%% file: thesis.tex
\documentclass[11pt,twoside]{uwthesis}

\usepackage{graphicx}
\usepackage{natbib}
\usepackage{aas_macros}
\usepackage{amsfonts}
\usepackage{amsbsy}
\usepackage{ulem}
\usepackage{color}

\graphicspath{{./Lens3D/writeup/}{./shear_KL/paper/}{./COSMOS/paper/fig/}{./figures/}}

\newcommand{\myvec}[1]{\boldsymbol{#1}}
\newcommand{\mymat}[1]{\boldsymbol{#1}}
\newcommand{\dd}{\mathrm{d}}
\newcommand{\Map}{\ensuremath{M_{\rm ap}}\ }
\newcommand{\naive}{na\"{i}ve\ }
\newcommand{\Noise}{\mymat{\mathcal{N}}}

\newcommand{\KL}{Karhunen-Lo\`{e}ve}
\newcommand{\rcom}{\xi}

\usepackage[bookmarks, plainpages=false, pdfpagelabels]{hyperref}

\hypersetup{%
  pdfauthor = {Jacob T Vanderplas},
  pdftitle = {Karhunen-Lo\`{e}ve Analysis for Weak Gravitational Lensing},
  pdfsubject = {Ph.D. Dissertation},
  pdfkeywords = {},
  pdfcreator = {University of Washington, Department of Astronomy},
  pdfproducer = {},
  bookmarksopen = {true},
  pdfpagelayout = {TwoColumnRight},
}

\begin{document}
 
% ==========   Preliminary pages
%
% ( revised 2012 for electronic submission )
%

\prelimpages
\include{prelim}

%
% ==========      Text pages
%

\textpages
\include{chapter1}
\include{chapter2}
\include{chapter3}
\include{chapter4}
\include{chapter5}
\include{chapter6}

% bibliography
%
% add to table of contents
\cleardoublepage
\addcontentsline{toc}{chapter}{Bibliography}

% now include it
\bibliographystyle{apj}
\bibliography{thesis}
 
% appendices
%
\appendix
\include{appendixA}
\include{appendixB}
\include{appendixC}

% curriculum vitae
\include{vita}

\end{document}

%% file: prelim.tex
%
% ----- copyright and title pages
%
\Title{Karhunen-Lo\`{e}ve Analysis for Weak Gravitational Lensing}
\Author{Jacob T. Vanderplas}
\Year{2012}
\Program{Department of Astronomy}

\Chair{Andrew Connolly}{Professor}{Department of Astronomy}
\Signature{Bhuvnesh Jain}
\Signature{Andrew Becker}

\copyrightpage
\titlepage

\setcounter{page}{-1}

\abstract{
In the past decade, weak gravitational lensing has become an important tool
in the study of the universe at the largest scale, giving insights into the
distribution of dark matter, the expansion of the universe, and the nature
of dark energy. This thesis research explores several applications
of Karhunen-Lo\`{e}ve (KL) analysis to speed and improve the comparison of
weak lensing shear catalogs to theory in order to constrain cosmological 
parameters in current and future lensing
surveys. This work addresses three related aspects of weak lensing analysis:

\begin{description}
   \item[Three-dimensional Tomographic Mapping:]
        \citep[Based on work published in][]{Vanderplas2011}
        We explore a new fast approach to three-dimensional mass mapping in
        weak lensing surveys.  The KL approach uses a KL-based filtering of
        the shear signal to reconstruct mass structures on the line-of-sight,
        and provides a unified framework to evaluate the efficacy of linear
        reconstruction techniques.  We find that the KL-based filtering leads
        to near-optimal angular resolution, and computation times which are
        faster than previous approaches.  We also use the KL formalism to
        show that linear non-parametric reconstruction methods are
        fundamentally limited in their ability to resolve lens redshifts.
   \item[Shear Peak Statistics with Incomplete Data]
        \citep[Based on work published in][]{Vanderplas2012}
        We explore the use of KL eigenmodes for interpolation across masked     
        regions in observed shear maps.  Mass mapping is an inherently
        non-local calculation, meaning gaps in the data can have a significant
        effect on the properties of the derived mass map.  Our KL mapping
        procedure leads to improvements in the recovery of detailed statistics
        of peaks in the mass map, which holds promise of improved cosmological
        constraints based on such studies.
   \item[Two-point parameter estimation with KL modes]
        The power spectrum of the observed shear can yield powerful cosmological
        constraints.  Incomplete survey sky coverage, however, can lead to
        mixing of power between Fourier modes, and obfuscate the cosmologically
        sensitive signal.  We show that KL can be used to derive an alternate
        orthonormal basis for the problem which avoids mode-mixing and allows
        a convenient formalism for cosmological likelihood computations.
        Cosmological constraints derived using this method are shown to be
        competitive with those from the more conventional correlation function
        approach.  We also discuss several aspects of the KL approach which
        will allow improved handling of correlated errors and redshift
        information in future surveys.
\end{description}

}
 
%
% ----- contents & etc.
%
\tableofcontents
\listoffigures
\listoftables
 
%
% ----- glossary 
%
%\chapter*{Glossary}      % starred form omits the `chapter x'
%\addcontentsline{toc}{chapter}{Glossary}
%\thispagestyle{plain}
%
%\begin{glossary}
%\item[item1] description description description description description
%  description description description description description description
%  description description description description description description
%\item[item2] description description description description description
%  description description description description description description
%  description description description description description description
%\item[item3] description description description description description
%  description description description description description description
%  description description description description description description 
%\end{glossary}
 
%
% ----- acknowledgments
%
\acknowledgments{% \vskip2pc
   {\narrower\noindent
   Thanks first to my wife Cristin for supporting me during the last five
   years, and understanding my late nights of working and writing.  Her
   genuine and deep love for the people in her life is a daily inspiration.

   Thanks to my family: especially to my mother Gretchen for her constant
   encouragement and support.  From collecting butterflies to growing   
   carnivorous plants to counting the stars, she instilled in me a desire
   to know the world around me.

   Thanks to my advisors, Andrew Connolly and Bhuvnesh Jain, for their
   patient mentorship.  Their thoughtfulness, humor, and expectation of
   excellence has given me a graduate school experience that exceeded even
   my highest expectations.

   Thanks to Andrew Becker for offering detailed and insightful comments
   on a draft of this work.

   I am indebted to several colleagues for helpful discussions through the
   course of this research, including Debbie Bard, Gary Bernstein, Anna Cabre,
   Joerg Dietrich, Mike Jarvis, Jan Kratochvil, Tim Schrabback, Patrick Simon,
   Andy Taylor, Vinu Vikram, Risa Wechsler, and many others.

   Support for this research was provided by DOE Grant DESC0002607,
   NSF Grant AST-0709394, and NASA Grant NNX07-AH07G.
   \par}
}

%
% ----- dedication
%

\dedication{
\begin{center}
To my father, Hugh Vander Plas, who lost his battle with cancer just weeks
before I presented this dissertation.  I will always be thankful for his
ceaseless support, encouragement, and willingness to listen.  His loyal,
loving, and faithful character will remain my highest example of how to live.
\end{center}
}

%
% end of the preliminary pages

%% file: chapter1.tex
\chapter{Brief Introduction to Cosmology}

By the late part of the 20th century, the standard model of Cosmology
seemed to rest on firm foundations.  The model, known as the Cold Dark Matter
(CDM) model, consisted of a uniformly
expanding universe, composed of baryonic matter, cold
non-baryonic (dark) matter, and radiation, with space-time evolving
according to the dynamics of Einstein's Theory of General Relativity.
This dynamical description, as was realized by Alexander Friedmann,
George Lemaitre, Howard Robertson, and Arthur Walker during the
1920s and 1930s,
predicts a dynamic universe, where space itself must expand or contract
under the influence of the energy within it.  The expansion of the universe,
first observed via the characteristic redshifts of distant galaxies
\citep{hubble1929}, was thought to be slowing under the
gravitational effect of the matter within it.

Uniform expansion of space-time lent support to the notion
that the early universe was filled with a hot, dense plasma
from which the constituents of chemical elements
formed.  This theory of {\it Big Bang Nucleosynthesis} \citep[BBN:][]{Alpher48}
was, and remains, extremely successful in explaining the relative
abundance of chemical elements in the universe.  Another important prediction
within the BBN model is that a Cosmic Background Radiation should be present,
due to the photons which free-streamed once the universe cooled enough for
the gas within it to no longer be ionized \citep{Alpher48b}.
The CBR signal had been observed in detail \citep{Smoot92},
resulting in a confirmation of the inflationary
hypothesis and the standard model of cosmology.

Nonetheless, the standard model had some cracks in its foundation:
cosmological probes and studies of galaxy clusters
yielded widely discrepant estimates of the
matter content in the universe.  The implied age of the universe in the CDM
model was younger than the age inferred for the oldest observed globular
clusters and white dwarfs.  Something needed to change.

Observations of type Ia supernovae by \citet{Riess98} and
\citet{Perlmutter99} offered a solution: these showed that distant supernovae
were fainter than expected in a standard CDM model.  This implied that the
expansion of the universe was accelerating, due to a cosmological component
dubbed {\it dark energy}.  This dark energy has an effective negative pressure,
which overcomes the gravitational attraction of matter and causes the
expansion of space to accelerate.  This additional energy component in the
universe solved many of the problems posed by CDM, and is now part of the
standard $\Lambda$-CDM cosmological model ($\Lambda$ refers to the cosmological
constant first proposed by Einstein).

Led by these supernova results, as well as new observations of the cosmic
microwave background \citep[CMB;][]{WMAP1}, the baryon acoustic oscillations
\citep[BAO;][]{Eisenstein05}, and other observational
campaigns, the last 15 years has seen a surge in precision cosmological
measurements.
The dark energy postulated to explain supernova distances is now known
to make up over 70\% of the energy density of the universe, and have
an equation of state consistent with it being due to
vacuum energy or a cosmological constant \citep{Kessler2009, WMAP7}.
Future surveys in many diverse areas of astronomy are seeking to place even
tighter constraints, allowing greater insight into the nature and evolution
of dark energy.

With such a wide and diverse field as Cosmology, we can't hope to offer
a complete introduction of the relevant theory in this work.
For a more complete discussion, there
are several very well-written books available; much of the material
discussed below is taken from formalism developed more fully in these works
\citep[see, e.g.][]{peebles1993principles, peacock1999cosmological,
  ryden2003cosmology, longair2008galaxy}.
This chapter will cover the basic physical and mathematical background of
the physical study of cosmology.  We will begin with a discussion of the
FLRW metric (named for Friedmann, Lemaitre,
Robertson, and Walker) which describes the geometry of space-time.
Next we'll move on to define the Friedmann Equations, which condense
the field equations of Einstein's General Relativity to the basic pieces
needed to describe the dynamics of a globally homogeneous and isotropic
universe.  We will then briefly discuss the relevant theory behind
gravitational structure formation within this model.
This paves the way to relate theory to data, using observations
including cluster counts, correlation functions, and Fourier power spectra.
Finally, we will develop the equations describing gravitational lensing
in the weak limit, and show how weak lensing observations can be used
to gain insight into the parameters of our cosmological model.
Throughout, we'll point out the relevant observational work which supports
and constrains these theories.

\section{FLRW Metric}
\label{sec:FLRW}
The physical study of cosmology in its classical form
is based on the fundamental assumption of symmetry:
that the universe on the largest scales is isotropic.
Homogeneity is an expression of translational symmetry: the appearance of
the universe does not depend on the location of the observer.  Isotropy
is an expression of rotational symmetry: the appearance of the universe
does not change with respect to the orientation of the observer.
These assumptions are clearly incorrect at small scales -- our galaxy
has a much higher density of stars in the central bulge than in the
outer halo, for example -- but they appear to hold at
the largest scales.
At distance scales larger than the size of typical superclusters
(about 50 Mpc or more), the distribution of
quasars and galaxies reflect the nearly homogeneous and isotropic
nature of large scale structure \citep{Yadav2005, Sarkar2009}.
More importantly, the Cosmic Microwave
Background appears homogeneous and isotropic to within one part in
$10^5$, giving evidence that our assumptions of homogeneity and isotropy
are well-founded for the universe as a whole \citep[For an interesting
discussion of the limits of this approach, however, see][]{Maartens2011}.

The most general metric for a homogeneous and isotropic space-time is due
to Howard Robertson and Arthur Walker, who showed that the space-time
distance $ds$ in spherical coordinates is given by
\begin{equation}
  \label{eq:FLRW_metric}
  ds^2 = -c^2 dt^2 + a(t)^2\left[dr^2 + S_\kappa^2(r)d\Phi^2\right]
\end{equation}
where $t$ is the time coordinate, $r$ and $\Phi$ are comoving spherical
coordinates,
$a(t)$ describes the distance scale (which may be an arbitrary function
of $t$), and $S_\kappa^2(r)$ is the curvature term.  The curvature term
depends on the curvature, $\kappa$, which may be either $+1$, $-1$, or $0$:
\begin{equation}
  \label{eq:FLRW_curvature}
  S_\kappa(r) = \left\{
  \begin{array}{ll}
    R\,\sin(r/R) & \kappa = +1\\
    r & \kappa = 0\\
    R\,\sinh(r/R) & \kappa = -1
  \end{array}
  \right.
\end{equation}
where $R$ is the radius of curvature today.  Often, the curvature
sign $\kappa$ and radius $R$ are compactly expressed in a single curvature
parameter $k$, such that $\kappa = k/|k|$ and $R = |k|^{-1/2}$.

Robertson and Walker derived the above metric
from purely geometric arguments.
An interesting aspect of this metric is the scale factor $a(t)$.  A general
homogeneous and isotropic universe is not necessarily static: it can be
expanding or contracting with time.  The detailed nature of this expansion
cannot be derived from purely geometric means: the description of the dynamics
of cosmic expansion comes from the field equations of Einstein's theory
of General Relativity.

\section{The Friedmann Equations}
\label{sec:friedmann}
The Robertson-Walker metric (eq.~\ref{eq:FLRW_metric})
is a purely geometric result,
where the scale factor $a(t)$ is arbitrary and unspecified.
Friedmann and Lemaitre had earlier independently derived this expression
from Einstein's field equations, with the addition of certain dynamical
constraints on the scale factor.  For this reason, the Robertson-Walker
metric is often referred to as the Friedmann-Robertson-Walker metric
or the Friedmann-Lemaitre-Robertson-Walker (FLRW) metric.
The general relativistic constraints on the scale factor $a(t)$
are compactly expressed by the Friedmann
equations\footnote{For a  derivation of the Friedmann
  equations from the field equations of general relativity,
  refer to \citet{peebles1993principles}}:
\begin{equation}
  \label{eq:friedmann_1}
  \left(\frac{\dot{a}}{a}\right)^2
  = \frac{8\pi G}{3c^2}\varepsilon
  + \frac{\Lambda}{3} - \frac{\kappa c^2}{a^2 R^2}
\end{equation}
\begin{equation}
  \label{eq:friedmann_2}
  \frac{\ddot{a}}{a}
  = -\,\frac{4\pi G}{3c^2}(\varepsilon + 3P) + \frac{\Lambda}{3}.
\end{equation}
where $G$ is the gravitational constant, and $c$ is the speed of light.
The scale factor $a$ is understood to be a function of time, with the
dots representing derivatives with respect to time.
By convention, the scale factor at the present day is
chosen to be $a(t_0) = 1$.  $\varepsilon$ and $P$ are
the energy density and pressure of the mass-energy in the universe, and
$\Lambda$ represents the cosmological constant.
Equations~\ref{eq:friedmann_1} and \ref{eq:friedmann_2} are the first and
second Friedmann equations, respectively.  The third Friedmann equation
can be easily derived from the first two:
\begin{equation}
  \label{eq:friedmann_3}
  \dot{\varepsilon} = -3\,\frac{\dot{a}}{a}\,(\varepsilon + P).
\end{equation}
This expression is equivalent to the first law of thermodynamics
expressed for the universe as a whole.

\subsection{Time Dilation and Redshift}
\label{sec:redshift}
General Relativity tells us
that light always travels along null geodesics, that is, the space time
interval in eqn.~\ref{eq:FLRW_metric} satisfies $ds = 0$.  For a light
beam with no angular deflection $d\Omega$, this gives
\begin{equation}
  dr = \frac{c}{a(t)} dt.
\end{equation}
If a beam of light is emitted at time $t_e$ and travels
a comoving distance $r$, the
time $t_o$ that the light is observed can be found by solving
\begin{equation}
  r = \int_{t_e}^{t_o} \frac{c}{a(t)} dt
\end{equation}
If a second photon is emitted a short time later at time $t_e + \Delta t_e$,
and arrives at time $t_o + \Delta t_o$, this gives
\begin{eqnarray}
  r &=& 
  \int_{t_e + \Delta t_e}^{t_o + \Delta t_o} \frac{c}{a(t)} dt \nonumber\\
  &\approx& \int_{t_e}^{t_o} \frac{c}{a(t)} dt + \frac{c\Delta t_o}{a(t_o)}
  - \frac{c\Delta t_e}{a(t_e)},
\end{eqnarray}
where we have used a first-order approximation.  Equating these
two expressions gives for small $\Delta t$:
\begin{equation}
  \label{eq:time_dialation}
  \Delta t_o = \Delta t_e \frac{a(t_o)}{a(t_e)}.
\end{equation}
In an expanding universe, the observed time interval is longer than
the time interval in the emitted frame.  This {\it time dilation} is
a general feature of space-time governed by Einstein's field equations.

The time dilation has an observable effect on emitted light:
if an atom emits light with a period 
$P_e = \Delta t_e = \lambda_e / c$, then the observed wavelength $\lambda_o$
and the emitted wavelength $\lambda_e$ are related by
\begin{equation}
  \lambda_o = \lambda_e \frac{a(t_o)}{a(t_e)}.
\end{equation}
The wavelength of light is lengthened due to the expansion of space.  For
historical reasons, this expansion is generally parametrized using the
redshift:
\begin{equation}
  1 + z \equiv \frac{a(t_o)}{a(t_e)}.
\end{equation}
Because we define $a(t_o) = 1$, we have
\begin{equation}
  a(t_e) = \frac{1}{1 + z}.
\end{equation}
Thus the redshift of a light source gives us a direct measurement of the
scale factor at the time that photon was emitted.  As such, it can be
substituted for $a$ as the dependent variable in the above equations
with a suitable change-of-variables; we will switch between these two
conventions depending on which is convenient.

\subsection{Equation of State}
The Friedmann equations can be further simplified
by relating the pressure $P$ and energy
density $\varepsilon$ in terms of a linear equation of state parameter
\begin{equation}
  \label{eq:w_EOS}
  w \equiv P / \varepsilon.
\end{equation}
Using this, the solution of eqn.~\ref{eq:friedmann_3} gives
\begin{equation}
  \label{eq:w_constant}
  \varepsilon = \varepsilon_0\, a^{-3(1 + w)}
\end{equation}
for $w$ constant in time.
Here $\varepsilon_0 = \varepsilon(t_0)$ is the energy density today,
and we have used the standard convention $a(t_0) = 1$.
Given this parametrization, we can now
separate the various contributions to the mass-energy of the universe
and re-write eqn.~\ref{eq:friedmann_1} in terms of the equation of
state for each:
\begin{equation}
  \label{eq:friedmann_1_split}
  \left(\frac{\dot{a}}{a}\right)^2 = \frac{8\pi G}{3c^2}
  \sum_w \varepsilon_{w, 0} \,\, a^{-3(1 + w)}
\end{equation}
where $\varepsilon_{w,0}$ is the energy density of each species at present.
The various possible contributions are:
\begin{description}
  \item[Vacuum energy/Cosmological Constant:]
    The vacuum energy or cosmological constant $\Lambda$ has
    energy density that does not change with time.
    So by Equation~\ref{eq:friedmann_3}, $P = -\varepsilon$ and $w = -1$.
  \item[General Quintessence:] Quintessence is defined as any sort of matter
    or energy field
    that can balance the gravitational attraction, leading to accelerated
    expansion.  By Equation~\ref{eq:friedmann_2}, $\ddot{a}/a > 0$ only
    if $w < - 1/3$.  We see that the cosmological constant is a form of
    quintessence.
  \item[Curvature:] Though it may seem strange to think about the curvature
    of space as having an energy density, in General Relativity the curvature
    is, in some sense, a stand-in for gravitational potential energy.
    Comparing eqns.~\ref{eq:friedmann_1} and \ref{eq:friedmann_1_split},
    the dependence of the curvature term on scale factor $k \propto a^{-2}$
    means it has an effective equation of state parameter $w = -1/3$.
    This makes it clear why curvature does
    not appear in the second Friedmann equation (eqn.~\ref{eq:friedmann_2}):
    for $w=-1/3$, $\varepsilon + 3P = 0$, and the presence of curvature
    cannot lead to a change in the expansion rate.
  \item[Non-relativistic matter:] Non-relativistic matter (often known
    as {\it cold matter}) has kinetic energy much less than its rest mass;
    in other words $P \sim kT \ll \varepsilon$.  This corresponds to
    $w \ll 1$, and we will often approximate this as simply $w=0$.
  \item[Relativistic Matter:] Relativistic matter 
    (known as {\it warm matter} or {\it hot matter}) has energy given by
    $E^2 = p^2c^2 + m^2 c^4$, where $p$ is the total momentum and $m$ is
    the rest-mass.  If $pc \ll mc^2$, we have the non-relativistic
    case above, and find $w \to 0$.  If $pc \gg mc^2$, then in analogy to
    the radiation case discussed below, we find $w \to 1/3$.
    For general relativistic matter, this leads to
    $0 \le w < 1/3$, with the exact value dependent on the energy density.
  \item[Radiation:] Radiation has energy per particle
    proportional to the momentum times the speed of light.  From basic
    electrodynamics, one can show that for an ideal photon gas, each spatial
    degree of freedom contributes equally to the energy, so that the pressure
    is $P = dp/dt = \varepsilon / 3$.  So relativistic mass-energy has
    $w = 1/3$.
\end{description}

\subsection{Evolving Equation of State}
In Equation~\ref{eq:w_constant} we show the solution of
eqn.~\ref{eq:friedmann_3} for constant $w$.  Another possibility
(especially applicable for general quintessence models) is that
the equation of state parameter $w$ evolves with time.
\begin{equation}
  \label{eq:w_general}
  \varepsilon(a) = \varepsilon_0\, \exp\left[-3\int_1^a
    \frac{1 + w(a^\prime)}{a^\prime}\dd a^\prime\right].
\end{equation}
Many parametrizations of the form of $w(a)$ have been proposed.  Perhaps
the simplest is a model which is simply linear in the redshift $z$, i.e.
\begin{equation}
  w(z) \approx w_0 + w_1 z.
\end{equation}
Though simple, this parametrization is encumbered by the fact that $z$ changes
very quickly with time, especially in the early universe.  A better 
choice is the CPL parametrization \citep{Chevallier01, Linder03a}, which is
of the form
\begin{equation}
  w(z) \approx w_0 + w_a\frac{z}{1 + z},
\end{equation}
or equivalently
\begin{equation}
  w(a) \approx w_0 + w_a(1 - a)
\end{equation}
which is linear in the scale factor rather than the redshift.  Adopting this
parametrization gives
\begin{equation}
  \label{eq:w_first_order}
  \varepsilon(a) = \varepsilon_0\, a^{-3(1 + w_0 + w_a)}e^{-3w_a(1-a)}.
\end{equation}
One of the main goals of future cosmological surveys is to put meaningful
constraints on the time-evolution of dark energy, often by placing constraints
on $w_1$ or $w_a$.  This will be discussed further in later sections.

\subsection{Hubble Parameter}
\label{sec:hubble_parameter}
The first Friedmann equation (eqn.~\ref{eq:friedmann_1}) is commonly expressed
in terms of dimensionless parameters via the generalization in
eqn.~\ref{eq:friedmann_1_split}.  If we define the Hubble parameter
\begin{equation}
  \label{eq:hubble_parameter}
  H \equiv \frac{\dot{a}}{a},
\end{equation}
and let $H_0$ be the value of the Hubble parameter today, then
eqn.~\ref{eq:friedmann_1_split} becomes
\begin{equation}
  \left(\frac{H}{H_0}\right)^2 = \frac{8\pi G}{3H_0^2c^2}
  \sum_w \varepsilon_{w, 0} \,\, a^{-3(1 + w)}.
\end{equation}
The constant in front of the sum has dimensions of inverse energy density:
this motivates the definition of the critical density
\begin{equation}
  \label{eq:critical_density}
  \varepsilon_c \equiv \frac{\rho_c}{c^2} \equiv \frac{3 H^2 c^2}{8\pi G},
\end{equation}
where, to be explicit, both the critical densities $\varepsilon_c$,
$\rho_c$,
and Hubble parameter $H$ are functions of time.
With this definition, and defining the dimensionless density parameter
\begin{equation}
  \label{eq:density_parameter}
  \Omega_w(t) \equiv \varepsilon_w(t) / \varepsilon_c(t)
\end{equation}
the Friedmann equation can be compactly expressed
\begin{equation}
  \left(\frac{H}{H_0}\right)^2
  = \sum_w \Omega_{w, 0}\,\, a^{-3(1 + w)},
\end{equation}
where the subscript $0$ indicates the value at present.
Alternatively, we can express the Friedmann equation as simply
\begin{equation}
  \sum_w \Omega_w(t) = 1.
\end{equation}
Notice that if the sum of all components $\Omega_w$ with the exception
of $\Omega_\kappa$ is unity, then
we must have curvature $\Omega_\kappa = 0$.  This shows the meaning of
the critical density (eqn.~\ref{eq:critical_density}): if the energy density
in the universe satisfies $\epsilon > \epsilon_c$ (i.e. $\Omega > 1$), then
$\kappa = +1$ and the universe is {\it spatially closed}.
If $\epsilon < \epsilon_c$ (i.e. $\Omega < 1$), then $\kappa = -1$ and
the universe is {\it spatially open}. If the energy density is
exactly equal to the critical density, then the curvature $\kappa = 0$
and the universe is {\it flat}.  Constraints from the Cosmic Microwave
Background show that our universe is flat to a very high precision
\citep{WMAP7}.  In addition, inflationary perturbation analysis shows that
if the Universe is close to flat, it probably is flat.
For this reason, in many of the below derivations we will
assume for simplicity that $\Omega_\kappa = 0$ and consequently
$S_\kappa(r) = r$.

For simplicity, we can limit our consideration to the principal contributors
to the density of the universe:
dark energy $(\Omega_\Lambda)$, matter $(\Omega_M)$, radiation $(\Omega_R)$,
and curvature $(\Omega_\kappa)$.  Neglecting other components gives the
familiar dimensionless form of the Friedmann Equation:
\begin{equation}
  \label{eq:friedmann_dimensionless}
  \left(\frac{H}{H_0}\right)^2
  = \Omega_{M,0}\,\,a^{-3} + \Omega_{R,0}\,\,a^{-4}
  + \Omega_{\kappa,0}\,\,a^{-2} + \Omega_\Lambda
\end{equation}

\section{Cosmological Distance Measures}
\label{sec:distances}
The FLRW metric of \S\ref{sec:FLRW} and the Friedmann equations of
\S\ref{sec:friedmann} lay the basic framework for the study of cosmology.
A large portion of
the history of $20^{\rm th}$ century cosmology surrounds various
attempts to understand the relative contributions of matter, radiation,
curvature, and dark energy to the Hubble parameter, which measures
the expansion rate of the universe.  The exact nature of these various
contributions has far-reaching consequences,
determining how, when, and where galaxies, clusters and other structure
form and evolve; determining the age of the universe and the character of
its evolution through time; determining cosmic abundances and 
the initial conditions of stellar evolution
and planet formation; and determining the nature of the universe's beginning,
and the possibility of its eventual end.
Observational measures of these consequences allow constraints of the
properties of the various $\Omega_w(z)$.

Eqn.~\ref{eq:friedmann_dimensionless} is simply a first-order differential
equation in $a$: For various choices of the density parameters $\Omega_w$, it
can be solved to yield a curve describing the scale factor $a$ as a function
of time $t$.  A straightforward route to placing
observational constraints on the densities of
various components, then, would require simply measuring the value of $a$ at
several times $t$ and performing a multidimensional fit to these observed
data points.

As discussed above in \S\ref{sec:redshift}, the redshift of
light offers a direct measurement of the scale factor at a given time.
In order to use eqn.~\ref{eq:friedmann_dimensionless} to gain information
about the cosmological densities $\Omega_w$, then, we must be able to
observe some property related to the time $t_e$ of emission of these photons.
To enable this, we'll introduce the concept of distances in Cosmology.

Distance measures in cosmology are a potentially confusing subject.  An
excellent resource describing these can be found in \citet{hogg1999distance}.
Here we'll briefly define four relevant distance measures: the comoving
distance, the proper distance, the angular diameter distance, and the
luminosity distance.

\begin{description}
  \item[Comoving Distance:] The comoving distance is the distance $r$ which
    enters into the FLRW metric, eqn.~\ref{eq:FLRW_metric}.  This distance is
    constant for two objects moving with the expansion of space.  Using the
    FLRW metric and setting $ds=0$, one can shown that the comoving
    distance for an object with redshift $z$ is given by
    \begin{equation}
      \label{eq:comoving_distance}
      r(z) = c\,\int_0^z \frac{dz^\prime}{H(z^\prime)}.
    \end{equation}
  \item[Proper Distance:] The proper distance $d_p(z)$ is the simultaneous
    separation between two objects.  From the FLRW metric with time interval
    $dt=0$, one can show that the proper distance is given by
    \begin{equation}
      \label{eq:proper_distance}
      d_p(z) = \frac{r(z)}{1 + z}.
    \end{equation}
  \item[Angular Diameter Distance:] The angular diameter distance $d_A(z)$
    is the
    ratio of the proper size to the observed angular size (in radians) of
    an extended source.  From the FLRW metric with $dt = dr = 0$, one
    can show that the angular diameter distance is given by
    \begin{equation}
      \label{eq:angular_diameter_distance}
      d_A(z) = \frac{S_\kappa(r)}{1 + z}.
    \end{equation}
  \item[Luminosity Distance:] The luminosity distance $d_L(z)$ relates the
    emitted flux of a source to the observed flux.  Taking into account
    the angular dilution of the flux, as well as the redshifted energy of
    each photon and time-delay of photon arrival leads to
    \begin{eqnarray}
      \label{eq:luminosity_distance}
      d_L(z) &=& (1 + z)^2\,d_A(z) \nonumber\\
      &=& (1 + z)\,S_\kappa(r)
    \end{eqnarray}
\end{description}
By measuring one of these distances as a function of redshift, we are
effectively measuring both the dependent and independent variable
in the dimensionless Friedmann equation, eqn.~\ref{eq:friedmann_dimensionless}.
In this way, it is possible to constrain the combinations of cosmological
parameters $\Omega_w$ that fit the data.

\section{Standard Candles: Cosmology via Luminosity Distance}
\label{sec:std_candles}
{\it Standard candles} have been one of the primary methods for measuring
cosmological parameters.  The earliest measure, from Edwin Hubble,
used Cepheid-type variable stars that have an absolute magnitude correlated
with their pulsation period \citep{Leavitt1912, hubble1929}.
Through observations of Cepheids
in nearby galaxies, Hubble found a positive correlation between the
luminosity distance to each galaxy and its apparent recessional velocity,
indicating a positive value for $H_0$.\footnote{
Although Hubble does not frame this measurement in terms of the Friedmann
equations, he cautiously mentions the potential relationship of the
observations to the ``de Sitter effect'', a predicted redshift within
a particular solution of
Einstein's equations which is a special case of the FLRW metric and
Friedmann equations (i.e., that with $\Omega_M = 0$, $\Omega_\kappa = 0$,
and $\Omega_\Lambda = 1$).}
Using our formalism above, we can show that the rate of change of the
proper distance to any object is
\begin{eqnarray}
  \frac{d}{dt}d_p(t) &=& H(t)\,\, d_p(t) \nonumber\\
                     &\approx& H_0\,\, d_L(t) \left[1 + \mathcal{O}(z)\right],
\end{eqnarray}
where in the second line we have separated corrections of order $z$.  Because
Hubble used a sample with $z <\sim 0.005$, the zeroth-order
linear fit is well within observed errors.
Thus Hubble's observations can be seen as a first attempt at constraining
cosmological parameters in Equation~\ref{eq:friedmann_dimensionless} through
the simultaneous measurement of redshift and luminosity distance.

In the 70 years after Hubble's discovery, there were many attempts to
confirm and improve upon his discovery and use it to derive tighter
constraints on the slope $H_0$ of the Hubble relation.  A large majority
of these have
been based on standard or standardizable candles.  These standardizable
candles have been based on several empirical relations, including
the period-luminosity relationship of Cepheid
variables \citep{Leavitt1912}; on the Tully-Fisher relationship between
brightness and rotation speed of spiral galaxies \citep{Tully77}; on
the Faber-Jackson/Fundamental Plane relationship between brightness,
velocity, and surface brightness for elliptical galaxies
\citep{Faber76, Djorgovski87}; and on the relationship between luminosity
and decay timescale for type Ia supernovae \citep{Philips93}.  Due to
their extreme intrinsic brightness and minimal scatter in
calibrated luminosity, type Ia supernovae have become a very important
tool for determining cosmological parameters, but they cannot be used
alone: independent measures are required to calibrate the distance
scale of type-I supernovae, making all the above approaches important.
Perhaps the most comprehensive study to date of these various standard
candles is the HST Key project \citep{Freedman01}, which combined
the above measures and others to arrive
at a value of $H_0 = 72 \pm 8$ km/s/Mpc.  This is consistent with the
tighter constraint from the WMAP 7-year CMB analysis, which gave
$H_0 = 70.4 \pm 2.5$ km/s/Mpc.

The common theme in the above methods is that they are based on the idea
of a standard candle: if we can determine the intrinsic brightness
of an object as well as its redshift, then we can compare this to the
apparent brightness and constrain the Hubble parameter $H(t)$.
Another path to this sort of constraint comes from standard {\it rulers}
rather than standard candles.  If we know the redshift as well as the
intrinsic size of an object, then we can use its apparent size to
constrain cosmological parameters.  One such standard ruler is given by the
characteristic scales of structure in the universe.

\section{The Growth of Structure}
\label{sec:growth}
There are several methods of cosmological parameter estimation that rely
on the idea of standard rulers.  Some examples are the observation of the
anisotropy scale in the Cosmic Microwave Background, and the observation
of the Baryon Acoustic Oscillation scale.
Another powerful method relies on detailed modeling of the growth of structure.
As we will see, along with offering the possibility of a standard ruler,
consideration of structure growth can offer other cosmologically interesting
observables.

The distribution of density throughout the universe can be expressed
in terms of the density contrast
\begin{equation}
  \label{eq:overdensity}
  \delta(\myvec{x}, t) = \frac{\rho(\myvec{x}, t) - \rho_b(t)}{\rho_b(t)},
\end{equation}
where the background matter density is
\begin{equation}
  \rho_b(t) \propto \frac{1}{a(t)^3}
\end{equation}
(i.e.~we assume the background consists of cold matter with $w=0$).
Rearranging this definition results in expressing the density field of the
universe as
\begin{equation}
  \rho(\myvec{x}, t) = \rho_b(t)[1 + \delta(\myvec{x}, t)].
\end{equation}

\subsection{Gravitational Instability}
We can proceed by treating matter as an ideal fluid with a
velocity field $\myvec{u}(\myvec{x}, t)$ and pressure $P(\myvec{x}, t)$
\citep{longair2008galaxy}.
In this case, it is governed by three equations: the
continuity equation, which describes conservation of mass,
\begin{equation}
  \label{eq:continuity}
  \left(\frac{\partial\rho}{\partial t}\right)_{\myvec{x}}
  + \rho\myvec{\nabla}_{\myvec{x}}\cdot\myvec{u} = 0,
\end{equation}
the Euler equation, which specifies conservation of momentum,
\begin{equation}
  \label{eq:euler}
  \left(\frac{\partial\myvec{u}}{\partial t}\right)_{\myvec{x}}
  + (\myvec{u}\cdot\myvec{\nabla}_{\myvec{x}})\myvec{u}
  + \frac{\myvec{\nabla}_{\myvec{x}}P}{\rho}
  = -\myvec{\nabla}_{\myvec{x}}\Phi,
\end{equation}
and the Poisson equation, which describes gravity in the Newtonian limit
\begin{equation}
  \nabla_{\myvec{x}}^2\Phi = 4\pi G\rho.
\end{equation}
Transforming to comoving coordinates $\myvec{r} = \myvec{x}/a(t)$,
defining comoving time derivatives
$d/dt \equiv \partial/\partial t + \myvec{u}\cdot\myvec{\nabla}$,
and expressing in terms of the dimensionless density contrast
$\delta(\myvec{r}, t)$ (eq.~\ref{eq:overdensity}) gives
\begin{equation}
  \label{eq:delta_evolution}
  \frac{d^2\delta}{d t^2} + 2 H \frac{d\delta}{d t}
  = \frac{c_s^2}{a^2}\nabla_{\myvec{r}}^2\delta + 4\pi G \rho_b\,\delta
\end{equation}
where $\nabla^2_{\myvec{r}}$ indicates the Laplacian with respect to comoving
coordinates \citep[for derivation see section 11.2 of][]{longair2008galaxy}.

We can gain insight by expressing this in terms of
wave-like solutions of the density contrast
$\delta_k(\myvec{r}, t) \propto \exp[i(\myvec{k}\cdot\myvec{r} - \omega t)]$,
where $\myvec{k}$ is the vector of spatial wave-numbers, and $\omega$
is the oscillation frequency.  In terms of these Fourier modes,
eqn.~\ref{eq:delta_evolution} becomes
\begin{equation}
  \label{eq:delta_evolution_k}
  \ddot{\delta}_k + 2 H \dot{\delta}_k - \delta_k(4\pi G\rho_b - k^2c_s^2) = 0.
\end{equation}
This differential equation describes an exponentially growing (or decaying)
density fluctuation, with a ``drag'' term governed by the Hubble parameter
$H(z)$.  The rate of growth of perturbations $\delta_k$ depends
on the balance between the gravitational force through $4\pi G\rho_b$ and the
pressure through $k^2c_s^2$.  The scale $\lambda_J = 2\pi/k_J$
where these forces balance is called the Jeans length:
\begin{equation}
  \label{eq:jeans_length}
  \lambda_J \equiv c_s \sqrt{\frac{\pi}{G\rho_b}}.
\end{equation}
This is the scale above which pressure cannot halt gravitational
collapse.  The length is directly proportional to the sound speed
$c_s = \sqrt{\partial P/\partial \rho}$ and so depends on the equation of
state of the total energy in the universe,
as well as the average density $\rho_b$.
The different components (radiation, matter, etc.) have different
equations of state (\S\ref{sec:friedmann}) and evolve
with different dependencies on the scale factor
$a$, and so the Jeans length also evolves through the course of cosmic
history.  Thus the scale of nonlinear
collapsed structure as a function of $z$ contains information which can
be used to place constraints on the components which make up the Universe.

In particular, we can consider two relevant regimes: radiation dominance
and matter dominance.  In the regime where radiation dominates the energy
density, the equation of state $w=1/3$ leads to $c_s^2 = c^2/3$.  In a flat
universe $\rho_b$ is the critical density (eqn.~\ref{eq:critical_density})
and the Jeans length for a radiation dominated universe can be expressed
\begin{equation}
  \label{eq:jeans_radiation}
  \lambda_J^{(R)} = \frac{\pi c \sqrt{8}}{3 H}.
\end{equation}
This scale is on the same order as that of the horizon scale,
$\lambda_s \approx 2c / (H\sqrt{3})$.  This means that sub-horizon modes
cannot collapse during the epoch when radiation dominates the energy
of the universe.

In the matter-dominated regime, there are two possibilities: if radiation and
matter are coupled, the pressure comes from the radiation
while the density is dominated by matter.  This gives
$c_s^2 \sim c^2 \Omega_r / \Omega_M \sim c^2 (1 + z)
\Omega_{r,0}/\Omega_{m, 0}$.
Putting in numbers from WMAP \citep{WMAP7}, we find approximately
$c_s \approx 10^{-2} c \sqrt{1 + z}$.
If radiation and matter are decoupled, the pressure comes from the nonzero
temperature of matter itself:
$c_s^2 \sim kT/m_p$ where $m_p$ is the proton mass.  Assuming the matter
is in thermal equilibrium with the CMB, then temperature goes as
$T \propto (1 + z)$.  Again using observational constraints from WMAP, we find
$c_s \approx 10^{-7} c \sqrt{1 + z}$.  Thus the Jeans length during the
matter-dominated epoch is given by
\begin{equation}
  \label{eq:jeans_matter}
  \lambda_J^{(M)} \approx \lambda_J^{(R)}\sqrt{1 + z}\,\, \times \left\{
  \begin{array}{ll}
    10^{-2} & \mathrm{before\ decoupling}\\
    10^{-7} & \mathrm{after\ decoupling}
  \end{array}
  \right.
\end{equation}
The redshift of decoupling is approximately $z \sim 1100$ \citep[for a physical
argument for this, see][]{ryden2003cosmology}.  This means that prior to
decoupling, growth below approximately sub-horizon scales is suppressed by
pressure.  After decoupling, the Jeans length shrinks by five orders of
magnitude, allowing linear structure on this scale to form.  The observable
effect of this sharp transition in the growth of structure will be discussed
further below.

A related question is that of the rate of structure growth
on scales larger than the Jeans length. 
This can be addressed by defining the linear growth
factor $D(t)$ such that
\begin{equation}
  \label{eq:linear_growth_def}
  \delta(\myvec{r}, t) = \delta_0(\myvec{r})D(t).
\end{equation}
Using $\rho_b = \Omega_M\rho_c$ and assuming negligible pressure (i.e. scales
above the Jeans length), we can rewrite
eqn.~\ref{eq:delta_evolution} as
\begin{equation}
  \label{eq:linear_growth_eqn}
  \ddot{D} + 2H\dot{D} - \frac{3}{2}\Omega_M H^2D = 0.
\end{equation}
This is a second-order differential equation, which will, in general,
admit a solution with a growing mode and a decaying mode:
\begin{equation}
  D(t) = A_1 D_1(t) + A_2 D_2(t).
\end{equation}
In a flat universe dominated by matter, the
Friedmann equation gives $H(t) = 2 / (3t)$, leading to solutions
\begin{equation}
  D(t) = A_1 t^{2/3} + A_2 t^{-1}.
\end{equation}
The first term quickly dominates the second, and structure grows as
$\delta \propto t^{2/3} \propto a$, where the last proportionality comes
from solving eqn.~\ref{eq:friedmann_dimensionless} for a matter dominated
universe.

In general, the growth factor for a flat universe is
\begin{equation}
  \label{eq:linear_growth}
  D(a) \propto \int_0^a \frac{da^\prime}{[a^\prime H(a^\prime)]^3},
\end{equation}
where the normalization is usually chosen such that $D(a) = 1$ at the
present day.

A radiation-dominated universe presents a more complicated case:
eqn.~\ref{eq:linear_growth_eqn} assumes the pressure is negligible compared
to the gravitational force.  This approximation breaks down in
cases when $\Omega_M \to 0$.  For these cases, we need a more involved
perturbative treatment.

\subsection{Perturbation Treatment}
To explore the growth rate in a universe where $\Omega_M$ is small, we
will perform a perturbation analysis of Friedmann's equations.
A full discussion of this treatment can be found in
\citet{peebles1993principles}.
Here we will briefly outline a schematic approach from \citet{kolb_turner}
which leads to the same results.

By Birkhoff's theorem \citep{birkhoff1923relativity},
a small spherical over-density can be treated as if
it were an independent homogeneous universe embedded within the background.
We'll assume the background is represented by a flat universe with
\begin{equation}
  H^2 = \frac{8\pi G}{3c^2}\varepsilon_0,
\end{equation}
and that a spherical perturbation has a small positive curvature
\begin{equation}
  H^2 = \frac{8\pi G}{3c^2}\varepsilon_1 - \frac{\kappa c^2}{a^2R_0^2}.
\end{equation}
The boundary requires that the expansion rate $H$ be equal between the
two; combining these we find
\begin{equation}
  \delta \equiv \frac{\varepsilon_1 - \varepsilon_0}{\varepsilon_0}
  = \frac{3\kappa c^4}{8\pi G R_0^2}\,\frac{1}{a^2 \varepsilon_0}.
\end{equation}
For a matter-dominated universe, $\varepsilon_0 \propto a^{-3}$
which gives $\delta \propto a$ as above.
For a radiation-dominated universe, $\varepsilon_0 \propto a^{-4}$
which gives $\delta \propto a^2$.

\subsection{Matter Power Spectrum}
In summary, the above results show that
\begin{itemize}
  \item In the radiation-dominated regime, fluctuations on scales above
    $\lambda_J^{(R)} = (\pi c \sqrt{8})/(3 H)$ grow as
    $\delta(a) \propto a^2$.
  \item In the matter-dominated regime, after decoupling, scales above
    $\lambda_J^{(M)}\approx 10^{-7} \lambda_J^{(R)} \sqrt{1 + z}$
    grow as $\delta(a) \propto a$.
\end{itemize}
The ratio of radiation density to matter density is
\begin{equation}
  \frac{\Omega_R(z)}{\Omega_M(z)} = (1 + z)\frac{\Omega_{R,0}}{\Omega_{M,0}}.
\end{equation}
So before the redshift of radiation-matter equality,
$z_{rm} \approx \Omega_{M,0}/\Omega_{R,0}$, radiation
dominates, while after this redshift matter dominates.
Thus the important scale is the horizon scale $\lambda_{rm}$ at
redshift $z=z_{rm}$.
Modes on length-scales $k > \lambda_{rm}$ will grow as $\delta \propto a^2$
for $a < a_{rm}$, and $\delta \propto a$ for $a > a_{rm}$.  Modes with
length-scales $k < \lambda_{rm}$ will grow as $\delta \propto a^2$ as long
as the horizon distance $d_{hor}(a) < k$, at which point the growth will
be suppressed by radiation pressure.  At $a > a_{rm}$, the Jeans length
shrinks by a factor of about $10^5$, and modes larger than $\lambda_J^{M}$
resume growth with $\delta \propto a$.

%\comment{add a picture of this here?}

Thus, density modes with $k > k_{rm}$ (that is, scales smaller than the
matter-radiation equality horizon scale) are suppressed by a factor of
$(a_k / a_{rm})^2 \propto k^{-2}$.  This motivates use of the power
spectrum of density fluctuations (for details see
Appendix~\ref{app:RandomFields}):
\begin{equation}
  \label{eq:power_spectrum}
  P_k \equiv \langle|\delta_k|^2\rangle,
\end{equation}
in theory a measurable quantity, which will have a distinct break at
$k = k_{rm}$ for the reasons discussed above.
In particular, if the power spectrum of primordial fluctuations
is a simple power law with $P_k \propto k^n$,
then after decoupling the power spectrum will be approximately
\begin{equation}
  P_k \propto \left\{
  \begin{array}{ll}
    k^{n-4}, & {\rm for\ } k > k_{rm} \\
    k^n, & {\rm for\ } k < k_{rm}.
  \end{array}
  \right.
\end{equation}
Under most inflationary scenarios, it is expected that the power spectral
index $n \approx 1$ \citep{peacock1999cosmological}.
The normalization of the power spectrum depends on the magnitude of the
primordial fluctuations, and is dependent on the cosmological model.
This normalization is typically expressed in terms of the parameter
$\sigma_8$, which measures the magnitude of average fluctuations within
a sphere of radius 8 Mpc.  For a more thorough discussion of power spectra
and the $\sigma_8$ normalization, see Appendix~\ref{app:RandomFields}.

\subsection{Putting it all together}
The sum of the above discussion paints a general picture of the growth of
structure within the universe presents several concepts with readily
observable consequences:

\begin{itemize}
  \item The size and mass/length scale of clusters as a function of redshift
    depends on the length-scale of gravitational instability $\lambda_J$
    (eqn.~\ref{eq:jeans_length}),
    which in turn depends on the densities of matter and radiation in the
    universe.  Clustering also depends on the linear growth rate
    $D(z)$ (eqn.~\ref{eq:linear_growth}) and its nonlinear extensions,
    which also depend on the relative cosmic densities as a function of $z$.
  \item The power spectrum of density fluctuations (eq.~\ref{eq:power_spectrum})
    has a turn-off at
    a length scale that is closely related to the horizon distance
    at the epoch of radiation-matter equality.  This scale acts as a standard
    ruler, such that the angular diameter distance can be estimated at a
    particular value of $z$, leading to cosmological constraints through the
    same means as the standard candle method discussed in
    \S\ref{sec:std_candles}.
  \item The linear growth factor (eqn.~\ref{eq:linear_growth}) affects the
    normalization of the power spectrum.  Therefore, measuring the power
    spectrum as a function of $z$ leads to cosmological constraints
    through the dependence of $D(z)$ on cosmological parameters.
\end{itemize}

So we see that there are powerful cosmological constraints that can be
obtained through the observation of the density fluctuations and clustering
of matter through the universe.  There are several caveats, however:
the above discussion focuses on the linear approximation (that is, we
discuss the behavior of perturbations of order $\delta$, while ignoring
$\delta^2$).  This is sufficient for small $\delta$, but not for when
$\delta$ is much larger than 1.  At small scales, structure is well beyond
the regime where this approximation holds: for example, $\delta$ for our
galaxy is approximately $10^5$!  Even moderate-sized galaxy clusters
(i.e. a hypothetical cluster 2Mpc across, containing 50 Milky-way
sized galaxies) have $\delta$ of order $10^2-10^3$.  Clearly, nonlinear
effects must be taken into account when measuring clustering.  These
effects can be estimated several ways; one of the more successful is the
halo model of \citet{Smith03}, which is calibrated using semi-analytic
results from N-body simulations.  Nonlinear effects lead to a significant
boosting of power on small scales.

A second caveat is that
the structure we are referring to here is that made up by the bulk
of the matter in the universe: collisionless dark matter.  Dark matter,
being non-luminous, cannot be observed directly through emitted light.
The power spectrum of luminous matter can be theoretically mapped to the
underlying mass power spectrum, but this mapping requires uncertain
corrections and introduces systematic errors that are difficult to
calibrate.

With careful accounting for the above two caveats, observations of the spatial
distribution of luminous matter have led to interesting cosmological results.
One of the most important surveys in this regard has been the Sloan Digital
Sky Survey (SDSS), which measured spectra of nearly a million sources across
over 8000 square degrees, leading to accurate photometric redshifts of
hundreds of thousands of galaxies across over a third of the sky.
Measurements of the angular power of these galaxies have been successfully
used to constrain cosmology both from the nonlinear power spectrum
\citep{Tegmark06} and the Baryon Acoustic Oscillation (BAO) signal
\citep{Eisenstein05}.

The BAO signal is another cosmological constraint based on the idea of a
standard ruler.  In the above discussion, we mention the role of pressure
in suppressing the growth of structure before radiation and matter are
decoupled.  This suppression by pressure leads to acoustic oscillations
in the plasma of the radiation and baryons.  When the radiation and matter
decouples, these oscillations freeze-out and form the seeds of baryonic
structure growth.  The remnants of this freeze-out can be observed in
baryonic structure today, and the characteristic length scale has been
used to place tight constraints on cosmological parameters
\citep{Eisenstein05}.

%\comment{ Should we also briefly mention WMAP anisotropies and the BAO signal?
%  They're essentially very accurate standard rulers at different epochs.
%  But I don't want to digress too much...}

\section{Gravitational Lensing}
\label{sec:gravitational_lensing}
In order to circumvent the astrophysical bias
involved with mapping luminous matter to
the underlying dark matter, it would be preferable to observe the dark
matter directly.  This is where gravitational lensing can make an
important contribution.
Einstein's theory of General Relativity predicts that photons will be deflected
in the presence of a gravitational field.  Under certain circumstances, this
deflection can be detected and used to learn about the nature of the
gravitating matter.

\subsection{Simplifying Assumptions}
\label{sec:lensing_simplification}
The propagation of light through a region of gravitational potential
$\Phi(\vec{r})$ is, in general, a very complicated\ problem, only analytically
solvable for potentials with various symmetries.  In cosmological contexts,
however, it is safe to assume that the universe is described by a
Robertson-Walker metric, with only small perturbations due to the density
fluctuations described by the potential $\Phi$.  In this case, the
gravitational deflection of a photon can be described by an effective
index of refraction given by
\begin{equation}
  n = 1-\frac{2}{c^2}\Phi 
\end{equation}
\citep[see][and references therein]{narayan1996lectures}.
As in conventional optics, light rays are deflected in proportion to the
perpendicular gradient of the refraction index, such that the deflection angle
$\hat{\alpha}$ is given by
\begin{equation}
  \label{eq:alpha-def}
  \hat{\alpha} = -\int_0^{D_S} \vec{\nabla}_\perp n \,\,\dd D
  = \frac{2}{c^2}\int_0^{D_S} \vec{\nabla}_\perp\Phi\,\,\dd D
\end{equation}
where $D_S$ is the distance from the observer to the photon source.  

For a point-mass located at a distance $D_L$ and an impact parameter $b$, with $D_L \gg b$ and $D_S \approx 2D_L$, equation \ref{eq:alpha-def} can be integrated to give
\begin{equation}
  \hat{\alpha} = \frac{4GM}{bc^2}\Bigg[1 - \frac{1}{2}\bigg(\frac{b}{D_L}\bigg)^2 + \mathcal{O}\bigg[\frac{b}{D_L}\bigg]^3 \Bigg]
\end{equation}
The first-order approximation is twice the deflection predicted by Newtonian gravity for a particle of arbitrary mass moving at a speed $c$.  It is important to note here that to first order, the deflection does not depend on the distance to the lens or source.  That is, for a mass distribution located at a distance $D_L \gg b$, equation \ref{eq:alpha-def} can be approximated
\begin{equation}
  \label{eq:alpha-def-approx}
  \hat{\alpha} \approx \frac{2}{c^2}\int_{D_L-\delta D}^{D_L+\delta D} \vec{\nabla}_\perp\Phi\,\dd D
\end{equation}
for $\delta D$ sufficiently greater than the size scale of the mass-distribution in question.
%Also note that for a large lens distance $D_L$, the contribution to $\hat{\alpha}$ becomes vanishingly small, and can be neglected.
So, to a very good approximation, the incremental deflection
$\delta \hat{\alpha}$ of a photon at a given point along its trajectory is
entirely due to an overdensity of matter with a thickness $2\,\delta D$,
oriented perpendicular to the unperturbed photon trajectory.  This is the
{\it thin lens} approximation.

\subsection{Lensing Geometry}

\begin{figure}
  \includegraphics[width=\textwidth]{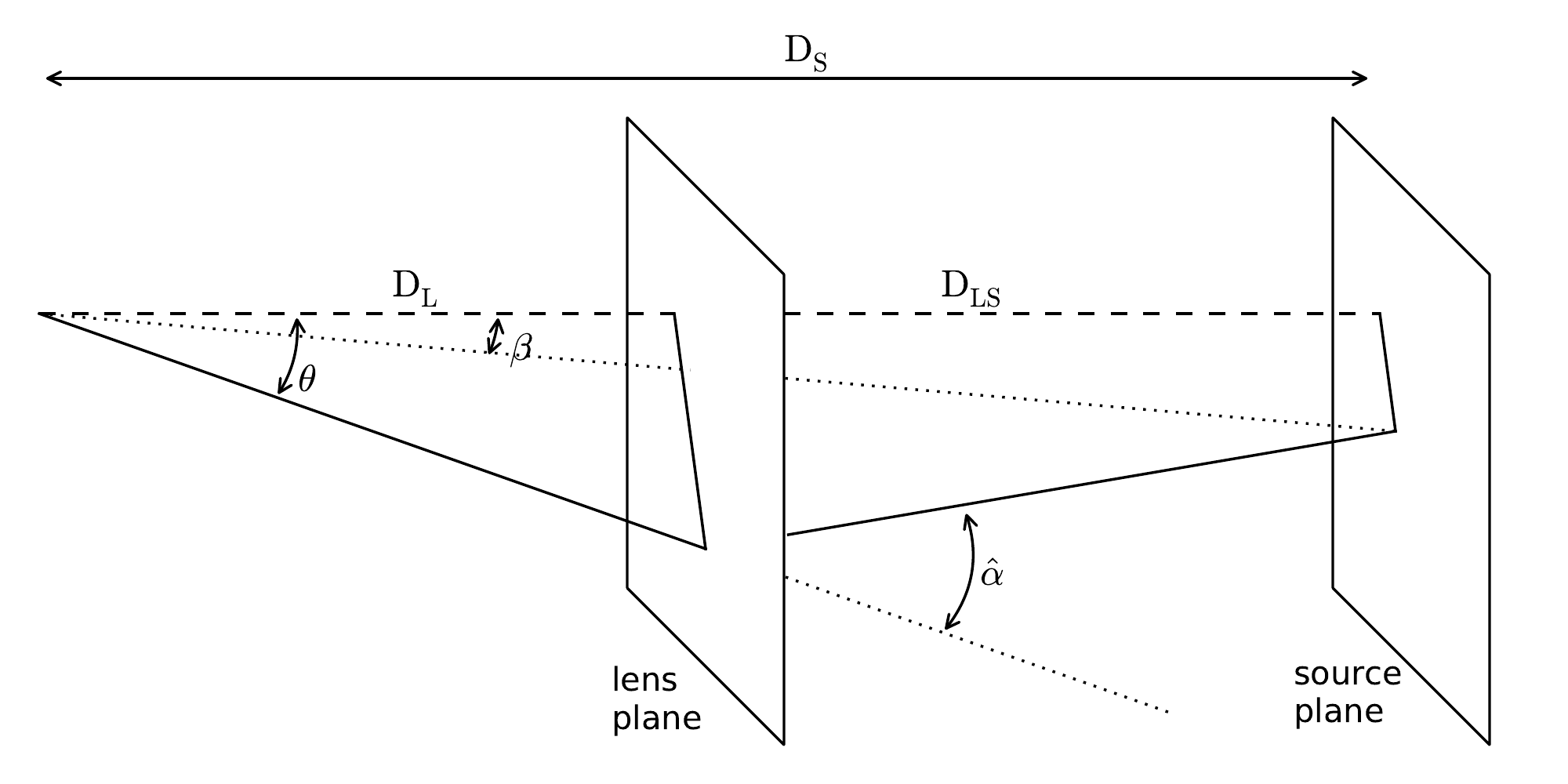}
  \caption{The geometry of gravitational lensing}
  \label{fig:lensing_geometry}
\end{figure}

For a mass-sheet located at a distance $D_L$, and a photon source located at a distance $D_S$ (with $D_{LS} = D_S - D_L$) geometric considerations in the small-angle approximation (see Figure~\ref{fig:lensing_geometry}) yield the relation
\begin{equation}
  \vec{\theta} = \vec{\beta} + \frac{D_{LS}}{D_S}\hat{\vec{\alpha}}
\end{equation}
where $\vec{\theta}$ and $\vec{\beta}$ are the observed and true positions of the source, respectively.  Rescaling $\hat{\vec{\alpha}}$ in more convenient units gives
\begin{equation}
  \label{eq:mapping}
  \vec{\theta} = \vec{\beta} + \vec{\alpha}
\end{equation}
where we have defined
\begin{equation}
  \label{eq:alpha-def2}
  \vec{\alpha} \equiv \frac{D_{LS}}{D_S}\hat{\vec{\alpha}}
\end{equation}

\subsection{Continuous Mass Distribution}
In the case of a continuous mass distribution, we can recall the remarks of section \ref{sec:lensing_simplification}, and define a surface-mass density for a mass-sheet located at a redshift $z_L$:
\begin{equation}
  \label{eq:sigma-def}
  \Sigma(\vec{\theta},z_L) 
  = \int_{D_L-\delta D}^{D_L+\delta D}\rho_M(\vec{\theta},D)\dd D 
  = \frac{1}{c^2}\int_{z_L - \delta z}^{z_L + \delta z} 
  \varepsilon_M(\vec{\theta},z)\frac{dD}{dz}\dd z
\end{equation}
where $\varepsilon_M \equiv \rho_M c^2$ is the energy density of matter,
$\vec{\theta}$ is the apparent angular position, and $z$ and $D(z)$ are the 
redshift and line-of-sight distance, respectively, with $D_S = D(z_s)$.  

A matter distribution $\rho_M(\vec{\theta},z)$, and its Newtonian potential $\Phi(\vec{\theta},z)$ are related by Poisson's equation:
\begin{equation}
  \label{eq:poisson}
  \nabla^2 \Phi(\vec{\theta},z) = 4\pi G \rho_M(\vec{\theta},z)
\end{equation}

It is convenient to define the unscaled lensing potential $\hat{\psi}$, 
given by
\begin{equation}
  \hat{\psi}(\vec{\theta},z_s) 
  = \int_{0}^{D(z_s)} \Phi(\vec{\theta},z(D))\,\,\dd D 
  = \int_{0}^{z_s}\Phi(\vec{\theta},z)\frac{dD}{dz}\,\,\dd z
\end{equation}
Using the approximation in equation \ref{eq:alpha-def-approx}, we can write this in terms of multiple mass-sheets, such that
\begin{equation}
  \hat{\psi} = \sum_i \delta \hat{\psi}_i 
  = \sum_i \int_{D_i-\delta D}^{D_i+\delta D}\Phi \dd D
\end{equation}
with $D_{i+1} = D_i + 2\delta D$.

The gradient of $\hat{\psi}$ with respect to 
$\vec{\xi} \equiv D_L\vec{\theta}$ is
\begin{equation}
  \vec{\nabla}_\xi\hat{\psi} 
  = \sum_i \vec{\nabla}_\xi\big(\delta\hat{\psi}_i\big) 
  = \sum_i \int_{D_i - \delta D}^{D_i+\delta D}\vec{\nabla}_\xi \Phi \dd D
\end{equation}
Comparing this with (\ref{eq:alpha-def-approx}) and (\ref{eq:alpha-def2}) gives the incremental deflection angle in terms of the lensing potential of a mass-sheet:
\begin{equation}
  \delta\vec{\alpha}_i = \frac{2}{c^2}\frac{D_{LS}}{D_S}\vec{\nabla}_\xi (\delta\hat{\psi}_i)
\end{equation}
Further simplification can be made by rescaling the lensing potential,
defining
\begin{equation}
  \delta\psi_i = \frac{2}{c^2}\frac{D_{LS}}{D_L D_S} \delta\hat{\psi}_i
\end{equation}
so that we are left with
\begin{equation}
  \label{eq:alpha-psi}
  \delta\vec{\alpha}_i(\vec{\theta},z_L) = \vec{\nabla}_\theta \Big( \delta\psi_i(\vec{\theta},z_L)\Big)
\end{equation}
Defining the total scaled lensing potential $\psi = \sum_i \delta\psi_i$, and the total deflection $\vec{\alpha} = \sum_i\delta\vec{\alpha}_i$, we obtain
\begin{equation}
  \vec{\alpha}(\vec{\theta},z_s) = \vec{\nabla}_\theta \psi(\vec{\theta},z_s)
\end{equation}

The Laplacian of $\delta\psi_i$ with respect to theta is given by
\begin{equation}
  \nabla_\theta^2 (\delta\psi_i) 
  = \frac{2}{c^2}\frac{D_{LS}D_L}{D_S}\int_{D_L-\delta D}^{D_L+\delta D} 
  \nabla_\xi^2\Phi \dd D
\end{equation}
Using (\ref{eq:sigma-def}) and (\ref{eq:poisson}) this becomes
\begin{equation}
  \label{eq:psi-sigma_init}
  \nabla_\theta^2(\delta\psi_i) = \frac{8\pi G}{c^2}\frac{D_{LS}D_L}{D_S}\Sigma(\vec{\theta},z_i)
\end{equation}

We now define the critical surface density,
\begin{equation}
  \Sigma_{c}(z) \equiv\frac{c^2 D_S}{4\pi G D_L(z) D_{LS}(z)}
\end{equation}
and the convergence
\begin{equation}
  \label{eq:kappa-sigma}
  \kappa(\vec{\theta},z_s) \equiv \sum \frac{\Sigma(\vec{\theta},z_i)}{\Sigma_c(z_i)}\ \forall\ z_i < z_s
\end{equation}
Now summing all the mass-sheets in (\ref{eq:psi-sigma_init}) gives the relation between the scaled lensing potential and the convergence
\begin{equation}
  \label{eq:psi-kappa-1}
  \nabla_\theta^2\psi(\vec{\theta},z_s) = 2\kappa(\vec{\theta},z_s)
\end{equation}

Solving this two-dimensional differential equation gives the effective potential in terms of the convergence:
\begin{equation}
  \label{eq:psi-kappa}
  \psi(\vec{\theta},z) 
  = \frac{1}{\pi}\int_{\mathbb{R}^2} \kappa(\vec{\theta}^\prime,z) 
  \ln|\vec{\theta} - \vec{\theta}^\prime|\dd^2\theta^\prime
\end{equation}

\section{Weak Gravitational Lensing}
\label{sec:weak_lensing_intro}

The local properties of the mapping in (\ref{eq:mapping}) are contained in its Jacobian matrix, given by
\begin{equation}
  \label{eq:Jacobian_def}
  \mathcal{A} \equiv \frac{\partial \vec{\beta}}{\partial \vec{\theta}} = \Big(\delta_{ij} - \frac{\partial \alpha_i}{\partial \theta_j} \Big) = \Big(\delta_{ij} - \frac{\partial^2 \psi}{\partial \theta_i \partial \theta_j} \Big),
\end{equation}
where $i,j$ index the two components of the angular position.

Introducing the abbreviation
\begin{equation}
  \label{eq:psi_ij}
  \psi_{ij} = \frac{\partial^2 \psi}{\partial \theta_i \partial \theta_j}
\end{equation}
We can then rewrite the convergence $\kappa$ (eqn \ref{eq:psi-kappa-1}) and define the complex shear $\gamma \equiv \gamma_1 + i\gamma_2$ of the mapping and write:
\begin{equation}
  \label{eq:gamma-def}
  \begin{array}{lcl}
    \kappa & = & (\psi_{11} + \psi_{22})/2\\
    \gamma_1 & = & (\psi_{11} - \psi_{22})/2\\
    \gamma_2 & = & \psi_{21} = \psi_{12}
  \end{array}
\end{equation}
The local Jacobian matrix (\ref{eq:Jacobian_def}) of the lens mapping can then be written
\begin{equation}
  \label{eq:jacobian_kappa_gamma}
  \mathcal{A} = \left(
  \begin{array}{cc}
    1 - \kappa - \gamma_1 & -\gamma_2\\
    -\gamma_2             & 1-\kappa+\gamma_1
  \end{array}\right)
\end{equation}

Equations \ref{eq:psi-kappa}, \ref{eq:psi_ij} and \ref{eq:gamma-def} can be combined and simplified to yield the following relationship between the convergence and the shear, where for simplicity we define the complex angle
$\theta \equiv \theta_1 + i\theta_2$:
\begin{equation}
  \label{eq:gamma-kappa}
  \gamma(\theta) 
  = \frac{-1}{\pi}\int_{\mathbb{R}^2} \mathcal{D}(\theta - 
  \theta^\prime)\kappa(\theta^\prime) \dd^2\theta^\prime
\end{equation}
where 
\begin{equation}
  \label{eq:scriptD}
  \mathcal{D}(\theta) 
  = \frac{\theta_1^2 - \theta_2^2 + 2i\theta_1\theta_2}{(\theta_1^2+\theta_2^2)^2}
  = \frac{\theta^2}{|\theta|^4}
\end{equation}
is the Kaiser-Squires kernel \citep{Kaiser93}.
The lens mapping in eqn.~\ref{eq:jacobian_kappa_gamma} describes an
image transformation consisting of a magnification with magnitude 
given by the real convergence $\kappa$
and a distortion with magnitude and orientation given by the complex shear
$\gamma = \gamma_1 + i\gamma_2$.  This distortion results in a measurable
effect, at least in principle.  If the intrinsic shape, size, or brightness
of a distant image were known, then the observed shape, size, or brightness
could be observed to determine the shear and convergence at that point.
Unfortunately, the intrinsic shape and size of a galaxy cannot be known
{\it a priori}, but using well-founded assumptions about the statistics
of the {\it distribution} of shapes and sizes of sources can lead to
useful estimates of the shear and/or convergence across the sky.

%\comment{add some references: measuring gamma \& kappa from galaxy shapes
%  and sizes, from number counts,
%  galaxy-galaxy lensing, quasar variability-luminosity relationship, etc.}

In the most common approach to weak lensing,
the ellipticities of source galaxies are measured, 
giving a noisy estimate of the {\it reduced shear}
\begin{equation}
  \gamma_r(\theta) = \frac{\gamma(\theta)}{1 - \kappa(\theta)}.
\end{equation}
In the weak limit where $\kappa(\theta) \ll 1$, analyses often assume
$\gamma_r(\theta) \approx \gamma(\theta)$, though with higher-precision
measurements, this second-order effect can introduce systematic
errors in mass maps and power spectra \citep{Dodelson06, Shapiro09, Krause10}.
Once the shear field is estimated, the
measurements can be utilized in a number of ways to learn about
fundamental physical principles; this work will focus on three areas:
\begin{description}
  \item[Direct mapping:] Having measured the shear $\gamma$ at locations
    across the sky, the convergence $\kappa$ can be estimated.  $\kappa$
    relates to the projected density via eqn.~\ref{eq:kappa-sigma}.  Thus
    the measured shear can be used to directly estimate a map of the
    distribution of dark matter in two dimensions.  Using redshift
    information for the lensed sources, there is the potential to extend
    this mapping to three dimensions. This is the subject of Chapter 3.
  \item[Peak statistics:] The two dimensional maps recovered as above
    represent a two-dimensional projection of the three-dimensional
    distribution of large scale structure, in particular massive galaxy
    clusters.  As discussed in \S\ref{sec:growth}, both the number of
    clusters and their mass distribution depend on the details of the
    geometry, expansion, and makeup of the universe.  By computing the
    statistics of observed lensing peaks to 
    that predicted by theory, it is possible to
    constrain cosmological parameters using the peaks alone.  This is
    one of the subjects reviewed and explored in Chapter 4.
  \item[Power spectrum:] The power spectrum of the shear is closely related
    to the power spectrum of the matter distribution that generates it.
    By measuring two point information of observed shear, it is possible
    to constrain cosmological parameters in a way that is complementary
    to the peak counts mentioned above.  This is the subject of Chapter 5.
\end{description}

To enable these three analyses, we will develop a bit further the basic
principles of weak lensing mappings and power spectra.

\subsection{Mapping with Weak Lensing}
\label{sec:lens_mapping_intro}
The tomographic approach to the 3D lensing mapping problem can be
computed using the following steps \citep[see, e.g.][]{Hu02, Simon09,
  Vanderplas2011}:
\begin{enumerate}
  \item From the measured ellipticities and redshifts of photometrically
    observed galaxies, obtain noisy estimates of the shear
    $\gamma_{obs}(\theta, z)$.
  \item Using eqn.~\ref{eq:gamma-kappa}, recover an estimate of
    $\kappa(\theta, z)$.  Note that due to the integral over the lensing
    kernel $\mathcal{D}(\theta)$, the convergence estimate is non-local:
    the value of $\kappa$ at a given location is related to the value of the
    $\gamma$ at {\it all other} locations.
  \item Using eqn.~\ref{eq:kappa-sigma}, determine the projected density
    $\Sigma(\theta, z)$.
  \item As a final step, it is possible in principle to use
    eqn.~\ref{eq:sigma-def} to recover the 3D mass density
    $\rho(\theta, z)$.  This is the subject of Chapter 3.
\end{enumerate}
To accomplish this, it is convenient to combine steps 3-4 and write 
the expression for $\kappa(\theta,z)$ in terms of $\rho(\theta,z)$
explicitly.  From (\ref{eq:sigma-def}) and (\ref{eq:kappa-sigma}),
approximating the sum as an integral, we find
\begin{equation}
  \kappa(\vec{\theta},z_s) 
  = 4\pi G \int_0^{z_s} 
  \frac{D^{(A)}(z)[D^{(A)}(z_s)-D^{(A)}(z)]}{D^{(A)}(z_s)} 
  \rho_M(\vec{\theta},z) \frac{dD^{(A)}(z)}{dz} \dd z
\end{equation}
The notation has been changed here to make clear that the distances in 
question are in fact angular diameter distance, the relevant distance 
in the context of lensing calculations (See \S\ref{sec:distances}).
Recall that angular diameter 
distance $D^{(A)}$ is related to the comoving distance $D$ by
\begin{equation}
  D^{(A)}(z) = a S_\kappa (D)
\end{equation}
where $S_\kappa(D) = D$ for a flat universe.  Assuming a flat universe,
converting to comoving distances, 
and writing this in terms of $\varepsilon = \rho c^2$, we find
\begin{equation}
  \label{kappa-epsilon-1}
  \kappa(\vec{\theta},z_s) 
  = \frac{4\pi G}{c^2} \int_0^{z_s} \dd z\frac{dD}{dz} 
  a^2\frac{D(D_S-D)}{D_S} \varepsilon_M(\vec{\theta},z),
\end{equation}
where we've used the shorthand $D \equiv D(z)$ and 
$D_S \equiv D(z_s)$.

To further progress, we can follow \S\ref{sec:growth} and write the
matter density $\varepsilon_M(\theta, z)$ in equation
\ref{kappa-epsilon-1} in terms of the density contrast $\delta$:
\begin{equation}
  \label{delta-def}
  \varepsilon_M(\vec{\theta},z) = \Omega_M(z) \varepsilon_c(z)\Big[1+\delta(\vec{\theta},z)\Big]
\end{equation}
where we have assumed a flat universe, such that the total density is
equal to the critical density $\varepsilon_c(z)$
(eqn.~\ref{eq:critical_density}).
We'll make use of two further algebraic substitutions:
from the definition of comoving distance (eq.~\ref{eq:comoving_distance}),
we can write
\begin{equation}
  \label{dDdz}
  \frac{dD}{dz} = \frac{c}{H(z)},
\end{equation}
and from the Friedmann equation (eqn.~\ref{eq:friedmann_dimensionless})
matter density fraction can be written
\begin{equation}
  \Omega_M(z) = \frac{H_0^2\Omega_{M,0}(1+z)^3}{[H(z)]^2}.
\end{equation}

Combining these equations gives
\begin{equation}
  \label{eq:kappa-delta}
  \kappa(z_s) = \frac{3cH_0^2\Omega_{M,0}}{2}\int_0^{z_s} \dd z \frac{(1+z)}{H(z)} \frac{D(D_S-D)}{D_S}\big[1+\delta(z)\big]
\end{equation}

Because of the mass-sheet degeneracy, $\kappa(z_s)$ can only be determined
up to an additive constant across a given redshift bin
\citep[see][for discussion]{seitz_schneider1996}.
Assuming the observed field is large enough to average the effects of cosmic
variance, the additive constant will be due simply to the background
matter distribution.
Defining $\bar{\kappa}(z_s)$ to be the convergence due to the background
matter distribution in matter-dominated growth,
and $\Delta(z) \equiv \delta(z)/a$ we find
\begin{equation}
  \label{eq:kappa-delta-2}
  \kappa(z_s) \equiv \hat{\kappa}(z_s)-\bar\kappa(z_s) = 
  \frac{3cH_0^2\Omega_{M,0}}{2}\int_0^{z_s} \dd z 
  \frac{1}{H(z)} \frac{D(D_S-D)}{D_S}\Delta(z)
\end{equation}
where, to be explicit,
\begin{equation}
  \bar\kappa(z_s) = \frac{3cH_0^2\Omega_{M,0}}{2}\int_0^{z_s} \dd z 
  \frac{(1+z)}{H(z)} \frac{D(D_S-D)}{D_S}
\end{equation}
To be clear, here, $D$ is the comoving distance to a redshift $z$,
and $D_S$ is the comoving distance to the redshift $z_s$ of the photon
source.  Equation \ref{eq:kappa-delta-2} defines the mapping from $\kappa(z_s)$
to $\Delta(z)$ for $z<z_s$.

\subsection{Power Spectra}
Mass mapping can lead to deep astrophysical and cosmological insights through
the comparison of dark and luminous matter distributions
\citep[e.g.][]{Clowe06}, through constraints on the mass profiles of
collapsed structures \citep[e.g.][]{Oguri2012}, or
through the comparison of observed mass peaks to
theoretical predictions (see Chapter 4).  Because of the noise inherent in
lensing observations, most of these localized analyses are limited to
very dense regions, far from the linear regime.

The linear regime, as well as the presence of nonlinear effects on small
scales, can be measured using power spectra of the weak lensing shear.  In
order to accomplish this, however, the power spectra of observed shear must
be related to the mass power spectra discussed in \S\ref{sec:growth}.

%In \S\ref{sec:growth}, we defined the power spectrum
%\begin{equation}
%  P_\delta(k) = \langle |\hat\delta_k|^2 \rangle,
%\end{equation}
%where

\subsubsection{E and B modes}
\label{sec:EBmode}
In this section, we will outline the basic results of \citet{Schneider02b}.
We will start by defining the E/B decomposition of the shear field $\gamma$.
If the shear $\gamma$ and convergence $\kappa$ can be expressed as shown
in eqn.~\ref{eq:gamma-def}, then the gradient of $\kappa$ can be written
\begin{equation}
  \label{eq:u_def}
  \nabla_\theta \kappa =
  \left(
  \begin{array}{l}
    \partial \kappa / \partial\theta_1\\
    \partial \kappa / \partial\theta_2 
  \end{array}
  \right) 
  =
  \left(
  \begin{array}{l}
    \partial \gamma_1 / \partial\theta_1 + \partial\gamma_2/\partial\theta_2\\
    \partial \gamma_2 / \partial\theta_1 - \partial\gamma_1/\partial\theta_2
  \end{array}
  \right)
  \equiv
  \myvec{u}.
\end{equation}
If $\kappa$ and $\gamma$ are due entirely to weak lensing, then the vector
$\myvec{u}$ should be a pure gradient field, as will every
quantity in the equality in eqn.~\ref{eq:u_def}.  This condition can be
compactly expressed by noting that the curl of a gradient is identically
zero:
\begin{equation}
  \label{eq:curl-free}
  \myvec{\nabla} \times \myvec{u} = 0.
\end{equation}
If, however, other effects are involved (e.g.~shot noise, second-order
lensing effects, intrinsic alignments, systematic errors, etc.)
then $\myvec{u}$ will not be a pure gradient field and will have
a nonzero curl.
With this in mind, we will use an analogy from electrodynamics and decompose
$\kappa$ into a curl-free ``E-mode'' $\kappa_E$ and a divergence-free
``B-mode'' $\kappa_B$ such that
\begin{eqnarray}
  \nabla^2 \kappa_E &=& \myvec{\nabla} \cdot \myvec{u}\\
  \nabla^2 \kappa_B &=& \myvec{\nabla} \times \myvec{u}.
\end{eqnarray}
We'll also define the E-mode and B-mode lensing potential following
eqn.~\ref{eq:psi-kappa-1}:
\begin{equation}
  \nabla^2 \psi_{E, B} = 2\kappa_{E, B}.
\end{equation}
This allows us to define the E and B modes of $\gamma$ via
eqn.~\ref{eq:gamma-def}.  Explicitly,
\begin{equation}
  \gamma_{E,B} = \left(\frac{1}{2}\left[\frac{\partial^2}{\partial \theta_1\partial\theta_1}
  - \frac{\partial^2}{\partial \theta_2\partial\theta_2}\right]
  + i\frac{\partial^2}{\partial \theta_1\partial\theta_2}\right)
  \psi_{E,B}.
\end{equation}
Combining the convergence E and B modes as a complex linear combination
$\kappa = \kappa_E + i\kappa_B$, we find in analogy to
eqn.~\ref{eq:gamma-kappa},
\begin{equation}
  \label{eq:gamma-kappa-EB}
  \left[\gamma_E(\theta) + i\gamma_B(\theta)\right]
  = \frac{-1}{\pi}\int_{\mathbb{R}^2} \mathcal{D}(\theta - 
  \theta^\prime)
  \left[\kappa_E(\theta^\prime) + i\kappa_B(\theta^\prime)\right]
  \dd^2\theta^\prime
\end{equation}
We can define the Fourier transform of the convergence
\begin{equation}
  \hat{\kappa}_{E,B}(\myvec\ell)
  = \int\dd^2\myvec{\theta}\,e^{i\myvec{\ell}\cdot\myvec{\theta}}
  \kappa_{E,B}(\myvec{\theta}),
\end{equation}
where $\myvec{\ell}$ is the angular Fourier variable.  We can then define
the power spectra \citep[e.g.][]{Schneider02b}
\begin{eqnarray}
  \langle\hat{\kappa}_{E, B}(\myvec{\ell})
  \hat{\kappa}^\ast_{E,B}(\myvec{\ell}^\prime)\rangle
  =(2\pi)^2\delta_D(\myvec{\ell} - \myvec{\ell}^\prime)P_{E,B}(\ell).
\end{eqnarray}
By the convolution theorem, the
Fourier transform of eqn.~\ref{eq:gamma-kappa-EB} gives
\begin{equation}
  \hat{\gamma}_{E,B}(\myvec{\ell})
  =\left(\frac{\ell_1^2 - \ell_2^2 + 2i\ell_1\ell_2}{|\myvec{\ell}|^2}\right)
  \hat{\kappa}_{E,B}(\myvec{\ell}).
\end{equation}
The factor relating $\hat{\gamma}$ and $\hat{\kappa}$ can be expressed
as a simple phase $e^{2i\beta}$, so that
\begin{equation}
  \langle\hat{\gamma}_{E, B}(\myvec{\ell})
  \hat{\gamma}^\ast_{E,B}(\myvec{\ell}^\prime)\rangle = 
  \langle\hat{\kappa}_{E, B}(\myvec{\ell})
  \hat{\kappa}^\ast_{E,B}(\myvec{\ell}^\prime)\rangle.
\end{equation}
Based on this equality, we can define the shear correlation function
and relate it to the convergence power spectra $P_{EB}$,
\begin{eqnarray}
  \label{eq:xi_plus}
  \xi_+(\theta)
  &\equiv& \langle\gamma(\myvec{0})\gamma^\ast(\myvec{\theta})\rangle\nonumber\\
  &=& \int_0^\infty\frac{\dd\ell\,\ell}{2\pi}J_0(\ell\,\theta)
     [P_E(\ell) + P_B(\ell)],
\end{eqnarray}
where $J_0(x)$ is a Bessel function of the first kind.
The ``+'' distinguishes this correlation measure from two other shear
correlations that can be defined, $\xi_-$ and $\xi_\times$
\citep[see][for details]{Schneider02}.
We will limit the discussion here to $\xi_+$, because this is the relevant
measure for our purposes (See discussion in \S\ref{sec:WhichCorrelation}).

The E-mode angular shear power spectrum $P_E(\ell)$ can be expressed as a
weighted line-of-sight integral over the matter power 
spectrum via eqn.~\ref{eq:kappa-delta}.  Taking into account the redshift
distribution of galaxies gives \citep[see][]{Takada04}
\begin{equation}
  P_E(\ell) = \int_0^{r_s}dr W^2(r)r^{-2}
  P_\delta\left(k=\frac{\ell}{r};z(r)\right)
\end{equation}
Here $r$ is the comoving distance, $r_s$ is the distance to the
source, and $W(r)$ is the lensing weight function,
\begin{equation}
  \label{eq:lens_weight}
  W(r) = \frac{3\Omega_{m,0}H_0^2}{2a(r)}\frac{r}{\bar{n}_g}
  \int_{z(r)}^{z(r_s)}dz\ n(z) \frac{r(z)-r}{r(z)}
\end{equation}
where $n(z)$ is the empirical redshift distribution of galaxies,
with $\bar{n}_g = \int_0^\infty n(z)dz$.
This allows us to predict an analytic relation between the 3D mass
fluctuation power spectrum $P_\delta(k, z)$ and the correlation function
of the shear signal $\xi_+(\ell, z)$.
The nonlinear mass fluctuation power spectrum $P_\delta(k, z)$ can be
predicted semi-analytically (e.g. \citet{Smith03}) and has a form dependent on
the assumed cosmological model.
The shear correlation function $\xi_+(\myvec{\ell})$
can be computed from observed data.  The important point is that this
relation provides a direct (if noisy) measure of the distribution of mass:
it requires no assumptions
about the mass to light ratio or how the non-baryonic matter distribution
relates to that of luminous matter. The observations can be related directly
to theoretical expectations for the nonlinear power spectrum.
For this reason, weak lensing studies provide a powerful
probe for constraining the cosmological parameters that describe
the geometry and dynamics of the Universe.

In the following sections we undertake a systematic study of gravitational
lensing applications enabled by \KL\ (KL) analysis.  We begin in Chapter 2
with a development of the mathematical theory behind KL analysis.  In
Chapter 3, we explore the use of a KL-based method for reconstruction of
three-dimensional mass maps from shear data.  In Chapter 4, we explore how
KL can be used to address incompleteness in shear surveys, allowing us to
interpolate the signal across masked regions.  In Chapter 5, we use KL as
a basis for computing parameter constraints from two-point statistics of a
shear field with incomplete sky coverage.

%% file: chapter2.tex
\chapter{Introduction to Karhunen-Lo\`{e}ve Analysis}

%\begin{itemize}
%  \item Introduce the notational formalism
%  \item KL as an eigenvalue problem
%  \item KL as a signal-to-noise ranking
%  \item KL as an optimal low-dimensional representation
%  \item KL on noisy data
%  \item KL with missing information
%\end{itemize}

%Also mention LLE and cite \cite{Vanderplas2009}, which uses a nonlinear
%alternative to PCA for dimensionalty reduction.

\KL\ (KL) analysis is a commonly used statistical tool
in a broad range of astronomical applications, from, e.g.~studies of
correlations in observed properties of galaxy photometry \citep{Efstathiou84}
and galaxy and quasar spectra \citep{Connolly95,Connolly99,Yip04a,Yip04b},
to  analysis of the spatial distribution of galaxies 
\citep{Vogeley96, Matsubara00, Szalay03, Pope04, Tegmark06},
to characterization of the 
expected errors in weak lensing surveys \citep{Kilbinger06, Munshi06}.
In this chapter, we will develop the formalism of KL that will form
the basis of the applications in the subsequent chapters.

The KL formalism requires the liberal
employment of algebra with vectors, scalars, matrices, and their
generalizations.
For clarity, we will begin by briefly specifying the notational
conventions used in this chapter and throughout this work.

\section{Notational Conventions}

It is important to clearly distinguish between vectors, matrices, and
scalars in the following formulation.  Vectors will be denoted by
bold lower-case symbols; e.g. $\myvec{x}$.  Matrices will be denoted by
bold upper-case symbols; e.g. $\mymat{X}$.  Scalars will be denoted by
non-bold symbols, either upper or lower-case.
All vectors are assumed to be column vectors, while a row-vector is
indicated by the transpose, $\myvec{x}^T$.  
Single elements of a given
vector or matrix are given with subscripts: $x_i$ is the $i^{\rm th}$
element of the vector $\myvec{x}$, and $X_{ij}$ is the element in the
$i^{\rm th}$ row and $j^{\rm th}$ column of the matrix $\mymat{X}$.
The vector making up the $j^{\rm th}$ column of $\mymat{X}$ is
indicated by $\myvec{x}^{(j)}$.  Note then, that by this convention,
the $(i, j)$ component of matrix $\mymat{X}$ can be equivalently expressed
$X_{ij}$ or $x_i^{(j)}$.

In algebraic expressions, the normal linear algebra rules are assumed.
For example, the expression 
\begin{equation}
  \myvec{y} = \mymat{M}\mymat{x} + \mymat{b}
\end{equation}
involves the vectors $\myvec{y}$, $\myvec{x}$, and $\myvec{b}$ and the
matrix $\mymat{M}$.  This expression is short-hand for the summation:
\begin{equation}
  y_i = \sum_{j=1}^{n} M_{ij} x_j + b_i
\end{equation}
Using these rules, we can define the magnitude of a vector
\begin{equation}
  |\myvec{x}| \equiv \sqrt{\myvec{x}^T \myvec{x}}
\end{equation}

\section{Basis function decomposition}
KL analysis is simply a basis function decomposition, where the basis
functions are derived based on the variance and covariance properties
of a class of functions.  We'll start by describing what is perhaps the
best-known basis function decomposition, the Fourier series.
We start here because it's a familiar concept that generalizes well to
the fundamental ideas of KL analysis.

\subsection{Fourier Series}
The Fourier Series is a means of expressing a bounded function in terms
of a certain class of oscillatory basis functions.  It is a discrete version
of the Fourier Transforms used in cosmological power spectrum analysis, and
discussed in \S\ref{sec:EBmode}.

We'll define a set of oscillatory basis functions 
\begin{equation}
  \label{eq:fourier_basis}
  \Phi_k(t) = \frac{1}{\sqrt{t_b - t_a}}
  \exp\left[\frac{2\pi i k (t-t_a)}{t_b - t_a}\right]
\end{equation}
where $t$ is the arbitrary dependent variable, $k$ is the wave-number,
and the function is defined in the region $t_a \le t \le t_b$.
We'll postulate that a function $f(t)$ can be expressed as a linear
combination of these basis functions:
\begin{equation}
  \label{eq:fourier_1D}
  f(t) = \sum_{k=-\infty}^\infty a_k\Phi_k(t).
\end{equation}
Here $a_k$ are an infinite set of complex coefficients.  Our claim is that
any piecewise continuous and square-integrable function $f(t)$ in the
interval $[t_a, t_b]$ can be represented this way.
A rigorous mathematical proof
of this statement can be found elsewhere, but below we will lend support
to this claim.

Given the claim that Equation~\ref{eq:fourier_1D} holds, how can we
compute the Fourier coefficients $a_k$ associated with 
a particular $f(t)$?  Though the expression is well-known, we'll briefly
derive it here because it illuminates some of the properties of
Fourier transforms that will generalize to KL transforms.

To begin, we'll multiply both sides of Equation~\ref{eq:fourier_1D} by the
complex conjugate $\Phi^\ast_{k^\prime}(x)$ of the basis function given
in Equation~\ref{eq:fourier_basis}, and integrate both sides over $t$
from $t_a$ to $t_b$:
\begin{equation}
  \int_{t_a}^{t_b} \Phi^\ast_{k^\prime}(t) f(t) \dd t
  = \int_{t_a}^{t_b} \Phi^\ast_{k^\prime}(t) \sum_{k=-\infty}^\infty a_k\Phi_k(x)\dd t
\end{equation}
On the right-hand side, we can exchange the order of integration and
summation to find 
\begin{equation}
  \label{eq:fourier_coef_der}
  \int_{t_a}^{t_b} \Phi^\ast_{k^\prime}(t) f(t) \dd t
  =  \sum_{k=-\infty}^\infty a_k \left[\int_{t_a}^{t_b} \Phi^\ast_{k^\prime}(t)\Phi_k(x)\dd t\right].
\end{equation}
Let's examine the term in the square brackets.  Plugging in the definition
of the basis functions from Equation~\ref{eq:fourier_basis}, we have
\begin{equation}
  \left[\int_{t_a}^{t_b} \Phi^\ast_{k^\prime}(t)\Phi_k(x)\dd t\right]
  = \left[\frac{1}{t_b-t_a}\int_{t_a}^{t_b} \exp\left(\frac{2\pi i (k - k^\prime) (t-t_a)}
    {t_b - t_a}\right)\dd t\right]
\end{equation}
This gives two distinct situations for the integral on the right-hand side:
when $k=k^\prime$, both the integrand 
and the term in the brackets is exactly $1$.  When $k\ne k^\prime$,
the integrand oscillates through an integer number of cycles between
$t_a$ and $t_b$ (remember that $k$ and $k^\prime$ here are integers),
and the result of the integral is exactly $0$.
So we see that the term in brackets is equal to simply 
the Kronecker delta function $\delta_{kk^\prime}$, defined as
\begin{equation}
  \delta_{ij} = \left\{
  \begin{array}{ll}
    1 & {\rm if}\ i = j\\
    0 & {\rm if}\ i \ne j.
  \end{array}
  \right.
\end{equation}
Putting this result into Equation~\ref{eq:fourier_coef_der}, only one
term of the sum remains and we find
\begin{equation}
  \label{eq:fourier_coef}
  a_k = \frac{1}{t_b - t_a}\int_{t_a}^{t_b} \Phi^\ast_k(t) f(t) \dd t
\end{equation}

Equation~\ref{eq:fourier_coef} shows how to compute the Fourier coefficients
$a_k$ for a given $f(t)$.  But one might wonder if this is a unique result.
Could there be several possible sets of valid Fourier coefficients for
a given function?

Let's assume that given a function $f(t)$, there are two valid sets of
Fourier coefficients $a_k$ and $a^\prime_k$ that satisfy
Equation~\ref{eq:fourier_1D}.  In this case, subtracting the two equations
gives
\begin{equation}
  f(t) - f(t) = \sum_{k=-\infty}^\infty (a_k - a^\prime_k)\Phi_k(t) = 0.
\end{equation}
In a similar manner to above, we can multiply by $\Phi^\ast_k(t)$, integrate
over $t$ from $t_a$ to $t_b$, and extract a Kronecker delta function to
yield
\begin{equation}
  \sum_{k=-\infty}^\infty (a_k - a^\prime_k) \delta_{kk^\prime} = 0
\end{equation}
Collapsing the sum, we find that $a_k = a^\prime_k$ for all $k$.  This
shows the uniqueness of the Fourier coefficients $a_k$ for a given function
$f(t)$ on an interval $[t_a, t_b]$.  Thus, given an orthonormal basis $\Phi_k$,
there is a single unique linear combination that reconstructs a function
$f(t)$ on the defined interval.

\subsection{Generalizing Orthonormal Bases}

Stepping back for a moment, we have shown that for a particular
class of basis functions $\Phi_k(t)$, one can find unique coefficients
$a_k$ such that one of the expansions of Equation~\ref{eq:fourier_1D} holds.
A key observation is that all the derivations above rested 
solely on two special properties of these basis functions:
\begin{enumerate}
  \item
    The basis functions are {\it orthonormal} on the interval $[a, b]$.
    That is, $\Phi_k(t)$ satisfies
    \begin{equation}
      \int_{t_a}^{t_b} \Phi_k(t) \Phi^\star_{k^\prime}(t) = \delta_{kk^\prime}
    \end{equation}
  \item
    The basis functions are {\it complete} on the interval $[t_a, t_b]$.
    That is, an arbitrary function $f(t)$ can be approximated by
    \begin{equation}
      f(t) = \sum_{k=1}^N a_k \Phi_k(t)
    \end{equation}
    and the mean square error satisfies
    \begin{equation}
      \label{eq:completeness}
      \lim_{N\to\infty} \int_{t_a}^{t_b}
      \left[f(t) - \sum_{k=1}^{N}a_k \Phi_k(t)\right]^2 \dd t = 0.
    \end{equation}
\end{enumerate}
As long as these two properties hold for a class of functions $\Phi_k(t)$,
we would be able to repeat the above derivations and express any $f(t)$
via Equation~\ref{eq:fourier_1D}.
This suggests the possible existence of other functions that fit these
criteria.  Some examples are the Legendre polynomials on the interval
$[-1, 1]$, the Laguerre polynomials on the interval $[0, \infty)$, and
the Hermite polynomials on the interval $(-\infty, \infty)$.
In fact, these different orthonormal basis functions are simply a
generalization of well-known geometric bases (such as the $x$ and $y$ axis
of a two dimensional vector space) into an abstract function space.
Just as there are an infinite number of possible orientations for
an $(x, y)$ axis in a two-dimensional vector space, there are an infinite
number of possible orthogonal function bases that work in the above
formalism.  Choosing the right basis can lead to a much easier analysis of
a given problem.

\section{\KL\ Analysis}
Because of the infinite number of possible orthogonal function classes,
one might wonder how to choose the optimal class for any particular problem.
\KL\ analysis seeks to answer this question in a very general case.

\subsection{Derivation of \KL\ theorem}
Imagine now that we have a random process $F_t$.  This can be thought of as
an arbitrarily large collection of functions $f^{(i)}(t)$ defined
on the interval
$t \in [a, b]$.  At a given location $t$, the expectation value of the
random process is given by
\begin{equation}
  E[F_t] = \lim_{N \to \infty} \frac{1}{N} \sum_{i=1}^N f^{(i)}(t)
\end{equation}
For simplicity, we'll assume that the random process $F_t$ is {\it
centered}; that is $E[F_t] = 0$.  A general random process can be
centered by subtracting the expectation value for each $t$.
A centered random process $F_t$ can be characterized
by its covariance function, which is defined as
\begin{eqnarray}
  \label{eq:corrfunc_def}
  \mathcal{C}_F(t, t^\prime) &\equiv& E[F_t F_{t^\prime}]
  \nonumber\\
  &=& \lim_{N\to\infty}\frac{1}{N} \sum_{i=1}^N
  f^{(i)}(t)f^{(i)}(t^\prime).
\end{eqnarray}
For an {\it uncorrelated} random process,
$\mathcal{C}_F(t, t^\prime) = \mathcal{V}_F(t) \delta_{t, t^\prime}$
where $\mathcal{V}_F(t) $ is the {\it variance} of $F_t$.

\subsection{Eigenfunctions}
We'll now introduce the {\it eigenfunctions} $e_k(t)$
of the covariance function $\mathcal{C}_F(t, t^\prime)$, which satisfy
\begin{equation}
  \label{eq:eigfunc_def}
  \int_{t_a}^{t_b} \mathcal{C}_F(t, t^\prime) e_k(t^\prime)\dd t^\prime
  = \lambda_k e_k(t)
\end{equation}
subject to the constraint that $\int_{t_a}^{t_b}|e_k(t)|^2\dd t \ne 0$
(i.e. $e_k(t)$ is not everywhere zero).
Here $\lambda_k$ is the {\it eigenvalue} associated with the
eigenfunction $e_k(t)$.

Now what are the properties of these eigenfunctions?  First of all, they
are orthogonal on the interval $[t_a, t_b]$.
We can show this by considering two arbitrary eigenfunctions
$e_k(t)$ and $e_{k^\prime}(t)$.  Consider the quantity
\begin{equation}
  \int_{t_a}^{t_b} \dd t \int_{t_a}^{t_b} \dd t^\prime\,\, \mathcal{C}_F(t, t^\prime)
  e_{k^\prime}(t^\prime) e_k(t)
\end{equation}
Because of the symmetry of the covariance,
i.e.~$\mathcal{C}_F(t, t^\prime) = \mathcal{C}_F(t^\prime, t)$, and
because the order of integration can be switched, this can be evaluated
two different ways, which must be equal:
\begin{eqnarray}
  \int_{t_a}^{t_b} \dd t\,\, e_k(t)
  \int_{t_a}^{t_b} \dd t^\prime\,\, \mathcal{C}_F(t, t^\prime) e_{k^\prime}(t^\prime) &=&
  \int_{t_a}^{t_b} \dd t^\prime\,\, e_k^\prime(t^\prime)
  \int_{t_a}^{t_b} \dd t\,\, \mathcal{C}_F(t^\prime, t) e_{k}(t)
  \nonumber\\
  \int_{t_a}^{t_b} \dd t\,\, \lambda_{k^\prime} e_k(t) e_{k^\prime}(t) &=&
  \int_{t_a}^{t_b} \dd t^\prime\,\, \lambda_k e_k(t^\prime) e_{k^\prime}(t^\prime)
\end{eqnarray}
Rearranging the bottom line leads to
\begin{equation}
  (\lambda_k - \lambda_{k^\prime})
  \int_{t_a}^{t_b} \dd t\,\, e_k(t) e_{k^\prime}(t) = 0
\end{equation}
So for $\lambda_k \ne \lambda_{k^\prime}$, then $e_k$ and $e_{k^\prime}$
must be orthogonal\footnote{In the degenerate case when
$\lambda_k = \lambda_{k^\prime}$, one can still construct orthogonal
vectors by linear combinations:
\begin{eqnarray}
  e_+(t) &=& \frac{e_k(t) + e_{k^\prime}(t)}{\sqrt{2}} \nonumber\\
  e_-(t) &=& \frac{e_k(t) - e_{k^\prime}(t)}{\sqrt{2}}. \nonumber
\end{eqnarray}
This leads to two new orthogonal eigenfunctions $e_+$ and $e_-$ with the
same eigenvalue $\lambda_k$.}.  From the definition in
Equation~\ref{eq:eigfunc_def}, we see that if $e_k(t)$ is an eigenfunction
with eigenvalue $\lambda_k$,
then for any arbitrary constant $K$,
$K\, e_k(t)$ is an eigenfunction with eigenvalue $\lambda_k$ as well.
To make the choice of eigenfunction more definite, we will assume all
eigenfunctions are normalized: that is
\begin{equation}
  \int_{t_a}^{t_b} \dd t e_k(t) e_{k^\prime}(t) = \delta_{kk^\prime}
\end{equation}
for all $k$.  This still allows any eigenfunction to have an arbitrary
phase: that is, an eigenfunction may be multiplied by $e^{i\theta}$ for
any theta, and still satisfy our orthogonality condition.  This fact
will become important later.

The net result is that the eigenfunctions $e_k(t)$ form an orthonormal
basis for the space of functions represented by the random process $F_t$.
In general, the eigenfunctions
also satisfy the completeness relation (see eqn.~\ref{eq:completeness}).
The proof of the completeness of eigenfunctions for a symmetric kernel
$\mathcal{C}_F(t, t^\prime)$ is rather involved, and
can be found in, e.g.~\citet{Courant1989}.

Let's now consider the expansion of the random process $F_t$ onto the
eigenvectors $e_k(t)$.  Analogously to the Fourier case discussed
above, we have
\begin{equation}
  \label{eq:Ft_decomp}
  F_t = \sum_{k=1}^\infty A_k e_k(t).
\end{equation}
where here $A_k$ can be thought of as a set of coefficients $a_k^{(i)}$
in the same way that the random process $F_t$ can be thought of as a
set of functions $f^{(i)}(t)$.
Multiplying both sides by $e_{k^\prime}(t)$, integrating, and using the
orthogonality of eigenvectors (this is analogous to the derivation
in Equations~\ref{eq:fourier_coef_der}-\ref{eq:fourier_coef}) leads to
\begin{equation}
  \label{eq:F_k_def}
  A_k = \int_{t_a}^{t_b} F_t e_k(t) \dd t
\end{equation}

Because of the fact that the random process is centered (i.e.~$E[F_t] = 0$),
it is straightforward to show that $E[A_k] = 0$ as well.  The more interesting
computation is that of the covariance of the eigenvectors,
$\mathcal{C}_A(k, k^\prime)$.
From Equations~\ref{eq:corrfunc_def} and \ref{eq:F_k_def}, we have
\begin{eqnarray}
  \mathcal{C}_A(k, k^\prime)
  &=& E[A_k A_{k^\prime}]\nonumber\\
  &=& E\left[\int_{t_a}^{t_b} \dd t \int_{t_a}^{t_b} \dd t^\prime
    {F}_t e_k(t)
    {F}_{t^\prime} e_{k^\prime}(t^\prime)\right] \nonumber\\
  &=& \int_{t_a}^{t_b} \dd t \int_{t_a}^{t_b} \dd t^\prime
    E[{F}_t {F}_{t^\prime}]
    e_{k^\prime}(t^\prime) e_k(t) \nonumber\\
  &=& \int_{t_a}^{t_b} \dd t \int_{t_a}^{t_b} \dd t^\prime
    \mathcal{C}_F(t, t^\prime)
    e_{k^\prime}(t^\prime) e_k(t) \nonumber
\end{eqnarray}
Substituting Equation~\ref{eq:eigfunc_def}, we find that this gives
\begin{eqnarray}
  \label{eq:cov_A}
  \mathcal{C}_A(k, k^\prime)
  &=& \int_{t_a}^{t_b} \dd t \lambda_{k^\prime} e_{k^\prime}(t) e_k(t) \nonumber\\
  &=& \lambda_k \delta_{kk^\prime}
\end{eqnarray}
So we see that projection of the centered random process $\tilde{F}_t$ onto
the eigenvectors of its covariance matrix yields coefficients which
are uncorrelated, with variance equal to the eigenvalues $\lambda_k$.
This result is the \KL\ theorem, and it has many ramifications that will
be discussed below.

\subsection{Partial Reconstructions}
\label{sec:partial_recons}
We have shown that \KL\ provides an orthonormal basis for a random field
with uncorrelated projection coefficients.  We can go further and show
that \KL\ provides the
optimal orthonormal basis for low-rank approximations of functions in
a random field.

Above, we expressed the completeness relation for a single function $f(t)$
(eqn.~\ref{eq:completeness}).
For a random process, an orthonormal basis
$\phi_k(t)$ is complete if and only if there
exists a random process $B_k$ such that
\begin{equation}
  \label{eq:completeness_rp}
  \lim_{N\to\infty} \int_{t_a}^{t_b}
  \left[F_t - \sum_{k=1}^{N}B_k \phi_k(t)\right]^2 \dd t = 0.
\end{equation}
A random process is {\it low-rank} if and only if there exists a complete
orthonormal
basis $\phi_k(t)$ such that $E[B_k^2] = 0$ for one or more values of $k$.
In other words, a random process is low-rank if for some basis $\phi_k(t)$,
some values of $k$ are not required for a perfect reconstruction of the
function.

Let us consider an arbitrary complete orthonormal basis
$\phi_k(t)$, with $F_t = \sum_{k=1}^\infty B_k\phi_k(t)$.
Given this basis, we'll define the low-rank approximation of $F_t$
\begin{equation}
  F^{(N)}_t = \sum_{k=1}^N B_k\phi_k(t)
\end{equation}
We'll seek to minimize the expectation value of the squared error
\begin{eqnarray}
  \mathcal{E}^2_N &=& \int_{t_a}^{t_b} \left[F_t - F_t^{(N)}\right]^2 \dd t
  \nonumber\\
  &=& \int_{t_a}^{t_b} \left[\sum_{k = N + 1}^\infty B_k \phi_k(t)\right]^2 \dd t.
\end{eqnarray}

Expanding the sum leads to
\begin{eqnarray}
  \mathcal{E}^2_N &=& \sum_{k=N + 1}^\infty
  \sum_{k^\prime=N + 1}^\infty \int_{t_a}^{t_b} 
  \left[B_k B_{k^\prime} \phi_k(t) \phi_{k^\prime}(t)\right] \dd t
  \nonumber\\
  &=& \sum_{k=N+1}^\infty \sum_{k^\prime=N+1}^\infty B_k B_{k^\prime}
  \delta_{kk^\prime}\nonumber\\
  &=& \sum_{k=N+1}^\infty B_k^2
\end{eqnarray}
Taking the expectation value and plugging in the equivalent of
Equation~\ref{eq:F_k_def} for $B_k$, we find
\begin{eqnarray}
  E[\mathcal{E}^2_N] &=&
  E\left[\sum_{k=N+1}^\infty \int_{t_a}^{t_b} \dd t \int_{t_a}^{t_b} \dd t^\prime
  F_tF_{t^\prime} \phi_m(t)\phi_m(t^\prime)\right] \nonumber\\
  &=& \sum_{k=N+1}^\infty \int_{t_a}^{t_b} \dd t \int_{t_a}^{t_b} \dd t^\prime
  C_F(t, t^\prime) \phi_m(t)\phi_m(t^\prime)
\end{eqnarray}

We'd like to minimize this expected error over the basis $\phi_m(t)$, subject
to the constraint that $\phi_m(t)$ are an orthonormal basis.  We'll accomplish
this by the method of Lagrange multipliers.  Our Lagrangian is
\begin{equation}
  \mathcal{L}(\{\phi(t)\}) = \sum_{k=N+1}^\infty
  \left[\int_{t_a}^{t_b} \dd t \int_{t_a}^{t_b} \dd t^\prime
  C_F(t, t^\prime) \phi_m(t)\phi_m(t^\prime)
  - \lambda_k\left(1 - \int_{t_a}^{t_b} \phi_m(t)^2 \dd t\right)
  \right]
\end{equation}
Minimizing this with respect to $\phi_k$ gives
\begin{equation}
  \frac{\partial\mathcal{L}}{\partial\phi_k(t)} = \int_{t_a}^{t_b} \dd t^\prime
  C_F(t, t^\prime) \phi_k(t^\prime)
  - \lambda_k \phi_k(t).
\end{equation}
The optimum for each $k$ is where this derivative equals zero; setting to
zero and solving recovers the original eigenvalue problem
 (Eq.~\ref{eq:eigfunc_def}) from which we
derived the KL basis $\{e_k(t)\}$.  By the uniqueness of the eigenvalue
decomposition, this shows that the KL basis is the optimal basis for low-rank
approximations of functions drawn from $F_t$.  Furthermore, for an
approximation using $N$ eigenvectors, the mean squared error is given by
\begin{eqnarray}
  E[\mathcal{E}^2_N] &=& \sum_{k=N+1}^\infty E[A_k^2]\nonumber\\
  &=& \sum_{k=N+1}^\infty \lambda_k
\end{eqnarray}
This is an interesting result: it says that in order to minimize the
expectation value of the reconstruction error $\mathcal{E}_N$ for all $N$,
we simply need to order the eigenvalues such that $\lambda_k \ge \lambda_{k+1}$
for all eigenvalue-eigenfunction pairs $(\lambda_k, e_k(t))$.

Because of this, throughout this work we will follow this convention when
ordering the eigenvalues in a KL decomposition.

\subsection{KL in the presence of noise}
\label{sec:whitening_chp2}
In practice, the observed random field $F_t$ is composed of the sum of
a signal $S_t$ and noise $N_t$.  We'll continue to assume that both of these
are centered.  The covariance matrix then becomes
\begin{equation}
  \mathcal{C}_F(t, t^\prime) = E[(S_t + N_t)(S_{t^\prime}+N_{t^\prime})]
\end{equation}
Under the assumption that the signal $S_t$ and noise $N_t$ are uncorrelated,
this can be simplified to
\begin{eqnarray}
  \mathcal{C}_F(t, t^\prime) &=& E[S_t S_{t^\prime}]
  + E[N_t N_{t^\prime}] \nonumber\\
  &=& \mathcal{C}_S(t, t^\prime) + \mathcal{C}_N(t, t^\prime)
\end{eqnarray}
The \KL\ eigenfunctions always diagonalize the full covariance
$\mathcal{C}_F(t, t^\prime)$.
In the case of uncorrelated ``white'' noise,
$\mathcal{C}_N(t, t^\prime) = \sigma^2 \delta(t - t^\prime)$
and both the signal and the noise become diagonalized.  In this case,
the noise per mode is a constant $\sigma^2$, and the ranking of the
eigenfunctions leads to modes which are ranked in signal-to-noise.
This is why KL modes are often referred to as ``signal-to-noise
eigenmodes'' \citep{Vogeley96}.

In cases when the noise is not white, we can still recover signal-to-noise
information by preprocessing with a whitening transformation.  The eigenvectors
$n_k(t)$ of the noise covariance, which satisfy
\begin{equation}
  \int_{t_a}^{t_b} \mathcal{C}_N(t, t^\prime) n_k(t^\prime) \dd t^\prime 
  = \sigma_i n_k(t)
\end{equation}
can be used to apply a whitening transform to both the signal and the
noise.  The whitened covariance is given by
\begin{eqnarray}
  \mathcal{C}_F^{(W)}(k, k^\prime) &\equiv& \int_{t_a}^{t_b}\dd t\int_{t_a}^{t_b}\dd t^\prime
  \mathcal{C}_F(t, t^\prime)
  \frac{n_k(t) n_{k^\prime}(t^\prime)}{\sigma_k\sigma_{k^\prime}}\nonumber\\
  &=& \mathcal{C}_S^{(W)}(k, k^\prime) + \delta_{kk^\prime}.
\end{eqnarray}
The signal is now expressed in the basis of the noise eigenmodes $n_k(t)$,
and the noise has been made white with unit variance.  The KL modes derived
from the whitened covariance $\mathcal{C}_F^{(W)}(k, k^\prime)$ will have
eigenvalues that represent the signal-to-noise in each mode.

\subsection{\KL: theory to practice}
\label{sec:KL_practice}
The abstract formalism presented above is interesting in itself, but one
might wonder what practical advantages can be gained from this discussion.
In practice, we don't deal with an abstract stochastic process $F_t$, but
with discrete, measured data.  For this reason, the continuous formalism
from above can be transformed into a discrete linear algebraic formalism,
as seen below.  In this section we will discuss the
practical computational aspects of KL analysis.

Imagine, for the moment, that an astronomer has observed the spectra of
$N$ galaxies.  After normalization and correction for redshift effects,
the spectra can be encoded as a series of $N$ real-valued functions
$f^{(i)}(\lambda)$ over some defined domain $\lambda \in [a, b]$.
In practice, we measure these spectra at a finite set of wavelengths
$\myvec{\lambda}^T = [\lambda_1, \lambda_2 \cdots \lambda_M]$ so that
our observations $\myvec{f}^{(i)}$ become $M$-dimensional vectors.  For
convenience, we'll store these spectra in a $N \times M$ matrix
$\mymat{F} = [\myvec{f}^{(1)}, \myvec{f}^{(2)} \cdots \myvec{f}^{(N)}]^T$,
where each row of the matrix represents one spectrum.
These series of spectra $\mymat{F}$ 
can be considered a finite realization of a particular
random process $F_\lambda$.

The expectation value $E[F_\lambda]$ can be approximated via the sample mean:
\begin{equation}
  E[F_\lambda] \approx \myvec{\bar{f}}
  = \frac{1}{N} \sum_{i=1}^N \myvec{f}^{(i)}
\end{equation}
and the covariance function can be approximated by the covariance matrix
\begin{equation}
  \label{eq:mat_corr_def}
  E[F_\lambda F_{\lambda^\prime}] \approx 
  \mymat{\mathcal{C}_F} = \frac{1}{N - 1} \mymat{\tilde{F}}^T \mymat{\tilde{F}}
\end{equation}
where we have defined the centered matrix
\begin{equation}
  \mymat{\tilde{F}} \equiv \mymat{F} - \myvec{1}_N\myvec{\bar{f}}^T.
\end{equation}
and $\myvec{1}_N$ is the length-$N$ vector of ones.  The $N-1$ in the
denominator of Equation~\ref{eq:mat_corr_def} is called the
{\it Bessel Correction}, and results from the reduced number of degrees of
freedom after the mean is subtracted.

We can approximate the eigenfunctions $e^{(i)}(\lambda)$ and
eigenvalues $\lambda_i$ via the diagonalization 
of $\mymat{\mathcal{C}_F}$, computed using standard linear algebra techniques.
The diagonalization of the covariance matrix is
\begin{equation}
  \label{eq:evd_def}
  \mymat{\mathcal{C}_F} = \mymat{V}\mymat{\Lambda}\mymat{V}^T,
\end{equation}
where the columns of the matrix $\mymat{V}$ are the eigenvectors
(such that $\mymat{V}^T \mymat{V} = \mymat{I}$, the identity matrix), and
$\mymat{\Lambda}$ is the diagonal matrix of eigenvalues, such that
$\Lambda_{ij} = \lambda_i\delta_{ij}$.  In practice, the eigenvalues and
eigenvectors can often be computed more efficiently via a singular value
decomposition:
\begin{equation}
  \label{eq:svd_def}
  \mymat{U}\mymat{\Sigma}\mymat{V}^T
  = \frac{1}{\sqrt{N - 1}}\mymat{\tilde{F}}
\end{equation}
where the orthogonal matrices $\mymat{U}$ and $\mymat{V}$ are called the
left and
right singular vectors, respectively, and $\mymat{\Sigma}$ is a diagonal matrix
of singular values.  One can quickly show from Equations~\ref{eq:mat_corr_def}
and \ref{eq:svd_def} that
\begin{eqnarray}
  \mymat{\mathcal{C}_F}
  &=& \mymat{V}\mymat{\Sigma}\mymat{U}^T
  \mymat{U}\mymat{\Sigma}\mymat{V}^T\nonumber\\
  &=& \mymat{V}\mymat{\Sigma}^2\mymat{V}^T.
\end{eqnarray}
Comparing to Equation~\ref{eq:evd_def}, we see that 
$\mymat{\Lambda} = \mymat{\Sigma}^2$, and the eigenvectors $\mymat{V}$ are
identical to the right singular vectors of Equation~\ref{eq:svd_def},
up to an arbitrary ordering of columns.  Ordering our eigenvalues
according to the rule in \S\ref{sec:partial_recons} takes care of
this uncertainty.

The columns of $\mymat{V}$ are the {\it eigenvectors}, and are the discrete
representation of the eigenfunctions $e^{(i)}(\lambda)$.  These eigenvectors
satisfy all the properties of the KL bases discussed above: they diagonalize
the sample correlation matrix, they provide the best possible low-rank
linear approximation to any spectrum from the sample, and in the presence of
uncorrelated noise, they allow an orthogonal decomposition onto a basis
with a natural ranking in signal-to-noise.

In the case of the spectrum example, we have no theoretical expectation for the
correlation matrix $\mathcal{C}(\lambda, \lambda^\prime)$, so we are forced
to approximate the matrix based on the sample correlation using
Equation~\ref{eq:mat_corr_def}.  If we had sufficient physical understanding
of every process at work in each galaxy, it might be possible to compute
that correlation matrix from theory alone.  The number of variables involved,
however, make this prospect near impossible.

There are other situations in astronomical measurement, however, when a
theoretical expectation of the correlation matrix is possible in practice.
We will see in the following sections how the correlation matrix of
particular cosmological observations can be approximated from theory alone.

\subsection{KL with missing data}
Along with discrete data samples, another challenge when applying KL to
real data is the presence of missing data: given a KL basis, how can
one derive the projected coefficients when data are missing?
Note that in this section we'll assume that the KL basis has been
obtained independently.  It is also possible to derive a KL eigenbasis
from incomplete data using an iterative approach:
see, e.g.~\citet{Connolly99, Yip04a}.

To begin, we'll assume that we have
an observed object represented by the $K$-dimensional vector 
$\myvec{x} = [x_1, x_2 \cdots x_K]^T$ and a set of normalized
KL basis functions 
$\mymat{V} = [\myvec{v_1}, \myvec{v_2} \cdots \myvec{v_K}]$.
arranged in order of decreasing eigenvalue $\lambda_i$.  We have shown above
that the best rank-$N$ linear approximation of $\myvec{x}$ is given by
\begin{equation}
  \myvec{x}^{(N)} = \sum_{i=1}^{N}a_i\myvec{v_i}.
\end{equation}
where the coefficients can be calculated as
\begin{equation}
  \label{eq:a_coeff_def}
  a_i = \myvec{v_i}^T\myvec{x}.
\end{equation}
When $\myvec{x}$ has missing data,
however, the question of how to compute $a_i$ is not as straightforward.
Simply setting the missing values to zero
will not work: the reconstruction will then faithfully recover those
zero values.  We desire instead to constrain the expected contribution
of each eigenvector to $x$ while {\it ignoring} the contribution of the
missing data.

A simple solution may be to simply truncate the vectors such that the dot
product is only computed over unmasked values.  It is easy to see that this
is identical to the  ``set to zero'' solution just
discussed.  Furthermore, there is the problem that in general, a set
of bases truncated in this way does not retain its orthogonality.

Another approach may be to derive a new basis which {\it is} orthogonal
over the truncated space.  This is similar in spirit to the method
explored by \citet{Gorski1994} in analysis of CMB data.
While this leads to a complete orthogonal basis for the observed portion
of the field, coefficients of these new modes have no simple relationship to
coefficients of modes covering the full field.  In general, a rank-deficient
transformation matrix must be inverted in order to convert between the two.

We'll use a different approach.  We have shown above that when noise is
not present, the KL vectors define the optimal basis for rank-$N$
reconstruction in the least-squares sense.  That is, for an arbitrary
truncated orthonormal basis $\mymat{\Phi}_{(N)}$, (where truncated means
we use only the first $N$ columns of $\mymat{\Phi}$, with $0 < N \le K$),
\begin{equation}
  \chi^2 = \left|\left(\myvec{x}
  - \mymat{\Phi}_{(N)}^T\myvec{a}^{(N)}\right)\right|^2
\end{equation}
is minimized on average when
$\mymat{\Phi} = \mymat{V}_{(N)}
= [\myvec{v_1}, \myvec{v_2} \cdots \myvec{v_N}]$, $N \le K$.
Here we flip the problem: we know the desired basis $\mymat{V}_{(N)}$,
and hope to find the optimal vector of coefficients $\myvec{a}^{(N)}$
which minimizes the $\chi^2$ in the presence of the missing data
\citep{Connolly99}.
We'll define a 
diagonal weight matrix $\mymat{W}$ such that $W_{ij} = w_i\delta_{ij}$,
where $w_i=1$ where $x_i$ is defined, and $w_i=0$ where $x_i$ is
missing.  Our expression to minimize then becomes
\begin{equation}
  \chi^2(\myvec{a}^{(N)}) = \left(\myvec{x}
  - \sum_{i=1}^N a^{(N)}_i \myvec{v^{(i)}}\right)^T
  \mymat{W}^2 \left(\myvec{x}
  - \sum_{i=1}^N a_{(N),i} \myvec{v^{(i)}}\right).
\end{equation}
To minimize this with respect to the coefficients $a_{(N),i}$, we differentiate
and find
\begin{equation}
  \frac{\partial\chi^2}{\partial a_{(N),i}} = -2 \myvec{x}^t\mymat{W}^2\myvec{v^{(i)}}
  + 2\sum_{j=1}^N a_{(N),j} \myvec{v^{(j)}}^T \mymat{W}^2 \mymat{v^{(i)}}.
\end{equation}
Setting the derivative to zero and combining terms gives
\begin{equation}
  \label{eq:masked_coeff}
  \myvec{x}^T\mymat{W}^2\myvec{v^{(i)}} =
  \sum_{j=1}^N a^{(N)}_j \myvec{v^{(j)}}^T \mymat{W}^2 \mymat{v^{(i)}}.
\end{equation}
If there are no areas of missing data, then $\mymat{W} = \mymat{I}$ and we
simply recover $a_i = \myvec{v_i}^T\myvec{x}$, our standard expression
for finding the KL coefficients with no missing data.
In the general case, however, because the inner-product of
$\myvec{v^{(i)}}$ and $\myvec{v^{(j)}}$ is modulated by $\mymat{W}$,
there is no delta function to collapse the sum.

We can simplify this notation by defining the correlation matrix of the
mask $\mymat{M}_N$, such that
\begin{equation}
  [\mymat{M}_N]_{ij} \equiv \myvec{v^{(j)}}^T \mymat{W}^2 \mymat{v^{(i)}}
\end{equation}
so that eq.~\ref{eq:masked_coeff} can be compactly written
\begin{equation}
  \myvec{x}^T \mymat{W}^2 \mymat{V}_{(N)}
  = {\myvec{a}^{(N)}}^T \mymat{M}_N.
\end{equation}
From this, we can quickly see that the optimal set of coefficients
$\myvec{a}^{(N)}$ is given by
\begin{equation}
  \label{eq:a_truncated}
  \myvec{a}^{(N)} = [\mymat{M}_N]^{-1}
  \mymat{V}_{(N)}^T \mymat{W}^2 \myvec{x}.
\end{equation}
This can be viewed as a generalized form of
the expression in eq.~\ref{eq:a_coeff_def}:
if $\mymat{W}$ is set equal to the identity matrix (indicating no
missing data), then $\mymat{M}_N$ is also the identity and we recover
eq.~\ref{eq:a_coeff_def} exactly.

If some of the diagonal entries in $\mymat{W}$ are zero, then the
correlation matrix $\mymat{M}_K$ for the full set of $K$ eigenvectors
is rank-deficient
and cannot be inverted as required for eq.~\ref{eq:a_truncated}.  For this
reason, it is essential to use the truncated eigenvectors $\mymat{V}_{(N)}$,
with $N \le {\rm rank}(\mymat{W})$.  It is not strictly necessary to
discard the eigenvectors corresponding to the smallest eigenvalues, but this
choice leads to the highest signal-to-noise result.
For the simple binary masking case
where $\mymat{W}$ is a diagonal matrix consisting of zeros and
ones, the rank of $\mymat{W}$ is equivalent to its trace, or the sum
of nonzero diagonal terms.

Once these approximate KL coefficients $\myvec{a}^{(N)}$ are determined,
it is straightforward to use these to approximate the unmasked vector
$\myvec{x}$:
\begin{equation}
  \myvec{x} \approx \sum_{i=0}^N a_i^{(N)} \myvec{v^{(i)}}.
\end{equation}
Here we have used the coefficients determined from the unmasked region of
the data to constrain the unobserved value in the masked regions.

This could be further generalized by allowing $\mymat{W}$ to be an arbitrary
matrix, for instance encoding the inverse of the noise covariance
associated with the observed vector $\myvec{x}$.  In unconstrained regions,
the noise is infinite and the inverse is zero.  This leads to very similar
results to those expressed here (in this case, $\mymat{W}^2$ is replaced
by $\mymat{W}^T\mymat{W}$).  This is equivalent to the whitening operation
discussed in \S\ref{sec:whitening_chp2}. We will not use this formalism
here, so we leave it only as a suggested extension.

\section{\KL\ Analysis and Bayesian Inference}
\label{sec:KL_bayes}
Because of the signal-to-noise optimality properties of KL, it can be
very useful within Bayesian parameter estimation.  
Given observations $D$ and prior information $I$, Bayes' theorem specifies the
posterior probability of a model described by the parameters $\{\theta_i\}$:
\begin{equation}
  \label{eq:bayes_chp2}
  P(\{\theta_i\}|DI) = P(\{\theta_i\}|I) \frac{P(D|\{\theta_i\}I)}{P(D|I)}
\end{equation}
The term on the left hand side
is the \textit{posterior} probability of the set of
model parameters $\{\theta_i\}$, which is the quantity we are interested in.

The first term on the right side of the equation is the \textit{prior}.
It quantifies how our prior
information $I$ affects the probabilities of the model parameters.  The 
prior is where information from other surveys (e.g. WMAP, etc) can be
included. The likelihood function for the observed coefficients $\myvec{a}$
enters into the numerator $P(D|\{\theta_i\}I)$.  The denominator $P(D|I)$
is essentially a normalization constant, set so that the sum of probabilities
over the parameter space equals unity.

KL is useful in the case where the model $\{\theta\}$ can be expressed in
terms of a covariance $\mymat{\mathcal{C}}_{\{\theta\}}$ \citep[see][]{Vogeley96}.
Given a data vector $\mymat{d}$ with observed noise covariance
$\mymat{\mathcal{N}}$, the KL vectors are the eigenvectors
$\mymat{V}_{\{\theta\}}$ of the whitened total covariance 
\begin{equation}
  \mymat{\mathcal{C}}_{\{\theta\}, W} =
  \mymat{\mathcal{N}}^{-1/2}
  \mymat{\mathcal{C}}_{\{\theta\}}
  \mymat{\mathcal{N}}^{-1/2}.
\end{equation}
These eigenvectors can be used to quickly compute the KL coefficients of
the observed data,
\begin{equation}
  \myvec{a}_{\{\theta\}} = \mymat{V}^\dagger_{\{\theta\}}
  \mymat{\mathcal{N}}^{-1/2}\myvec{d}.
\end{equation}
For a given model $\{\theta_i\}$, we can predict the expected
distribution of coefficients $\myvec{a}_{\{\theta_i\}}$:
\begin{eqnarray}
  \mymat{X}_{\{\theta_i\}}
  & \equiv & \langle\myvec{a}_{\{\theta_i\}}
  \myvec{a}_{\{\theta_i\}}^\dagger\rangle\nonumber\\
  &=& \mymat{V}_{\{\theta\}}^\dagger \mymat{\mathcal{N}}^{-1/2} 
  \mymat{\mathcal{C}}_{\{\theta_i\}}\mymat{\mathcal{N}}^{-1/2}\mymat{V}_{\{\theta\}}
  + \mymat{I}.
\end{eqnarray}
If the length of the data vector $\myvec{d}$ is $N$, then the full
analysis results in $\mymat{X}_{\{\theta_i\}}$ being an $N\times N$ matrix.
Alternatively, one can truncate the eigenvectors to $n < N$ terms:
this leads to $\mymat{X}_{\{\theta_i\}}$ being an $n\times n$ matrix, and
is equivalent to working with an optimal low-rank approximation of the
data $\myvec{d}$.
Using this, the measure of departure from the model $\{\theta\}$
is given by the quadratic form
\begin{equation}
  \chi_{\{\theta\}}^2 = \myvec{a}^\dagger\mymat{X}_{\{\theta_i\}}^{-1}\myvec{a}
\end{equation}
The likelihood is then given by
\begin{equation}
  \label{eq:likelihood_chp2}
  P(D|\{\theta\}I) =
  \mathcal{L}(\myvec{a}|\{\theta_i\}) = 
  (2\pi)^{n/2} |\det(X_{\{\theta_i\}})|^{-1/2}
  \exp(-\chi_{\{\theta\}}^2/2)
\end{equation}
where $n$ is the number of degrees of freedom: in most cases, $n = N$, the
number of eigenmodes included in the analysis.  The likelihood given by
Equation~\ref{eq:likelihood_chp2} enters into Equation~\ref{eq:bayes_chp2}
when computing the posterior probability.  This sort of approach will be
applied to observed shear in Chapter 5.

\section{\KL\ Analysis of Shear}
\label{sec:KL_shear}
In the case of shear observations, the observed vector $\myvec{\gamma}$
consists of the $N$ ellipticity observations within the series of window
functions $A_i(\myvec{\theta}, z)$, with $i = 1...N$.  In general, these
window functions can overlap, though in most cases this would lead to
correlated noise which makes the analysis more difficult.
The expected correlation matrix of the
observed shear $\myvec{\gamma}$ is given by
\begin{equation}
  [\mymat{\mathcal{C}_\gamma}]_{ij}
  = \int \dd^2\theta A_i(\myvec{\theta}, z)
  \int \dd^2\theta^\prime A_j(\myvec{\theta}^\prime, z)
  \xi_+(|\myvec{\theta} - \myvec{\theta}^\prime|) + \mymat{\mathcal{N}}_{ij}.
\end{equation}
where the matrix $\mymat{\mathcal{N}}_{ij}$ is the noise covariance between
bins.  The shear correlation function $\xi_+$ can be computed from the
theoretical 3D mass power spectrum, using the results from the previous
chapter (eqs.~\ref{eq:xi_plus}-\ref{eq:lens_weight}).

The noise matrix $\mymat{\mathcal{N}}$ can be estimated from the measurement
process: in the simplest case where noise is due to shot-noise only and
the windows $A_i$ are non-overlapping,
$\mymat{\mathcal{N}} \propto \sigma_i^2 \mymat{I}$
where $\sigma_i = \sigma_{int} / \sqrt{N_i}$ is the shot noise, given
the intrinsic ellipticity $\sigma_{int}$ and the number of galaxies $N_i$
in the bin described by $A_i$.

Once the theoretical correlation matrix is computed, the KL basis can be
determined using linear algebraic methods, and the basis functions can
be employed in any of the variety of ways suggested above.  In this work,
we will explore three applications of this procedure.  In Chapter 3, we
will use the properties of the theoretical covariance as a basis for
a filter for constructing 3D mass maps from weak lensing shear observations.
In Chapter 4, the KL modes will be used to correct for masking in a realistic
shear map, and the resulting maps will be analyzed using shear peak statistics.
In Chapter 5, the KL modes will be used for cosmological parameter estimation
using shear catalogs from the COSMOS weak lensing survey.  In all three cases,
the ability to theoretically compute the covariance matrix of the observations
leads to an analysis optimally tuned to the noise properties of the observed
signal.

%% file: chapter3.tex
\chapter{3D weak lensing maps with KL}

The material in this chapter is adapted from \citet{Vanderplas2011}.
In it, we present a new method for constructing three-dimensional
mass maps from gravitational lensing shear data.  We solve the lensing
inversion problem using truncation of singular values
(within the context of generalized
least squares estimation) without a priori assumptions about the
statistical nature of the signal.   This singular value framework 
allows a quantitative
comparison between different filtering methods: we evaluate our method
beside the previously explored Wiener filter approaches.
Our method yields near-optimal angular resolution of the lensing
reconstruction and allows cluster sized halos to be de-blended robustly.
It allows for mass reconstructions which are
2-3 orders-of-magnitude faster than the Wiener
filter approach; in particular, we estimate that an all-sky
reconstruction with arcminute resolution could be performed
on a time-scale of hours on a current workstation.
We find however that linear, 
non-parametric reconstructions have a fundamental limitation in the
resolution achieved in the redshift direction.

This chapter was originally published in collaboration with Andrew Connolly,
Bhuvnesh Jain, and Mike Jarvis in the February 2011 edition of the
Astronomical Journal \citep[][ApJ, Vol. 727, p. 118; \copyright ~2011 by
the American Astronomical Society]{Vanderplas2011} and is reproduced below
with permission of the American Astronomical Society.

\section{Introduction}

\citet{Taylor01}, \citet[][hereafter HK02]{Hu02} and 
\citet{Bacon03} first looked at
non-parametric 3D mapping of a gravitational potential.  HK02
presented a linear-algebraic method for \textit{tomographic mapping}
of the matter distribution -- splitting the sources and lenses into
discrete planes in redshift.  They found that the inversion along each
line-of-sight is ill-conditioned, and requires regularization through 
\textit{Wiener filtering}.  Wiener filtering reduces reconstruction noise
by using the expected statistical properties of the signal as a prior: 
for the present problem, this prior is the nonlinear mass power spectrum.  
\citet[][hereafter STH09]{Simon09}
made important advances to this method by constructing an efficient
framework in which the inversions for every line-of-sight are computed
simultaneously, allowing for greater flexibility in the
type of filter used.  They introduced two types of Wiener filters: 
a ``radial Wiener filter'', based on the HK02 method, 
and a ``transverse Wiener filter'', 
based on the Limber approximation to
the 3D mass power spectrum.
They showed that the use of a generalized form of either
filter leads to a biased result -- the filtered reconstruction of the
line-of-sight matter distribution for a localized lensing mass is both
shifted and spread-out in redshift.

One issue with the Wiener filter approach is the assumption of
Gaussian statistics in the reconstructed signal.  In reality, the matter 
distribution at relevant scales can be highly non-Gaussian.  
It is possible that the redshift bias found in STH09 is not inherent
to nonparametric linear mapping, but rather a result of this deficiency
in the Wiener filtering method.

In this chapter, we develop an alternate noise-suppression scheme for
tomographic mapping that,  unlike Wiener filtering, has no dependence
on assumptions about the signal.   Our goal is to explore improvements
in the reconstruction and examine, in particular, the recovery of
redshift information using the different methods. We begin in 
Section~\ref{Method} 
by discussing the tomographic weak-lensing model developed by HK02
and STH09 and presenting our estimator for the density parameter, $\delta$.  
In Sections~\ref{Results} and \ref{Conclusions} we implement this method
for a simple case, and compare the results with those of the STH09 transverse
and radial Wiener filters.

\section{Method}
\label{Method}

For tomographic weak lensing, we are concerned with three quantities:  the 
complex-valued shear $\gamma(\vec\theta,z)$, the real-valued convergence
$\kappa(\vec\theta,z)$, and the dimensionless density parameter 
$\delta(\vec\theta,z)$.  As discussed in \S\ref{sec:weak_lensing_intro},
the relationship between $\gamma$ and $\kappa$ is given
by a convolution over all angles $\vec\theta$, and the density $\delta$ is
related to $\kappa$ by a line-of-sight integral over the lensing efficiency
function, $W(z,z_s)$.  The key observation is that in the weak lensing regime,
each of these operations is linear: if the variables are discretized, 
they become systems of linear equations, 
which can in principle be solved using standard matrix methods.

\subsection{Linear Mapping}
\label{LinearMapping}

Details of the weak lensing formalism are covered in
\S\ref{sec:lens_mapping_intro}.  Here we'll briefly review the
most relevant pieces.
To compute 3D mass maps with weak lensing, we begin by
creating a common pixel binning of the observed field of 
$N_x$ by $N_y$ equally sized square pixels
of angular width $\Delta\theta_x = \Delta\theta_y$.  
Within each of the $N_x N_y \equiv N_{xy}$ individual
lines of sight, we bin $\gamma$ into $N_s$ source-planes,
and bin $\delta$ into $N_l$ lens-planes, $N_l \le N_s$.  
Thus we have two 1D data vectors, which are concatenations of
the line-of-sight vectors within each pixel:
 $\myvec\gamma$, of length $N_{xy} N_s$; 
and $\myvec\delta$, of length $N_{xy} N_l$.  
(Note that throughout this section, boldface denotes a vector quantity.)  
As a result of this binning, we can write the discretized lensing 
equations in a particularly simple form:
\begin{equation}
  \label{M_gd}
  \myvec\gamma = M_{\gamma\delta}\myvec\delta + \myvec{n_\gamma}
\end{equation}
where $\myvec\gamma$ is the vector of binned shear observations with noise
given by $\myvec{n_\gamma}$, and $\myvec\delta$ is the vector 
of binned density parameter.  For details on the form of 
the matrix $M_{\gamma\delta}$, refer to Appendix~\ref{appB}.

The linear estimator $\myvec{\hat{\delta}}$ of the signal is found by 
minimizing the quantity
\begin{equation}
  \chi^2 = (\myvec\gamma-\mymat{M_{\gamma\delta}}\myvec{\delta})^\dagger 
  \mymat{\mathcal{N}_{\gamma\gamma}}^{-1} (\myvec\gamma-\mymat{M_{\gamma\delta}}\myvec{\delta})
\end{equation}
where $\dagger$ indicates the conjugate transpose, and 
$\mymat{\mathcal{N}_{\gamma\gamma}} \equiv 
\langle \myvec{n_\gamma}\myvec{n_\gamma}^\dagger\rangle$
is the noise covariance of the measurement $\myvec{\gamma}$,
and we assume $\langle \myvec{n_\gamma}\rangle = 0$.  The best linear
unbiased estimator for this case is due to \citet{Aitken34}: 
\begin{equation}
  \label{Aitken_estimator}
  \myvec{\hat{\delta}_A} \equiv
  \left[\mymat{M_{\gamma\delta}}^\dagger
    \mymat{\mathcal{N}_{\gamma\gamma}}^{-1}\mymat{M_{\gamma\delta}}\right]^{-1}
  \mymat{M_{\gamma\delta}}^\dagger
  \mymat{\mathcal{N}_{\gamma\gamma}}^{-1}\myvec{\gamma}
\end{equation}
The noise properties of this estimator can be made clear by
defining the matrix
$\mymat{\widetilde{M_{\gamma\delta}}}\equiv 
\mymat{\mathcal{N}_{\gamma\gamma}}^{-1/2}\mymat{M_{\gamma\delta}}$ and
computing the singular value decomposition (SVD)
$\mymat{\widetilde{M_{\gamma\delta}}} 
\equiv \mymat{U}\mymat{\Sigma}\mymat{V}^\dagger$.
Here $\mymat{U}^\dagger \mymat{U} = \mymat{V}^\dagger \mymat{V} = \mymat{I}$ 
and $\Sigma$ is the square diagonal matrix of singular values 
$\sigma_i\equiv\Sigma_{ii}$, ordered such that $\sigma_i\ge\sigma_{i+1}$, 
$i\ge 1$.  Using these properties, the Aitken estimator can be equivalently 
written
\begin{equation}
  \label{Aitken_SVD}
  \myvec{\hat{\delta}_A} = \mymat{V} \mymat{\Sigma}^{-1} 
  \mymat{U}^\dagger \mymat{\mathcal{N}_{\gamma\gamma}}^{-1/2} \myvec{\gamma}
\end{equation}
It is apparent in this expression that the presence of small singular values 
$\sigma_i \ll \sigma_1$ can lead to extremely large diagonal entries 
in the matrix $\mymat{\Sigma^{-1}}$, which in turn amplify the errors in the
estimator $\myvec{\hat{\delta}_A}$. 
This can be seen formally by expressing the noise covariance
in terms of the components of the SVD:
\begin{equation}
  \label{Ndd_decomp}
  \mymat{\mathcal{N}_{\delta\delta}} = 
  \mymat{V}\mymat{\Sigma}^{-2}\mymat{V}^\dagger.
\end{equation}
This makes clear the connection with KL, as discussed in Chapter 2.
The columns of the matrix $\mymat{V}$ are eigenvectors of
$\mymat{\mathcal{N}_{\delta\delta}}$, with eigenvalues $\sigma_i^{-2}$.
When many small singular values are present, the noise will dominate
the reconstruction, and it is necessary to 
use a more sophisticated estimator to recover the signal.

\subsection{KL Filtering}
\label{sing_val_formalism}
One strategy that can be used to reduce this noise is to add a penalty function
to the $\chi^2$ that will suppress the large spikes in signal.  This
is the Wiener filter approach explored by HK02 and STH09.
A more direct noise-reduction method, which does not require knowledge 
of the statistical properties of the signal, involves approximating 
the SVD in Equation~\ref{Aitken_SVD} to remove the contribution of the 
high-noise modes. We choose a cutoff value
$\sigma_{\rm cut}$, and determine $n$ such that 
$\sigma_n > \sigma_{\rm cut} \ge \sigma_{n+1}$.
We then define the truncated matrices 
$\mymat{U_n}$, $\mymat{\Sigma_n}$, and $\mymat{V_n}$,
such that $\mymat{U_n}$ ($\mymat{V_n}$) contains the first $n$ columns of 
$\mymat{U}$ ($\mymat{V}$),
and $\mymat{\Sigma}_n$ is a diagonal matrix of the largest $n$ singular 
values, $n \le n_{\rm max}$.
To the extent that $\sigma_{\rm cut}^2 \ll \sum_{i=1}^n\sigma_i^2$, 
the truncated matrices satisfy
\begin{equation}
  \mymat{U_n}\mymat{\Sigma_n}\mymat{V_n}^\dagger \approx 
  \mymat{U}\mymat{\Sigma}\mymat{V}^\dagger = \mymat{\widetilde{M_{\gamma\delta}}}
\end{equation}
and the signal estimator in Equation~\ref{Aitken_estimator} can be approximated
by the SVD estimator:
\begin{equation}
  \label{SVD_estimator}
  \myvec{\hat{\delta}}_{\rm svd}(n) \equiv 
  \mymat{V}_n \mymat{\Sigma}_n^{-1} 
  \mymat{U}_n^\dagger \mymat{\mathcal{N}_{\gamma\gamma}}^{-1/2} \myvec{\gamma}
\end{equation}
This approximation is optimal in the sense that it preferentially
eliminates high-noise orthogonal components in $\myvec{\delta}$ 
(cf.\ equation~\ref{Ndd_decomp}), leading to an estimator which is much
more robust to noise in $\myvec{\gamma}$.

SVDs are often used in the context of KL or Principal Component Analysis, 
where the square of the singular value is equal to the variance described 
by the corresponding principal component (see
\S\ref{sec:KL_practice}).  The variance can be
thought of, roughly, as a measure of the information contributed by the
vector to the matrix in question.  It will be useful for
us to think about SVD truncation in this way.  To that end, 
we define a measure of the truncated variance for a given value of $n$:
\begin{equation}
  \label{v_cut}
  v_{\rm cut}(n) = 1 - \frac{\sum_{i=1}^n\sigma_i^2}
  {\sum_{i=1}^{n_{\rm max}}\sigma_i^2}
\end{equation}
such that $0\le v_{\rm cut}\le 1$.  If $v_{\rm cut} = 0$, then $n = n_{\rm max}$ and we are 
using the full Aitken estimator.  As $v_{\rm cut} \rightarrow 1$, we are increasing the amount
of truncation.

In practice, taking the SVD of the transformation matrix 
$\mathcal{N}_{\gamma\gamma}^{-1/2}M_{\gamma\delta}$
is not entirely straightforward: the matrix is of size 
$(N_{xy}N_s) \times (N_{xy}N_l)$.
With a $128 \times 128$-pixel field, 20 lens-planes, and 25 source-planes,
the matrix contains $1.3\times 10^{11}$ mostly nonzero complex entries, 
amounting to 2TB in memory (double precision).  
Computing the SVD for a non-sparse matrix of this size is far from trivial.

We have developed a technique to speed-up this process, which involves decomposing
the matrices $\mathcal{N}_{\gamma\gamma}$ and $M_{\gamma\delta}$ into tensor
products, so that the full SVD can be determined through computing SVDs of
two smaller matrices: an $N_s\times N_l$ matrix, 
and an $N_{xy}\times N_{xy}$ matrix.
The second of these individual SVDs can be approximated using the
Fourier-space properties of the $\gamma\to\kappa$ mapping.  The result is
that the entire SVD estimator can be computed very quickly.  The details
of this method are described in Appendix~\ref{appB}.

\begin{figure*}[t] 
 \centering
 \includegraphics[width=0.8\textwidth]{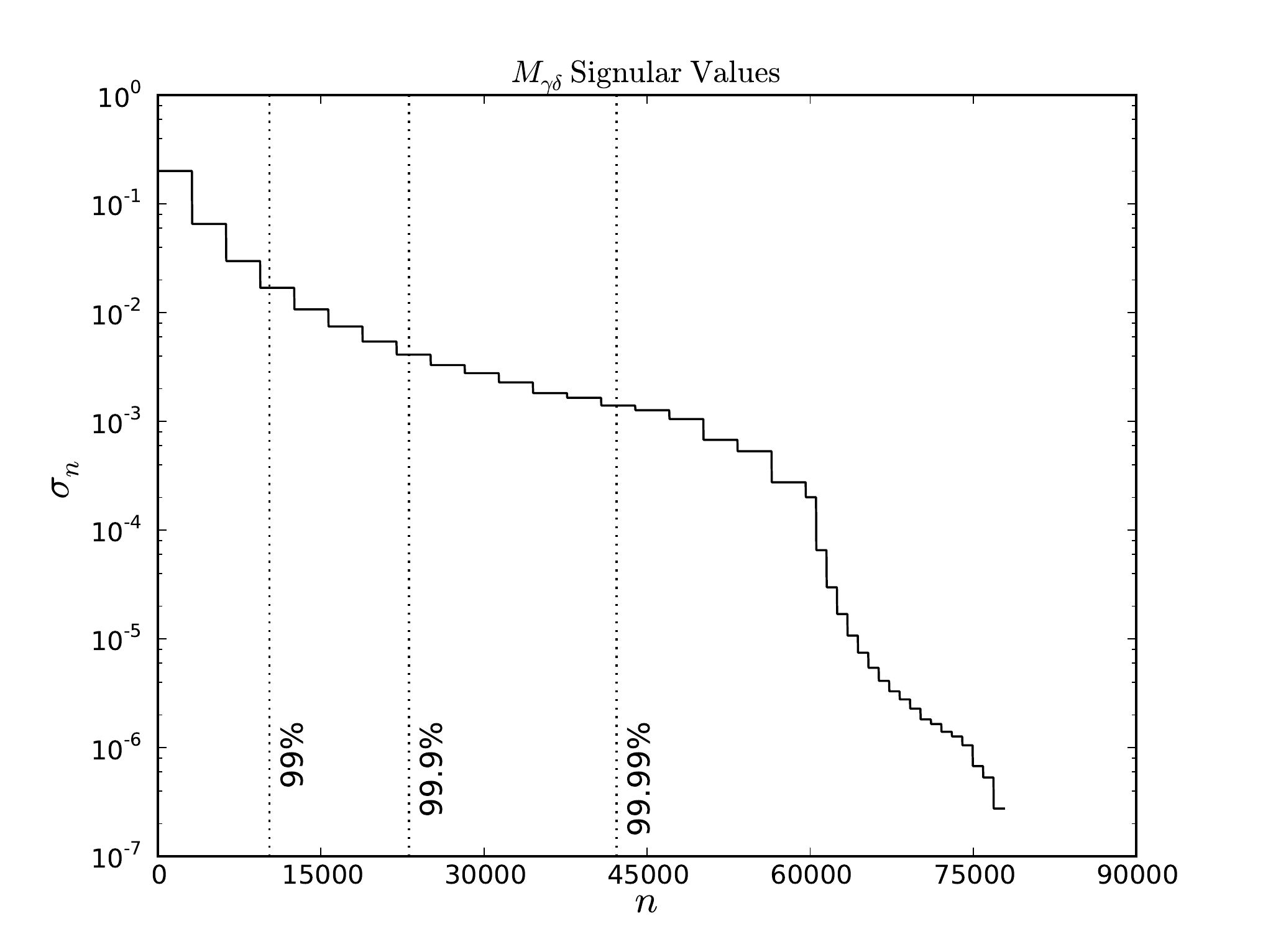}
 \caption[Ordered singular values]{
   Ordered singular values of the matrix
   $\mymat{\widetilde{M_{\gamma\delta}}}$.
   The dotted lines show the values of $n$ 
   such that 99\%, 99.9\%, and 99.99\% of the variance is preserved.
   The sharp drop-off near $n=60,000$ is due to the $10^{-3}$ 
   deweighting of border pixels.
   \label{fig_sing_vals}}
\end{figure*}

\section{Results}
\label{Results}
\label{Parameters}
Using the above formalism, we can now explore the
tomographic weak lensing problem using the 
techniques of Section~\ref{Method}.
For the following discussion, we will use a field of approximately 
one square degree: a $64 \times 64$ grid of
$1^\prime \times 1^\prime$ pixels, with 25 source redshift
bins ($0\le z\le 2.0$, $\Delta z = 0.08$) and 20 lens redshift bins
($0\le z\le 2.0$, $\Delta z = 0.1$).  This binning approximates the 
expected photometric redshift errors of future surveys.
We suppress edge effects by increasing
the noise of all pixels within $4^\prime$ of the field border
by a factor of $10^3$, effectively
deweighting the signal in these pixels (cf.\ STH09).  The noise
for each redshift bin is set to $n_i = \sigma_\gamma/\sqrt{N_i}$, where 
$\sigma_\gamma$ is the intrinsic ellipticity dispersion, and
$N_i$ is the number of galaxies in the bin.  We assume $\sigma_\gamma = 0.3$
\citep[based on the Hubble Deep Field image][]{Mellier99},
and 70 galaxies per square arcminute, with a redshift distribution given by
\begin{equation}
  \label{gal_z_dist}
  n(z) \propto z^2\exp{\left[-(z/z_0)^{3/2}\right]},
\end{equation}
with $z_0 = 0.57$.  We assume a flat cosmology with 
$h=0.7$, $\Omega_M = 0.27$  and $\Omega_\Lambda = 0.73$ at the present day.

\subsection{Singular Values}
The singular values of the transformation matrix for this configuration 
are depicted in Figure~\ref{fig_sing_vals}.  The step pattern visible in
this plot is due to the fact that the noise across 
each source plane is identical,
aside from the $4^\prime$ deweighted border.  
It is apparent from this figure that the large
majority of the singular values are very small: 99.9\% of the variance
in the transformation is contained in less than $1/3$
of the singular values.  The large number of very small
singular values will, therefore, dominate in the Aitken estimator 
(Equation~\ref{Aitken_SVD}), leading to the very noisy unfiltered 
results seen in HK02.  

\subsection{Evaluation of the SVD Estimator}
\label{SVD_Evaluation}
To evaluate the performance of the SVD filter, we first create a
field-of-view containing a single halo at redshift $z = 0.6$.
One well-supported parametrization of halo shapes is the
NFW profile \citep{NFW97}.  We use the analytic form of the
shear and projected density due to an NFW profile, given by equations
13-18 in
\citet{Takada03}.

\begin{figure*}[p]
 \centering
 \includegraphics[width=0.8\textwidth]{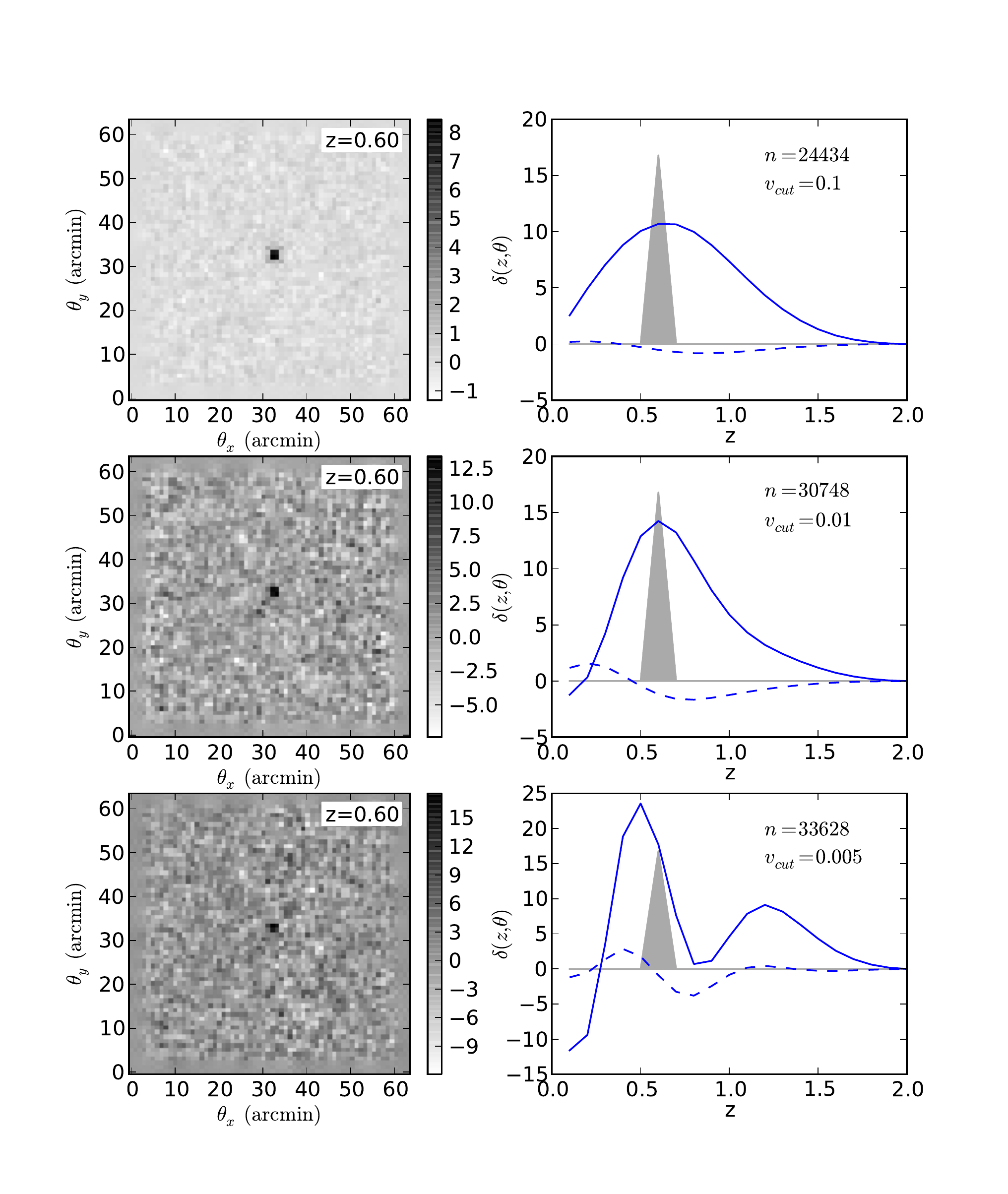}
 \caption[The effect of SVD truncation on a single $z=0.6$ NFW halo]{
   The effect of SVD truncation on a single $z=0.6$ NFW halo
   in the center of the field, for three different levels of filtering.
   \textit{left column:}
   reconstructed density parameter $\delta(\theta)$ in the $z=0.6$ lens-plane.
   The true matter distribution is represented by a tight ``dot'' in the 
   center of the plot.
   \textit{right column:}
   line-of-sight profile at the central pixel.  The gray shaded area 
   shows the input density parameter.  The solid line shows the E-mode signal, 
   while the dashed line shows the B-mode signal.
   $n$ gives the number of singular values
   used in the reconstruction (out of a total $n_{\rm max}=81920$), 
   and $v_{\rm cut}$ gives the amount of variance cut by the truncation 
   (Equation~\ref{v_cut}); the level of filtration decreases from the top
   panels to the bottom panels.  
   The bottom panels show a case of under-filtering:
   for small enough $v_{\rm cut}$, the noise overwhelms the signal and creates
   spurious peaks along the line-of-sight.
   \label{fig_los_plot}} 
\end{figure*}

\begin{figure*}[t]
 \centering
 \includegraphics[width=\textwidth]{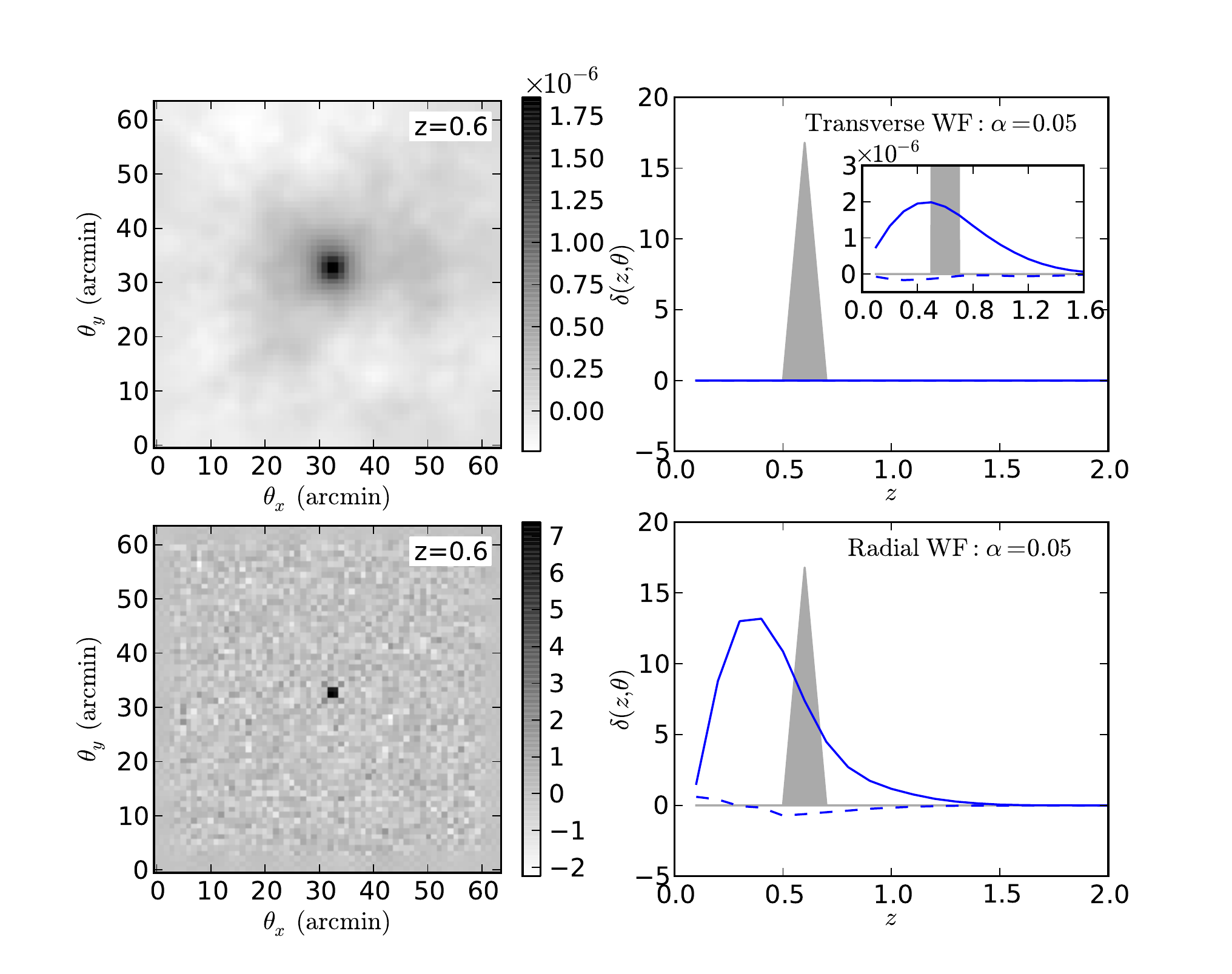}
 \caption[The effect of Wiener filtering]{
   The effect of Wiener filtering on the same input as 
   Figure~\ref{fig_los_plot}. Here we have used both transverse 
   \textit{(top panels)}
   and radial \textit{(bottom panels)} Wiener filtering, both down-tuned 
   by $\alpha = 0.05$ (the value recommended by STH09).  
   The transverse Wiener filter suppresses the response by several 
   orders of magnitude; a closer view of the line-of-sight peak is shown 
   in the inset plot.  The radial Wiener filter gives similar angular 
   results to the SVD filter, but takes much longer to compute.
   \label{fig_los_plot_ST}} 
\end{figure*}

We reconstruct the density map using the SVD filter 
(Figure~\ref{fig_los_plot}) with the above survey parameters.  
We show the results for three different values of $v_{\rm cut}$: 
0.1, 0.01, and 0.005.  In all three cases, the halo is easily 
detected at its correct location (left panels),
although as $v_{\rm cut}$ decreases, there is more noise in the 
surrounding field.  The right panels show the computed density profile
along the line of sight for the central pixel. 
The peak of this curve is close to the correct redshift, 
but there is a significant spread in redshift, as well as a bias.
As the level of SVD filtering (measured by $v_{\rm cut}$) decreases, 
the magnitude of these effects decreases, but the increased noise 
leads to spurious peaks. 

Similar plots for the transverse Wiener filter recommended by STH09 are
shown in the upper panels of Figure~\ref{fig_los_plot_ST}, using their 
recommended value of $\alpha = 0.05$. 
The response shows a significant spread in angular space, and 
the signal is seen to be suppressed by six orders-of-magnitude along with
a similar suppression of the noise. 
These effects worsen, 
in general, as the filtering level $\alpha$ increases.
Mathematically it is apparent why the transverse filter performs so poorly:
the small singular values primarily come from the line-of-sight 
part of the mapping, and the this filter has no effect along the line-of-sight.

The effect of the radial Wiener filter is shown in the bottom panels of 
Figure~\ref{fig_los_plot_ST}.
It shares the positive aspects of the SVD filter, having very 
little signal suppression or angular spread.
However, this filter uses some priors on the statistical form of 
the signal that are not as physically well-motivated as those for the 
transverse Wiener filter.  In contrast, the SVD filter does not make 
any prior assumptions about the signal. In this way, the SVD reconstruction
can be thought of as even more non-parametric than the Wiener filter
reconstructions.

\subsection{Comparison of Estimators}
\label{Comparison}

\begin{figure*}[t]
 \centering
 \includegraphics[width=0.8\textwidth]{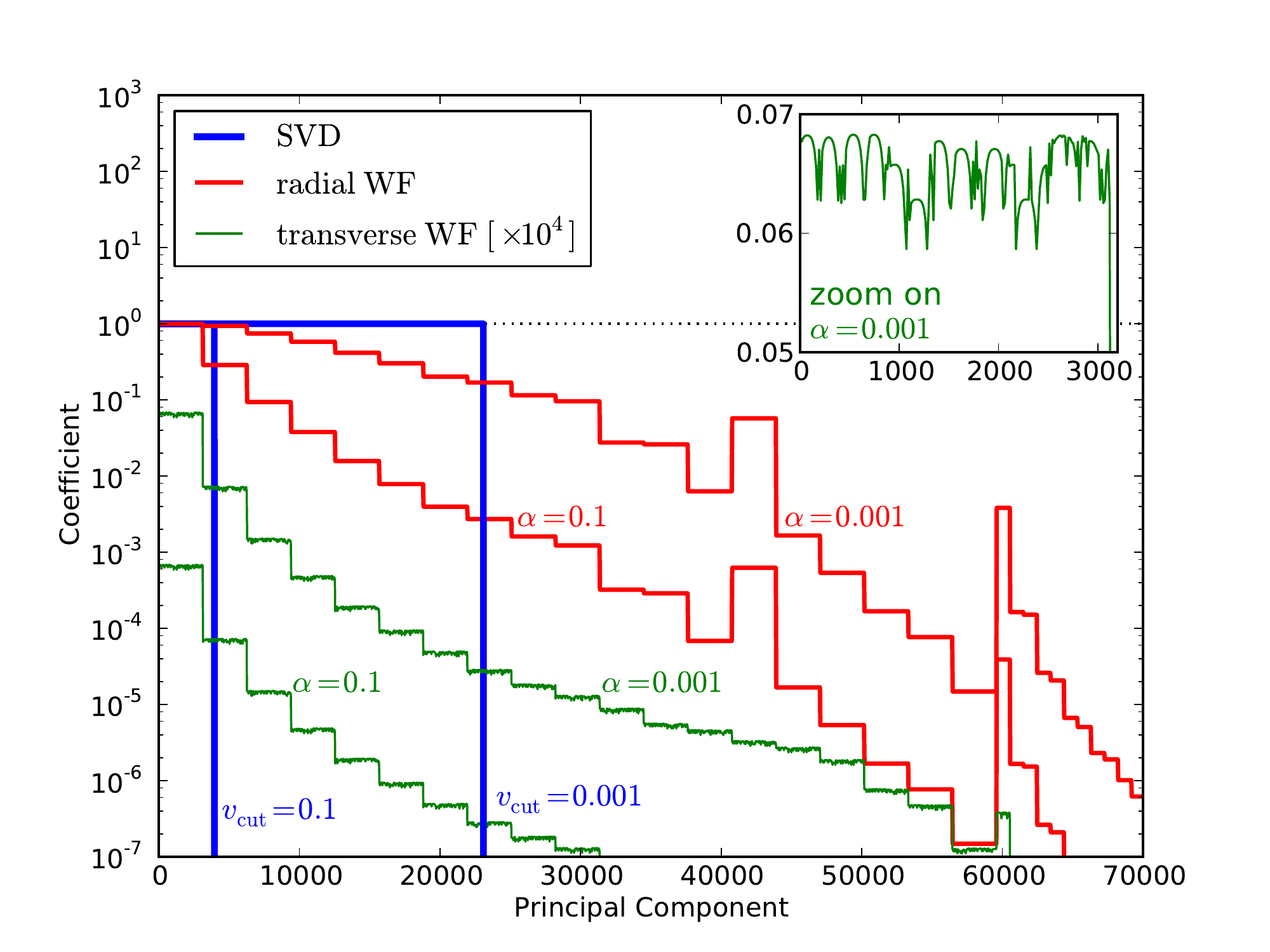} 
 \caption[Contribution of each shear mode to the reconstruction]
 {Contribution of each shear mode to the reconstruction
 for three different filters.  The dotted line at $10^0$ represents the
 unfiltered result.  Each filtering method leads to a different
 weighting of the shear modes.  The SVD filter, by design, completely 
 removes higher-order modes beyond a given cutoff, while the Wiener 
 filter deweights modes in a more gradual fashion.
 Note that the transverse Wiener filter deweights
 all modes by up to seven orders of magnitude; it has been scaled by
 a factor of $10^4$ for this plot.  The inset plot shows a
 closeup of the fluctuations within each ``step'' of the transverse
 filter.  These fluctuations lead to angular spread in the response
 (see discussion in Section~\ref{Comparison})
 \label{fig_mode_comparison} }
\end{figure*}

The SVD framework laid out in Section~\ref{sing_val_formalism} can be used
to quantitatively compare the behavior of different estimators.  A general
linear estimator has the form
\begin{equation}
  \myvec{\hat\delta_R} = \mymat{R}\myvec\gamma
\end{equation}
for some matrix $\mymat{R}$.
This general estimator can be expressed in terms of the components of
the unbiased estimator (Equation~\ref{Aitken_SVD}):
\begin{equation}
  \mymat{R} = \mymat{V_R} \mymat{\Sigma}^{-1} 
  \mymat{U}^\dagger \mymat{\mathcal{N}_{\gamma\gamma}}^{-1/2}.
\end{equation}
Here the matrices $\mymat\Sigma$, $\mymat{U}$ and 
$\mymat{\mathcal{N}_{\gamma\gamma}}$ are defined as in 
Equation~\ref{Aitken_SVD}, and we have defined the matrix
\begin{equation}
  \mymat{V_R} \equiv \mymat{R}\mymat{\mathcal{N}_{\gamma\gamma}}^{1/2}
  \mymat{U}\mymat{\Sigma}
\end{equation}
The rows of the matrix $\mymat{\Sigma}^{-1} \mymat{U}^\dagger 
\mymat{\mathcal{N}_{\gamma\gamma}}^{-1/2}$ provide a convenient basis
in which to work: they are the weighted principal components of the shear, 
ordered with decreasing signal to noise.  The norm of
the $i^{\rm th}$ column of $\mymat{V_R}$ measures the contribution of
the $i^{\rm th}$ mode to the reconstruction of $\delta$.  For the unfiltered
estimator, $\mymat{V_R}=\mymat{V}$ and all the norms are unity.
This leads to a very intuitive comparison between different filtering schemes.
Figure~\ref{fig_mode_comparison} compares the column-norms of $V_R$ 
for the SVD filter with those of the radial and
transverse Wiener filters.
  
The steps visible in the plot originate the same way as the steps in
Figure~\ref{fig_sing_vals}: the flatness of each step comes from the 
assumption of uniform noise in each source plane.  This plot shows 
the tradeoff between noise and bias.  The flat line at norm=$10^0$ 
represents a noisy but unbiased estimator.  Any departure
from this will impose a bias, but can increase signal-to-noise.
There are two important observations from this figure.
First, because each step on the plot is relatively flat for
the SVD filter and radial Wiener filter, we don't expect much bias 
\textit{within} each lens plane.  
The transverse filter, on the other hand, has fluctuations
at the $10\%$ level within each step (visible in the inset of 
Figure~\ref{fig_mode_comparison}), which will lead to a noticeable bias
within each lens plane, resulting in the degraded angular resolution
of the reconstruction seen in Figure~\ref{fig_los_plot_ST}.
Second, the transverse Wiener filter deweights even the highest 
signal-to-noise modes by many orders of magnitude, resulting 
in the signal suppression seen in Figure~\ref{fig_los_plot_ST}.
The SVD filter and radial Wiener filter, on the other hand,
have weights near unity for the highest signal-to-noise modes.
These two observations show why the SVD filter and radial Wiener filter 
are the more successful noise reduction techniques for the present problem.

\subsection{Noise Properties of Line-of-Sight Modes}
\label{SN_modes}
As seen in equation \ref{Ndd_decomp}, the columns of $\mymat{V}$ provide
a natural orthogonal basis in which to express the signal $\myvec{\delta}$.  
It should be emphasized that this eigenbasis is 
valid for any linear filtering scheme: the untruncated
SVD is simply an equivalent re-expression of the original transformation.
Examining the characteristics of these eigenmodes can yield insight
regardless of the filtering method used.

\begin{figure*}[t]
 \centering
 \includegraphics[width=0.45\textwidth]{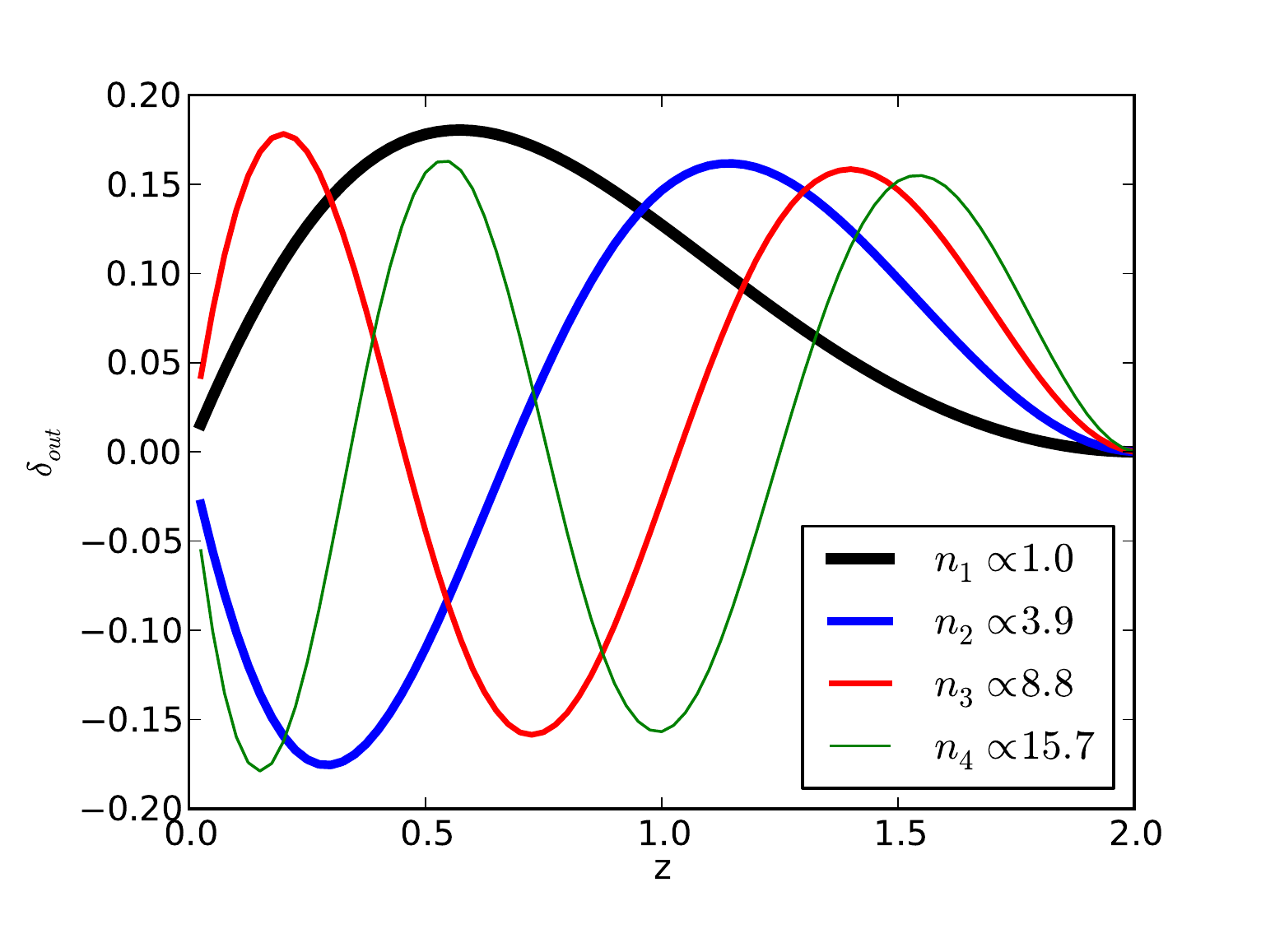}
 \includegraphics[width=0.45\textwidth]{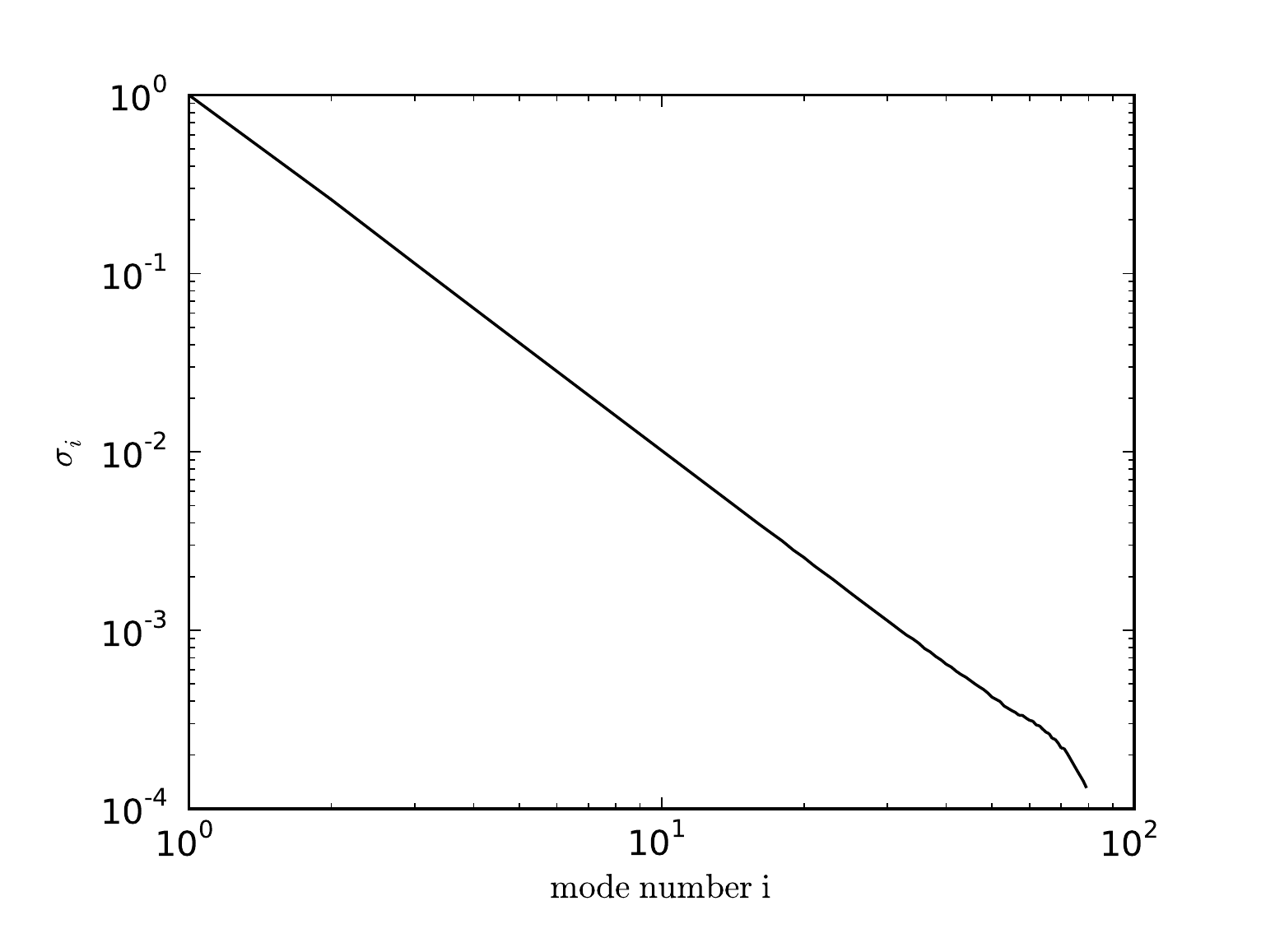}  
 \caption[The radial components and singular values of the filtering]
 {\textit{left panel:} 
   The radial components of the first four
   columns of the matrix $\mymat{V}$ (see section~\ref{LinearMapping}).
   This is calculated for 100 equally spaced redshift bins $(0\le z\le 2.5)$ 
   in $\myvec{\gamma}$, and 80 bins $(0\le z\le 2.0)$ in $\myvec{\delta}$
   These orthogonal eigenmodes are analogous to radial Fourier modes.  Each
   is labeled by its relative noise level, $n_i = (\sigma_i/\sigma_1)^{-1}$.
   \textit{right panel:}
   The singular values $\sigma_i$ associated with the 80 radial eigenmodes.
 \label{fig_radial_modes} }
\end{figure*} 

The radial components of the first four eigenmodes are plotted in 
figure~\ref{fig_radial_modes}.  Each is labeled by its normalized 
noise level, $n_i \equiv (\sigma_i/\sigma_1)^{-1}$.
The total number of modes will be equal to the number of output redshift 
bins; here, for clarity, we've used 80 equally-spaced bins out to redshift 2.0.
As the resolution is lessened, the overall shape and relative noise level of 
the lower-order modes is maintained.
These radial modes are analogous to angular Fourier modes,
and are related to the signal-to-noise KL modes discussed in HK02.
It is clear from this plot that any linear, non-parametric
estimator will be fundamentally limited in its redshift resolution: 
the noise level of the $i^{th}$ mode approximately scales as
\begin{equation}
  n_i\ \widetilde{\propto}\ i^2
\end{equation}
The signal-to-noise level for any particular halo will depend on
its mass and redshift. The magnitude of the signal
scales linearly with mass (see discussion in STH09), 
but the redshift dependence is more complicated: it is
affected by the lensing efficiency function, which depends on the redshift
of the lensed galaxies.  Using the above survey parameters, with an NFW halo 
of mass $M_{200} = 10^{15} \rm{M_\odot}$ and redshift $z=0.6$, 
the signal-to-noise ratio 
of the central pixel for the fundamental radial mode is $\sim 5.9$,
consistent with the results for Wiener filtered reconstructions of
singular isothermal halos explored in STH09.
This means that for even the largest halos, with a very deep survey,
only the first few modes will contribute significantly 
to the reconstructed halo.  Adding higher-order modes can in theory 
provide redshift information, but at the cost of increasingly high 
noise contamination. This is a general result which will apply to all 
nonparametric linear reconstruction algorithms.

This lack of information in the redshift direction leads directly 
to an inability to accurately determine halo masses:
the lensing equations relate observed shear $\gamma$ 
to density parameter $\delta$, which is related to mass in a
redshift-dependent way.  This is a fundamental limitation on the
ability of linear nonparametric methods to determine halo masses
from shear data. Indeed, even moving to fully parametric models, 
line-of-sight effects can lead to halo mass errors of 
20\% or more \citep{Hoekstra03, dePutter05}.

\subsection{Reconstruction of a Realistic Field}
\label{Realistic_Field}

\begin{figure*}%[t] 
 \centering
 \includegraphics[width=\textwidth]{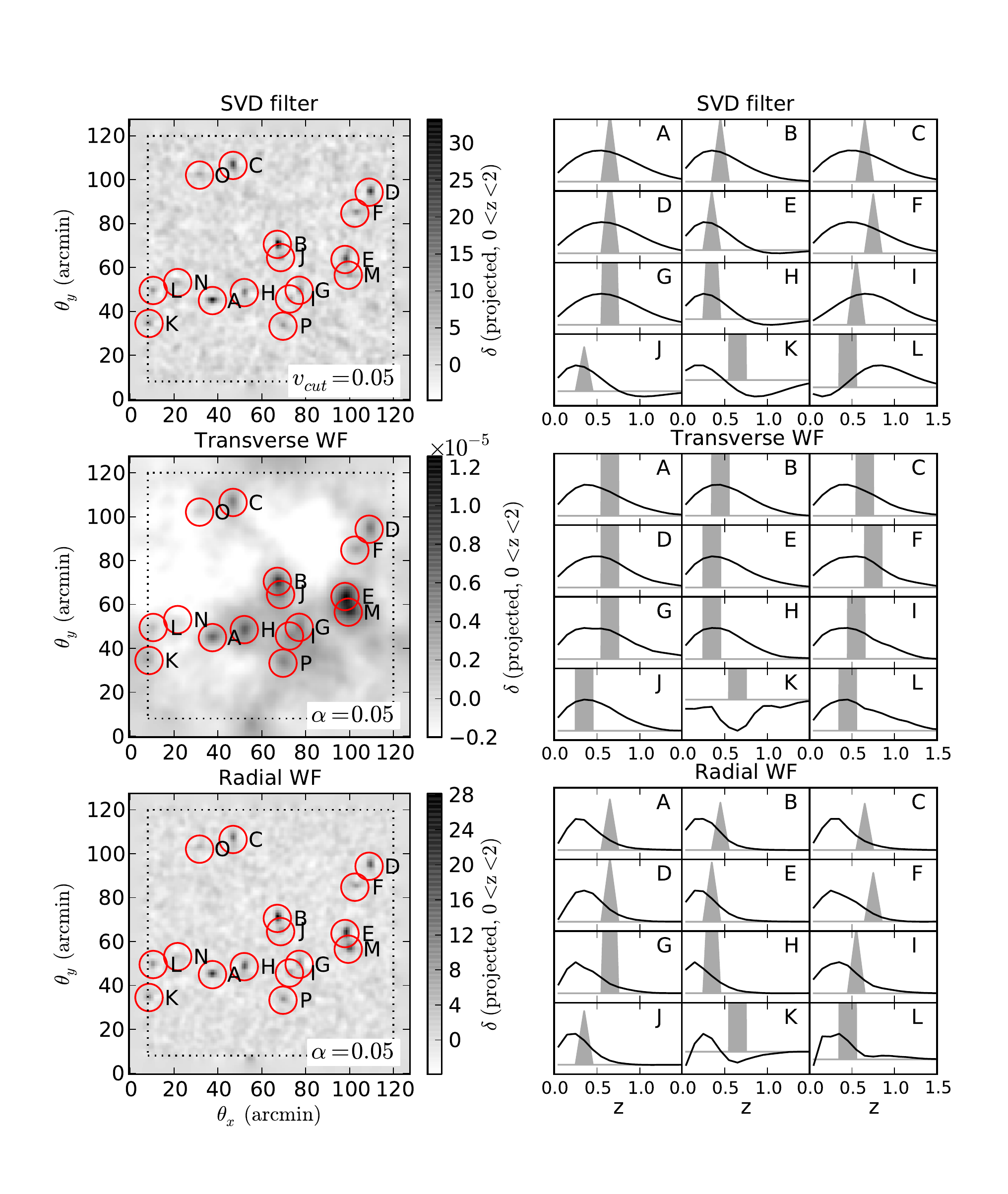} 
 \caption[Reconstruction of an artificial shear field with the SVD filter]{
   Reconstruction of an artificial shear field with the 
   SVD filter \textit{(top panels)}, 
   Transverse Wiener filter \textit{(middle panels)}, 
   and Radial Wiener filter \textit{(bottom panels)}.  
   The left column shows the projected density reconstruction
   across the field using each method, all
   smoothed with a 1-pixel wide Gaussian filter.  
   Red circles indicate the true locations of the input halos.
   The right column shows the line-of-sight distributions of the
   twelve most massive NFW halos, labeled A-L.  The masses and redshifts of
   the halos are listed in Table~\ref{halo_table}.
   The signal suppression of the transverse Wiener filter seen in 
   Figure~\ref{fig_los_plot_ST} is apparent in the color-bar scaling
   of the middle panels. The anomalous results seen in halo K are due to
   its proximity to the deweighted border.  As suggested by the discussion
   in Section~\ref{SN_modes}, none of the three methods succeed 
   in recovering precise redshifts of the halos.
  \label{fig_many_halos} }
\end{figure*}

To compare the performance of the three filtering methods 
for a realistic field, we
create a 4 square degree field with approximately 20 halos between masses of
$2\times 10^{14}$ and $8\times 10^{14} {\rm M_\odot}$ with a mass 
distribution approximating the cluster mass function of \citet{Rines07},
and a redshift distribution given by Equation~\ref{gal_z_dist},
adding a hard cutoff at $z=1.0$.  
These parameters are chosen to approximate the true 
distribution of observable halos in a field this size.
The results of the reconstruction are shown in 
Figure~\ref{fig_many_halos} 

The red circles are the locations of the input halos, not the result 
of some halo-detection algorithm.  However, it is clear that, 
for at least most of the mass range, we are able to produce a map 
for which any reasonable detection algorithm should detect the halos
in the correct locations.  A few of the lower mass halos would 
certainly be missed though, since they are not significantly 
different from the noise peaks in the image.

In practice, one may vary the parameter $v_{\rm cut}$ as in 
Figure~\ref{fig_los_plot}
to trade-off robustness of detecting peaks with resolution in angle and in 
redshift. As shown in Section~\ref{Comparison}, we expect
filtering to introduce very little bias in angular resolution, 
so large values of $v_{\rm cut}$ lead to the most robust angular results. 
On the other hand, as shown in Section~\ref{SN_modes},
filtering introduces an extreme bias along the line-of-sight.
The effects of this bias can be seen qualitatively
in the right column of Figure~\ref{fig_los_plot}.
Optimal redshift resolution requires choosing a 
filtering level which balances the effects of noise and bias,
and may require some form of bias correction.  In future work, we will
explore in detail the ways in which the SVD method allows
for a near optimal reconstruction of projected mass maps
and halo redshifts from data on galaxy shapes and photometric redshifts. 

\subsection{Scalability}
\label{Scalability} 

As we look forward to future surveys, it becomes important to consider methods 
that will scale upward with increasing survey volumes. Present weak lensing
surveys cover fields on the order of a few square degrees 
\citep[e.g.\ COSMOS,][]{Massey07}.  Future surveys will increase the field
size exponentially: up to $\sim\!\!20,000$ square degrees for LSST
\citep{LSST09}.  Though the flat-sky approximation used in this 
work is not appropriate for such large survey
areas, the weak lensing formalism can be modified to 
account for spherical geometry \citep[see, e.g.][]{Heavens03}. 

The main computational cost for both SVD and Wiener filtering is the
Fast Fourier Transform (FFT) required to implement the mapping from
$\gamma$ to $\kappa$.  For an $N \times N$ pixel field, the FFT algorithm 
performs in $\mathcal{O}[N\log N]$ in each dimension, 
meaning that the 2D FFT takes 
$\mathcal{O}[(N\log N)^2] \approx \mathcal{O}[N^2]$.  The Wiener
filter method, however, requires the inversion of a very large matrix 
using, for example, a conjugate-gradient method.  
The exact number of iterations
depends highly on the condition number of the matrix to be inverted;
STH09 finds that up to 150 iterations are required for this problem.
We find that \textit{each iteration} takes over 3 times longer than the 
entire SVD reconstruction. The net result is that both algorithms
scale nearly linearly with the area of the field (for constant pixel scale),
though the SVD estimator is computed up to 500 times faster 
than the Wiener filter. 

Extrapolating this scaling, the appropriately scaled 
SVD filter will allow reconstruction
of the entire $\sim\!\!20,000$ square-degree 
LSST field in a few hours on a single $\sim$GHz processor, given
enough memory.  On the same computer, the Wiener-filter method would take 
over a month, depending on the amount and type of filtering and 
assuming that the required number of iterations stays constant with 
increasing field size. For the SVD-filtered reconstruction of this 
large field, the real challenge will not be computational time, 
but memory constraints: the complex shear
vector itself for such a field will require $\sim\!\!30$ GB of memory,
with the entire algorithm consuming approximately three times this.
The memory requirements for the Wiener filter will be comparable.
This is within reach of current high-end workstations as well as 
shared-memory parallel clusters.

\begin{table}
\centering
\caption{Masses and redshifts of halos in Figure~\ref{fig_many_halos}.}
\label{halo_table}
\begin{tabular}{lllll}
\hline
& $\theta_x$ & $\theta_y$ & $z$ & $M/M_\odot$ \\
\hline
  A & 37.5 & 44.9 & 0.60 & $7.2\times 10^{14}$ \\
  B & 67.1 & 70.5 & 0.47 & $6\times 10^{14}$ \\
  C & 46.9 & 106.5 & 0.63 & $5.5\times 10^{14}$ \\
  D & 108.9 & 94.3 & 0.63 & $5.4\times 10^{14}$ \\
  E & 97.9 & 63.6 & 0.39 & $4.9\times 10^{14}$ \\
  F & 102.4 & 84.8 & 0.70 & $3.8\times 10^{14}$ \\
  G & 77.0 & 49.6 & 0.58 & $3.2\times 10^{14}$ \\
  H & 52.0 & 48.5 & 0.36 & $3.2\times 10^{14}$ \\
  I & 72.6 & 45.6 & 0.78 & $2.9\times 10^{14}$ \\
  J & 68.6 & 64.5 & 0.68 & $2.5\times 10^{14}$ \\
  K & 8.6 & 34.5 & 0.32 & $2.3\times 10^{14}$ \\
  L & 10.5 & 49.5 & 0.51 & $2.3\times 10^{14}$ \\
  M & 99.4 & 56.5 & 0.22 & $2.3\times 10^{14}$ \\
  N & 21.7 & 53.1 & 0.76 & $2.3\times 10^{14}$ \\
  O & 31.6 & 102.1 & 0.69 & $2.2\times 10^{14}$ \\
  P & 69.7 & 33.2 & 0.39 & $2.2\times 10^{14}$ \\
\hline
\end{tabular}
\end{table}

\section{Conclusion}
\label{Conclusions}

We have presented a new method for producing tomographic maps of dark matter
through weak lensing, using truncation of singular values.  We have
tested and compared our method to the Wiener filter based
method of STH09, which is the first three-dimensional mass mapping
approach that is applicable to large area surveys. Our reconstruction
shares many of the aspects of the Wiener filter reconstruction,
in the sense that it massively reduces the noise inherent in the problem.
Our SVD method may be considered even more non-parametric
than the Wiener filter method, since it does not rely on any a priori 
assumptions of the statistical properties of the signal: all of the noise 
reduction is derived from the observed noise properties of the data.

The SVD framework allows a unique quantitative comparison between the
different filtering methods and filtering strengths.  Using the
coefficients of the weighted principal components contained in the SVD,
we have compared the three filtering methods, and have found that 
the radial Wiener filter of HK02 and SVD filter of this work are 
less-biased noise reduction techniques than the transverse Wiener filter
of STH09.  These authors have recently implemented the radial Wiener 
filter and obtain results consistent with our findings (P. Simon and 
A. Taylor, private communication).

The angular resolution of the SVD-reconstructed mass maps seems to 
be significantly better than that of the transverse Wiener filter method,
the method chosen in the STH09 analysis.
This allows for more robust separation of pairs of halos into two 
separate halos rather than blurring them into a single mass peak.  
We discuss how our reconstruction method provides a scheme for
optimizing the 3D reconstruction of projected mass maps by
balancing the goals of robustness of detecting specific structures
and improved redshift resolution. 

The SVD method can compute the three-dimensional mass maps
rapidly provided sufficient computational memory is available. 
This allows for the possibility of solving the full-sky tomographic
lensing inversion on the scale of hours, rather than months, which
makes it readily applicable to upcoming surveys. 

On the other hand, the redshift resolution with the SVD method is not
significantly  better than that of either Wiener filter method.  
This was a problem identified by STH09, and unfortunately
the SVD method does not significantly improve the situation. 
Our analysis of the noise characteristics of radial modes 
indicates that linear, non-parametric reconstruction methods are
fundamentally limited in this regard.

%% file: chapter4.tex
\chapter{Shear Peak Statistics with KL}

This chapter will cover the results from \citet{Vanderplas2012}. In it,
we explore the utility of Karhunen Lo\`{e}ve (KL) analysis in 
solving practical problems in the analysis of gravitational
shear surveys, with a specific application to cosmological constraints
from shear peak statistics.
Shear catalogs from large-field weak lensing
surveys will be subject to many systematic limitations, notably
incomplete coverage and pixel-level masking due to foreground sources.  
We develop a method to use two dimensional KL eigenmodes of 
shear to interpolate noisy shear measurements across masked regions.  
We explore the results of this method with simulated shear catalogs, 
using statistics of high-convergence regions in the resulting map.  
We find that the KL procedure not only
minimizes the bias due to masked regions in the field, it also reduces
spurious peak counts from shape noise by a factor of $\sim 3$ in the
cosmologically sensitive regime.  This indicates that KL reconstructions 
of masked shear are not only useful for creating robust convergence maps
from masked shear catalogs, but also offer promise of improved parameter
constraints within studies of shear peak statistics.

This chapter was originally published in collaboration with Andrew Connolly,
Bhuvnesh Jain, and Mike Jarvis in the January 2012 edition of the
Astronomical Journal \citep[][ApJ, Vol. 744, p. 180; \copyright ~2012 by
the American Astronomical Society]{Vanderplas2012} and is reproduced below
with permission of the American Astronomical Society.

\section{Introduction}
Currently, a new generation of wide-field weak lensing surveys are
in the planning and construction stages.  
Among them are the the Dark Energy Survey (DES), 
the Panoramic Survey Telescope \& Rapid
Response System (PanSTARRS), the Wide Field Infrared Survey Telescope (WFIRST),
and the Large Synoptic Survey Telescope (LSST), to name a few.
These surveys, though not as deep as small-field space-based lensing surveys,
will cover orders-of-magnitude more area on the sky: up to $\sim 20,000$
square degrees in the case of LSST.

This is a fundamentally different regime 
than early weak lensing reconstructions
of single massive clusters: the strength of the shear signal is only
$\sim 1$\%, and is dominated by $\sim 30$\% intrinsic shape noise.  
This, combined with source galaxy densities of only 
$n \sim 20-50\ \mathrm{arcmin}^{-2}$
(compared with $n > 100\ \mathrm{arcmin}^{-2}$ for deep, space-based surveys)
and atmospheric PSF effects leads to a situation where the signal is 
very small compared to the noise.  
Additionally, in the wide-field regime,
the above-mentioned priors cannot be used.  
Nevertheless, many methods have been developed to extract useful
information from wide-field cosmic shear surveys, including
measuring the N-point power spectra and correlation functions 
\citep{Schneider02,Takada04,Hikage10}, 
performing log transforms of the convergence field 
\citep{Neyrinck09,Neyrinck10,Scherrer10,Seo11},
analyzing statistics of convergence and aperture mass peaks 
\citep{Marian10,Dietrich10,Schmidt10,Kratochvil10,Maturi11}.
Another well-motivated application of wide-field weak lensing 
is using wide-field mass reconstructions to minimize the effect
mass-sheet degeneracy in halo mass determination.

Many of the above applications require reliable recovery of 
the projected density, 
either in the form of the convergence $\kappa$, or filter-based quantities 
such as aperture mass \citep{Schneider98}.  
Because each of these amounts to a non-local filtering of the shear, 
the presence of masked regions can lead to a bias across significant
portions of the resulting maps.  Many of these methods have been demonstrated 
only within the context of idealized surveys, with exploration of the 
complications of real-world survey geometry left for future study.  
Correction for masked pixels has been studied within the context 
of shear power spectra \citep{Schneider10,Hikage10}
but has not yet been systematically addressed
within the context of mapmaking and the associated statistical methods
\citep[see, however,][for some possible approaches]{Padmanabhan03,Pires09}.
We propose to address this missing data problem through 
Karhunen-Lo\`{e}ve (KL) analysis.

In Section~\ref{KL_Intro} we summarize the theory of KL analysis in the
context of shear measurements, including the use of KL for interpolation
across masked regions of the observed field.
In Section~\ref{Testing_Reconstruction} we show the shear eigenmodes for
a particular choice of survey geometry, and use these eigenmodes to
interpolate across an artificially masked region in a simulated shear catalog.
In Section~\ref{Shear_Peaks} we discuss the nascent field of 
``shear peak statistics'',
the study of the properties of projected density peaks, and propose this
as a test of the possible bias imposed by KL analysis of shear.
In Section~\ref{Discussion} we utilize simulated shear catalogs 
in order to test the effect of KL interpolation on
the statistics of shear peaks.

\section{Karhunen-Lo\`{e}ve Analysis of Shear}
\label{KL_Intro}
As discussed in Chapter 2, KL analysis is a commonly used statistical tool
in a broad range of astronomical applications, from, e.g.~studies of 
galaxy and quasar spectra \citep{Connolly95,Connolly99,Yip04a,Yip04b}, to 
analysis of the spatial distribution of galaxies 
\citep{Vogeley96,Matsubara00,Pope04}, to characterization of the 
expected errors in weak lensing surveys \citep{Kilbinger06, Munshi06}.
A full description of KL analysis is presented in Chapter 2; here we
will briefly review the points relevant to this chapter.

In general, any set of $N$-dimensional data can be represented as a sum of 
$N$ orthogonal basis functions: this amounts to a rotation and scaling of 
the $N$-dimensional coordinate axis spanning the space in which the data live.
KL analysis seeks a set of orthonormal basis functions which can optimally
represent the dataset.  The sense in which the KL basis is optimal will be
discussed below.  For the current work, the data we wish to represent are the 
observed gravitational shear measurements across the sky.  
We will divide the survey 
area into $N$ discrete cells, at locations $\myvec{x}_i,\ 1\le i \le N$.  
From the ellipticity of the galaxies within each cell, 
we infer the observed shear $\gamma^o(\myvec{x}_i)$, which we assume
to be a linear combination of the true underlying shear, $\gamma(\myvec{x}_i)$
and the shape noise $n_\gamma(\myvec{x}_i)$.\footnote{
Throughout this chapter, we assume we are in the regime where the convergence
$\kappa \ll 1$ so that the average observed ellipticity in a 
cell is an unbiased estimator of shear; see \citet{Bartelmann01}}
In general, the cells may be of any shape (even overlapping) 
and may also take into account the redshift of sources.
In this analysis, the cells will be square pixels across the locally 
flat shear field, with no use of source redshift information.  
For notational clarity, we will represent quantities with a vector notation,
denoted by bold face: i.e. $\myvec{\gamma} = [\gamma_1,\gamma_2\cdots]^T$; 
$\gamma_i = \gamma(\myvec{x}_i)$. 

\subsection{KL Formalism}
\label{KL_Formalism}
As discussed in Chapter 2, KL analysis provides an optimal framework
such that our measurements $\myvec\gamma$ 
can be expanded in a set of $N$ orthonormal basis functions 
$\left\{ \myvec{\Psi}_j(\myvec{x}_i),\ j=1,N\right\}$, via a vector of
coefficients $\myvec{a}$.  
In matrix form, the KL projection of the observed shear can be written
\begin{equation}
  \myvec\gamma = \myvec\Psi\myvec{a}
\end{equation}
where the columns of the matrix $\myvec\Psi$ are the basis vectors 
$\myvec\Psi_i$.  Orthonormality is given by the condition 
$\myvec\Psi_i^\dagger\myvec\Psi_j = \delta_{ij}$, so that the coefficients
can be determined by
\begin{equation}
  \myvec{a} = \myvec{\Psi}^\dagger\myvec{\gamma}
\end{equation}
A KL decomposition is optimal in the sense that it seeks basis 
functions for which the 
coefficients are statistically orthogonal;\footnote{Note that statistical
orthogonality of coefficients is conceptually distinct from the 
geometric orthogonality of the basis functions themselves; 
see \citet{Vogeley96} for a discussion of this property.}
that is, they satisfy
\begin{equation}
  \langle a_i^* a_j \rangle = \langle a_i^2 \rangle \delta_{ij}
\end{equation}
where angled braces $\langle\cdots\rangle$ denote averaging over all 
realizations.  This definition leads to several important properties
\citep[see][for a thorough discussion \& derivation]{Vogeley96}:
\begin{enumerate}
\item \textbf{KL as an Eigenvalue Problem:} 
  Defining the correlation matrix 
  $\myvec{\xi}_{ij} = \langle \gamma_i\gamma_j^*\rangle$, 
  it can be shown that the KL vectors $\myvec{\Psi}_i$ are eigenvectors 
  of $\myvec{\xi}$ with eigenvalues $\lambda_i = \langle a_i^2\rangle$.
  For clarity, we'll order the eigenbasis such that 
  $\lambda_i \ge \lambda_{i+1}\ \forall\ i\in(1,N-1)$.  We define the
  diagonal matrix of eigenvalues $\mymat{\Lambda}$, such that
  $\mymat{\Lambda}_{ij} = \lambda_i\delta_{ij}$
  and write the eigenvalue decomposition in compact form:
  \begin{equation}
    \mymat{\xi} = \mymat{\Psi}\mymat{\Lambda}\mymat{\Psi}^\dagger
  \end{equation}

\item \textbf{KL as a Ranking of Signal-to-Noise}
  It can be shown that KL vectors of a whitened covariance matrix (see
  Section~\ref{Adding_Noise})
  diagonalize both the signal and the noise of the problem, with the
  signal-to-noise ratio proportional to the eigenvalue.  This is
  why KL modes are often called ``Signal-to-noise eigenmodes''.

\item \textbf{KL as an Optimal Low-dimensional Representation:}
  An important consequence of the signal-to-noise properties of KL modes  
  is that the optimal rank-$n$ representation of the data is 
  contained in the KL vectors corresponding to the $n$ largest eigenvalues:
  that is,
  \begin{equation}
    \label{eq_truncation}
    \myvec{\hat\gamma}^{(n)}
    \equiv \sum_{i=1}^{n<N} a_i\myvec{\Psi}_i
  \end{equation}
  minimizes the reconstruction error between $\myvec{\hat\gamma}^{(n)}$ and 
  $\myvec\gamma$ for reconstructions using $n$ orthogonal basis vectors.
  This is the theoretical basis of Principal Component Analysis (sometimes
  called Discrete KL), and leads to a common application of KL 
  decomposition: filtration of noisy signals.  For notational compactness,
  we will define the truncated eigenbasis $\mymat{\Psi}_{(n)}$ and truncated
  vector of coefficients $\myvec{a}_{(n)}$ such that 
  Equation~\ref{eq_truncation} can be written in matrix form:
  $\myvec{\hat\gamma}^{(n)} = \mymat{\Psi}_{(n)}\myvec{a}_{(n)}$.
\end{enumerate}
 
\subsection{KL in the Presence of Noise}
\label{Adding_Noise}
When noise is present in the data, the above properties do not 
necessarily hold.
To satisfy the statistical orthogonality of the KL coefficients $\myvec{a}$ 
and the resulting signal-to-noise properties of the KL eigenmodes, 
it is essential that the noise in the covariance matrix be ``white'': 
that is, $\Noise_\gamma \equiv 
\langle \myvec{n}_\gamma\myvec{n}_\gamma^\dagger \rangle \propto \mymat{I}$.  
This can be accomplished through a judicious
choice of binning, or by rescaling the covariance with a whitening 
transformation.  We take the latter approach here.

Defining the noise covariance matrix $\Noise_\gamma$ as above,
the whitened covariance matrix can be written 
$\myvec{\xi}_W = \Noise_\gamma^{-1/2} \myvec{\xi} \Noise_\gamma^{-1/2}$.  Then 
the whitened KL modes become
$\myvec{\Psi}_W\myvec{\Lambda}_W\myvec{\Psi}_W^\dagger \equiv \myvec{\xi}_W$.
The coefficients $\myvec{a}_W$ are calculated from the noise-weighted signal,
that is
\begin{equation}
  \myvec{a}_W = \myvec{\Psi}_W^\dagger\Noise_\gamma^{-1/2}
  (\myvec{\gamma}+\myvec{n}_\gamma)
\end{equation}
For the whitened KL modes, if signal and noise are uncorrelated, this leads to 
$\langle\myvec{a}_W\myvec{a}_W^\dagger\rangle = \myvec{\Lambda}_W + \myvec{I}$:
that is, the coefficients $\myvec{a}_W$ are statistically orthogonal.
For the remainder of this work, we will drop the subscript ``$_W$'' and assume
all quantities to be those associated with the whitened covariance.

\subsection{Computing the Shear Correlation Matrix}
\label{Shear_Correlation}
The KL reconstruction of shear requires knowledge of the
form of the pixel-to-pixel correlation matrix $\myvec\xi$.  
In many applications of KL
\citep[e.g.~analysis of galaxy spectra,][]{Connolly95} 
this correlation matrix is
determined empirically from many realizations of the data (i.e.~the set
of observed spectra).  In the case of weak lensing shear, we generally
don't have many realizations of the data, so this approach is not
tenable.  Instead, we compute this correlation matrix analytically.  The 
correlation of the cosmic shear signal between two regions of the sky
$A_i$ and $A_j$ is given by
\begin{eqnarray}
  \label{xi_analytic}
  \myvec{\xi}_{ij} 
  &=& \langle\gamma_i\gamma_j^*\rangle + 
  \langle n_in_j^*\rangle\nonumber\\
  &=& \left[\int_{A_i}d^2x_i\int_{A_j}d^2x_j 
    \xi_+(|\myvec{x_i}-\myvec{x_j}|)\right]
  + \delta_{ij}\frac{\sigma_\epsilon^2}{\bar{n}}
\end{eqnarray}
where $\sigma_\epsilon$ is the intrinsic shape noise (typically assumed to 
be $\sim 0.3$), $\bar{n}$ is the average galaxy count per pixel, and 
$\xi_+(\theta)$ is the ``+'' shear correlation function \citep{Schneider02}. 
$\xi_+(\theta)$ can be expressed as an integral over the shear power spectrum:
\begin{equation}
  \label{xi_plus_def}
  \xi_+(\theta) 
  = \frac{1}{2\pi} \int_0^\infty d\ell\ \ell P_\gamma(\ell) J_0(\ell\theta)
\end{equation}
where $J_0$ is the zeroth-order Bessel function of the first kind.  The 
shear power spectrum $P_\gamma(\ell)$ can be expressed as an 
appropriately weighted line-of-sight integral over the 3D mass power 
spectrum \citep[see, e.g.][]{Takada04}:
\begin{equation}
  \label{P_gamma}
  P_\gamma(\ell) = \int_0^{\chi_s}d\chi W^2(\chi)\chi^{-2}
  P_\delta\left(k=\frac{\ell}{\chi};z(\chi)\right)
\end{equation}
Here $\chi$ is the comoving distance, $\chi_s$ is the distance to the
source, and $W(\chi)$ is the lensing weight function,
\begin{equation}
  \label{Lensing_Weight}
  W(\chi) = \frac{3\Omega_{m,0}H_0^2}{2a(\chi)}\frac{\chi}{\bar{n}_g}
  \int_{z(\chi)}^{z(\chi_s)}dz\ n(z) \frac{\chi(z)-\chi}{\chi(z)}
\end{equation}
where $n(z)$ is the redshift distribution of galaxies.  We assume a
DES-like survey, where $n(z)$ has the approximate form
\begin{equation}
  \label{Number_Distribution}
  n(z) \propto z^2 \exp[-(z/z_0)^{1.5}]
\end{equation}
with $z_0 = 0.5$, where $n(z)$ is normalized to the observed galaxy density
$\bar{n}_g = 20\ {\rm arcmin}^{-2}$.

The 3D mass power spectrum $P_\delta(k,z)$ in Equation~\ref{P_gamma}
can be computed theoretically.  
In this work we compute $P_\delta(k,z)$ using the halo model of 
\citet{Smith03}, and compute the correlation matrix $\myvec\xi$ using 
Equations~\ref{xi_analytic}-\ref{Number_Distribution}.
When computing the double integral of Equation~\ref{xi_analytic},
we calculate the integral in two separate regimes:
for large separations ($\theta > 20$ arcmin), 
we assume $\xi_+(\theta)$ doesn't change appreciably over the area 
of the pixels, so that only a single evaluation of
the $\chi_+(\theta)$ is necessary for each pixel pair.  
For smaller separations, 
this approximation is insufficient, and we evaluate $\myvec\xi_{ij}$ using
a Monte-Carlo integration scheme.  Having calculated the 
theoretical correlation matrix $\myvec\xi$ for a given field, 
we compute the KL basis directly using an eigenvalue decomposition.

\subsection{Which Shear Correlation?}
\label{sec:WhichCorrelation}
Above we note that the correlation matrix of the measured shear 
can be expressed
in terms of the ``+'' correlation function, $\xi_+(\theta)$.  This is not
the only option for measurement of shear correlations 
\citep[see, e.g.][]{Schneider02}.  So why use $\xi_+(\theta)$ rather 
than $\xi_-(\theta)$?  The answer lies in the KL formalism itself.  The KL
basis of a quantity $\myvec\gamma$ is constructed via its correlation
$\langle\myvec\gamma \myvec\gamma^\dagger\rangle$.  Because of the complex
conjugation involved in this expression, the only relevant correlation 
function for KL is $\xi_+(\theta)$ \textit{by definition}.  Nevertheless,
one could object that by neglecting $\xi_-$, 
KL under-utilizes the theoretical information available 
about the correlations of cosmic shear. However, in the absence of noise, 
the two correlation functions contain identical information: 
either function can be determined from the other.  
In this sense, the above KL formalism
uses all the shear correlation information that is available.

One curious aspect of this formalism is that the theoretical covariance
matrix and associated eigenmodes are real-valued, while the shear we are
trying to reconstruct is complex-valued.  This can be traced
to the computation of the shear correlation:
\begin{equation}
  \xi_+ \equiv \langle\gamma \gamma^*\rangle
  = \langle \gamma_t\gamma_t\rangle + \langle \gamma_\times\gamma_\times\rangle
  + i[\langle\gamma_t\gamma_\times\rangle - \langle\gamma_\times\gamma_t\rangle]
\end{equation}
By symmetry, the imaginary part of this expression
is zero. At first glance, this might seem a bit strange:
how can a complex-valued data vector be reconstructed from a
real-valued orthogonal basis?  The answer lies in the complex KL coefficients
$a_i$: though each KL mode contributes only a single phase across the field
(given by the phase of the associated $a_i$), the reconstruction has a 
plurality of phases due to the varying magnitudes of the contributions 
at each pixel (given by the elements of each basis vector $\myvec{\Psi}_i$).

An important consequence of this observation is that the KL modes 
themselves are not sensitive by construction to the E-mode (curl-free) 
and B-mode (divergence-free) components of the shear field. As we will
show below, however, the signal-to-noise properties of KL modes lead to 
some degree of sensitivity to the E and B-mode information in a given
shear field (See Section~\ref{Discussion}).

\subsection{Interpolation using KL Modes}
\label{KL_Interpolation}
Shear catalogs, in general, are an incomplete and inhomogeneous
tracer of the underlying shear field, and some regions of the field may 
contain no shear information.  This sparsity of data poses a problem,
because the KL modes are no longer orthogonal over the incomplete field.
\citet{Connolly99} demonstrated how this missing-information problem can be 
addressed for KL decompositions of galaxy spectra. 
This application is discussed in more detail in Chapter 2; here we will
briefly summarize the results using the notation of shear studies.
First we define the weight function $w(\myvec{x}_i)$.
The weight function can be defined in one of two ways: 
a binary weighting convention where
$w(\myvec{x}_i)=0$ in masked pixels and $1$ elsewhere, or a continuous 
weighting convention where $w(\myvec{x}_i)$ scales inversely with the noise 
$[\Noise_\gamma]_{ii}$.
The binary weighting convention treats the noise is part of the data, 
and so the measurements should be whitened as outlined in 
Section~\ref{Adding_Noise}.  
The continuous weighting convention assumes the noise is part of the mask, 
so data and noise are not whitened.
We find that the two approaches lead to qualitatively similar results, 
and choose to use the binary weighting convention for the simplicity of 
comparing masked and unmasked cases.

Let $\myvec{\gamma}^o$ be the observed data vector, 
which is unconstrained where
$w(\myvec{x}_i)=0$.  Then we can obtain the KL coefficients $a_i$ by 
minimizing the reconstruction error of the whitened data
\begin{equation}
  \label{chi2_min}
  \chi^2 = ( \Noise_\gamma^{-1/2}\myvec{\gamma}^o
  - \myvec{\Psi}_{(n)}\myvec{a}_{(n)} )^\dagger 
  \myvec{W}
  ( \Noise_\gamma^{-1/2}\myvec{\gamma}^o
  - \myvec{\Psi}_{(n)}\myvec{a}_{(n)} )
\end{equation}
where we have defined the diagonal weight matrix 
$\myvec{W}_{ij} = w(\myvec{x}_i) \delta_{ij}$.
Minimizing Equation~\ref{chi2_min} with respect to $\myvec{a}$ leads to
the optimal estimator $\myvec{\hat{a}}$, which can be expressed
\begin{equation}
 \myvec{\hat{a}}_{(n)} = 
 \mymat{M}_{(n)}^{-1} 
 \myvec{\Psi}_{(n)}^\dagger \myvec{W} \Noise_\gamma^{-1/2}\myvec{\gamma}^o
\end{equation}
Where we have defined the mask convolution matrix 
$\mymat{M}_{(n)} \equiv \mymat{\Psi}_{(n)}^\dagger\mymat{W}\mymat{\Psi}_{(n)}$.
These coefficients $\myvec{\hat{a}}_{(n)}$
can then be used to construct an estimator for the unmasked shear field:
\begin{equation}
  \label{shear_recons}
  \myvec{\hat{\gamma}}^{(n)} = \Noise_\gamma^{1/2}\myvec{\Psi}_{(n)}\myvec{\hat{a}}_{(n)}
\end{equation}
In cases where the mask convolution matrix $\mymat{M}_{(n)}$  
is singular or nearly singular, the estimator in Equation~\ref{shear_recons}
can contain unrealistically large values within the reconstruction 
$\myvec{\hat{\gamma}}^{(n)}$.
This can be addressed either by reducing $n$, or by 
adding a penalty function to the right side of Equation~\ref{chi2_min}.  
One convenient form of this penalty is the generalized
Wiener filter \citep[see][]{Tegmark97}, which penalizes results which 
deviate from the expected correlation matrix.  Because the correlation
matrix has already been computed when determining the KL modes, 
this filter requires very little extra computation.
With Wiener filtering, Equation~\ref{chi2_min} becomes
\begin{eqnarray}
  \label{chi2_min_WF}
  \chi^2 & = & ( \Noise_\gamma^{-1/2}\myvec{\gamma}^o - \myvec{\Psi}_{(n)}\myvec{a}_{(n)} )^\dagger 
  \myvec{W} 
  (\Noise_\gamma^{-1/2}\myvec{\gamma}^o - \myvec{\Psi}_{(n)}\myvec{a}_{(n)} )
  \nonumber\\
  && + \alpha\ \myvec{a}_{(n)}^\dagger\myvec{C}_{a(n)}^{-1}\myvec{a}_{(n)}
\end{eqnarray}
where $\myvec{C}_{a(n)}\equiv\langle 
\myvec{a}_{(n)}\myvec{a}_{(n)}^\dagger\rangle$ 
and $\alpha$ is a tuning parameter which lies in the range $0\le\alpha\le 1$. 
Note that for $\alpha=0$, the result is the same as in the unfiltered case.
Minimizing Equation~\ref{chi2_min_WF} with respect to $\myvec{a}$ gives the
filtered estimator
\begin{equation}
  \label{a_WF}
  \myvec{\hat{a}}_{(n,\alpha)} = 
  \mymat{M}_{(n,\alpha)}^{-1} 
  \myvec{\Psi}_{(n)}^\dagger \myvec{W} \Noise_\gamma^{-1/2}\myvec{\gamma}^o
\end{equation}
where we have defined $\mymat{M}_{(n,\alpha)} 
\equiv [\myvec{\Psi}_{(n)}^\dagger\myvec{W}\myvec{\Psi}_{(n)} 
  + \alpha\myvec{\Lambda}_{(n)}^{-1}]$, and
$\myvec{\Lambda}_{(n)}$ is the truncated diagonal matrix of 
eigenvalues associated with $\myvec{\Psi}_{(n)}$.

\begin{figure*}
 \centering
 \includegraphics[width=\textwidth]{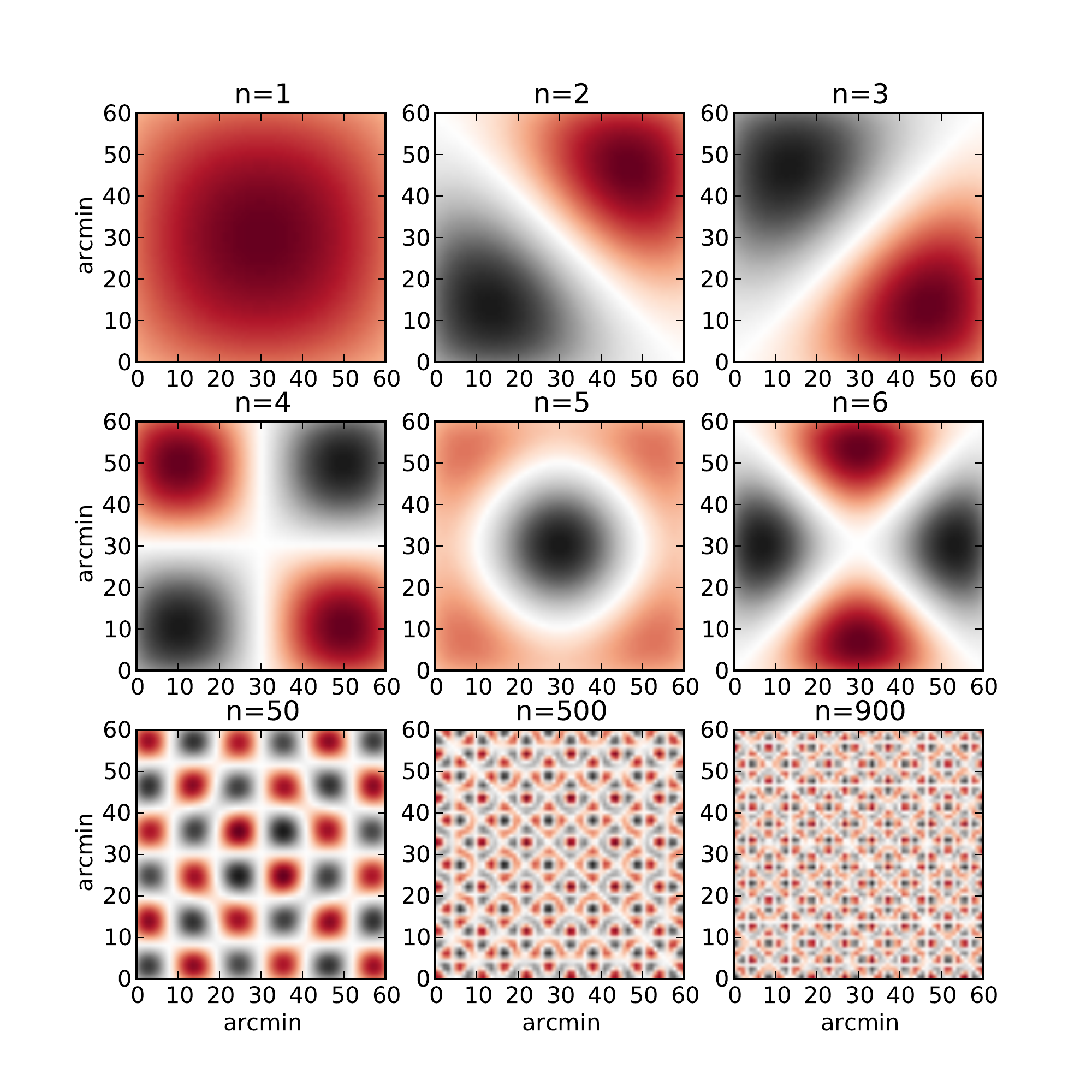} 
 %created by ../fig_code/fig01_eigenmodes.py
 \caption[ A sample of nine of the 4096 KL eigenmodes]{
   A sample of nine of the 4096 KL eigenmodes 
   of a $1^\circ\times 1^\circ$ patch of the sky partitioned into
   $64\times 64$ pixels.  Black is positive, red is negative, and each mode
   has unit norm. The modes are calculated from the theoretical
   shear correlation function (see Section~\ref{Shear_Correlation}).  
   As a consequence of the isotropy of the cosmic shear field,
   the covariance matrix -- and thus the associated eigenmodes --
   are purely real (see Section~\ref{Testing_Shear_KL}).
   \label{fig_KL_modes} }
\end{figure*} 

\section{Testing KL Reconstructions}
\label{Testing_Reconstruction}
In this section we show results of the KL analysis of shear fields for
a sample geometry.  In Section~\ref{Testing_Shear_KL} we discuss the general
properties of shear KL modes for unmasked fields, while in
Section~\ref{Testing_Interpolation} we discuss KL shear reconstruction 
in the presence of masking. 

\subsection{KL Decomposition of a Single Field}
\label{Testing_Shear_KL}
To demonstrate the KL decomposition of a shear field, we assume a square field
of size $1^\circ\times 1^\circ$, divided into $64\times 64$ pixels.  We assume
a source galaxy density of $20\ \mathrm{arcmin}^{-2}$ -- appropriate for a
ground-based survey such as DES -- and calculate the
KL basis following the method outlined in Section~\ref{KL_Formalism}.
For the computation of the nonlinear matter power spectrum, we assume a flat 
$\Lambda$CDM cosmology with $\Omega_M=0.27$ at the present day, with
the power spectrum normalization given by $\sigma_8=0.81$.

\begin{figure*}
 \centering
 \includegraphics[width=0.8\textwidth]{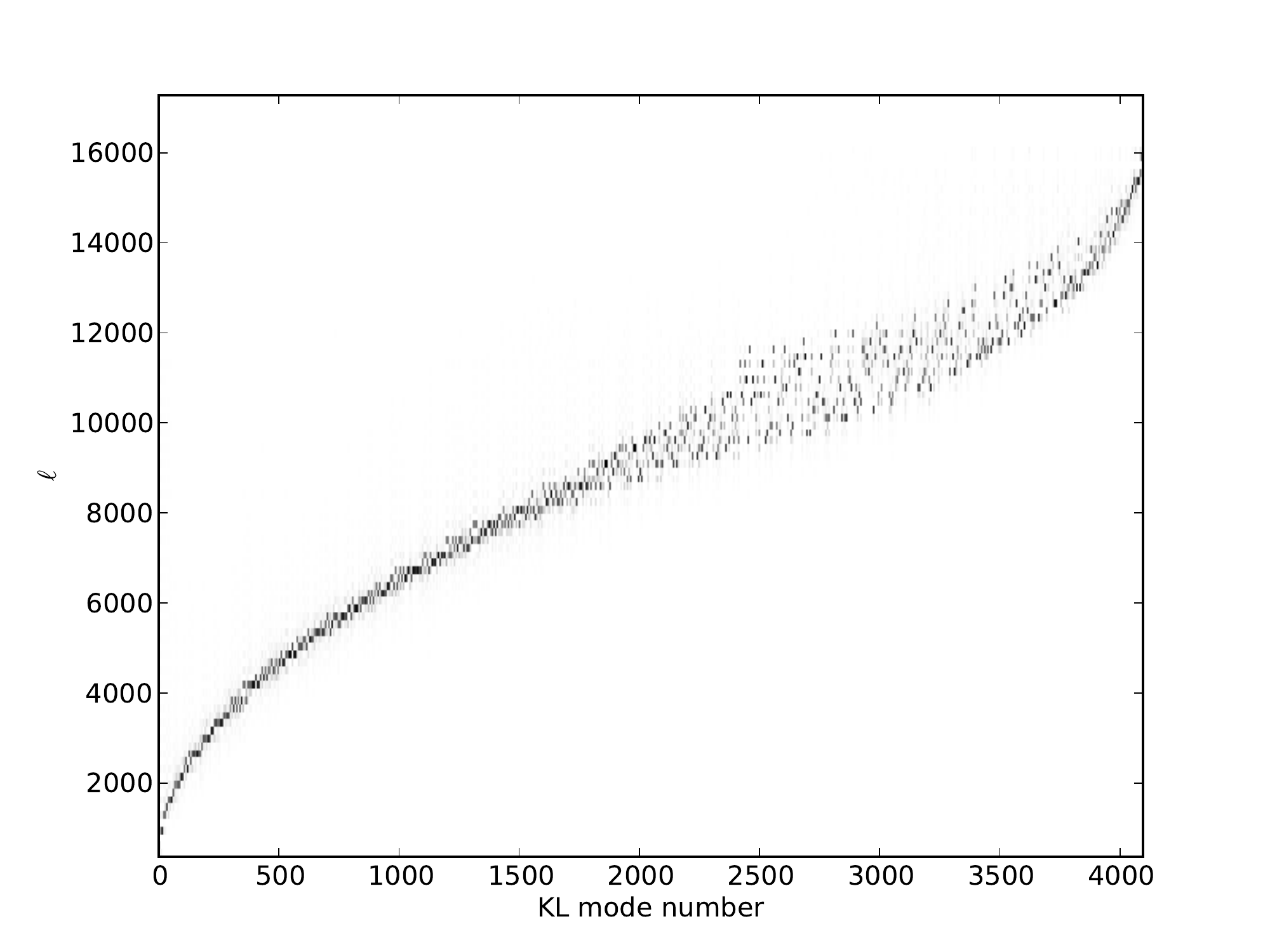}
 %created by ../fig_code/fig02_bandpower.py
 \caption[The normalized power spectrum of each KL mode]
 {The normalized power spectrum of each KL mode.  For constant
   mode number, the figure represents a histogram of the power in that KL
   mode, normalized to a constant total power.  KL modes represent a 
   linear combination of Fourier modes, so that the power in each KL 
   mode is spread over a range of $\ell$ values.  Nevertheless,
   the general trend is clear: larger mode numbers are associated with
   larger wave numbers, and thus smaller length scales.
   \label{fig_bandpower} }
\end{figure*}

Figure~\ref{fig_KL_modes} shows a selection of
nine of the 4096 shear eigenmodes within
this framework.  The KL modes are reminiscent of 2D Fourier modes, with
higher-order modes probing progressively smaller length scales.  
This characteristic length scale of the eigenmodes
can be seen quantitatively in Figure~\ref{fig_bandpower}.  
Here we have computed the rotationally averaged
power spectrum $C_\ell$ for each individual Fourier mode, and plotted the
power vertically as a density plot for each mode number.  
Because the KL modes are
not precisely equivalent to the 2D Fourier modes, each contains power at
a range of values in $\ell$.  But the overall trend is clear: larger modes
probe smaller length scales, and the modes are very close to Fourier in
nature.

\begin{figure*}
 \centering
 \includegraphics[width=0.8\textwidth]{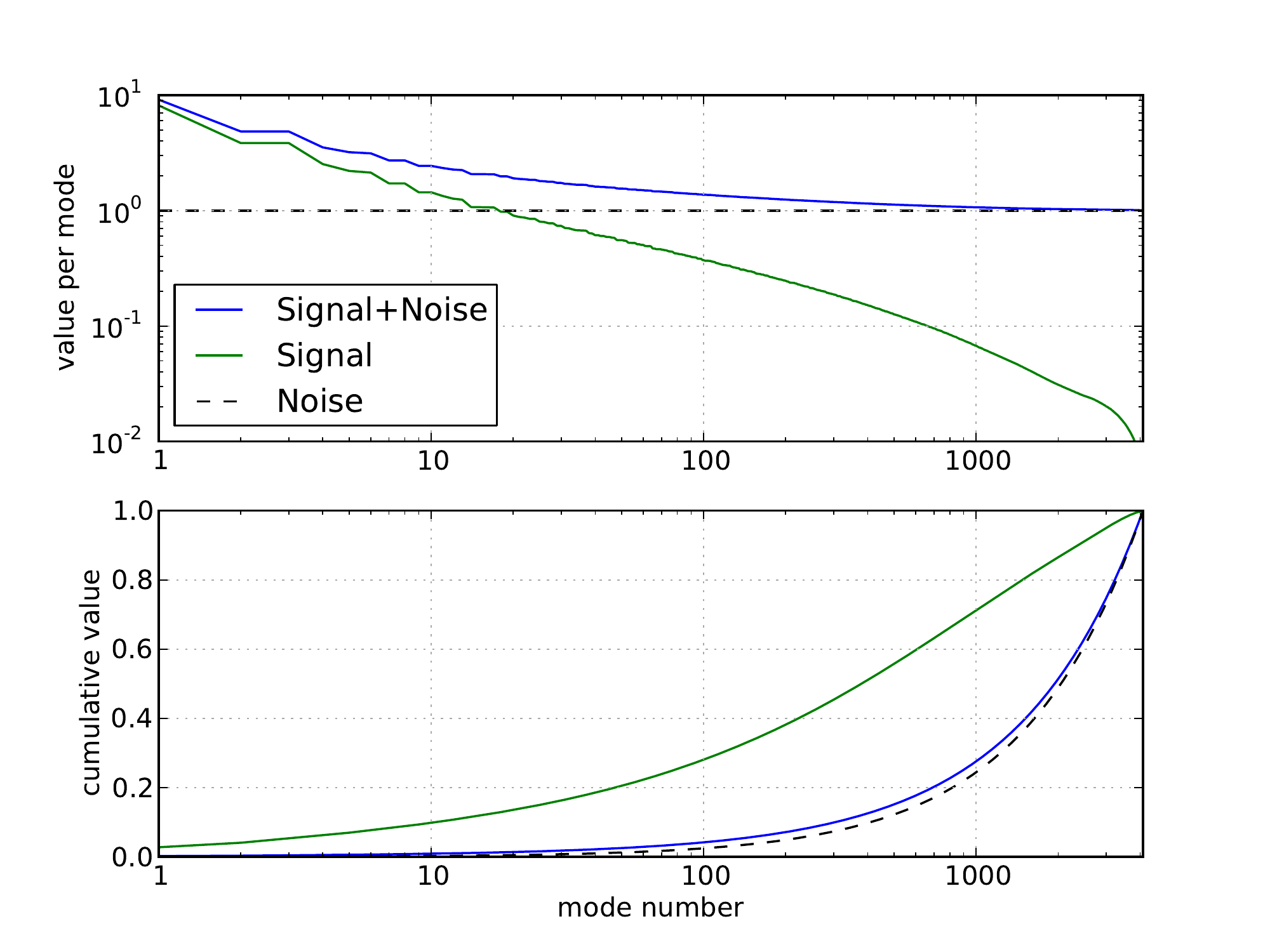}
 %created by ../fig_code/fig03_eigenvalues.py
 \caption[The eigenvalues associated with the eigenmodes]{
   The eigenvalues associated with the eigenmodes discussed in
   Figure~\ref{fig_KL_modes}.  By construction, the eigenvalue is 
   proportional to the sum of signal and
   noise within each mode.  The upper figure shows the value per mode,
   while the lower figure shows the normalized cumulative value.
   The stepped-pattern evident in the upper panel is due to the presence
   of degenerate eigenmodes which have identical eigenvalues
   (e.g. modes $n=2$ and $n=3$, related by parity as 
   evident in Figure~\ref{fig_KL_modes}).
   Because the eigenmodes are computed from a whitened covariance matrix 
   (see Section~\ref{Adding_Noise}), the noise contribution within each
   mode is equal to 1.  Subtracting this contribution leads to the plot of
   signal only: this shows that the signal-to-noise ratio is above unity 
   only for the first $17$ modes.  Still, as seen in the lower panel,
   higher modes are required: the first $17$ modes account for only $12\%$
   of the signal on average.
   To recover $70\%$ of the signal in a particular reconstruction requires
   about 1000 modes.  At this point, each additional mode has a signal-to-noise
   ratio of less than $0.1$.  Such a small signal-to-noise ratio is a
   well-known aspect of cosmic shear studies.
   \label{fig_eigenvalues} }
\end{figure*}

As noted in Section~\ref{KL_Intro}, one useful quality of a KL decomposition
is its diagonalization of the signal and noise of the problem.
To explore this property, we plot in the upper panel of 
Figure~\ref{fig_eigenvalues} the eigenvalue profile 
of these KL modes. By construction, higher-order modes 
have smaller KL eigenvalues.  What is more,
because the noise in the covariance matrix is whitened 
(see Section~\ref{Adding_Noise}), the expectation of the noise covariance
within each mode is equal to 1.  Subtracting this noise from each eigenvalue 
gives the expectation value of the signal-to-noise ratio: 
thus we see that the expected
signal-to-noise ratio of the eigenmodes is above unity only for the first
17 of the 4096 modes.

At first glance, this may seem to imply that only the first 17 or so modes
are useful in a reconstruction.  On the contrary: as seen in the lower panel
of Figure~\ref{fig_eigenvalues}, these first 17 modes contain only a
small fraction of the total information in the shear field (This is not 
an unexpected result: cosmic shear measurements have 
notoriously low signal-to-noise ratios!)
About 900 modes are needed to preserve an average of 70\% of the total signal, 
and at this level, each additional mode has a signal-to-noise ratio 
of below $1/10$.  The noisy input shear field can be exactly recovered
by using all 4096 modes: in this case, though, the final few modes 
contribute two orders-of-magnitude more noise than signal.

\begin{figure*}
 \centering
 \includegraphics[width=\textwidth]{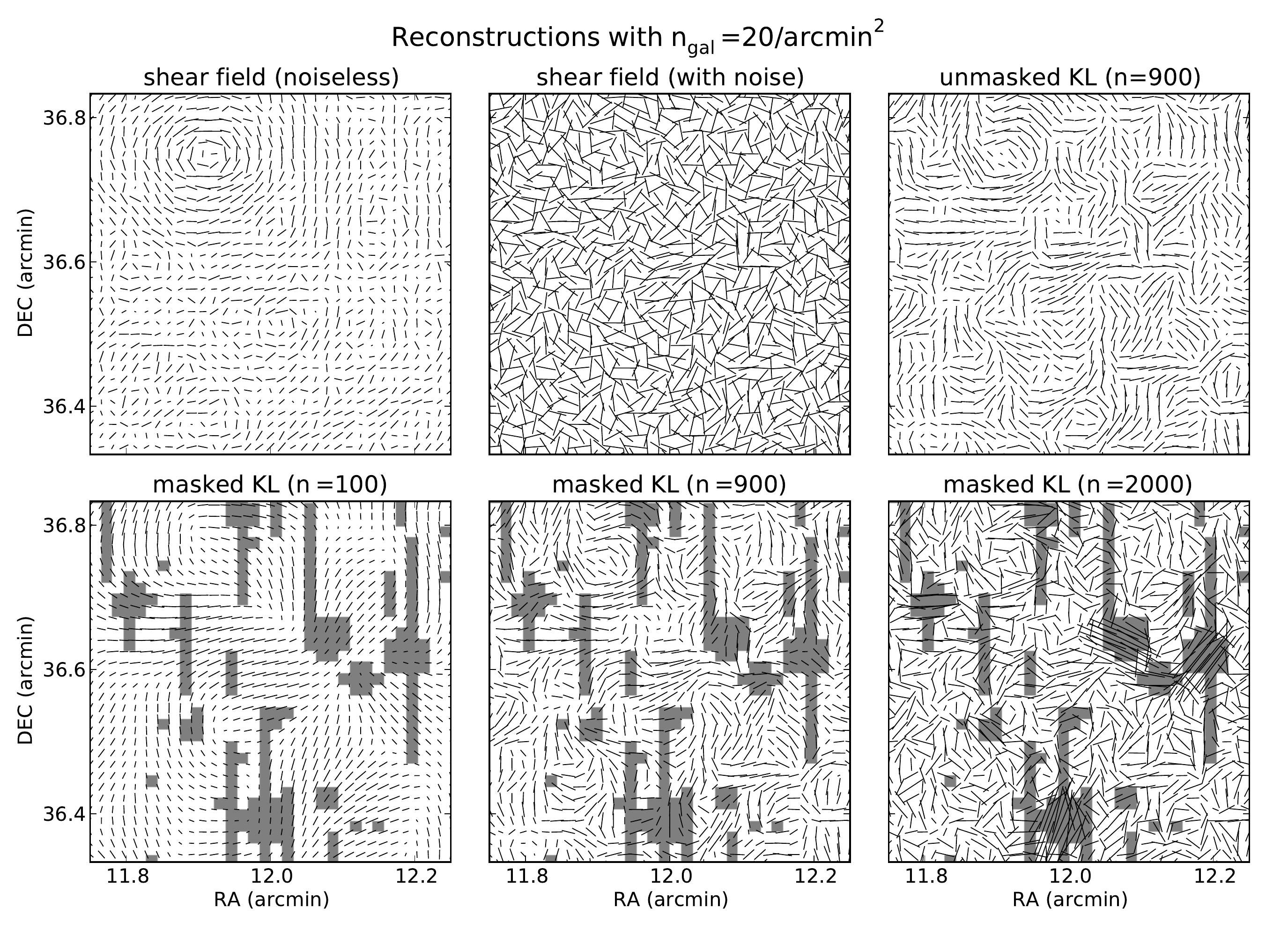}
 %created by ../fig_code/fig04_reconstruction.py
 \caption[reconstruction of a small patch of masked shear from simulated shear catalog]{
   This figure illustrates the reconstruction of a small patch of masked
   shear from simulated shear catalog.
   \textit{upper panels:} 
   The underlying noiseless shear signal \textit{(left)},
   the observed, noisy shear signal \textit{(middle)},
   and the unmasked reconstruction with 900 modes and $\alpha=0.15$.
   The amplitude of the noise is calculated using an intrinsic ellipticity 
   $\sigma_\epsilon = 0.3$, with an average number density of 
   $n_{gal}=20\ \mathrm{arcmin}^{-2}$.  The large peak in the upper 
   portion of the figure is well-recovered by the KL reconstruction.
   \textit{lower panels:} The KL reconstruction of the shear in the
   presence of 20\% masking, with increasing number of modes $n$.
   The mask is represented by the shaded regions in panels: 
   within these regions, the value of the shear is
   recovered through KL interpolation (see Section~\ref{KL_Interpolation}).
   We see in this progression the effect of the KL cutoff choice:
   using too few modes leads to loss of information, 
   while using too many modes leads to over-fitting within the masked regions 
   (See Appendix~\ref{Choosing_Params} for a discussion of the choice of
   number of modes).
   \label{fig_reconstruction} }
\end{figure*}

\begin{figure*}
 \centering
 \includegraphics[width=\textwidth]{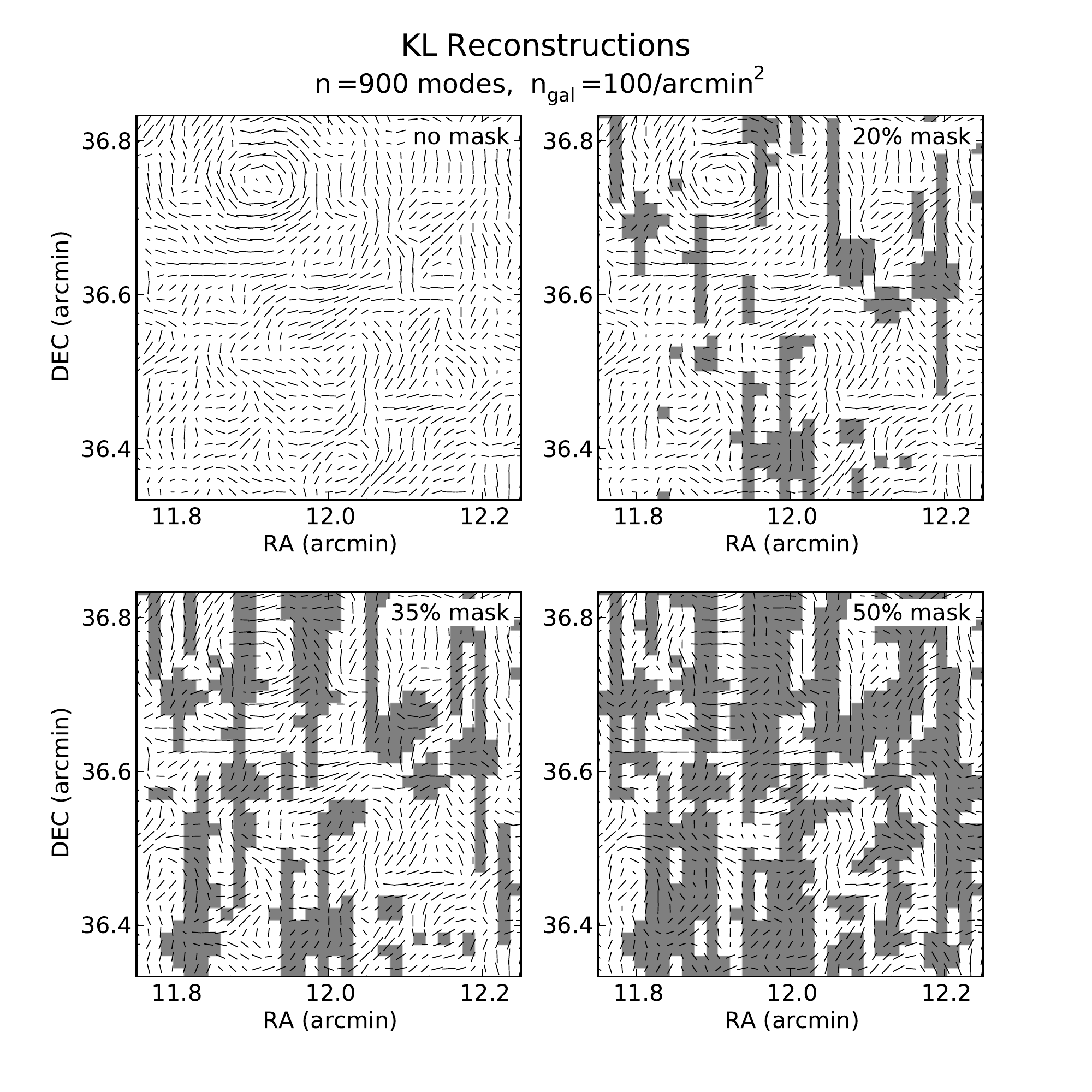}
 %created by ../fig_code/fig05_reconstruction_2.py
 \caption[reconstruction using $n=900$ modes]{
   Here we show the same field as in Figure~\ref{fig_reconstruction}, 
   reconstructed using $n=900$ modes, with increasing levels of mask
   coverage.  The density of source galaxies has been increased to 
   100 arcmin$^{-2}$, typical of a space-based weak lensing survey.  
   At this noise
   level, smaller halos can be detected within the unmasked KL
   reconstruction (upper-left panel).  Even at a 50\% masking level, 
   the large peak at (RA,DEC) = (11.9,36.75) is adequately recovered.
   \label{fig_reconstruction_2} }
\end{figure*}
 
%\begin{figure*}
% \centering
% \includegraphics[width=\textwidth]{fig06_kappa_multires.pdf}
% %created by ../fig_code/fig06_kappa_multires.py
% \caption{The large-field convergence map from the noiseless input
%   shear \textit{(upper panels)} and the 900-mode, $\alpha=0.15$
%   reconstruction of the noisy shear, with 20\% of pixels masked out 
%   \textit{(lower panels)}.  
%   The rightmost plots cover one square degree.
%   Maps are smoothed with gaussian filters of width 4, 2 and 1 arcmin,
%   as labeled.  This shows that even with 20\% masking of the input signal, 
%   the KL interpolation procedure recovers the most significant peaks.
%   Note that the spurious structure in the KL plots is primarily due 
%   to the shape noise of the input shear, not to the KL procedure itself.
%   \label{fig_kappa_reconstruction} }
%\end{figure*} 

\begin{figure*}
 \centering
 \includegraphics[height=0.8\textheight]{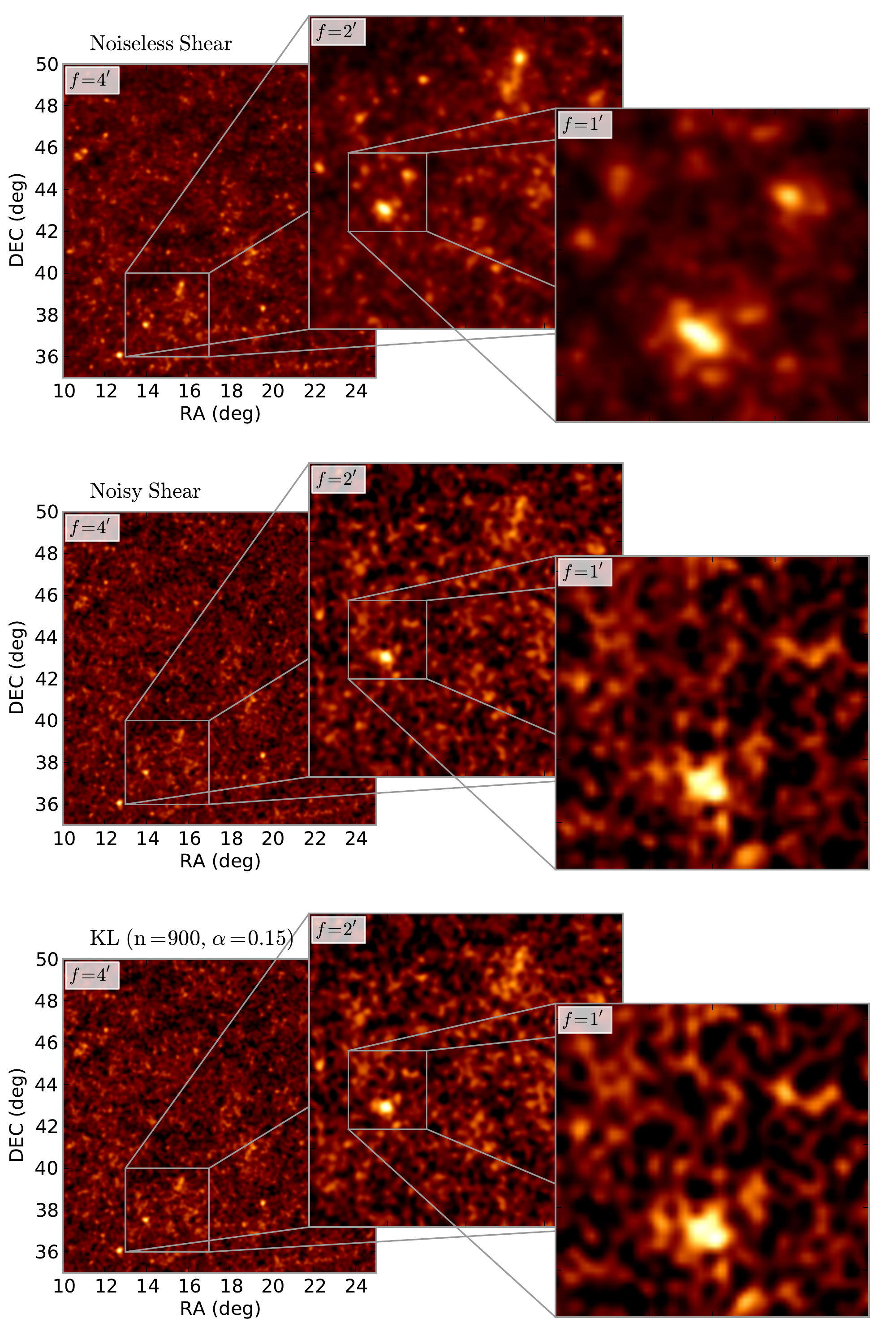}
%created by ../fig_code/fig06_kappa_multires.py
 \caption[The large-field convergence map]{The large-field convergence map derived using the method of
   \citet{Kaiser93} from the
   noiseless input shear \textit{(upper panels)}, 
   the noisy input shear \textit{(middle panels)},
   and the 900-mode, $\alpha=0.15$ reconstruction of the noisy shear, 
   with 20\% of pixels masked out \textit{(lower panels)}.  
   The rightmost plots of each row cover one square degree.
   The larger-field maps are smoothed with Gaussian filters of width 
   4 and 2 arcmin, while the smaller fields are unsmoothed.
   These plots show that even with 20\% masking of the input signal, 
   the KL interpolation procedure recovers the most significant peaks,
   and offers improvement over the results derived from unmasked,
   observed shear. 
   This improvement will be discussed more quantitatively below.
   \label{fig_kappa_reconstruction} }
\end{figure*}

\subsection{Testing KL Interpolation}
\label{Testing_Interpolation}
To test this KL interpolation technique, we use simulated shear 
catalogs\footnote{The simulated shear catalogs were kindly made available 
to us by R. Wechsler, M Busha, and M. Becker.}. 
These catalogs contain 220 square degrees of simulated shear maps, computed 
using a ray-tracing grid through a cosmological N-body simulation of the 
standard $\Lambda$-CDM model. The shear signal is computed at the locations 
of background galaxies with a median redshift of about 0.7.  Galaxies are 
incorporated in the simulation using the ADDGALS algorithm 
\citep[][Wechsler \textit{et al.} in preparation]{Wechsler04}, 
tuned to the expected observational characteristics of the DES mission. 

We pixelize this shear field using the same pixel size as above: 
$64\times 64$ pixels per square degree.
To perform the KL procedure on the full field with this angular 
resolution would lead to a data vector containing over $10^6$ elements, 
and an associated covariance matrix containing $10^{12}$ entries.  
A full eigenvalue decomposition of such a matrix is 
computationally infeasible, so we reconstruct the field in 
$1^\circ\times 1^\circ$ tiles, each $64\times 64$ pixels in size.  To reduce
edge effects between these tiles, we use only the central 
$0.5^\circ\times 0.5^\circ$ region of each, so that covering the 300 square 
degree field requires 1200 tiles. 

In order to generate a realistic mask over the field area, 
we follow the procedure outlined in
\citet{Hikage10} which generates pixel-level masks characteristic of  
point-sources, saturation spikes, and bad CCD regions.  We tune the mask
so that 20\% of the shear pixels have no data.  The geometry of the mask
over a representative patch of the field can be seen in the lower 
panels of Figure~\ref{fig_reconstruction}, where we also show the result of
the KL interpolation using 900 out of 4096 modes, with $\alpha = 0.15$
(For a discussion of these parameter choices, see 
Appendix~\ref{Choosing_Params}).

The upper panels of Figure~\ref{fig_reconstruction} give a qualitative view
of the difficulty of cosmic shear measurements.  The top left panel shows the
noiseless shear across the field, while the top right panel shows the
shear with shape noise for a DES-type survey ($\sigma_\epsilon=0.3$, 
$\bar{n}_{gal} = 20\ \mathrm{arcmin}^{-2}$).  To the eye, the signal seems
entirely washed out by the noise.  Nevertheless, the shear signal is there,
and can be fairly well-recovered using the first 900 KL modes 
(middle-left panel). For masked data, we must resort to the 
techniques of Section~\ref{KL_Interpolation} to fill-in the missing data.
The middle-right panel shows this reconstruction, with gray shaded
regions representing the masked area.  A visual comparison of the masked 
and unmasked $n=900$ panels of Figure~\ref{fig_reconstruction} confirms
qualitatively that the KL interpolation is performing as desired.  This is
especially apparent near the large cluster located at (RA,DEC)=(11.9,36.7).  
The remaining two lower panels of Figure~\ref{fig_reconstruction} 
show cases of over-fitting and under-fitting
of the shear data. If too few KL modes are used, the structure of the input 
shear field is lost.  If too many KL modes are used, the masked regions 
are over-fit,
causing the interpolated shear values to become unnaturally large.  This
observation suggests one rubric by which the ideal number of modes can be
chosen; see the discussion in Appendix~\ref{Choosing_Params}.

It is interesting to explore the limits of this interpolation algorithm.
Figure \ref{fig_reconstruction_2} shows the KL reconstruction with
increasing masked fractions, using a noise level typical of space-based
lensing surveys $(n_{gal}=100\ {\rm arcmin}^{-2})$  Though the quality of
the reconstruction understandably degrades, the lower panels show that
large features can be recovered even with up to 50\% of the pixels masked.

In Figure~\ref{fig_kappa_reconstruction} we provide a comparison of the
convergence maps generated from the noiseless shear (upper panels) and
the KL-reconstructed noisy shear with 20\% of pixels masked (lower panels).  
The convergence maps are smoothed by a Gaussian filter to 
ease comparison with the quantitative
results of Sections~\ref{Shear_Peaks}-\ref{Discussion}, where we explore
the distribution of peaks through an aperture mass filter.  The aperture
mass filter amounts to a particular smoothing function over the 
convergence field (see Section~\ref{Aperture_Mass}, below).
Comparison of the upper and lower panels of 
Figure~\ref{fig_kappa_reconstruction} give a qualitative indication of
the performance of KL: high-convergence regions are recovered remarkably
well, while convergence peaks of lower magnitude are obscured by the 
background noise: as we show in the following sections, 
this obscuration is largely the result of shape noise 
in the simulated shear measurements.

For a quantitative analysis of the effectiveness of the KL interpolation 
in convergence mapping, and the potential biases it introduces, 
a large-scale statistical measure is most appropriate.
In the following sections, we test the utility of this KL interpolation scheme
within the framework of shear peak statistics.

\section{Shear Peak Statistics}
\label{Shear_Peaks}
It has long been recognized that much useful cosmological information 
can be deduced from the masses and spatial distribution of galaxy clusters 
\citep[e.g.][]{Press74}.  
Galaxy clusters are the largest gravitationally bound objects in the universe,
and as such are exponentially sensitive to cosmological parameters
\citep{White93}.  The spatial distribution of clusters and redshift evolution
of their abundance and clustering is sensitive to both geometrical effects
of cosmology, as well as growth of structure.  Because of this, cluster
catalogs can be used to derive constraints on many interesting 
cosmological quantities, including the matter density $\Omega_M$ 
and power spectrum normalization $\sigma_8$ \citep{Lin03}, 
the density and possible evolution of dark energy \citep{Linder03,
 Vikhlinin09b},
primordial non-gaussianities \citep{Matarrese00,Grossi07}, 
and the baryon mass fraction \citep{Lin03,Giodini09}.

Various methods have been developed to measure the mass and 
spatial distribution of galaxy clusters, and each are subject to 
their own difficult astrophysical and observational biases.
They fall into four broad categories: optical or infrared richness, 
X-ray luminosity and surface brightness, 
Sunyaev-Zeldovich decrement, and weak lensing shear.

While it was long thought that weak gravitational lensing studies would lead
to robust, purely mass-selected cluster surveys, it has since become 
clear that shape noise and projection effects limit the usefulness 
of weak lensing in determining the 3D cluster mass function 
\citep{Hamana04,Hennawi05,Mandelbaum10,Vanderplas2011}.  
The shear observed in weak lensing is non-locally related to the convergence,
a measure of \textit{projected} mass along the line of sight.  
The difficulty in deconvolving the correlated and uncorrelated projections
in this quantity leads to difficulties in relating these projected peak
heights to the masses of the underlying clusters in three dimensions.
Recent work has shown, however, that this difficulty 
in relating the observed quantity 
to theory may be overcome through the use of statistics of
the projected density itself.

\citet{Marian09,Marian10} first explored the extent to which 2D 
projections of the
3D mass field trace cosmology.  They found, rather surprisingly, that the
statistics of the projected peaks closely trace the statistics of the 3D
peak distribution: in N-body simulations, both scale 
with the \citet{Sheth99} analytic
scaling relations.  The same correlated projections which bias
cluster mass estimates contribute to a usable signal: 
statistics of projected mass alone can provide useful cosmological 
constraints, without the need for bias-prone conversions from peak 
height to cluster mass.

A host of other work has explored diverse aspects of these
shear peak statistics,
including tests of these methods with ensembles of N-body simulations
\citep{Wang09,Kratochvil10,Dietrich10},
the performance of various filtering functions
and peak detection statistics \citep{Pires09,Schmidt10,Kratochvil11},
exploration of the spatial correlation of noise with signal
within convergence maps \citep{Fan10},
and exploration of shear-peak constraints on primordial non-gaussianity
\citep{Maturi11}.  The literature has yet to converge
on the ideal mapping procedure: convergence maps, Gaussian filters,
various matched filters, wavelet transforms, and more novel filters
are explored within the above references.  There is also variation in
how a ``peak'' is defined: simple local maxima, ``up-crossing'' criteria, 
fractional areas above a certain threshold, connected-component labeling, 
hierarchical methods, and Minkowski functionals are all shown to be useful.  
Despite diverse methodologies, all the above work confirms that there 
is useful cosmological information within the projected peak 
distribution of cosmic shear fields, and that this information adds to 
that obtained from 2-point statistics alone.

\subsection{Aperture Mass Peaks}
\label{Aperture_Mass}
Based on this consensus, we use shear peak statistics to explore
the possible bias induced by the KL interpolation method outlined above.  
We follow the aperture mass methodology of \citet{Dietrich10}:
The aperture mass magnitude at a point $\vec\theta_0$ is given by
\begin{equation}
  \label{Map_def}
  M_{\rm ap}(\vec{\theta_0})
  = \int_\Omega d^2\theta Q_{\rm NFW}(\vartheta=|\vec\theta-\vec\theta_0|) 
  \gamma_t(\vec\theta;\vec\theta_0)
\end{equation}
where $\gamma_t(\vec\theta;\vec\theta_0)$ is the component of the shear at 
location $\vec\theta$ tangential to the line $\vec\theta-\vec\theta_0$, 
and $Q_{\rm NFW}$ is the 
NFW-matched filter function defined in \citet{Schirmer07}:
\begin{equation}
  Q_{\rm NFW}(x;x_c) \propto \frac{1}{1+e^{6-150x}+e^{-47+50x}}
  \frac{\tanh(x/x_c)}{x/x_c}
\end{equation}
with $x = \vartheta/\vartheta_{\rm max}$ and $x_c$ a free parameter. We follow
\citet{Dietrich10} and set $\vartheta_{\rm max}=5.6\ \mathrm{arcmin}$ and
$x_c=0.15$.  The integral in Equation~\ref{Map_def} is over the whole sky, 
though the filter function $Q$ effectively cuts this off at a radius
$\vartheta_{\rm max}$.  
In the case of our pixelized shear field, the integral is converted to 
a discrete sum over all pixels, with
$\vartheta$ equal to the distance between the pixel centers:
\begin{equation}
  \label{Map_def_discrete}
  M_{\rm ap}(\vec{\theta}_i) 
  = \sum_j Q_{\rm NFW}(\vartheta_{ij}) 
  \gamma_t(\vec{\theta}_i;\vec{\theta}_j) 
\end{equation}
where we have defined $\vartheta_{ij} \equiv |\vec\theta_i-\vec\theta_j|$.
%Figure~\ref{fig_Map_reconstruction} shows the aperture mass calculated
%from the four shear fields shown in Figure~\ref{fig_reconstruction}.   

We can similarly compute the \textit{B-mode} aperture mass, by substituting
$\gamma_t \to \gamma_\times$ in Equations~\ref{Map_def}-\ref{Map_def_discrete}
\citep{Crittenden02}.  For pure gravitational
weak lensing with an unbiased shear estimator, the B-mode signal is 
expected to be negligible, though second-order effects such as 
source clustering and intrinsic 
alignments can cause contamination on small angular scales 
\citep{Crittenden02,Schneider02b}
These effects aside, the B-mode signal can be used as a rough estimate of the
systematic bias of a particular analysis method.

For our study, the aperture mass is calculated with the same resolution as the
shear pixelization: $64^2$ pixels per square degree.  A pixel is defined to be
a peak if its value is larger than that of the surrounding eight pixels:
a simple local maximum criterion. 

\begin{figure*}
 \centering
 \includegraphics[width=\textwidth]{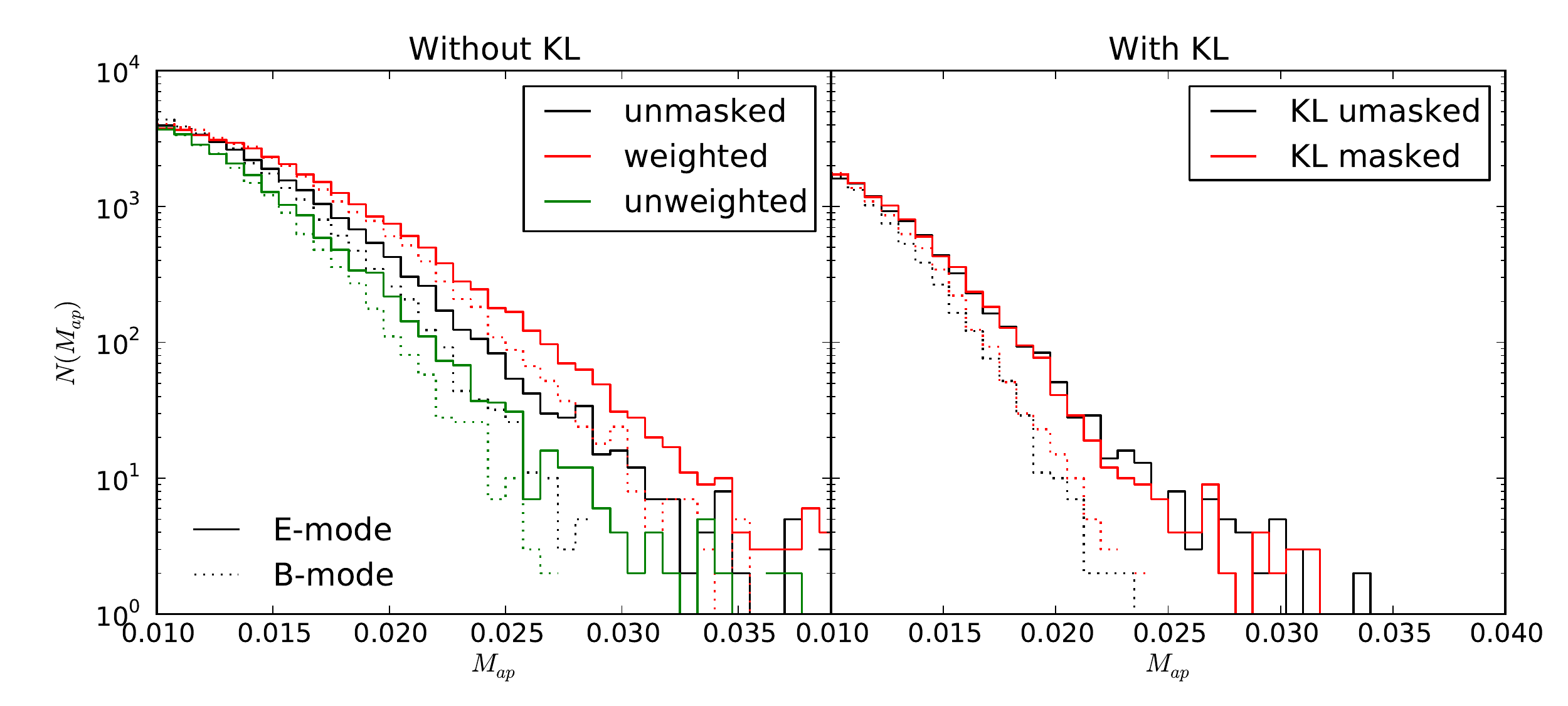}
 %created by ../fig_code/fig07_mask_nomask.py
 \caption[Comparison of the masked and unmasked peak distributions.]{
   Comparison of the masked and unmasked peak distributions.
   \textit{left panel:} the peak distributions without the use of KL.
   The black line is the result with no masking, while the red and green
   lines show the two \naive methods of correcting for the mask (see
   Section~\ref{MaskingEffects}).
   \textit{right panel:}  the masked and unmasked peak distributions
   after applying KL. 
   Neither \naive method of mask-correction adequately recovers 
   the underlying peak distribution.  It is evident, however, that 
   the KL-based interpolation procedure recovers a mass map with 
   a similar peak distribution to the unmasked KL map.
   It should be noted that the unmasked peak distribution 
   (black line, left panel) is not identical to the unmasked peak distribution
   after application of KL (black line, right panel).  This difference is
   addressed in Figure~\ref{fig_num_peaks}.
   \label{fig_mask_nomask} 
 }
\end{figure*} 

\subsection{The Effects of Masking}
\label{MaskingEffects}
When a shear peak statistic is computed across a field with masked regions,
the masking leads to a bias in the peak height distribution
(see Section~\ref{Discussion} below).  Moreover, due
to the non-local form of the aperture mass statistic, a very large region is
affected: in our case, a single masked pixel biases the aperture mass 
measurement of an area of size
 $\pi(2\vartheta_{\rm max})^2 \approx 400\ {\rm arcmin}^2$.  
There are two \naive approaches one could use when measuring the 
aperture mass in this situation:

\begin{description}
  \item[Unweighted:] Here we simply set the shear value within each 
    masked pixel to zero, and apply Equation~\ref{Map_def_discrete}.  
    The shear within the masked regions do not contribute to the peaks, so the
    height of the peaks will be underestimated.
  \item[Weighted:] Here we implement a weighting scheme which
    re-normalizes the filter $Q_{\rm NFW}$ to reflect the
    reduced contribution from masked pixels.  The integral in 
    equation \ref{Map_def} is replaced by the normalized sum:
    \begin{equation}
      M_{\rm ap}(\vec\theta_i) 
      = \frac{\sum_j 
        Q_{\rm NFW}(\vartheta_{ij})
        \gamma_t(\vec\theta_j)w(\vec\theta_j)}
      {\sum_j 
        Q_{\rm NFW}(\vartheta_{ij})w(\vec\theta_j)}
    \end{equation}
    where $w(\vec\theta_j) = 0$ if the pixel is masked, and 
    $1$ otherwise.  This should correct for the underestimation of peak
    heights seen in the unweighted case.
\end{description}
In order to facilitate comparison between this weighted definition of \Map 
and the normal definition used in the unmasked and unweighted cases, 
we normalize the latter by $\sum_j Q_{\rm NFW}(\vec\theta_{ij})$, 
which is a constant normalization across the field.

\begin{figure}
 \centering
 \includegraphics[width=0.8\textwidth]{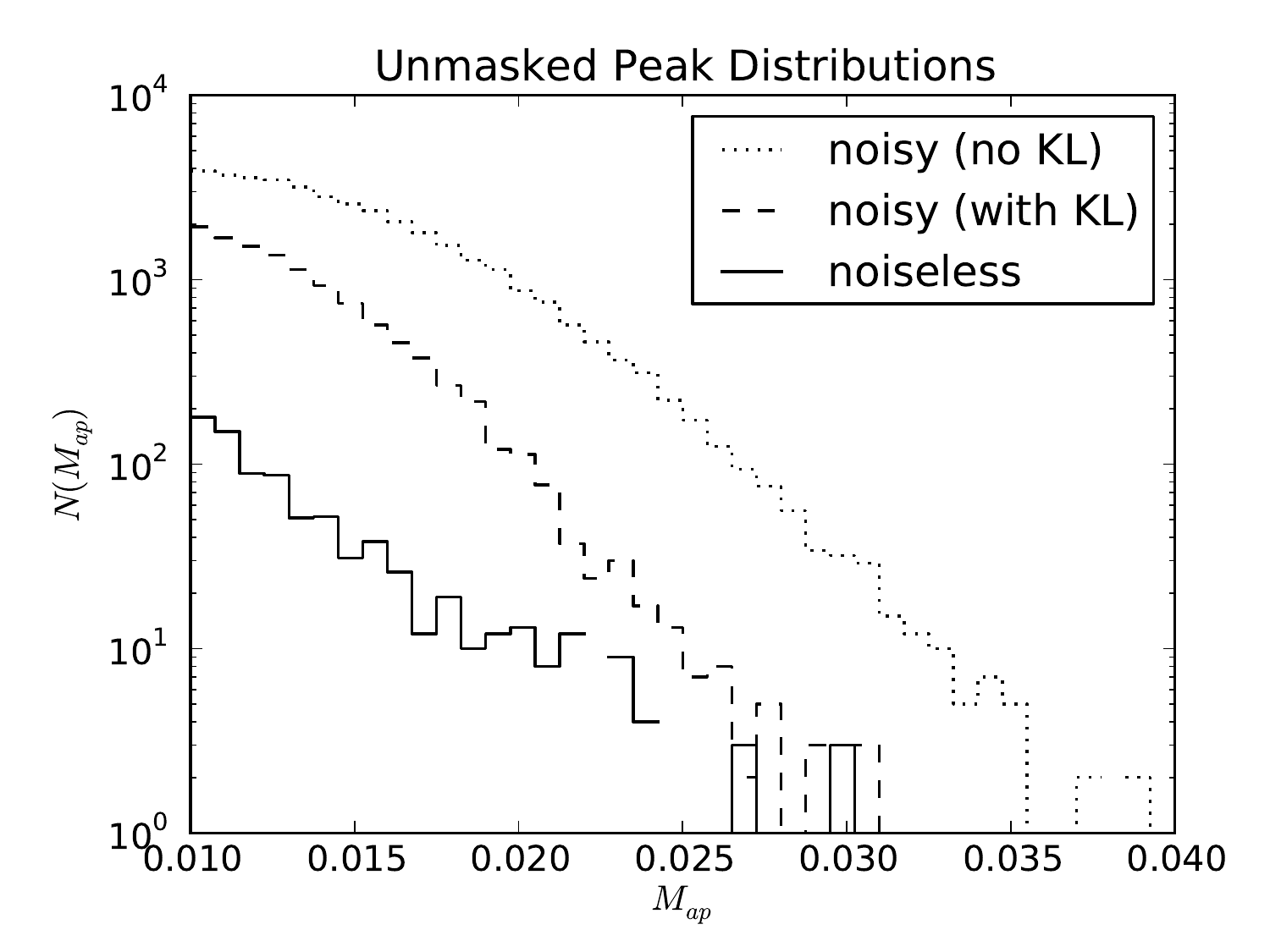}
 %created by ../fig_code/fig08_num_peaks.py
 \caption[Comparison of the distribution of \Map peaks for unmasked shear]{
   Comparison of the distribution of \Map peaks for unmasked shear,
   before and after filtering
   the field with KL (dotted line and dashed line, respectively).  The
   peak distribution in the absence of noise is shown for comparison
   (solid line).  It is clear that the addition of shape noise leads to 
   many spurious \Map peaks: noise peaks outnumber true peaks by nearly a
   factor of 10 for smaller peak heights.  Filtering by KL reduces these
   spurious peaks by about a factor of 3, and for larger peaks leads
   to a distribution similar in scale to that of the noiseless peaks.
   \label{fig_num_peaks}  
 } 
\end{figure}

Note that in both cases, it is the \textit{shear} that is masked, not the
\Map peaks.  Aperture mass is a non-local measure, so that the value can
be recovered even within the masked region.  This means that
masking will have a greater effect on the observed magnitude of the peaks
than it will have on the count.  In particular, on the small end of the
peak distribution, where the peaks are dominated by shape noise, the
masking of the shear signal is likely to have little effect on the
distribution of peak counts.  This can be seen in Figure~\ref{fig_num_peaks}.

\subsection{\Map Signal-to-Noise}
It is common in shear peak studies to study signal-to-noise peaks 
rather than directly study aperture-mass or convergence
peaks \citep[e.g.~][]{Wang09,Dietrich10,Schmidt10}.
We follow this precedent here.  
The aperture mass (Eqn.~\ref{Map_def_discrete}) is defined in terms of the
tangential shear.  Because we assume that the shear measurement is dominated
by isotropic, uncorrelated shape noise, the noise covariance of 
\Map can be expressed
\begin{eqnarray}
  [\Noise_M]_{ij} &\equiv& \langle M_{\rm ap}(\vec\theta_i) 
  M_{\rm ap}(\vec\theta_j)\rangle  \nonumber\\
  & = & \frac{1}{2}\sum_{k}
  Q^2_{\rm NFW}(\vartheta_{ik}) [\Noise_\gamma]_{kk} \delta_{ij}
\end{eqnarray}
where we have used the fact that shape noise is uncorrelated:
$[\Noise_\gamma]_{ij} = \langle n_i n_j\rangle \propto \delta_{ij}$.

In the case of a KL-reconstruction of a masked shear field, the reconstructed
shear has non-negligible correlation of noise between pixels.  
From Equations~\ref{shear_recons}-\ref{a_WF}, it can be shown that
\begin{eqnarray}
  \Noise_{\hat\gamma} 
  & \equiv  &
  \langle\myvec{\hat\gamma} \myvec{\hat\gamma}^\dagger\rangle\nonumber\\
  & = & \left[\Noise_\gamma^{1/2}\myvec\Psi \mymat{M}^{-1} \mymat{\Psi}^\dagger\right]
  \mymat{W}^2 
  \left[\mymat{\Psi} \mymat{M}^{-1} \mymat{\Psi}^\dagger \Noise_\gamma^{1/2}\right]
\end{eqnarray}
The covariance matrix $\Noise_{\hat\gamma}$ is no longer diagonal, but
the noise remains isotropic under the linear transformation, so that
$\Noise_{\hat{\gamma}_t} = \Noise_{\hat{\gamma}}/2$.
The aperture mass noise covariance can thus be calculated in a similar 
way to the non-KL case:
\begin{equation}
  [\Noise_M]_{ij} = \frac{1}{2}\sum_k\sum_\ell Q_{\rm NFW}(\vartheta_{ik})
  Q_{\rm NFW}(\vartheta_{j\ell}) [\Noise_{\hat\gamma}]_{k\ell}.
\end{equation}
This expression can be computed through standard linear algebraic techniques.  
The aperture mass signal-to-noise in each pixel is given by 
\begin{equation}
  [S/N]_i = M_{\rm ap}(\vec\theta_i)/\sqrt{[\mymat{\Noise_M}]_{ii}}
\end{equation}

\begin{figure}
 \centering
 \includegraphics[width=0.8\textwidth]{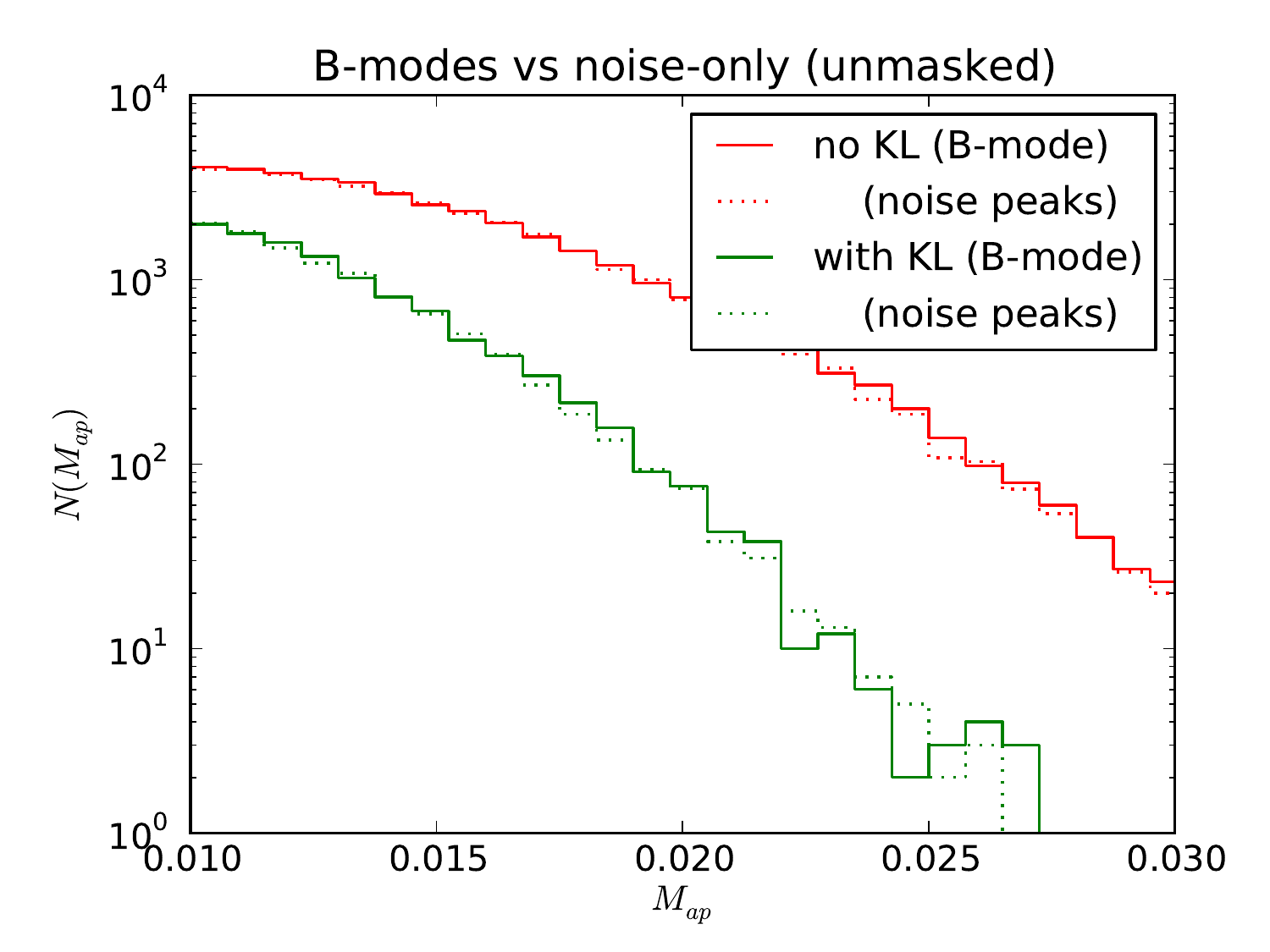}
 %created by ../fig_code/fig09_B_noise.py
 \caption[B-mode peak distributions]
 {The comparison between B-mode peak distributions and the
   peak distributions for a shear field composed entirely of noise.
   As expected, the B-mode peak distributions are largely consistent
   with being due to noise only.  Because of this, we can use B-mode
   peaks as a rough proxy for the noise.
   \label{fig_B_noise} 
 }
\end{figure}

\begin{figure} 
 \centering
 \includegraphics[width=0.8\textwidth]{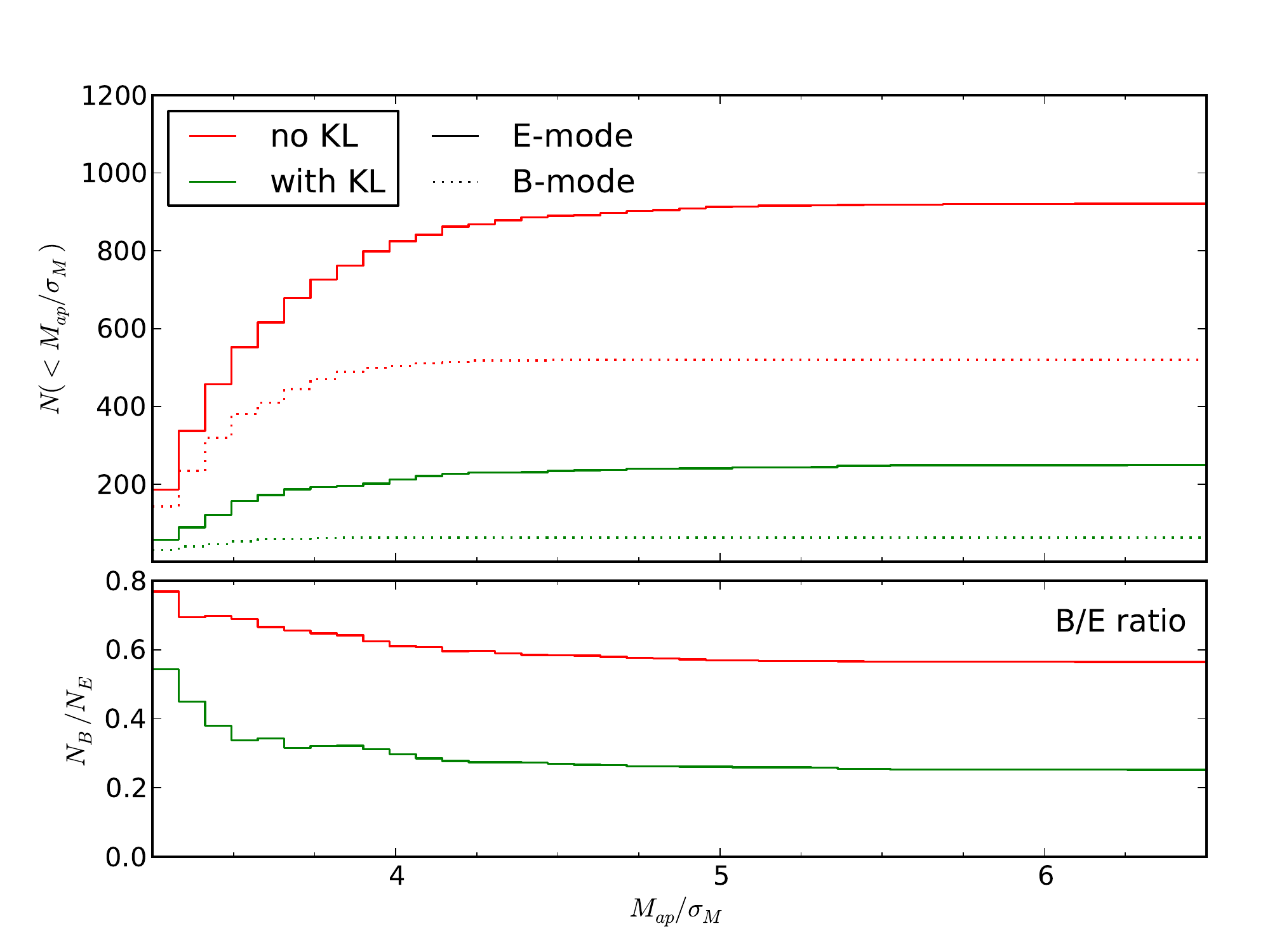}
 %created by ../fig_code/fig10_EB_comp.py
 \caption[Distributions of peaks with and without KL]{
   \textit{top panel:} The cumulative distributions of peaks in 
   signal-to-noise, for peaks with $M_{\rm ap}/\sigma_M > 3.25$.
   This is the statistic used by \citet{Dietrich10} to discriminate
   between cosmological models.  \textit{bottom panel:}  The ratio
   of B-mode to E-mode peak distributions.  Filtration by KL reduces
   the relative number of B-mode peaks by about 1/3.  Because B-mode
   peaks are a proxy for contamination by shape noise
   (see Figure~\ref{fig_B_noise}), this indicates that KL-filtration
   results in peak distributions less affected by statistical errors.
   KL also reduces the total number of both E and B peaks by about 2/3; 
   this effect can also be seen in Figure~\ref{fig_num_peaks}.
   \label{fig_EB_comp} 
 }
\end{figure}

\section{Discussion}
\label{Discussion}
\subsection{\Map Peak Distributions}
In Figures~\ref{fig_mask_nomask}-\ref{fig_EB_comp}
we compare the peak distribution obtained with and without KL.
We make three broad qualitative observations which point to the
efficacy of KL in interpolation of masked shear fields, and in
the filtration of shape-noise from these fields.  We stress the
qualitative nature of these results: quantifying 
these observations in a statistically rigorous way would require
shear fields from an ensemble of cosmology simulations, which is beyond the
scope of this work.  These results nevertheless point to the
efficacy of KL analysis in this context.

\textbf{KL filtration corrects for the bias due to masking.}
Figure~\ref{fig_mask_nomask} compares the effect of masking on the
resulting peak distributions with and without KL.  The left panel
shows the unmasked noisy peak distribution, and the masked
peak distributions resulting from the weighted and unweighted
approaches described in Section~\ref{MaskingEffects}.  
Neither method of accounting for the masking accurately 
recovers the unmasked distribution of peak heights. The unweighted
approach (green line) leads to an underestimation of peak heights.  
This is to be expected, because it does not correct for the missing 
information in the masked pixels.  The weighted approach, on the other
hand, over-estimates the counts of the peaks.  We suspect
this is due to an analog of Eddington bias: the lower signal-to-noise 
ratio of the weighted peak statistic leads to a larger scatter in peak heights.
Because of the steep slope of the peak distribution, this scatter 
preferentially increases the counts of larger peaks.  This suspicion is
confirmed by artificially increasing the noise in the unmasked peak function.
Increasing $\sigma_\epsilon$ from 0.30 to 0.35 in the unmasked case
results in a nearly identical peak function to the weighted masked case.

The right panel of Figure~\ref{fig_mask_nomask} shows that when 
KL is applied to the shear field, the distribution of the masked and
unmasked peaks is very close, both for E-mode and B-mode peaks.
This indicates the success of the KL-based interpolation outlined in
Section~\ref{KL_Interpolation}.  Even with 20\% of the pixels masked,
the procedure can recover a nearly identical peak distribution as 
from unmasked shear.

\textbf{KL filtration reduces the number of noise peaks.}
Comparison of the unmasked lines in the left and right panels of 
Figure~\ref{fig_mask_nomask} shows that application of KL to a shear
field results in fewer peaks at all heights.  This is to be expected:
when a reconstruction is performed with fewer than the total number
of KL modes, information of high spatial frequency is lost.  In this
way, KL acts as a sort of low-pass filter tuned to the particular
signal-to-noise characteristics of the data.  Figure~\ref{fig_num_peaks}
over-plots the KL and non-KL peak distributions with the noiseless
peak distribution.  From this figure we see that the inclusion of
shape noise results in nearly an order-of-magnitude more peaks than 
the noiseless case.  The effect of noise on peak counts lessens slightly 
for higher-\Map peaks: this supports the decision of \citet{Dietrich10} 
to limit their distributions to peaks with a signal-to-noise 
ratio greater than 3.25: the vast majority of peaks are lower 
magnitude, and are overwhelmed by the effect of shape noise.

Omission of higher-order KL modes of shear field reduces the number of
these spurious peaks by a factor of 3 or more.  For low-magnitude peaks,
$(M_{\rm ap}<\sim 0.02)$, KL still produces peak counts which are 
dominated by noise. For higher-magnitude peaks, the number of observed 
KL peaks more closely approaches the number of peaks in the noiseless case.

\textbf{KL filtration reduces the presence of B-modes.}
To first order, weak lensing shear is expected to consist primarily of
curl-free, E-mode signal.  Because of this, the presence of B-modes can
indicate a systematic effect.  It is not obvious that filtration by KL
will maintain this property: as noted in Section~\ref{sec:WhichCorrelation},
KL modes individually are agnostic to E-mode and B-mode information.  
E\&B information is only recovered within a 
complex-valued linear combination of the set KL modes.

Figure~\ref{fig_B_noise} shows a comparison between the unmasked
B-mode peak functions from Figure~\ref{fig_mask_nomask} and the associated
E-mode peak functions due to shape-noise only.  For both the non-KL version
and the KL version, the B-mode peak distributions closely follow the
distributions of noise peaks.  This supports the use of B-mode peaks as
a proxy for the peaks due to shape noise, even when truncating higher-order
KL modes.

The near-equivalence of B-modes and noise-only peaks shown in 
Figure~\ref{fig_B_noise} suggests a way of recovering the true peak function,
by subtracting the B-mode count from the E-mode count as a proxy for the 
shape noise. This approach has one fatal flaw: because it involves computing
the small difference between two large quantities, the result has
extremely large uncertainties.  It should be
noted that this noise contamination of small peaks is not an impediment to
using this method for cosmological analyses: the primary information
in shear peak statistics is due to the high signal-to-noise peaks.

In the top panel of Figure~\ref{fig_EB_comp}, we show the cumulative 
distribution of peaks in signal-to-noise, for peaks with $S/N > 3.25$:
the quantity used as a cosmological discriminant in \citet{Dietrich10}.
The difference in the total number of E-mode peaks in the KL and non-KL
approaches echoes the result seen
in Figure~\ref{fig_num_peaks}: truncation of higher-order KL modes acts as
a low-pass filter, reducing the total number of peaks by a factor of $\sim 3$.
More interesting is the result shown in the lower panel of 
Figure~\ref{fig_EB_comp}, where the ratio of B-mode peak counts to E-mode 
peak counts is shown.  
Before application of KL, the B-mode contamination is above 
30\%.  Filtration by KL reduces this contamination by a factor of $\sim 3$,
to about 10\%.  This indicates that the truncation of higher-order KL modes
leads to a preferential reduction of the B-mode signal, which traces
the noise.  This is a promising observation: the counts of
high signal-to-noise peaks, 
which offer the most sensitivity to cosmological parameters 
\citep{Dietrich10}, are significantly less contaminated by noise after 
filtering and reconstruction with KL.  This is a strong indication that 
the use of KL could improve the cosmological 
constraints derived from studies of shear peak statistics.

Note that in Figure~\ref{fig_EB_comp} we omit the masked 
results for clarity.  The masked cumulative signal-to-noise peak 
functions have B/E ratios comparable to the unmasked versions, so 
the conclusions here hold in both the masked and unmasked cases.

\subsection{Remaining Questions}
The above discussion suggests that KL analysis of masked shear fields holds
promise in constraining cosmological parameters of shear peaks in both
masked and unmasked fields.  KL greatly
reduces the number of spurious noise peaks at all signal-to-noise levels.
It minimizes the bias between masked and unmasked constructions, and leads
to a factor of 3 suppression of the B-mode signal, which is a proxy
for the spurious signal introduced through shape noise.

The question remains, however, how much cosmological information is contained
in the KL peak functions.  The reduction in level of noise peaks is promising,
but the omission of higher-order modes in the KL reconstruction leads to 
a smoothing of the shear field on scales smaller than the cutoff mode.
This smoothing could lead to the loss of cosmologically useful information.
In this way, the choice of KL mode cutoff can be thought of as a balance 
between statistical and systematic error.
The effect of these competing properties on cosmological
parameter determination is difficult to estimate.  Quantifying this effect
will require analysis within a suite of synthetic shear maps, similar to the
approach taken in previous studies \citep[e.g.][]{Dietrich10,Kratochvil10}, 
and will be the subject of future work.

Another possible application of KL in weak lensing is to use KL to 
directly constrain 2-point information in the measured shear data.  
In contrast to the method outlined in the current work, 
KL basis functions can be computed for the unmasked region only.
The projection of observed data onto this basis can be used to 
directly compute cosmological parameters via the 2-point function, 
without ever explicitly calculating the power spectrum.  
This is similar to the approach taken for galaxy counts
in \citet{Vogeley96}.  This approach is the subject of Chapter 5.

%% file: chapter5.tex
\chapter{Application to COSMOS lensing data}

This chapter will cover the application of KL parameter estimation to
lensing data from the COSMOS survey, a $\sim 1.5$ square degree field
observed by the Hubble Space Telescope. We use KL to express the data
within the optimal orthonormal basis dictated by the 
survey geometry, and use the Bayesian inference framework developed in
\S\ref{sec:KL_bayes} to perform a simple parameter estimation using
the KL basis in place of the usual Fourier basis.  From this, we obtain
parameter constraints on $\Omega_M$ and $\sigma_8$ which are similar to
those from conventional angular correlation analyses, with a framework that
is free from the systematic errors associated with incomplete sky coverage
and irregular survey geometry.

This chapter represents a first exploration of this problem; the results
consider a simple two dimensional analysis for two cosmological parameters.
KL can naturally be extended to 3D tomographic approaches with any number of
parameters; this will be the subject of future work.

\section{Introduction}
\label{sec:introduction}

In this chapter we explore the evaluation of cosmological
likelihoods using KL analysis of shear fields.  
In Chapter 4 we explored the use of KL analysis in shear surveys,
focusing on the ability of
KL modes to help fill-in missing information within the context of weak
lensing convergence mapping and studies of the peak statistics of the
resulting mass maps.
Here we follow a different approach:
we use KL analysis to aid in the calculation of cosmological likelihoods
using two-point statistics within a Bayesian framework.
This draws upon similar work done
previously to constrain cosmological parameters using number counts of
galaxy surveys \citep{Vogeley96, Pope04}.

In \S\ref{sec:lensing_intro} we review and discuss the strengths and
weaknesses of constraining cosmological quantities using two-point shear
statistics. In \S\ref{sec:kl_intro} we review KL analysis and its application
to shear surveys.  In \S\ref{sec:data} we describe the COSMOS shear data
used in this analysis, and we discuss these results in \S\ref{sec:results}.

\section{Two-point Statistics in Weak Lensing}
\label{sec:lensing_intro}
As noted and outlined in Chapter 1, the large-scale structure
of the Universe provides a powerful probe of
cosmological parameters.  Through gravitational instability, the initial
matter fluctuations have grown to the nonlinear structure we see today.
This happens in a hierarchical manner, with the smallest structures
collapsing before the largest.  One of the most powerful probes of this
structure is the redshift-dependent power spectrum of matter density
fluctuations, $P_k(z)$, which gives the amplitude of the
Fourier mode with wave-number $k$ at a redshift $z$. 
This approach has often been used to measure cosmological parameters
through optical tracers of the underlying dark matter structure
\citep[e.g.][]{Tegmark06}.  In this chapter we explore the use of
weak lensing measurements of the matter power spectrum.
Recent work has shown the power of this lensing-based
approach \citep{Ichiki09, Schrabback10}.

The are two approaches to measuring two-point information are 
mathematically equivalent: the power spectrum $\mathcal{P}(\ell)$,
and its Fourier transform $\xi(\theta)$.  In practice,
the most commonly used method of measuring two-point information
is through correlation functions \citep[see][]{Schneider02}.
The main advantage of correlation functions is their ease of measurement:
they can be straightforwardly estimated from the positions and shapes of
galaxies, even in very complicated survey geometries.
Their disadvantage is that the
signal is highly correlated between different scales.  Accounting for this 
correlation is very important when computing cosmological likelihoods,
and often requires large suites of simulations.

Shear power spectra, on the other hand, have a number of nice properties.
Compared to correlation functions, they provide a simpler mapping
to theoretical expectations.  They have
weaker correlations between different multipoles: on the largest scales,
where structure is close to Gaussian, the scales are expected to be
statistically independent.
Even on small scales where non-Gaussianity leads to correlated errors,
these correlations have a relatively small effect on derived cosmological
constraints \citep{Takada09}.
The disadvantage of shear power spectra as direct cosmological probes is
the difficulty of measuring them from data.  In particular, survey geometry
effects such as incomplete sky coverage
and masking can lead to mixing of power on all angular scales.
This mode-mixing is a direct result of the loss of orthogonality:
spherical harmonics are orthogonal over the entire sky, but are not
necessarily orthogonal over the incomplete patch of the sky represented by
lensing surveys.  Even in the case of future all-sky surveys, the masking from
foreground sources will pose a problem.  This means
that the spherical harmonic decomposition on which power spectra are based
is not unique for realistic surveys.  
It may be possible to construct
a survey in order to limit the magnitude of these effects
\citep[see][for some approaches]{Kilbinger04, Kilbinger06}.
There have also been a few attempts to correct for this
difficulty through direct deconvolution of the survey geometry from
the correlation signal \citep{Brown03, Hikage11}, but because of the
computational difficulty involved with these methods,
results based on correlation function measures remain more common.
Here we explore an alternate approach which relies on constructing
a new set of orthogonal modes for the observed survey geometry.
Because the new modes are orthogonal by construction,
one can avoid the difficulties associated with mode mixing. We propose to
take this latter approach using Karhunen-Lo\'{e}ve (KL) analysis.

\section{KL for Parameter Estimation}
\label{sec:kl_intro}
As discussed more fully in Chapter 2,
KL analysis and the related Principal Component Analysis are well-known
statistical tools which have been applied in a wide variety of astrophysical
situations, from e.g. analysis of the spatial power of galaxy counts
\citep{Vogeley96, Szalay03, Pope04}
to characterization of stellar, galaxy, and QSO spectra
\citep{Connolly95, Connolly99, Yip04a, Yip04b},
to studies of noise properties of weak lensing surveys
\citep{Kilbinger06, Munshi06}, and a host of other situations too numerous
to mention here.  Informally, the power of KL/PCA rests in the fact that 
it allows a highly efficient representation of a set of data, highlighting
the components that are most important in the dataset as a whole.
Though the framework is discussed more completely in Chapter 2, we will
review the most important points here.
The discussion of KL analysis below derives largely from \citet{Vogeley96},
reexpressed for application in cosmic shear surveys.

Any $D$-dimensional data point may be completely represented
as a linear combination of $D$ orthogonal basis functions: this is
a geometrical property, closely linked to the free choice of coordinate axes
used to represent points in a $D$-dimensional space.
For example, the data may be $N$ individual galaxy spectra, each with flux
measurements in $D$ wavelength bins.  Each spectrum can be thought of as a
single point in $D$-dimensional parameter space, where each axis corresponds
to the value within a single wavelength bin.  
Geometrically, there is nothing special about
this choice of axes: one could just as easily rotate and translate the axes
to obtain a different but equivalent representation of the same data.

In the case of of a shear survey, our single data vector is the set of
cosmic shear measurements across the sky.  We will divide the sky into $N$
cells in angular and redshift space, at coordinates
$\myvec{x}_i = (\theta_{x,i}, \theta_{y,i}, z_i)$
These cells may be spatially distinct, or they may overlap.
From the ellipticity of the galaxies within each cell, we
estimate the shear
$\gamma_i \equiv \gamma^o(\myvec{x}_i) = 
\gamma(\myvec{x}_i) + n_\gamma(\myvec{x}_i)$
where $\gamma(\myvec{x}_i)$ is the true underlying shear,
and $n_\gamma(\myvec{x}_i)$ is the measurement noise.
Our data vector is then
$\myvec{\gamma} = [\gamma_1, \gamma_2 \cdots \gamma_N]^T$.

We seek to express our set of measurements $\myvec{\gamma}$
as a linear combination of $N$ (possibly complex) 
orthonormal basis vectors
$\{\myvec{\Psi}_j(\myvec{x}_i, j=1,N)\}$ with complex coefficients
$a_j$:
\begin{equation}
  \label{eq:gamma_decomp}
  \gamma_i = \sum_{j=1}^{N} a_j \Psi_j(\myvec{x}_i)
\end{equation}
For conciseness, we'll create the matrix $\mymat{\Psi}$ whose columns are
the basis vectors $\myvec{\Psi}_j$, so that the above equation can be
compactly written $\myvec\gamma = \mymat\Psi\myvec{a}$.  Orthonormality
of the basis vectors leads to the property
$\mymat\Psi^\dagger\mymat\Psi = \mymat{I}$, where $\mymat{I}$ is the identity
matrix: that is, $\mymat\Psi$ is a unitary matrix with
$\mymat\Psi^{-1} = \mymat\Psi^\dagger$.  Observing this, we can easily compute
the coefficients for a particular data vector:
\begin{equation}
  \myvec{a} = \mymat\Psi^\dagger \myvec\gamma.
\end{equation}
We will be testing the likelihood of a particular set of coefficients
$\myvec{a}$.  
The statistical properties of these coefficients can be written in terms of
the covariance of the observed shear:
\begin{equation}
  \label{eq:a_cov}
  \left\langle \myvec{a}\myvec{a}^\dagger \right\rangle 
  =  \mymat\Psi^\dagger
  \left\langle \myvec\gamma\myvec\gamma^\dagger \right\rangle 
  \mymat\Psi
  \equiv \mymat\Psi^\dagger \myvec{\xi}  \mymat\Psi
\end{equation}
where we have defined the observed shear correlation matrix 
$\myvec{\xi} \equiv \left\langle 
\myvec\gamma\myvec\gamma^\dagger \right\rangle$, and angled braces
$\langle\cdots\rangle$ denote expectation value or ensemble average
of a quantity.

In order to perform a likelihood analysis on the coefficients
$\myvec{a}$, we will require that $\myvec{a}$ be statistically orthogonal:
\begin{equation}
  \label{eq:a_cov_2}
  \left\langle \myvec{a}\myvec{a}^\dagger \right\rangle_{ij}
  = \left\langle a_i^2 \right\rangle \delta_{ij}
\end{equation}
Comparing Equations \ref{eq:a_cov} \& \ref{eq:a_cov_2} we see that the desired
basis functions are the solution of the eigenvalue problem
\begin{equation}
  \mymat\xi \myvec\Psi_j = \lambda_j \myvec\Psi_j
\end{equation}
where the eigenvalue $\lambda_j = \left\langle a_i^2 \right\rangle$.
Comparison of this to the KL framework outlined in Chapter 2 shows that
the unique basis with these properties is given by the KL decomposition
of the shear field $\gamma$, represented by the correlation matrix
of observations $\myvec{\xi}$.
By convention, we'll again order the eigenvalue/eigenvector pairs such that
$\lambda_i \ge \lambda_{i+1} \forall i\in(1, N-1)$.
Expansion of the data $\myvec\gamma$ into this basis is the discrete form
of KL analysis.

In chapter 2 we discussed the Uniqueness, Efficiency, and Signal-to-noise
optimality of KL modes.  In particular, we showed that
if signal and noise are uncorrelated, then the covariance of the observed
shear can be decomposed as
\begin{equation}
  \mymat{\xi} = \mymat{\mathcal{S}} + \mymat{\mathcal{N}}
\end{equation}
where $\mymat{\mathcal{S}}$ is the covariance of the signal, and
$\mymat{\mathcal{N}}$ is the covariance of the noise.
Because the noise covariance $\mymat{\mathcal{N}} \equiv 
\langle\myvec{n_\gamma}\myvec{n_\gamma}^\dagger\rangle$ is proportional
to the identity by assumption, diagonalization of $\mymat{\xi}$ results
in a simultaneous diagonalization of both the signal $\mymat{\mathcal{S}}$
and the noise $\mymat{\mathcal{N}}$.  Based on this signal-to-noise
optimization property, KL modes can be proven to be the optimal basis
for testing of spatial correlations \citep[see Appendix A of][]{Vogeley96}.

\subsection{Shear Noise Properties}
\label{sec:whitening}
The signal-to-noise properties of shear mentioned above are based on the 
requirement that noise be ``white'', that is, the noise covariance is
$\mymat{\mathcal{N}} \equiv 
\langle\myvec{n_\gamma}\myvec{n_\gamma}^\dagger\rangle
= \sigma^2 \mymat{I}$.  Noise in measured shear is affected mainly by the
intrinsic ellipticity and source density, but can also be prone to systematic
effects that lead to noise correlations between pixels.  When the survey
geometry leads to shear with more complicated noise characteristics, a
whitening transformation can be applied.

Given the measured data $\myvec\gamma$ and noise covariance
$\mymat{\mathcal{N}}$, we can define the whitened shear
\begin{equation}
  \myvec{\gamma}^\prime = \mymat{\mathcal{N}}^{-1/2} \myvec{\gamma}
\end{equation}
With this definition, the shear covariance matrix becomes
\begin{eqnarray}
  \mymat{\xi}^\prime 
  &=& \left\langle \myvec{\gamma}^\prime 
  \myvec{\gamma}^{\prime\dagger}\right\rangle \nonumber\\
  &=& \mymat{\mathcal{N}}^{-1/2}\mymat{\xi}
  \mymat{\mathcal{N}}^{-1/2} \nonumber\\
  &=& \mymat{\mathcal{N}}^{-1/2}\left[
    \mymat{\mathcal{S}} + \mymat{\mathcal{N}}
    \right]\mymat{\mathcal{N}}^{-1/2} \nonumber\\
  &=& \mymat{\mathcal{N}}^{-1/2}\mymat{\mathcal{S}}\mymat{\mathcal{N}}^{-1/2} + \mymat{I}
\end{eqnarray}
We see that the whitened signal is $\mymat{\mathcal{S}}^\prime = 
\mymat{\mathcal{N}}^{-1/2}\mymat{\mathcal{S}}\mymat{\mathcal{N}}^{-1/2}$
and the whitened noise is $\mymat{\mathcal{N}}^\prime = \mymat{I}$, the
identity matrix. This transformation whitens the data covariance,
so that the noise in each bin is constant and uncorrelated.  Given the
whitened measurement covariance $\mymat{\xi}^\prime$, we can find the KL
decomposition that satisfies the eigenvalue problem
\begin{equation}
  \mymat{\xi}^\prime \myvec{\Psi^\prime}_j = 
  \lambda^\prime_j \myvec{\Psi^\prime}_j
\end{equation}
With KL coefficients given by
\begin{equation}
  \myvec{a}^\prime = \mymat{\Psi}^{\prime\dagger}
  \mymat{\mathcal{N}}^{-1/2}\myvec\gamma
\end{equation}
Note that because $\langle\myvec\gamma\rangle = 0$,
the expectation value of the KL coefficients is
\begin{eqnarray}
  \langle\myvec{a}^\prime\rangle 
  &=& \mymat{\mathcal{N}}^{-1/2}\langle\myvec\gamma\rangle\nonumber\\
  &=& 0
\end{eqnarray}
For the remainder of this chapter, it will be assumed that we are working with
whitened quantities.  The primes will be dropped for notational simplicity.

\subsection{Constructing the Covariance Matrix}
In many applications, the data covariance matrix can be estimated
empirically, using the fact that
\begin{equation}
  \tilde{\myvec\xi} = \lim_{N\to\infty} \sum_{i=1}^N 
  \myvec{\gamma}_i \myvec{\gamma}_i^\dagger
\end{equation}
Unfortunately, in surveys of cosmic shear, we have only a single sky to
observe, so this approach does not work.  Instead, we can construct the
measurement covariance analytically by assuming a theoretical form of the
underlying matter power spectrum.

The measurement covariance $\mymat{\xi}_{ij}$ between two regions of the
sky $A_i$ and $A_j$ is given by
\begin{eqnarray}
  \label{eq:xi_analytic}
  \myvec{\xi}_{ij} 
  &=& \mymat{\mathcal{S}}_{ij} + \mymat{\mathcal{N}}_{ij} \nonumber\\
  &=& \left[\int_{A_i}d^2x_i\int_{A_j}d^2x_j 
    \xi_+(|\myvec{x_i}-\myvec{x_j}|)\right]
  + \mymat{\mathcal{N}}_{ij}
\end{eqnarray}
where $\xi_+(\theta)$ is the ``+'' shear correlation function. 
$\xi_+(\theta)$ is expressible as an integral over the shear power spectrum
weighted by the zeroth-order Bessel function
\citep[see, e.g.][]{Schneider02}:
\begin{equation}
  \label{eq:xi_plus_def}
  \xi_+(\theta) 
  = \frac{1}{2\pi} \int_0^\infty d\ell\ \ell P_\gamma(\ell) J_0(\ell\theta)
\end{equation}
The angular shear power spectrum $P_\gamma(\ell)$ can be expressed as a
weighted line-of-sight integral over the matter power
\begin{equation}
  P_\gamma(\ell) = \int_0^{\chi_s}d\chi W^2(\chi)\chi^{-2}
  P_\delta\left(k=\frac{\ell}{\chi};z(\chi)\right)
\end{equation}
Here $\chi$ is the comoving distance, $\chi_s$ is the distance to the
source, and $W(\chi)$ is the lensing weight function,
\begin{equation}
  \label{eq:lensing_weight}
  W(\chi) = \frac{3\Omega_{m,0}H_0^2}{2a(\chi)}\frac{\chi}{\bar{n}_g}
  \int_{\chi}^{\chi_s}dz\ n(z) \frac{\chi(z)-\chi}{\chi(z)}
\end{equation}
where $n(z)$ is the empirical redshift distribution of galaxies.
The nonlinear mass fluctuation power spectrum $P_\delta(k, z)$ can be
predicted semi-analytically: in this work we use the halo model of
\citet{Smith03}.  With this as an input, we can analytically
construct the measurement covariance matrix $\mymat\xi$ using 
Equations~\ref{eq:xi_analytic}-\ref{eq:lensing_weight}.

\subsection{Cosmological Likelihood Analysis with KL}
The cosmological analysis with KL consists of the following steps:
from the survey geometry and galaxy ellipticities, we measure the
shear $\myvec\gamma$, estimate the noise covariance
$\mymat{\mathcal{N}}$ (see \S\ref{sec:bootstrap}) and derive
the whitened covariance matrix $\mymat\xi$. 
From $\mymat\xi$ we compute the KL basis $\mymat\Psi$ and $\myvec\lambda$.
Using the KL basis, we compute the coefficients
$\myvec{a} = \mymat{\Psi}^\dagger \mymat{\mathcal{N}}^{-1/2} \myvec\gamma$.
Given these KL coefficients $\myvec{a}$, we use a Bayesian framework to
compute the posterior distribution of our cosmological parameters.

The problem of Bayesian inference with KL was discussed in
\S\ref{sec:KL_bayes}.  Here we will briefly outline the portions relevant
to this chapter.
Given observations $D$ and prior information $I$, Bayes' theorem specifies the
posterior probability of a model described by the parameters $\{\theta_i\}$:
\begin{equation}
  \label{eq:bayes}
  P(\{\theta_i\}|DI) = P(\{\theta_i\}|I) \frac{P(D|\{\theta_i\}I)}{P(D|I)}
\end{equation}
The term on the left of the equality
is the \textit{posterior} probability of the set of
model parameters $\{\theta_i\}$, which is the quantity we are interested in.
The likelihood function for the observed coefficients $\myvec{a}$
enters into the numerator $P(D|\{\theta_i\}I)$.  The denominator $P(D|I)$
is essentially a normalization constant, set so that the total probability
over the parameter space equals unity.

For a given model $\{\theta_i\}$, we can predict the expected distribution of model KL
coefficients $\myvec{a}_{\{\theta_i\}} \equiv \mymat{\Psi}^\dagger
\mymat{\mathcal{N}}^{-1/2}\myvec{\gamma}$:
\begin{eqnarray}
  \mymat{C}_{\{\theta_i\}}
  & \equiv & \langle\myvec{a}_{\{\theta_i\}}
  \myvec{a}_{\{\theta_i\}}^\dagger\rangle\nonumber\\
  &=& \mymat{\Psi}^\dagger \mymat{\mathcal{N}}^{-1/2} 
  \mymat{\xi}_{\{\theta_i\}}\mymat{\mathcal{N}}^{-1/2}\mymat{\Psi}
\end{eqnarray}
Using this, the measure of departure from the model $m$ is given by the
quadratic form
\begin{equation}
  \chi^2 = \myvec{a}^\dagger\mymat{C}_{\{\theta_i\}}^{-1}\myvec{a}
\end{equation}
The likelihood is then given by
\begin{equation}
  \label{eq:likelihood}
  \mathcal{L}(\myvec{a}|\{\theta_i\}) = 
  (2\pi)^{n/2} |\det(C_{\{\theta_i\}})|^{-1/2}
  \exp(-\chi^2/2)
\end{equation}
where $n$ is the number of degrees of freedom: that is, the number
of eigenmodes included in the analysis.  The likelihood given by
Equation~\ref{eq:likelihood} enters into Equation~\ref{eq:bayes} when
computing the posterior probability.

\section{COSMOS data}
\label{sec:data}
\begin{figure}
  \centering
  \includegraphics[width=0.8\textwidth]{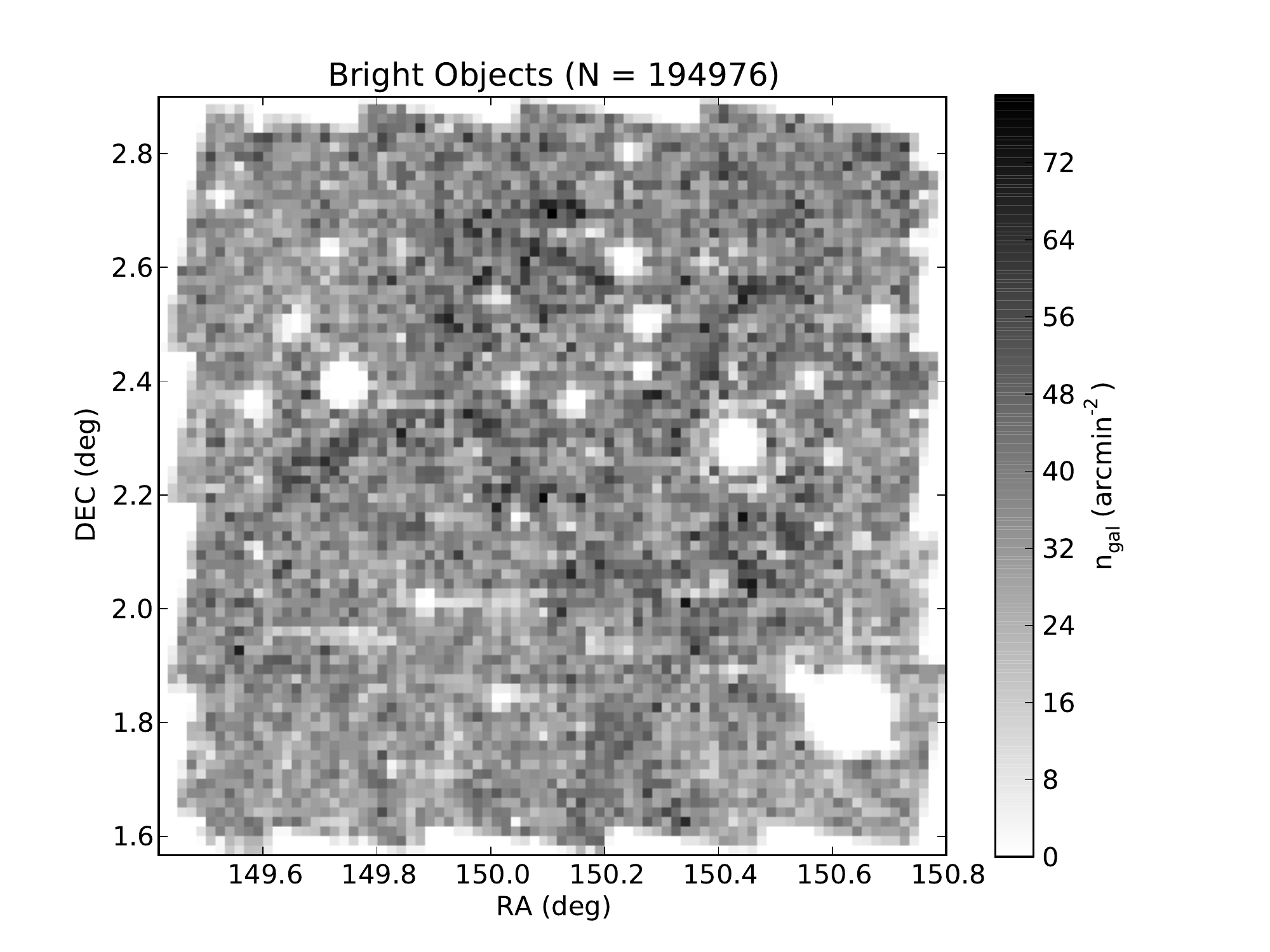}
  \includegraphics[width=0.8\textwidth]{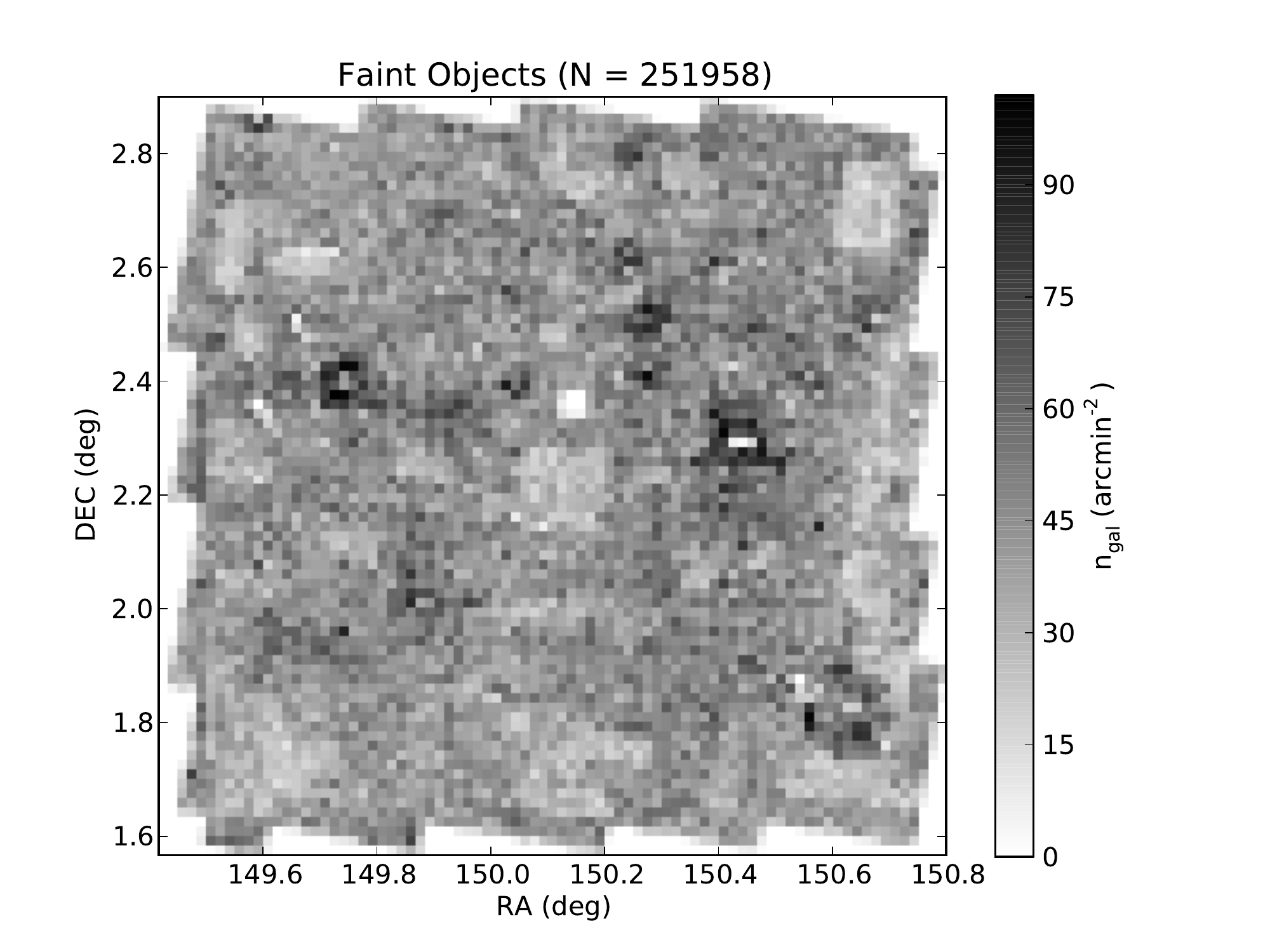}
  \caption[Angular locations of COSMOS galaxies]
  {Angular locations of the 194,976 COSMOS galaxies with photometric
    redshift measurements (top panel)
    and the 251,958 COSMOS galaxies without photometric redshift estimates
    (bottom panel).  The masking due to bright foreground sources is evident
    in both panels.}
  \label{fig:COSMOS_locations}
\end{figure}

\begin{figure}
  \centering
  \includegraphics[width=0.8\textwidth]{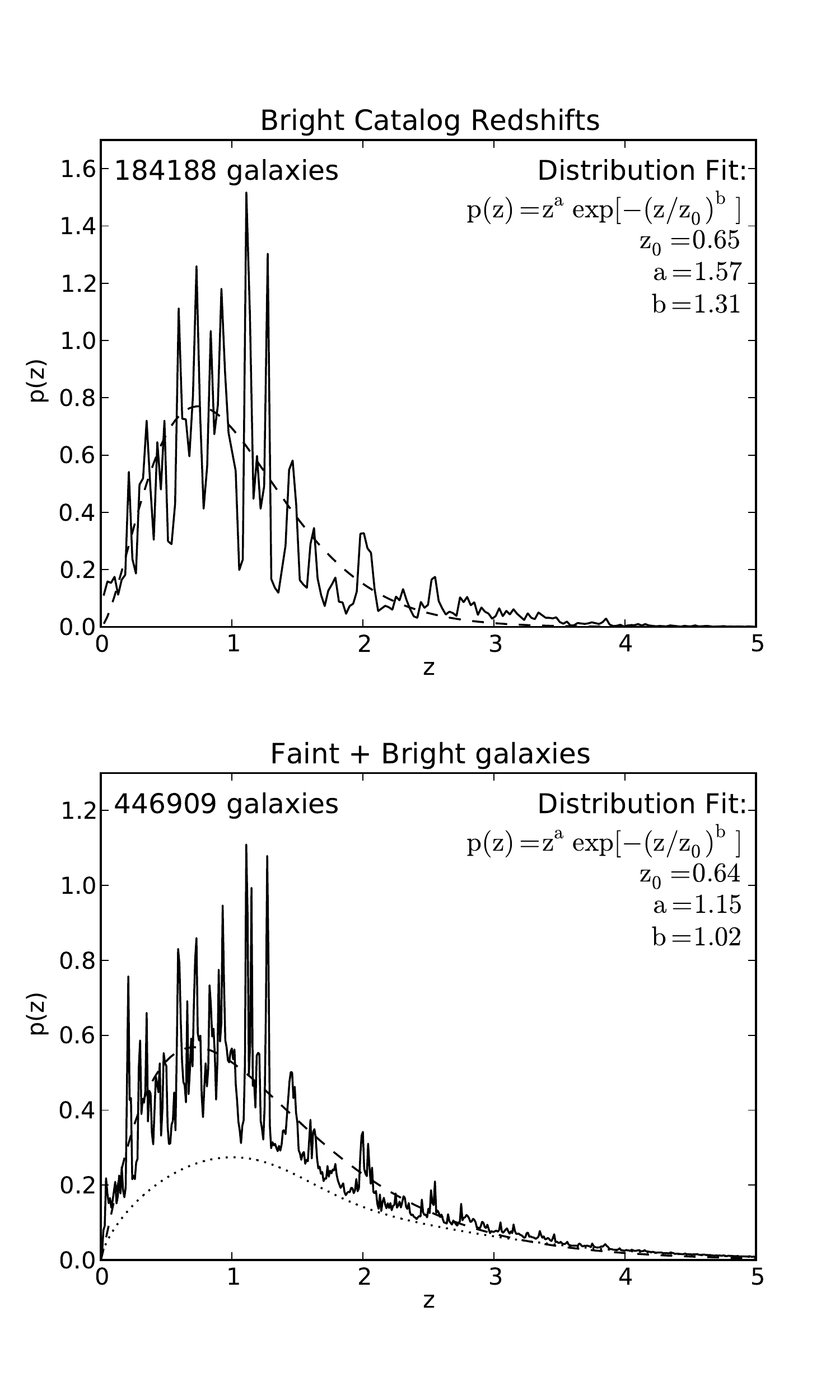}
  \caption[Redshift distributions of the COSMOS data]
  {Redshift distributions of the COSMOS data.  The top panel shows the
    distribution of photometric redshifts of the shear catalog cross-matched
    with the COSMOS30 photometric redshift catalog \citep{Ilbert09}, while
    the bottom panel includes the inferred redshift distribution of the
    remaining faint galaxies.}
  \label{fig:COSMOS_zdist}
\end{figure}

To test the KL likelihood formalism, we use a shear catalog derived from the
COSMOS survey\footnote{We are grateful to Tim Schrabback et al. for making 
these data available to us}.  A full description of this catalog and detailed 
tests of its systematics are presented in
\citet[][hereafter S10]{Schrabback10}; we will summarize some relevant
details here.
The catalog contains shape measurements of 446,934 source galaxies 
in a 1.64 square-degree field.
The ``bright'' sample of 194,976 galaxies are those with well-behaved
photometric redshifts drawn from the COSMOS30 pipeline
\citep[][S10]{Hildebrandt2009},  The angular distribution of these
galaxies is shown in the upper panel of Figure~\ref{fig:COSMOS_locations},
and the redshift distribution is shown in the upper panel of
Figure~\ref{fig:COSMOS_zdist}.  The redshift distribution is marked by
spikes indicating the presence of clusters of galaxies at the same redshift.

The remaining 446,909 galaxies are too faint to have been included
in the reference catalog \citep[the COSMOS30 redshifts are limited to
$i^+ < 25$; See][]{Ilbert09}, and S10 estimates their redshift distribution
using the empirical relationship between redshift and absolute $i$-band
magnitude.
The spatial and redshift distributions of this ``faint'' galaxy
sample are shown in the lower panels of Figures~\ref{fig:COSMOS_locations}
and \ref{fig:COSMOS_zdist}, respectively.

In addition, S10 identifies potential catastrophic outliers in the redshifts.
Photometric redshifts gain significant leverage from broad spectral features
such as the Lyman and Balmer spectral breaks.  The Balmer limit is
$\sim 364$nm, while the Lyman limit is $\sim 91$nm, so that the Balmer limit
of a galaxy at redshift $z_0$ is at the same observed wavelength as the
Lyman limit at redshift $z_0 + 3$.  This can lead to a degeneracy in redshift
determination that results in catastrophic outliers -- that is, low redshift
galaxies identified as high redshift, or high redshift galaxies identified as
low redshift.  In shear studies, the former acts to dilute the high-$z$
shear signal,
while the latter acts to add spurious signal at low redshift.  To prevent
the latter effect from affecting results, we follow S10 and remove galaxies
with $i^+ > 24$ and redshifts $z < 0.6$.  S10 provides several tests which
show that this cut does not generate appreciable systematic error.

S10 performs a classical shear correlation function analysis to find
constraints on $\sigma_8$ and $\Omega_M$ that are consistent with
those derived from WMAP: for a flat $\Lambda$CDM cosmology, they
find with 63\% confidence 
$\sigma_8(\Omega_M / 0.3)^{0.51} = 0.79 \pm 0.09$.  S10 performs both a
two-dimensional analysis and a three dimensional analysis: the constraints
from each are consistent, with a slightly better figure of merit for the
$\Omega_M$ vs. $\sigma_8$ constraint when the analysis is computed within
several redshift bins.  The strength of the
3D treatment comes when we drop assumptions about flatness or the dark
energy equation of state: for a
$\Lambda$CDM cosmology with varying dark energy equation of state $w$,
S10 finds at 90\% confidence that $w < -0.41$.
Though these constraints offer only a slight improvement over prior
information from WMAP constraints from measurements of the CMB, we must
note that they are derived from just over
1 square degree of observations, while the
WMAP constraints use the full sky.  Future wide-field
lensing surveys will be able to place much more competitive constraints
on these parameters.

Here we will not duplicate all the various analyses of S10: instead we
will use the KL-based estimation formalism of \S\ref{sec:KL_bayes} with
shear eigenbases computed for the observed field via the formalism
of \S\ref{sec:KL_shear}.  This will enable us to constrain two-point
information using the KL formalism.

\subsection{Intrinsic Ellipticity estimation}
\label{sec:bootstrap}
In order to apply the KL analysis techniques discussed
above and in Chapter 2, we require an accurate
determination of the noise for the observed shear.  Assuming systematic
errors are negligible, shape noise should be dominated by shot noise,
which scales as $\mymat{\mathcal{N}}_{ii} = \hat{\sigma}_\epsilon^2 / n_i$,
with $n_i$ representing the number of galaxies in bin $i$.

\begin{figure*}
 \centering
 \includegraphics[width=0.8\textwidth]{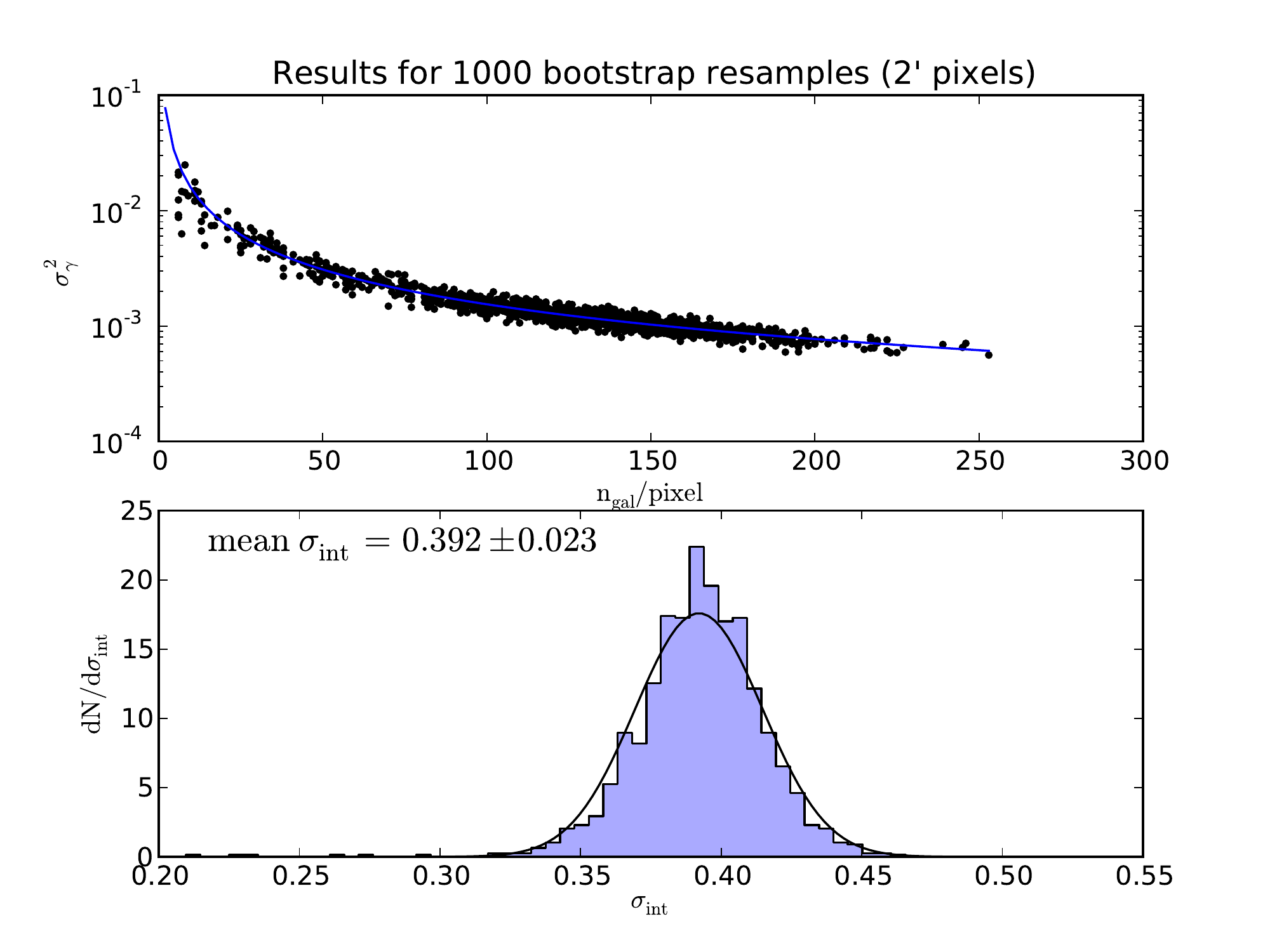}
 \caption[Bootstrap estimates of the shape noise]
 {Bootstrap estimates of the shape noise for each pixel.  The estimates
   reflect an intrinsic ellipticity of $0.393 \pm 0.013$.
   \label{fig:bootstrap}}
\end{figure*}

To test this assumption, we perform a bootstrap resampling of the observed
shear in square pixels that are two arcminutes on a side.
Generating 1000 bootstrap samples within each
pixel, we compute the variance in each pixel.
From Poisson statistics, we would expect the variance in each pixel to
scale inversely with the number of measurements within each pixel: with this
in mind we plot in Figure~\ref{fig:bootstrap}
the variance vs the number of galaxies and fit a curve of the form 
\begin{equation}
  \sigma_\gamma^2 = \frac{\sigma_{int}^2}{n_{\rm gal}},
\end{equation}
where $\sigma_{int}$ is the intrinsic ellipticity dispersion of the population.
As shown in the upper panel of Figure~\ref{fig:bootstrap}, the best-fit
curve has $\sigma_{int} = 0.393$.  The residual distribution, shown in the
lower panel of the figure, is close to Gaussian as expected.

In this figure, we see that the pixel-to-pixel fluctuation in shape noise
is only a few percent.  For the analysis below, we use for each pixel
the noise estimates derived the bootstrap resampling within each pixel.
Because bootstrapping is inaccurate
for pixels with a small number of galaxies, if a pixel has fewer than 10
sources we use the best-fit estimate for the noise,
$\mymat{\mathcal{N}}_{ii} = \hat{\sigma}_\epsilon^2 / n_i$
with $\hat{\sigma}_\epsilon = 0.392$.  Pixels with zero galaxies (i.e.
masked pixels) are treated using the techniques developed in
Section~\ref{sec:kl_intro}.

\subsection{Whitened KL modes}
Using the pixel-by-pixel noise estimates from the previous section, we can
now follow the formalism of \S\ref{sec:kl_intro} and construct the optimal
orthonormal basis across the survey window defined by the selection function
of the bright galaxies from the COSMOS survey.  We use pixels that are
two arcminutes on a side, in a grid of 40 $\times$ 41 = 1640 total
pixels.  We whiten the theoretical correlation matrix according to the noise
properties of the observed data, and compute the eigenvalue decomposition
of the resulting correlation matrix.

\begin{figure*}
 \centering
 \includegraphics[width=\textwidth]{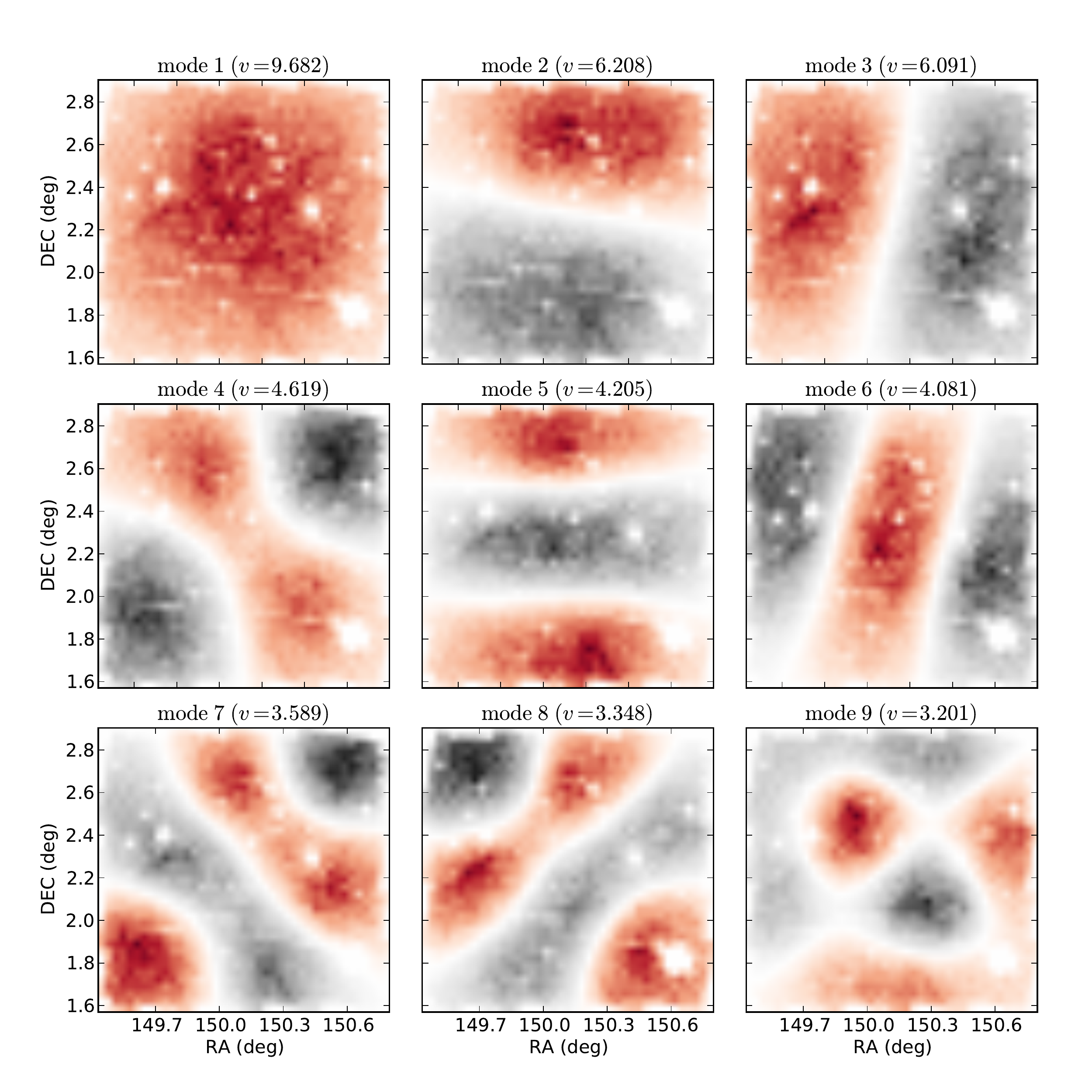}
 \caption[KL signal-to-noise eigenmodes for COSMOS data]{
   The first nine 2D KL signal-to-noise eigenmodes
   for the COSMOS bright objects.  This uses square pixels that are
   two arcminutes on a side, leading to $41 \times 40 = 1640$ pixels
   over the entire field.  Compare these results to the KL modes shown
   in Figure~\ref{fig_KL_modes}.}
   \label{fig:eigenmodes}
\end{figure*}

The first nine eigenmodes for the bright galaxy sample
are shown in Figure~\ref{fig:eigenmodes}.
It is interesting to compare these to the modes shown in
Figure~\ref{fig_KL_modes}, which are derived under the assumption that
each pixel has the same number of sources, and thus the same noise properties.
The window function of the COSMOS survey is clearly present, as can be
seen by comparing the
masking apparent in the eigenmodes to that of the galaxy distribution
shown in the upper panel of Figure~\ref{fig:COSMOS_locations}.
Moreover, the asymmetry of the mask acts as a perturbation that destroys
the rotational symmetry evident in the idealized eigenmodes of
the previous chapter (See Figure~\ref{fig_KL_modes}).

\begin{figure*}
 \centering
 \includegraphics[width=0.8\textwidth]{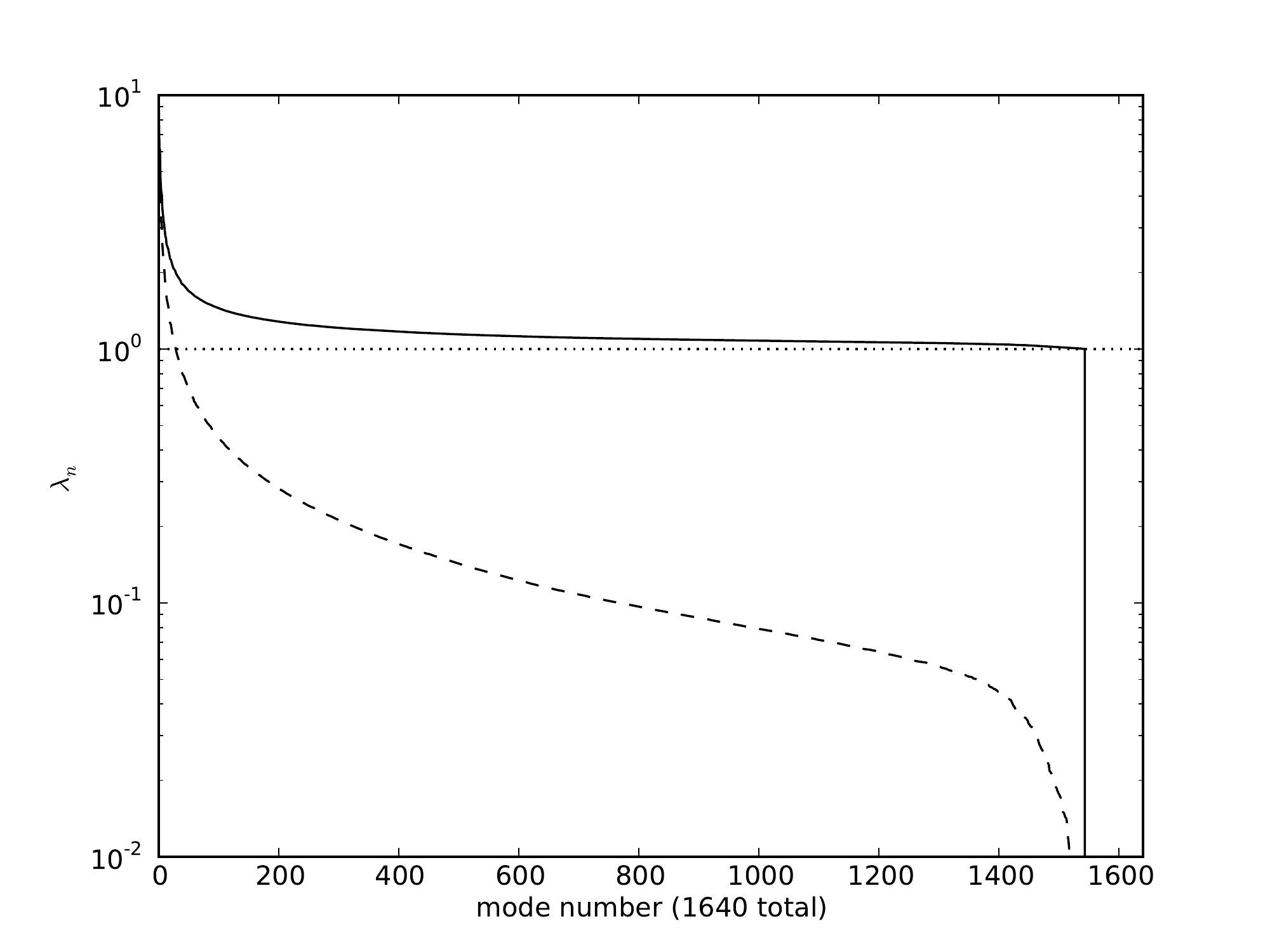}
 \caption[distribution of KL eigenvalues for COSMOS data]{
   The distribution of KL eigenvalues for the eigenmodes shown
   in Figure~\ref{fig:eigenmodes}.  There are $41 \times 40 = 1640$
   pixels, but approximately 90 of these contain no sources and are part of
   the mask.  This is reflected in the fact that the final 90 KL modes have
   zero eigenvalue.}
   \label{fig:eigenvalues}
\end{figure*}

The eigenvalues of the whitened correlation matrix are shown in
Figure~\ref{fig:eigenvalues}.  Because the covariance matrix is
whitened, the noise is normalized to $1$ within each mode.
Similar to the results seen in the previous chapter, only a very small
number of modes have signal-to-noise greater than 1.
This figure also shows that the signal drops to zero at just over 1500
modes.  This is due to the survey mask: approximately 120 of the 1640
modes are completely masked, such that they have no signal and do not
contribute to the correlation matrix.

As discussed above, an advantage of KL is its ability to yield an optimal
low-rank approximation of a set of observations, by truncating the low
signal-to-noise modes in a reconstruction.  The choice of which modes to
truncate for a reconstruction or other analysis is not straightforward:
as discussed in Chapters 3-4, this decision amounts to a tradeoff between
systematic bias and statistical error.  Below we impose a cutoff for
modes with signal-to-noise ratios of less than $\sim 1/10$, corresponding to
mode number 800.  This lies approximately at the inflection point of the
signal-to-noise curve.

\subsection{Is our shear Gaussian?}

\begin{figure*}
 \centering
 \includegraphics[width=0.8\textwidth]{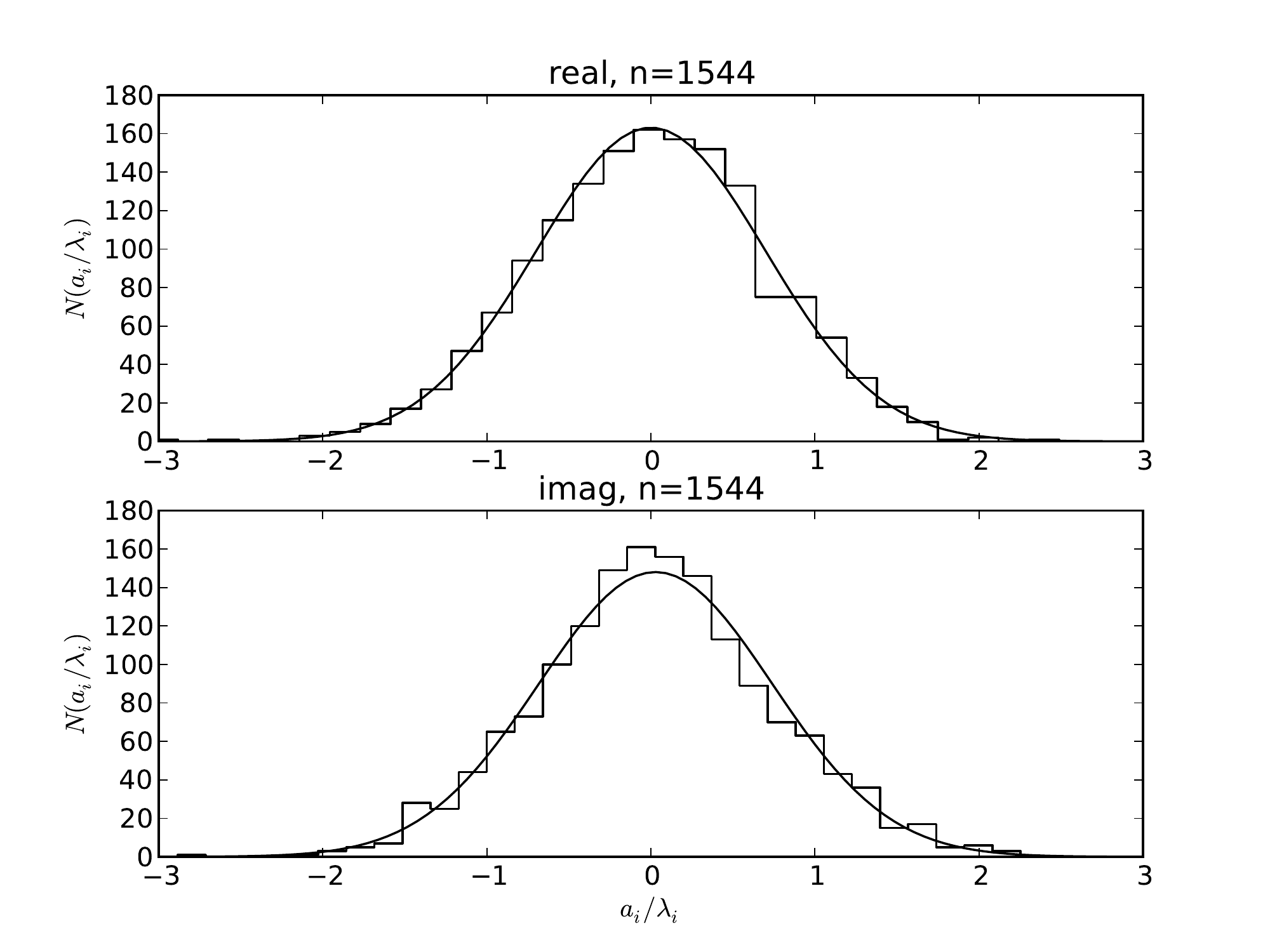}
 \caption[histogram of normalized coefficients for COSMOS data]{
   The histogram of normalized coefficients $a_i / \sqrt{\lambda_i}$.
   If the shear is truly a Gaussian random field, this distribution should
   be a Gaussian with unit variance.
   \label{fig:coeff_hist}}
\end{figure*}

The KL formalism for shear analysis assumes that the shear is well-described
by a Gaussian random field described by a covariance matrix, with mean
zero.  If this is the case, then (by the arguments of Chapter 2) we would
expect the observed KL coefficients of the whitened signal to be Gaussian
distributed with zero mean and variance equal to the associated eigenvalue.
Figure~\ref{fig:coeff_hist} shows a histogram of the observed coefficients
scaled by the corresponding eigenvalue.  As is evident, both the real
part and the imaginary part of the scaled coefficients are consistent with
being drawn from a standard normal distribution.  This is consistent with
our assumption that the shear is drawn from a Gaussian random field, and
that the noise properties estimated in \S\ref{sec:bootstrap} are accurate.

\subsection{Relationship to Power Spectrum}
\begin{figure*}
 \centering
 \includegraphics[width=0.8\textwidth]{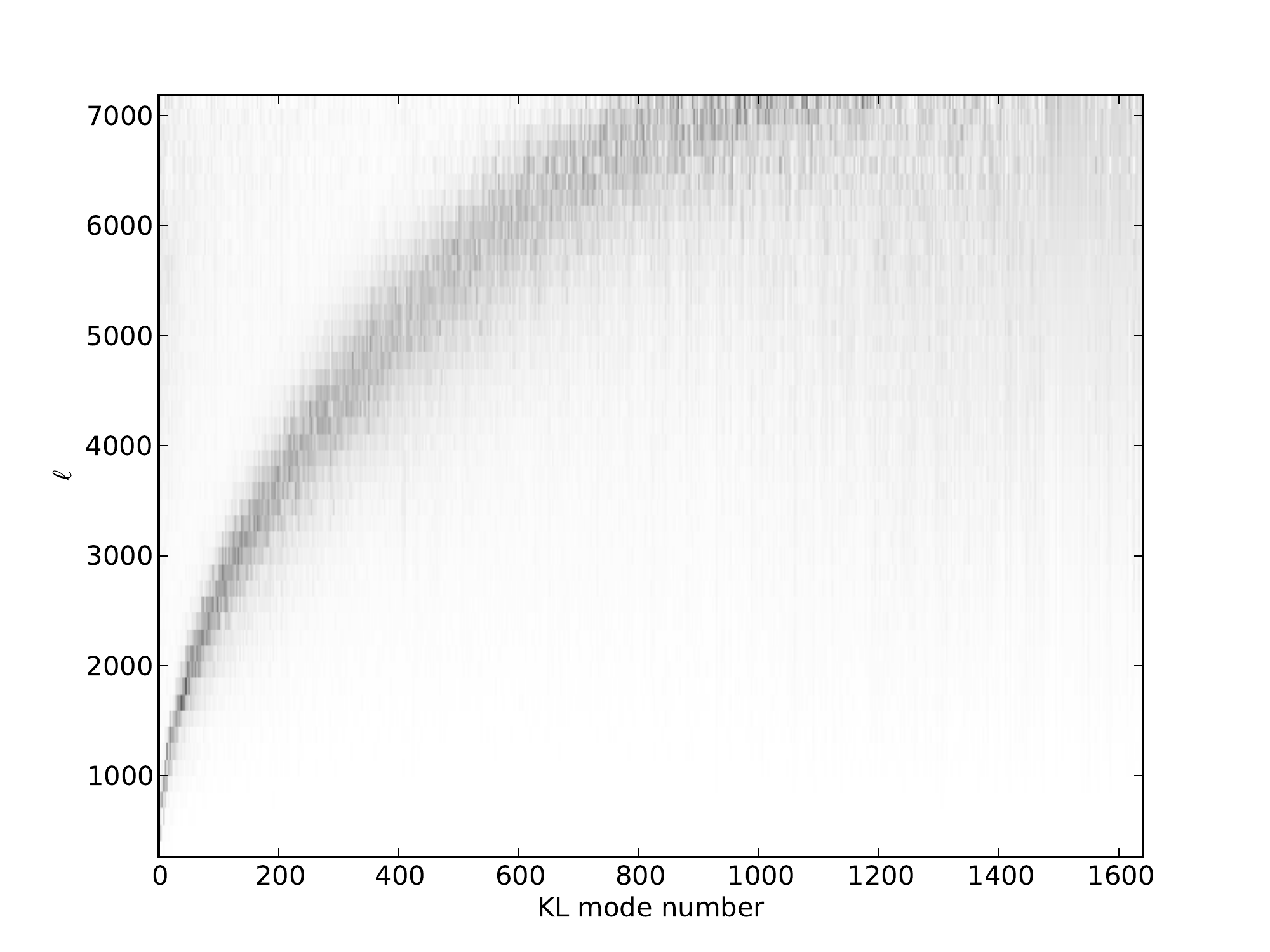}
 \caption[The Fourier power represented by each KL mode]{
   The Fourier power represented by each KL mode.  For each KL mode number,
   the vertical band shows the distribution of power with angular wavenumber
   $\ell$.  In general, the larger KL modes correspond to larger values of
   $\ell$, though there is significant mode mixing.
   \label{fig:bandpower_masked}}
\end{figure*}

As we did in Figure~\ref{fig_bandpower}, for each KL mode we compute the
associated two-dimensional power spectrum to determine the relationship
between each mode and the Fourier power it represents.  For the unmasked
KL modes explored in the previous chapter, this relationship displayed a
fairly tight scatter between KL mode number and Fourier mode number.
As seen in Figure~\ref{fig:bandpower_masked}, however, we see that the
masked KL modes have a much larger scatter in associated Fourier modes,
especially for higher KL mode numbers.

The analysis reflected in this plot can help in the choice of which KL
modes to truncate: the pixel scale is two arcmin, which corresponds to
a dimensionally-averaged Nyquist frequency of
\begin{equation}
  \ell = \left(\frac{1}{\sqrt{2}}\right)
  \left(\frac{2\pi\, {\rm rad}}{2\ {\rm arcmin}}\right) \approx 7200.
\end{equation}
So modes which do not have significant Fourier power on angular scales
$\ell < 7200$ are likely to be limited in their usefulness for parameter
estimation from shear on this grid.  This scale does not tell the whole story,
however.  The KL analysis tells us that the smaller scales probed by the
higher-order modes have progressively smaller signal-to-noise ratios.  For
this reason, we choose the mode cutoff at scales less than 7200, which
corresponds to $n \sim 800$ modes.  As was the case in Chapter 4, the
optimal choice of mode cutoff is hard to quantify precisely, and represents
a fundamental tradeoff between statistical and systematic errors.
Modes larger than our cutoff of 800 have an expected signal-to-noise of
less than $1/10$.

\section{Results}
\label{sec:results}

\begin{figure*}
 \centering
 \includegraphics[width=0.8\textwidth]{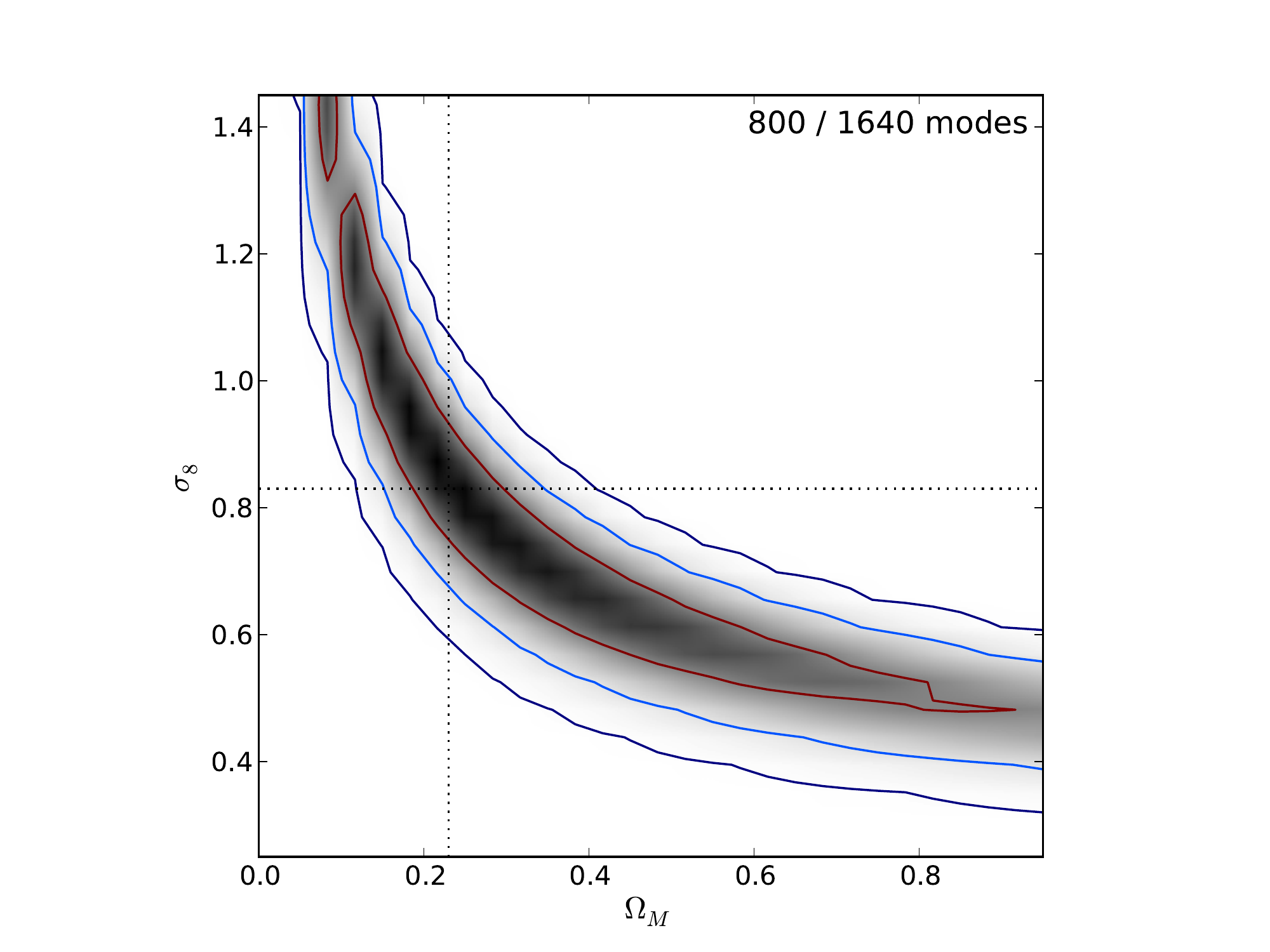}
 \caption[The posterior distribution in the $(\Omega_M, \sigma_8)$]
 {The posterior distribution in the $(\Omega_M, \sigma_8)$ plane
   for a 2D analysis of the bright galaxy sample.  This uses 800 of the
   1640 modes, such that the truncated modes have average signal-to-noise
   ratios of $<\sim 1/10$, and an approximate angular scale of
   $\ell \sim 7000$, which corresponds to 3 arcmin or 1.5 pixel-lengths.}
 \label{fig:posterior}
\end{figure*}

The result of the KL-based Bayesian inference for cosmological parameter
estimation is shown in Figure~\ref{fig:posterior}.  To compute the
eigenmodes, we assume a flat $\Lambda$CDM cosmology with
$\Omega_M = 0.27$, $\Omega_L = 0.73$, $\sigma_8 = 0.812$, and $h=0.71$.
These KL modes are computed for the angular and redshift distribution
of the bright galaxy sample, and the KL-based Bayesian inference is
performed assuming a flat cosmology.  We use only a single redshift bin in
this case, which is comparable to the first analysis performed in
S10.

This leads to a best-fit cosmology $\Omega_M = 0.23 \pm 0.06$,
$\sigma_8 = 0.83 \pm 0.09$, where the error bars represent 1$\sigma$
deviations about the maximum {\it a priori} value.
This does not capture the entire story, however, as there
is a strong degeneracy between the parameters (note that WMAP data offers
complementary constraints in this plane, and can be used to break this
degeneracy: see S10).  Following S10, we describe this degeneracy by
computing a power-law fit to the posterior distribution, to find
\begin{equation}
  \sigma_8 (\Omega_M / 0.3) ^ {1.03} = 0.88 \pm 0.14.
\end{equation}
This should be compared to the S10 result for the 2D analysis,
$\sigma_8 (\Omega_M / 0.3) ^ {0.62} = 0.62 \pm 0.11$.

Compared to S10, our results show a 50\% broader constraint on $\sigma_8$,
as well as a stronger degeneracy between $\sigma_8$ and $\Omega_M$ 
(reflected in the exponent of the relation).  This discrepancy is likely
due to the fact that we use only the bright galaxies in this analysis,
while the S10 results use both bright and faint galaxies, as well as the
fact that S10 marginalizes over nuisance parameters (the Hubble parameter
and redshift systematic corrections) while we fix these at the  expected
values.

S10 notes that the low value of $\sigma_8$ seen in their 2D analysis is likely
an artifact of cosmic variance: the strongest contributions to lensing
signals in COSMOS are from $z > 0.7$, which boosts the shear signal for higher
redshift sources but leads to a lower signal at intermediate redshifts.
Using the full 3D analysis, S10 is able to separate these regions, leading
to results consistent with those from WMAP.

Here we have limited the analysis to two dimensions, but this is by no means a
fundamental limitation of KL.  As long as we can sufficiently estimate the
correlation of signal and noise, KL can be used to analyze an arbitrary
geometry: in future work we will extend the present analysis to three
dimensions, fully taking into account the redshifts of the sources.

\section{Next Steps}
The above analysis presents a firm basis for further exploring the ability of
KL to provide a natural basis for extracting two-point information from
weak lensing surveys.  Further work is needed to fully understand the effect
of the mode truncation on results, as well as other effects such as the
pixel size, the assumptions of noise, and the effect of the assumed fiducial
cosmology.

There are also some potentially interesting and useful features of the
algorithm: first of all, the KL framework naturally extends from two dimensions
to three.  Unlike the rigidly tomographic approach used in conventional
correlation function studies, KL allows each pixel to have its redshift
distribution individually specified, potentially leading to a more robust
use of source redshift information.  For this reason, KL is a promising
technique for a full 3D analysis, and will give insight into the density and
perhaps equation of state of dark energy.

Second, assumptions about noise and bias can be built-in to the KL model.
For example, S10 does a careful job of correcting for the shape of the HST PSF
before computing the observed correlation function.  With KL, we could instead
account for these biases in the KL basis itself, allowing us to perform our
cosmological analysis one step closer to the observed data.

Third, there is the question of how this approach can scale from the one
square degree of COSMOS to the 20,000 square degrees of LSST.  The LSST
weak lensing analysis has potential to give very tight constraints on
cosmological parameters, especially the possible evolution of dark energy.
It will be increasingly important to address and explore how this KL
framework can be scaled to the size of future surveys.

%% file: chapter6.tex
\chapter{Conclusion}
In the above chapters, we have developed the \KL\ analysis as a useful tool
for several aspects of the analysis of present and future weak lensing
surveys.  In Chapter 2, we discussed the details of the KL formalism.
We showed that KL is a powerful technique that enables data to be represented
as a linear combination of orthogonal modes that are constructed such that
the modes are optimal representations of the signal-to-noise ratio.

In Chapter 3, we demonstrated that KL can be used to construct an optimal
linear framework for the mapping of three dimensional structure from weak
lensing surveys.  The KL filtering leads to an algorithm which is orders of
magnitude faster than previously studied approaches, and allows quantitative
constraints on the effectiveness of mapping for given survey depths and
geometries.

In Chapter 4, we demonstrated that KL can be used to address a practical problem
for two and three dimensional mass mapping: the interpolation of shear signal
across masked regions of a given survey.  The reconstruction takes into
account theoretical expectations of the shear correlation, and results in
peak counts that are more consistent with those of the underlying
distribution.  The KL approach also results in a natural filtration of
low-magnitude noisy peaks, which has the potential to increase the
performance of cosmological likelihood calculations from peak statistics
of shear.

In Chapter 5, we show how KL can be used directly as a tool to derive
cosmological parameter constraints from two-point information within
a Bayesian inference framework. Because KL can naturally account for 
arbitrary survey masks and geometries, it allows for robust determination
of likelihoods without the need for computationally expensive calibration
against N-body simulations.
As a proof-of-concept, we perform a
two-dimensional likelihood analysis to derive constraints on $\sigma_8$
and $\Omega_M$ which are consistent with those derived from conventional
correlation-function approaches using the same data.

In these three important areas of weak lensing analysis, the KL approach
has proven valuable in addressing the practical problems associated with
the science of weak lensing.  KL's robust, computationally efficient approach
has the potential to be very useful in many areas of future weak lensing
science.

This thesis opens nearly as many questions as it answers.  In the future,
we would like to explore more deeply how the shear KL basis can be used to
address real problems in data analysis.  Chapters 4 and 5 end with discussions
of some remaining questions: can the KL analysis of projected shear peaks
lead to robust cosmology constraints from realistic datasets?  Can the 
KL power spectrum approach be extended to 3D and make use of our knowledge
about the correlated statistical and systematic errors in real weak lensing
data?

Beyond that, we can ask other questions: is KL the best basis to use for these
sorts of studies?  In chapter 2, we show that KL gives the optimal low-rank
reconstruction and noise filtering among all possible linear, orthonormal
bases.  This does not, however, exclude the possibility of using other
non-standard representations of the data, such as non-orthonormal or
overcomplete bases.  Such approaches are common in the fields of sparse
coding and compressed sensing, and may also be fruitful methods within the
field of weak lensing.

%% file: appendixA.tex
\chapter{Random Fields, Correlation Functions, and Power Spectra}
\label{app:RandomFields}

In this appendix we discuss some of the details of the mathematics behind
random fields, and their correlation functions and power spectra.
In Appendix~\ref{app:sec:masspower} we
apply this to the cosmological density field introduced in
Chapter 1, and from this define the power spectrum normalization
$\sigma_8$.
Some common window functions and their Fourier transforms are listed
in Appendix~\ref{app:sec:Wtransforms}.

\section{Background on Gaussian random fields}
\label{app:sec:background}
Consider a field $g(\vec x)$ in $n$ dimensions. We'll enforce a 
few restrictions on this field to make it easier to work with.
Note that $\langle\cdot\rangle$ denotes a volume-average:
\begin{enumerate}
  \item vanishing: $\langle g(\vec x)\rangle=0$ 
    for all $\vec x$.
  \item homogeneous: $g(\vec x + \vec y)$ is statistically equivalent to
    $g(\vec x)$ for all $\vec x$ and $\vec y$.
  \item isotropic: $g(\mathbf{R}\vec x)$ is statistically equivalent to
    $g(\vec x)$ for all $\vec x$ and any unitary rotation matrix $\mathbf{R}$.
\end{enumerate}
These conditions become very useful when we study the 
(auto) correlation function, defined as
\begin{equation}
  \label{Cgg}
  C_{gg}(\vec r) \equiv \left\langle g(\vec x) g^*(\vec x+\vec{r})\right\rangle
\end{equation}
which for a homogeneous and isotropic field depends only on the 
distance $r = |\vec r|$.  It becomes useful to decompose $g$ into
orthogonal Fourier components:\footnote{
  Note that the Fourier transform convention in eqns \ref{ft}-\ref{ift}
  is useful in that it leads to a particularly simple form of the 
  convolution theorem, without any gratuitous factors of $\sqrt{2\pi}$:
  \begin{displaymath}
    h(\vec x) = \int\dd^nx^\prime 
    f(\vec x^\prime)g(\vec x-\vec x^\prime)
    \ \ \ \ \ \Longleftrightarrow\ \ \ \ \ 
    \hat h(\vec k) = \hat f(\vec k)\hat g(\vec k)
  \end{displaymath}
}
\begin{eqnarray}
  \label{ft}
  g(\vec x) = \int \frac{\dd^nk}{(2\pi)^n} \hat{g}(\vec k)
  e^{-i\vec k\cdot\vec x}\\
  \label{ift}
  \hat{g}(\vec k) = \int \dd^nx g(\vec x)e^{i\vec x\cdot\vec k}
\end{eqnarray}

From these, we can see that the n-dimensional Dirac delta 
function can be written
\begin{equation}
  \label{ddelta_form}
  \delta^n_D(\vec x-\vec x^\prime) = \frac{1}{(2\pi)^n}\int \dd^nke^{-i\vec k\cdot(\vec x-\vec x^\prime)}
\end{equation}
such that 
\begin{equation}
  \label{ddelta_def}
  \int \dd^nx f(\vec x)\delta^n_D(\vec x-\vec x^\prime) = f(\vec x^\prime)
\end{equation}

We now define the Power Spectrum of $g$ to be the Fourier transform of the 
auto-correlation function, which, due to isotropy, depends only on the 
magnitude of $\vec k$:
\begin{eqnarray}
  \label{pspec}
  P_g(k) = \int \dd^nx e^{-i\vec x\cdot\vec k}C_{gg}(x)\nonumber\\
  C_{gg}(x) = \frac{1}{(2\pi)^n}\int\dd^nke^{i\vec x\cdot\vec k}P_g(k)
\end{eqnarray}
A bit of math shows that the Power Spectrum is proportional to the 
Fourier-space correlation function:
\begin{equation}
  \label{pspec_corr}
  \hat C_{gg}(\vec k-\vec k^\prime) 
  \equiv \left\langle\hat g(\vec k)\hat g^*(\vec k^\prime) \right\rangle 
  =  (2\pi)^n \delta^n_D(\vec k -\vec k^\prime) P_g(|\vec k|).
\end{equation}
Along with isotropy and homogeneity, this result implies
\begin{equation}
  P_g(k) \propto \left\langle | \hat g(k)|^2 \right\rangle.
\end{equation}
The proportionality constant is finite only for a discrete Fourier series 
(i.e. a finite averaging volume).

\subsection{Smoothing of Gaussian Fields}
\label{smoothing}
When measuring a realization of Gaussian field, we often make the measurement
within a region defined by a window function $W(\vec x/R)$.
By convention, we express window functions in terms of $\vec x/R$,
where $R$ is a characteristic length scale of the window.
This window may reflect a sharp boundary in space (e.g. a spherical tophat 
function) or perhaps an observation efficiency in space (e.g. a 3D Gaussian).
In either case, our observed overdensity is given by
\begin{equation}
  g_W(\vec x) = \int \dd^n x^\prime 
  W\left(\frac{\vec{x}^\prime-\vec{x}}{R}\right) g(\vec{x}^\prime)
\end{equation}
where $W(\vec x/R)$ is normalized such that
\begin{equation}
  \label{W_normalization}
  \int \dd^nx W(\vec{x}/R) = 1.
\end{equation}
This can be simplified if we can define the Fourier transform pair 
of a window function in the convention
of equations \ref{ft} and \ref{ift} (cf. Liddle \& Lyth 2000):
\begin{eqnarray}
  \label{W-transform}
  \widetilde{W}(\vec{k}R) = \int W(\vec{x}/R) e^{i\vec{k}\cdot\vec{x}}\dd^nx 
  \nonumber\\
  W(\vec{x}/R) = \frac{1}{(2\pi)^n}\int 
  \widetilde{W}(\vec{k}R) e^{-i\vec{k}\cdot\vec{x}}\dd^nk 
\end{eqnarray}
This definition is convenient because, when combined with equation \ref{ft},
some straightforward algebra leads to the convolution theorem:
\begin{equation}
  \hat g_W(\vec k) = \widetilde W(\vec k R)\hat g(\vec k)
\end{equation}
It is also useful to calculate the cross-correlation between two windows,
\begin{equation}
  \label{W_correlation}
  \langle g_{W_1}(\vec x_1)g^*_{W_2}(\vec x_2)\rangle = 
  \int \dd^n x 
  W_1\left(\frac{\vec{x}-\vec{x}_1}{R}\right) 
  \int \dd^n x^\prime 
  W_2\left(\frac{\vec{x}^\prime-\vec{x}_2}{R}\right) 
  \left\langle g(\vec{x})g^*(\vec{x}^\prime)  \right\rangle
\end{equation}
Using equations \ref{pspec} and \ref{W-transform}, we can re-express
equation \ref{W_correlation} as a single integral over the
wave number:
\begin{equation}
  \label{W_cov_simp}
  \langle g_{W_1}(\vec x_1)g^*_{W_2}(\vec x_2)\rangle = 
  \frac{1}{(2\pi)^n} \int \dd^nk
  P_g(k)\widetilde{W}_1(\vec{k}R)\widetilde{W}_2(\vec{k}R)
  e^{i\vec k\cdot(\vec x_1-\vec x_2)}
\end{equation}
Appendix \ref{app:sec:Wtransforms} lists a few common window functions and
their Fourier transforms.

\section{Cosmological Mass Power Spectrum}
\label{app:sec:masspower}
In studies of the cosmological distribution of matter, we are interested
in the comoving matter density $\rho(\vec{x})$, which defines the mass 
density at every comoving point $\vec{x}$ in the universe.  In order
to take advantage of the preceding formalism, we can subtract the mean
cosmological density $\bar{\rho}(\vec{x})$, and define a dimensionless
density contrast $\delta(\vec{x})$, such that
\begin{equation}
  \delta(\vec{x}) = \frac{\rho(\vec{x}) - \bar{\rho}(\vec{x})}{\bar{\rho}(\vec{x})}.
\end{equation}
By the assumptions of the Cosmological Principle, for small
deviations $\delta(\vec x)$ is an isotropic, homogeneous random field.  
We can better understand the distribution of $\delta(\vec x)$ by looking
at the mean square deviation
\begin{eqnarray}
  \label{Cdd0}
  \left\langle|\delta(\vec x)|^2\right\rangle 
  &=& C_{\delta\delta}(0)\nonumber\\
  &=& \frac{1}{(2\pi)^3}\int \dd^3 kP_\delta(k)\nonumber\\
  &=& \frac{1}{2\pi^2}\int k^2\dd k P_\delta(k)
\end{eqnarray}
The power spectrum of density contrast given by equation \ref{pspec}
can be an inconvenient quantity to work with, because it has dimensions
of volume.  We can take the lead from equation \ref{Cdd0} and define
a dimensionless form of the power spectrum
\begin{equation}
  \label{P_conversion}
 \Delta^2(k) = \frac{k^3}{2\pi^2} P_\delta(k)
\end{equation}
This is constructed so that equation \ref{Cdd0} can be written in a
simple form:
\begin{equation}
  \left\langle|\delta(\vec x)|^2\right\rangle 
  = \int_0^\infty \Delta^2(k)\dd(\ln k)
\end{equation}
This convention is due to Peacock (1999).
For mathematical convenience, we'll continue to work with 
the $P_\delta(k)$ convention,
with the understanding that we can switch back and forth any time using
equation \ref{P_conversion}. 

\subsection{Power Spectrum Normalization}
In practice, the functional form of the power spectrum is determined only
up to a proportionality constant, such that
\begin{equation}
  P_\delta(k) = P_0 P^\prime_\delta(k)
\end{equation}
where $P^\prime_\delta(k)$ is the unnormalized form.
For historical reasons, the normalization constant $P_0$ is commonly 
expressed in terms of the parameter $\sigma_8$, 
which is defined as the mean density 
fluctuation within a sphere of radius 8 Mpc.  To compute this, we use a 
top-hat window function:
\begin{equation}
\label{top-hat}
  W_T(\vec{x}/R) = \left\{
    \begin{array}{ll}
      1, & |\vec x|/R \le 1 \\
      0, & |\vec x|/R > 1
    \end{array}
    \right.
\end{equation} 
The density fluctuation within this window is found using equation \ref{W_cov_simp}:
\begin{equation}
  \label{powerspec-sigma-def}
  \sigma_R^2 = \frac{1}{(2\pi)^3}\int \dd^3\vec{k} 
  P_\delta(k) [\widetilde{W}_T(\vec{k}R)]^2
\end{equation}
where the window function is assumed to be shallow enough that there is no
cosmological evolution of the signal.

For the top-hat window function of equation \ref{top-hat}, with 
$k = |\vec{k}|$,
the Fourier transform of equation \ref{top-hat} (cf. eqn. \ref{W-transform}) is
\begin{equation}
  \label{top-hat-f}
  \widetilde{W}_T(kR) = \frac{3}{(kR)^3}\left[\sin(kR) - kR\cos(kR) \right].
\end{equation}
$\sigma_8$ can be calculated using equation \ref{powerspec-sigma-def}
and \ref{top-hat-f} with $R=8$Mpc for a given $P_\delta(k)$. 
The WMAP 7-year measurement gives $\sigma_8 = 0.816 \pm 0.024$ 
\citep{WMAP7}.  Using this value, the correct normalization 
can be computed for any functional form of the power spectrum.

\subsection{Window functions and Measurement Covariance}
In a 3D lensing analysis, we are searching for a signal within a series
of windows defined as 
\begin{equation}
  W_{ij}(\vec{x}) = W_{ij}(\rcom,\vec{\theta}) 
  = q_i(\rcom) \cdot F_j(\vec\theta)
\end{equation}
where $\rcom$ is the radial comoving distance, and $\vec{\theta}$ is the
angular position on the sky.  To convert between angle on the sky and 
projected comoving separation, we multiply by the transverse comoving 
distance (eqn. 16 in Hogg 1999), given by
\begin{equation}
f_\kappa(\rcom) = \left\{
\begin{array}{ll}
  \kappa^{-1/2} \sin (\kappa^{1/2}\rcom) & (\kappa > 0) \\
  \rcom  & (\kappa = 0)\\
  (-\kappa)^{-1/2} \sinh [(-\kappa)^{1/2}\rcom] & (\kappa < 0)
\end{array}\right.
\end{equation}
Our observed overdensity in window $W_{ij}$ is given by
\begin{eqnarray}
  \delta_{ij} 
  & = & \int \dd^3x W_{ij}(\vec x)\delta(\vec x)\nonumber\\
  & = & \int \dd\rcom \int \dd^2\theta \left[ f_\kappa(\rcom)\right]^2F_j(\vec{\theta})   q_i(\rcom) \delta\left(f_\kappa(\rcom)\vec{\theta},\rcom\right)
\end{eqnarray}
$F_j$ is the function describing the shape of the $j^{th}$ pixel,
while $q_i$ is the function describing the $i^{th}$ redshift bin.
These window functions should be normalized as in equation \ref{W_normalization},
such that
\begin{equation}
  \label{q_normalization}
  \int \dd\rcom [f_\kappa(w)]^2 q_i(\rcom) = 1
\end{equation}
and
\begin{equation}
  \int \dd^2\theta F_j(\vec{\theta}) = 1
\end{equation}
We are usually concerned with the covariance matrix of the signal, given by
\begin{eqnarray}
  \label{Sdd}
  \left[S_{\delta\delta}\right]_{nm}
  & = & \left\langle \delta_{i_nj_n}\delta^*_{i_mj_m}\right\rangle \nonumber\\
  & = & \int \dd^2\theta F_{j_n}[\vec{\theta}] 
  \int \dd^2\theta^\prime F_{j_m}[\vec{\theta}^\prime] \nonumber\\
  & & \times
  \int \dd\rcom \left[f_\kappa(\rcom)\right]^2 q_{i_n}[\rcom]
  \int \dd\rcom^\prime \left[f_\kappa(\rcom^\prime)\right]^2 
  q_{i_m}[\rcom^\prime]\nonumber\\
  & &\times
  \left\langle
  \delta\left(f_\kappa(\rcom)\vec{\theta},
  \rcom\right)
  \delta^*\left(f_\kappa(\rcom^\prime)\vec{\theta^\prime},
  \rcom^\prime\right)
  \right\rangle
\end{eqnarray}
This can be simplified using the Limber approximation.

\subsection{The Limber Approximation}%\footnote{The following logic comes from \citet{Bartelmann01}.}}
Consider a projection of the density field along a certain radial direction
\begin{equation}
  g_i(\vec{\theta}) = \int \dd\rcom p_i(\rcom)\delta\left(f_\kappa(\rcom)\vec{\theta},\rcom\right).
\end{equation}
The cross correlation is
\begin{eqnarray}
  \label{Sgg}
  S_{g_ig_j}(\vec\theta-\vec\theta^\prime) 
  &=& \left\langle g_i(\vec\theta)g^*_j(\vec\theta^\prime)\right\rangle \nonumber\\
  &=& \int \dd\rcom p_i(\rcom)
  \int \dd\rcom^\prime p_j(\rcom^\prime)
  \left\langle\delta[f_\kappa(\rcom)\vec{\theta},\rcom]
  \delta^*[f_\kappa(\rcom^\prime)\vec{\theta}^\prime,\rcom^\prime]
  \right\rangle
\end{eqnarray}
Let's express $\delta(\vec x)$ in terms of the Fourier integral, equation \ref{ft}.  This gives
\begin{eqnarray}
  S_{g_ig_j}(\vec\theta-\vec\theta^\prime) 
  &=& \int \dd\rcom p_i(\rcom)
  \int \dd\rcom^\prime p_j(\rcom^\prime)
  \int \frac{\dd^3k}{(2\pi)^3}\int \frac{\dd^3k^\prime}{(2\pi)^3}
  \left\langle\hat\delta(\vec k,\rcom)
  \hat\delta^*(\vec k^\prime, \rcom^\prime)
  \right\rangle\nonumber\\
  &&\times
  \exp\left[-i f_\kappa(\rcom)\vec{\theta}\cdot \vec k_\perp -i k_\parallel w\right]
  \exp\left[i f_\kappa(\rcom^\prime)\vec{\theta}^\prime \cdot \vec k^\prime_\perp  + i k^\prime_\parallel w^\prime\right].
\end{eqnarray}
Here $\vec k_\perp$ is the 2-dimensional projection of $\vec k$ perpendicular to 
the line of sight, and $k_\parallel$ is the projection of $\vec k$ along 
the line of sight.
The second argument of $\hat\delta(\vec k,\rcom)$ parametrizes evolution with
time via $|c\cdot dt| = a\cdot d\rcom$.

Because the power spectrum $P_\delta(k)$ decreases linearly with $k$ as 
$k\to0$, there must be a coherence scale $L_c$ such that the
correlation is near zero for 
$|\rcom-\rcom^\prime| \equiv \Delta\rcom > L_c$.  
The first part of the Limber approximation is to assume that $S_{gg}$ vanishes
at these distances.
Next we make the assumption that $p_i(\rcom)$ and 
$p_j(\rcom^\prime)$ do not vary appreciably over the small range where
$S_{g_ig_j}$ is non-vanishing, and that this range is small enough that 
$f_\kappa(\rcom) \approx f_\kappa(\rcom^\prime)$. 
This allows us to rewrite the above expression in a simpler way:
\begin{eqnarray}
  S_{g_ig_j}(\vec\theta-\vec\theta^\prime) 
  &=& \int \dd\rcom p_i(\rcom)p_j(\rcom)
  \int \frac{\dd^3k}{(2\pi)^3}\int \frac{\dd^3k^\prime}{(2\pi)^3} 
  \left\langle\hat\delta(\vec k,\rcom)
  \hat\delta^*(\vec k^\prime, \rcom)
  \right\rangle\nonumber\\
  &&\times
  \exp\left[-i f_\kappa(\rcom) 
    \left(\vec{\theta} \cdot \vec k_\perp 
    - \vec{\theta}^\prime \cdot k^\prime_\perp \right) 
    -i \rcom k_\parallel \right]
  \int \dd\rcom^\prime\exp\left(i\rcom k^\prime_\parallel\right)
\end{eqnarray}
The integral over $\rcom^\prime$ is 
simply $2\pi\delta_D(k^\prime_\parallel)$
via equation \ref{ddelta_form}, and the Fourier space correlation function
is proportional to $P_\delta(k)\delta^3_D(\vec k-\vec k^\prime)$ 
via equation \ref{pspec_corr}:
\begin{eqnarray}
  S_{g_ig_j}(\vec\theta-\vec\theta^\prime) 
  &=& \int \dd\rcom p_i(\rcom)p_j(\rcom)
  \int \frac{\dd^3k}{(2\pi)^2}\int \dd^3k^\prime
  \delta_D^3(\vec k-\vec k^\prime)\delta_D(k_\parallel^\prime)P_\delta(k)\nonumber\\
  &&\times
  \exp\left[-i f_\kappa(\rcom) 
    \left(\vec{\theta} \cdot \vec k_\perp 
    - \vec{\theta}^\prime \cdot k^\prime_\perp \right) 
    -i \rcom k_\parallel \right]
\end{eqnarray}
Carrying out the integrals over the two delta functions we see
\begin{eqnarray}
  \label{Limber}
  S_{g_ig_j}(\vec\theta_{ij}) 
  &=& \int \dd\rcom p_i(\rcom)p_j(\rcom)
  \int \frac{\dd^2k_\perp}{(2\pi)^2}
  P_\delta(k_\perp,\rcom)
  \exp\left[-i f_\kappa(\rcom) 
    \vec \theta_{ij} \cdot \vec k_\perp 
    \right]\\
  \label{Limber2}
  &=& \int \dd\rcom p_i(\rcom)p_j(\rcom)
  \int \frac{k\dd k}{2\pi}P_\delta(k,\rcom)
  J_0\left[f_\kappa(\rcom)\theta k\right]
\end{eqnarray}
where we have defined $\vec \theta_{ij} = \vec{\theta} -\vec{\theta}^\prime$, 
and $J_0(x)$ is a Bessel function of the first kind, which comes from 
the angular integral via
\begin{equation}
  \label{bessel_j}
  J_n(x) = \frac{1}{2\pi}\int_{-\pi}^{\pi} e^{-i(n\tau-x sin\tau)}\dd\tau
\end{equation}
We see that all $k_\parallel$ terms have vanished, which leads to the main
result of the Limber approximation: there is no correlation between the 
density contrast in windows that do not overlap 
in redshift. This is quickly seen from the leading integral in equation 
\ref{Limber}.  If the windows $p_i(\rcom)$ and $p_j(\rcom)$ do 
not overlap, then the expression integrates to zero.  

Now that we have a simple expression for the 2-dimensional projected 
correlation functions, we can use equation \ref{pspec} to define the 
2-dimensional cross power spectrum of the measured covariance,
\begin{equation} 
  \label{cross_pspec}
  P_{g_ig_j}(\ell) = \int \dd^2\theta e^{i\vec\theta\cdot\vec\ell}S_{g_ig_j}(\vec\theta).
\end{equation}
Combining equations \ref{Limber} and \ref{cross_pspec}, we have
\begin{eqnarray}
  \label{cross_pspec2}
  P_{g_ig_j}(\ell) 
  &=& \int \dd^2\theta
  \int \dd\rcom p_i(\rcom)p_j(\rcom)
  \int \frac{\dd^2k_\perp}{(2\pi)^2}
  P_\delta(k_\perp,\rcom)
  \exp\left[i\vec\theta\cdot(\vec\ell- f_\kappa(\rcom) 
    \vec k_\perp)\right]\nonumber\\
  &=&\int \dd\rcom p_i(\rcom)p_j(\rcom)\int \dd^2k_\perp
  P_\delta(k_\perp,\rcom)\delta_D^2(\vec\ell- f_\kappa(\rcom)\vec k_\perp)
  \nonumber\\
  &=& \int \dd\rcom\frac{p_i(\rcom)p_j(\rcom)}{[f_\kappa(\rcom)]^2}P_\delta\left(\frac{\ell}{f_\kappa(\rcom)},\rcom\right).
\end{eqnarray}

\subsection{Applying the Limber Approximation}
Examining equation \ref{Sdd}, we see that the integrals over $\rcom$ and 
$\rcom^\prime$ appear in equation \ref{Sgg}, with 
$p_i(\rcom)\to q_{i_n}(\rcom)[f_\kappa(\rcom)]^2$ and 
$p_j(\rcom^\prime)\to 
q_{i_m}(\rcom^\prime)[f_\kappa(\rcom^\prime)]^2$.
Thus we can rewrite equation \ref{Sdd} using the Limber approximation:
\begin{equation}
  \label{Sdd_Limber}
  \left[S_{\delta\delta}\right]_{nm}
  = \int \dd^2\theta F_{j_n}[\vec{\theta}] 
  \int \dd^2\theta^\prime F_{j_m}[\vec{\theta}^\prime] 
  S_{nm}(\vec\theta -\vec\theta^\prime)
\end{equation}
Now using the procedure from section \ref{smoothing}, 
we can write this approximation as an integral in Fourier 
space over the cross power spectrum given by equation
\ref{cross_pspec2}.  We'll consider the simple case where 
each pixel has the same angular shape described by $F(\vec\theta)$,
so that $F_{j_n}(\vec\theta) = F(\vec\theta-\vec\theta_n)$ and
$F_{j_m}(\vec\theta) = F(\vec\theta-\vec\theta_m)$.  Defining
$\vec\theta_{nm} = |\vec\theta_n - \vec\theta_m|$,
\begin{eqnarray}
  \label{Sdd_final}
  \left[S_{\delta\delta}\right]_{nm} &=& 
  \frac{1}{(2\pi)^2}
  \int \dd^2\ell\, \widetilde{F}_{j_n}(\vec\ell)
  \widetilde{F}_{j_m}(\vec\ell)P_{nm}(\ell)
  e^{i\vec\ell\cdot\vec\theta_{nm}}\nonumber\\
  &=&\frac{1}{2\pi}
  \int \ell\dd\ell\, \widetilde{F}_{j_n}(\ell)
  \widetilde{F}_{j_m}(\ell)P_{nm}(\ell)J_0\left(\ell\theta_{nm}\right)
\end{eqnarray}
where the second line holds if the window functions are circularly symmetric, 
with the Bessel function given by equation \ref{bessel_j}.
The projected power spectrum $P_{nm}(\vec\ell)$ is given by 
equation \ref{cross_pspec2} 
with appropriate substitution for $p_{i,j}$:
\begin{equation}
  P_{nm}(\ell) = \int \dd\rcom[f_\kappa(\rcom)]^2 q_{i_n}(\rcom)q_{i_m}(\rcom)P_\delta\left(\frac{\ell}{f_\kappa(\rcom)},\rcom\right).
\end{equation}
Note the factor of $[f_\kappa(\rcom)]^2$ in the numerator, which has 
its root in the spherical coordinate differential 
$d^3x \to r^2dr \cdot d\Omega$.

Our main application of this formalism will involve discrete 
non-overlapping redshift bins with uniform weighting.  That is,
\begin{equation}
  q_i(\rcom) = \left\{
  \begin{array}{ll}
    A_i, & \rcom^{(i)}_{\mathrm{min}} 
    < \rcom < \rcom^{(i)}_{\mathrm{max}}\\
    0, & \mathrm{otherwise}
  \end{array}
  \right.
\end{equation}
with the normalization constant computed via the condition in equation 
\ref{q_normalization}:
\begin{equation}
  A_i = \left[\int_{\rcom^{(i)}_{\mathrm{min}}}^{\rcom^{(i)}_{\mathrm{max}}}
  [f_\kappa(\rcom)]^2 \dd\rcom\right]^{-1}.
\end{equation}
To summarize, the correlation between two windows $n$ 
and $m$ becomes, with $i_{n,m}$
indexing the redshift window, and $n,m$ indexing the angular window,
\begin{eqnarray}
  [S_{\delta\delta}]_{nm} 
  &=& \delta^K_{i_ni_m}\omega(|\theta_{nm}|)\nonumber\\
  \omega(\theta)
  &=&\frac{1}{2\pi}\int_0^\infty\ell\dd\ell\, 
  |\widetilde{F}(\ell)|^2
  P_{nm}(\ell)J_0(\ell\theta)\nonumber\\
  P_{nm}(\ell) &=& 
  \frac{1}{A^2}
  \int_{\rcom^{(i)}_\mathrm{min}}^{\rcom^{(i)}_\mathrm{max}} 
  \dd\rcom \left[f_\kappa(\rcom)\right]^2
  P_\delta\left(\frac{\ell}{f_\kappa(\rcom)},\rcom\right).\nonumber\\
  A &=& \int_{\rcom^{(i)}_\mathrm{min}}^{\rcom^{(i)}_\mathrm{max}} 
  \dd\rcom \left[f_\kappa(\rcom)\right]^2
\end{eqnarray} 
where $\delta^K_{ij}$ is the Kronecker delta. This result should be 
compared  to equations 39-41 of \citet{Simon09}.  The only difference is
that we have correctly accounted for the normalization of the redshift
bin.  Note that if we assume $f_\kappa(\rcom)$ is constant across
each redshift bin, the two formulations are equivalent.

For our analysis, we will use angular pixels with radius $\theta_s$, 
so that the Fourier transform of the window function is given by equation 
\ref{tophat2D} and the equation giving the signal covariance becomes
\begin{equation}
  \left[S_{\delta\delta}\right]_{nm}
  = \frac{2}{\pi\theta_s^2} 
  \int \dd\rcom \frac{q_{i_n}[\rcom]q_{i_m}[\rcom]}
  {f_\kappa(\rcom)^2} 
  \int  \frac{\dd\ell}{\ell}
  P_\delta\left(\frac{\ell}{f_\kappa(\rcom)},\rcom\right)
  [J_1(\theta_s\ell)]^2 J_0(\theta_{nm}\ell)
\end{equation}

\section{Window Functions and their Fourier Transforms}
\label{app:sec:Wtransforms}
Here we list a few common window functions and their Fourier transforms
\subsection{Gaussian Window Functions}
The $n$-dimensional Gaussian window function is defined as
\begin{equation}
  W(\vec x/R) = \frac{1}{(2\pi R^2)^{n/2}}
  \exp\left(\frac{-|\vec x|^2}{2R^2}\right)
\end{equation}
The Fourier transform is straightforward because the dimensions decouple
and we're left with $n$ 1-dimensional Gaussian integrals.  The resulting
window function is
\begin{equation}
  \widetilde{W}(\vec k R) 
  = \exp\left(\frac{-|\vec k|^2R^2}{2}\right)
\end{equation}
which itself is a Gaussian.
\subsection{Top-hat Window Functions}
The $n$-dimensional tophat window function is given by
\begin{equation}
W(\vec x/R) = A_n
\times\left\{
\begin{array}{ll}
  1, & |\vec x|\le R\\
  0, & |\vec x|>R
\end{array}
\right.
\end{equation}
with
\begin{equation}
  A_n = \frac{\Gamma(n/2+1)}{(R\sqrt\pi)^n}
\end{equation}
where $\Gamma(y)$ is the gamma function.  The normalization is simply the
inverse of the volume of an $n$-sphere of radius $R$.   
For $n=2$ and $n=3$, the normalizations are the familiar 
$1/(\pi R^2)$ and $3/(4\pi R^3)$, respectively.
For the top-hat window function, there is no
simple expression for the Fourier transform for arbitrary $n$. 
Here we compute three special cases:
\begin{description}
  \item{$n=1$}:
    \begin{eqnarray}
      \label{tophat1D}
      A_1 &=& (2R)^{-1}\nonumber\\
      \widetilde W(\vec k R) &=& A_1\int_{-R}^{R}\dd r e^{ikr} \nonumber\\
      &=&\frac{\sin(kR)}{kR}
    \end{eqnarray}
  \item{$n=2$}:
    \begin{eqnarray}
      \label{tophat2D}
      A_2 &=& (\pi R^2)^{-1}\nonumber\\
      \widetilde W(\vec k R) &=& A_2\int_0^R r\dd r\int_0^{2\pi} \exp[ikr\cos\phi]\dd\phi\nonumber\\
      &=& \frac{2}{R^2}\int_0^R r J_0(kr)\dd r \nonumber\\
      &=& \frac{2J_1(kR)}{kR}
    \end{eqnarray}
  \item{$n=3$}: 
    \begin{eqnarray}
      \label{tophat3D}
      A_3 &=& (4\pi R^3/3)^{-1}\nonumber\\
      \widetilde W(\vec k R) &=& A_3\int_0^Rr^2\dd r\int_0^\pi \sin(\theta)\dd\theta\int_0^{2\pi}\exp[ikr\cos\phi]\dd\phi\nonumber\\
      &=& \frac{3}{R^3}\int_0^R r^2 J_0(kr)\dd r\nonumber\\
      &=& _1F_2(\frac{1}{2};\ 1,\frac{5}{2};\ \frac{-k^2R^2}{4})
    \end{eqnarray}
\end{description}
where, $J_n(x)$ are Bessel functions of the first kind (see eqn \ref{bessel_j})
and the last line is a generalized hypergeometric function, 
$_nF_m(a_0\cdots a_n;b_0\cdots b_m;x)$.  

%% file: appendixB.tex
\chapter{Efficient Implementation of the SVD Estimator}
\label{appB}
As noted in Section~\ref{sing_val_formalism}, taking the SVD of the 
transformation matrix $\widetilde{M}_{\gamma\delta} \equiv 
\mathcal{N}_{\gamma\gamma}^{-1/2}M_{\gamma\delta}$
is not trivial for large fields.  This appendix will first give a rough
outline of the form of $M_{\gamma\delta}$, then describe our tensor 
decomposition method which enables quick calculation of the singular
value decomposition.  For a more thorough review of the lensing results, 
see e.g.~\citet{Bartelmann01}.

Our goal is to speed the computation of the SVD by writing \
$\widetilde{M}_{\gamma\delta}$
as a tensor product $\mymat{A} \otimes \mymat{B}$.  Here ``$\otimes$''
is the Kronecker product, defined such that, if $\mymat{A}$ is a matrix
of size $n \times m$, $B$ is a matrix of arbitrary size,
\begin{equation}
  \mymat{A}\otimes\mymat{B} \equiv \left(
  \begin{array}{cccc}
    A_{11}B & A_{12}B & \cdots & A_{1m}B \\
    A_{21}B & A_{22}B & \cdots & A_{2m}B \\
    \vdots  & \vdots & \ddots & \vdots  \\
    A_{n1}B & A_{n2}B & \cdots & A_{nm}B 
  \end{array}\right)
\end{equation}
In this case, the singular value decomposition
$A\otimes B = U_{AB}\Sigma_{AB}V^\dagger_{AB}$
satisfies
\begin{eqnarray}
  \label{AB_SVD}
  U_{AB} &=& U_A\otimes U_B \nonumber\\
  \Sigma_{AB} &=& \Sigma_A \otimes \Sigma_B\nonumber\\
  V_{AB} &=& V_A \otimes V_B
\end{eqnarray}
where $U_A\Sigma_AV^\dagger_A$ is the SVD of $A$, 
and   $U_B\Sigma_BV^\dagger_B$ is the SVD of $B$.
Decomposing $\widetilde{M}_{\gamma\delta}$ in this way can 
greatly speed the SVD computation.

\subsection{Angular and Line-of-Sight Transformations}
The transformation from shear to density, encoded in $M_{\gamma\delta}$,
consists of two steps: an angular integral relating shear $\gamma$ to
convergence $\kappa$, and a line-of-sight integral relating the convergence
$\kappa$ to the density contrast $\delta$.

The relationship between $\gamma$ and $\kappa$ is a convolution over
all angular scales,
\begin{equation}
  \label{gamma_integral}
  \gamma(\myvec\theta,z_s) \equiv \gamma_1 + i\gamma_2 = \int \dd^2\theta^\prime
  \ \mathcal{D}(\myvec\theta^\prime-\myvec\theta)\kappa(\myvec\theta^\prime,z_s),
\end{equation}
where $\mathcal{D}(\myvec\theta)$ 
is the Kaiser-Squires kernel \citep{Kaiser93}.  
This has a particularly simple form in Fourier space:
\begin{equation}
  \label{gamma_fourier}
  \hat\gamma(\myvec\ell,z_s) 
  = \frac{\ell_1 + i\ell_2}{\ell_1 - i\ell_2}\hat\kappa(\myvec\ell,z_s).
\end{equation}
where $\hat\gamma$ and $\hat\kappa$ are the Fourier transforms of $\gamma$
and $\kappa$ and $\myvec{\ell}\equiv(\ell_1,\ell_2)$ is the angular wavenumber.

The relationship between $\kappa$ and $\delta$ is an integral along each
line of sight:
\begin{equation}
  \label{kappa_integral}
  \kappa(\myvec\theta,z_s) = 
  \int_0^{z_s}\dd z\ W(z,z_s)\delta(\myvec\theta,z)
\end{equation}
where $W(z,z_s)$ is the lensing efficiency function at redshift $z$ 
for a source located at redshift $z_s$ 
(refer to STH09 for the form of this function).

Upon discretization of the quantities $\gamma$, $\kappa$, and $\delta$
(described in Section~\ref{LinearMapping}), 
the integrals in Equations~\ref{gamma_integral}-\ref{kappa_integral} 
become matrix operations.  The relationship between the data vectors
$\myvec{\gamma}$ and $\myvec{\kappa}$ can be written
\begin{equation}
  \label{P_gk}
  \myvec\gamma = [\mymat{P}_{\gamma\kappa} \otimes \mathbf{1}_s]\myvec\kappa 
  + \myvec{n}_\gamma
\end{equation}
where $\mathbf{1}_s$ is the $N_s \times N_s$ identity matrix and 
$\mymat{P}_{\gamma\kappa}$ is the matrix representing the linear 
transformation in Equations~\ref{gamma_integral}-\ref{gamma_fourier}.  
The quantity $[\mymat{P}_{\gamma\kappa} \otimes \mathbf{1}_s]$ 
simply denotes that $\mymat{P}_{\gamma\kappa}$ operates on each of the $N_s$ 
source-planes represented within the vector $\myvec\kappa$.
Similarly, the relationship between the vectors $\myvec{\kappa}$ and
$\myvec{\delta}$ can be written
\begin{equation}
  \label{Q_kd}
  \myvec\kappa = [\mathbf{1}_{xy} \otimes \mymat{Q}_{\kappa\delta}]\myvec\delta
\end{equation}
where $\mathbf{1}_{xy}$ is the $N_{xy} \times N_{xy}$ 
identity matrix, and the tensor product signifies that the operator 
$Q_{\kappa\delta}$ operates on each of the $N_{xy}$ lines-of-sight in
$\myvec\delta$.  $Q_{\kappa\delta}$ is the $N_s \times N_l$ matrix which
represents the discretized version of equation \ref{kappa_integral}.
Combining these representations allows us to decompose the matrix 
$\mymat{M}_{\gamma\delta}$ in Equation~\ref{M_gd} into a tensor product:
\begin{equation}
  \mymat{M}_{\gamma\delta} = 
  \mymat{P}_{\gamma\kappa} \otimes \mymat{Q}_{\kappa\delta}.
\end{equation}

\subsection{Tensor Decomposition of the Transformation}
We now make an approximation that the noise covariance 
$\mymat{\mathcal{N}}_{\gamma\gamma}$ can be written as a
tensor product between its angular part $\mymat{\mathcal{N}_P}$ 
and its line of sight part $\mymat{\mathcal{N}_Q}$:
\begin{equation}
  \label{noise_decomp}
  \mymat{\mathcal{N}}_{\gamma\gamma} 
  = \mymat{\mathcal{N}_P} \otimes \mymat{\mathcal{N}_Q}.
\end{equation}
Because shear measurement error comes primarily from shot noise, this 
approximation is equivalent to the statement that source galaxies are drawn 
from a single redshift distribution, with a different normalization along 
each line-of-sight.  For realistic data, this approximation will break down
as the size of the pixels becomes very small.  We will assume here for 
simplicity that the noise covariance is diagonal, but the following results
can be generalized for non-diagonal noise.  
Using this noise covariance approximation, we can compute the 
SVDs of the components of $\widetilde{M}_{\gamma\delta}$:
\begin{eqnarray}
  \mymat{U}_P\mymat{\Sigma}_P\mymat{V}_P^\dagger = \mathcal{N}_P^{-1/2} \mymat{P}_{\gamma\kappa}\nonumber\\
  \mymat{U}_Q\mymat{\Sigma}_Q\mymat{V}_Q^\dagger = \mathcal{N}_Q^{-1/2} \mymat{Q}_{\kappa\delta}
\end{eqnarray}

In practice the SVD of the matrix $\mymat{P}_{\gamma\kappa}$ 
need not be computed explicitly.  
$\mymat{P}_{\gamma\kappa}$ encodes the discrete linear operation expressed
by Equations~\ref{gamma_integral}-\ref{gamma_fourier}: 
as pointed out by STH09, in the large-field limit $P_{\gamma\kappa}$ 
can be equivalently computed in either real or Fourier space.
Thus to operate with $P_{\gamma\kappa}$ on a shear vector, 
we first take the 2D Fast Fourier Transform (FFT) of each
source-plane, multiply by the kernel $(\ell_1+i\ell_2)/(\ell_1-i\ell_2)$,
then take the inverse FFT of the result.  This is orders-of-magnitude
faster than a discrete implementation of the real-space convolution.
Furthermore, the conjugate transpose of this operation can be computed
by transforming $\ell \to -\ell^*$, so that
\begin{equation}
  \mymat{P}_{\gamma\kappa}^\dagger\mymat{P}_{\gamma\kappa} = \mymat{I}
\end{equation}
and we see that $P_{\gamma\kappa}$ is unitary in the wide-field limit.  This
fact, along with the tensor product properties of the SVD, allows us to
write $\widetilde{M}_{\gamma\delta} = U\Sigma V^\dagger$ where
\begin{eqnarray}
  U &\approx& \mathbf{1}_{xy} \otimes U_Q \nonumber\\
  \Sigma &\approx& \mathcal{N}_P^{-1/2} \otimes \Sigma_Q \nonumber\\
  V^\dagger &\approx& P_{\gamma\kappa} \otimes  V_Q^\dagger
\end{eqnarray}
The only explicit SVD we need to calculate is that of 
$\mathcal{N}_Q^{-1/2}\mymat{Q}_{\kappa\delta}$,
which is trivial in cases of interest.  
The two approximations we have made are the
applicability of the Fourier-space form of the $\gamma\to\kappa$ mapping
(Eqn.~\ref{gamma_fourier}), and the tensor
decomposition of the noise covariance (Eqn.~\ref{noise_decomp}).

%% file: appendixC.tex
\chapter{Choice of KL Parameters}
\label{Choosing_Params}
The KL analysis outlined in Section~\ref{KL_Intro}
has only two free parameters: the number of
modes $n$ and the Wiener filtering level $\alpha$.  Each of these parameters
involves a trade-off: using more modes increases the amount of information
used in the reconstruction, but at the expense of a decreased signal-to-noise
ratio.  Decreasing the value of $\alpha$ to $0$ reduces the smoothing effect 
of the prior, but can lead to a nearly singular convolution matrix 
$\mymat{M}_{(n,\alpha)}$, 
which results in unrealistically large shear values in the poorly-constrained 
areas areas of the map (i.e.~masked regions).

To inform our choice of the number of modes $n$, we recall the trend of 
spatial scale with mode number seen in Figure~\ref{fig_bandpower}.  Our 
purpose in using KL is to allow interpolation in masked regions.  To this 
end, the angular scale of the mask should inform the choice of angular
scale of the largest mode used.  An eigenmode which probes scales much smaller
than the size of a masked region will not contribute meaningful information to
the reconstruction within that masked region.  Considering the pixels within
our mask, we find that 99.5\% of masked pixels are within 2 pixels of a
shear measurement.  This corresponds to an angular scale of $\ell=6140$.
Consulting Figure~\ref{fig_bandpower}, we see that modes larger than
about $n=900$ out of 4096 will probe length scales significantly 
smaller than the mask scale.  
Thus, we choose $n=900$ as an appropriate cutoff for our reconstructions.

To inform our choice of the Wiener filtering level $\alpha$, we examine the
agreement between histograms of \Map peaks for a noise-only DES field
with and without masking (see Section~\ref{Shear_Peaks}).  
We find that for large (small) values of $\alpha$, the number
of high-\Map peaks is underestimated (overestimated) in the masked 
case as compared to the unmasked case.  
Empirically, we find that the two agree at $\alpha = 0.15$; 
we choose this value for our analysis.  Note that this 
tuning is done on noise-only reconstructions, 
which can be generated for observed data by assuming that
shape noise dominates: 
\begin{equation}
  [\Noise_\gamma]_{ij} = \frac{\sigma_\epsilon}{n_i^2}\delta_{ij}.
\end{equation}
The $\alpha$-tuning can thus be performed on artificial noise realizations 
which match the observed survey characteristics.

We make no claim that $(n,\alpha) = (900,0.15)$ is the optimal
choice of free parameters for KL: determining this would involve a more
in-depth analysis.  They are simply well-motivated choices which we use to
make a case for further study.

%% file: vita.tex
\vita{
\begin{center}
\noindent Jacob T. Vanderplas
\end{center}

\section*{Education}
\begin{tabular}{lll}
  2012: & PhD in Astronomy & {\it University of Washington}, Seattle WA\\
  2008: & MSc in Astronomy & {\it University of Washington}, Seattle WA\\
  2003: & BSc in Physics   & {\it Calvin College}, Grand Rapids, MI
\end{tabular}

\section*{Professional Experience}
\begin{tabular}{ll}
  2006-2012: & Graduate research assistant.
               {\it University of Washington}, Seattle WA\\
             & Advised by:\\
             & \hspace{1cm} Andrew Connolly: 2008-2012\\
             & \hspace{1cm} Bhuvnesh Jain (U. Penn): 2009-2012\\
             & \hspace{1cm} Andrew Becker: 2006-2008\\
             & \hspace{1cm} Craig Hogan: 2006-2007\\
  2010-2012: & Planetarium Digitization Project Coordinator\\
             & \hspace{1cm} {\it University of Washington}, Seattle, WA\\
  2008-2010: & Planetarium Educational Outreach Coordinator\\
             & \hspace{1cm} {\it University of Washington}, Seattle, WA\\
  2006-2008: & Graduate teaching assistant, introductory Astronomy\\
             & \hspace{1cm} {\it University of Washington}, Seattle, WA\\
  2004-2006: & Experiential Science Educator\\
             & \hspace{1cm} {\it Mount Hermon Outdoor Science School},
                            Santa Cruz, CA\\
  2004-2006: & Mountaineering Instructor\\
             & \hspace{1cm} {\it Summit Adventure}, Bass Lake, CA
\end{tabular}

\section*{Volunteer Experience}
\begin{tabular}{ll}
   2009-2012: & Science Communication Fellow\\
              & \hspace{1cm} {\it Pacific Science Center}, Seattle WA\\
   2007-2012: & Trip Leader for Inner-city Youth\\
              & \hspace{1cm} {\it Sierra Club Inner City Outings}, Seattle WA
\end{tabular}

\section*{Non-technical and Public Talks}
\begin{tabular}{ll}
  2009-2012:     & ``Scientist Spotlight'' and ``Portal to the Public''
                    events,\\ & (3-4 times per year)\\
                 & \hspace{1cm} Pacific Science Center, Seattle WA\\
  April 2012:    & Colloquium: {\it Dark Matter, Dark Energy,
                    and the Fate of the Universe}\\
                 & \hspace{1cm} {\it Calvin College}, Grand Rapids, MI\\
  November 2011: & Invited talk: Gravity, Lensing the Universe\\
                 & \hspace{1cm}  {\it KCTS9 Science Cafe}, Seattle WA\\
  June 2010 - Nov. 2011: & Invited World Wide Telescope demonstrations
                           (Microsoft)\\
                 & \hspace{1cm} {\it Supercomputing 2011}, Seattle WA\\
                 & \hspace{1cm} {\it Partners in Learning Global Forum 2011},
                    Washington DC\\
                 & \hspace{1cm} {\it Popular Mechanics Breakthrough Awards
                    2011}, New York NY\\
                 & \hspace{1cm} {\it ISTE 2010}, Denver CO\\
  March 2011:    & Invited Talk: Understanding the Dark Side of the Universe\\
                 & \hspace{1cm} {\it Science with a Twist: Star Wars}\\
                 & \hspace{1cm} Pacific Science Center, Seattle WA\\
  February 2011: & Talk: Interconnection in Art and Cosmology\\
                 & \hspace{1cm} Astronomy-inspired art show,
                   University of Washington\\
  May 2009:      & Dark Matter, Gravitational Lensing, and Cosmology\\
                 & \hspace{1cm} {\it Battle Point Astronomical Society},
                   Bainbridge Island WA
\end{tabular}

\section*{Technical Talks \& Presentations}
\begin{tabular}{ll}
   July 2012:     & Tutorial: Machine Learning in Python\\
                  & Talk: {\tt astroML}: Machine Learning for Astronomy\\
                  & \hspace{1cm} {\it Scipy 2012}, Austin TX\\
   March 2012:    & Invited tutorial: Scientific Machine Learning with
                    {\tt scikit-learn}\\
                  & \hspace{1cm} {\it PyData 2012}, Mountain View CA\\
   December 2011: & Poster: Shear Mapping with Karhunen-Loeve Analysis\\
                  & \hspace{1cm} {\it NIPS 2011}, Grenada, Spain\\
   June 2011:     & Invited talk: Digital Planetariums for the  Masses\\
                  & \hspace{1cm} {\it AstroViz 2011}, Seattle WA\\
   April-May 2011:& Talk: KL Interpolation of Weak Lensing Shear\\
                  & \hspace{1cm} {\it INPA Seminar},
                    Lawrence Berkeley National Laboratory\\
                  & \hspace{1cm} {\it Cosmology Seminar}, UC Davis\\
                  & \hspace{1cm} {\it KIPAC Cosmology Seminar},
                    Stanford University/SLAC\\
   February 2011: & Talk: Weak Lensing Peak Statistics\\
                  & \hspace{1cm} {\it Cosmology Seminar},
                    University of Pennsylvania\\
   January 2011:  & Poster: Finding the Odd-one Out in Spectroscopic Surveys\\
                  & Poster: 3D Reconstruction of the Density Field\\
                  & \hspace{1cm} {\it 217$^{\rm th}$ AAS meeting}, Seattle WA\\
   July 2010:     & Talk: A New Approach to Tomographic Mapping\\
                  & \hspace{1cm} {\it Ten Years of Cosmic Shear},
                    Edinburgh, UK\\
   April 2010:    & Talk: New Ideas for 3D Mapping with Cosmic Shear\\
                  & \hspace{1cm} {\it Cosmology Seminar},
                    University of Pennsylvania\\
   October 2009:  & Talk: Locally Linear Embedding of Astronomy Data\\
                  &  \hspace{1cm} Microsoft Research, Redmond, WA\\
   November 2007: & Invited talk: 
                    SALT-2 Light-curve Fitting for SDSS Supernovae\\
                  & \hspace{1cm} {\it SDSS Collaboration Meeting}, Fermilab\\
\end{tabular}

\section*{Publication List}
\begin{enumerate}
   \item {\bf VanderPlas, J. T.}; Connolly, A. J.; Jain, B.; Jarvis, M.\\
         {\it Interpolating Masked Weak-lensing Signal
          with Karhunen-Loève Analysis}.\
         ApJ 744:180 (2012)
   \item Pedregosa, Fabian; Varoquaux, Gaël; Gramfort, Alexandre;
         Michel, Vincent; Thirion, Bertrand; Grisel, Olivier;
         Blondel, Mathieu; Prettenhofer, Peter; Weiss, Ron; Dubourg, Vincent;
         {\bf Vanderplas, Jake;} Passos, Alexandre; Cournapeau, David;
         Brucher, Matthieu; Perrot, Matthieu; Duchesnay, Édouard.\\
         {\it Scikit-learn: Machine Learning in Python}.
         JMLR, 12:2825 (2011)
   \item Xiong, L.; Poczos, B.; Schneider, J.; Connolly, A.;
         {\bf VanderPlas, J.}\\
         {\it Hierarchical Probabilistic Models for Group Anomaly Detection}.
         AISTATS (2011)
   \item Jain, Bhuvnesh; {\bf VanderPlas, Jake}.\\
         {\it Tests of modified gravity with dwarf galaxies}
         JCAP 10:32 (2011)
   \item Daniel, Scott F.; Connolly, Andrew; Schneider, Jeff;
         {\bf Vanderplas, Jake}; Xiong, Liang.\\
         {\it Classification of Stellar Spectra with Local Linear Embedding}.
         AJ 142:203 (2011)
   \item {\bf VanderPlas, J. T.}; Connolly, A. J.; Jain, B.; Jarvis, M.\\
         {\it Three-dimensional Reconstruction of the Density Field:
              An SVD Approach to Weak-lensing Tomography}.
         ApJ 727:118 (2011)
   \item Lampeitl, H.; Nichol, R. C.; Seo, H.-J.; Giannantonio, T.;
         Shapiro, C.; Bassett, B.; Percival, W. J.; Davis, T. M.; Dilday, B.;
         Frieman, J.; Garnavich, P.; Sako, M.; Smith, M.; Sollerman, J.;
         Becker, A. C.; Cinabro, D.; Filippenko, A. V.; Foley, R. J.;
         Hogan, C. J.; Holtzman, J. A.; Jha, S. W.; Konishi, K.; Marriner, J.;
         Richmond, M. W.; Riess, A. G.; Schneider, D. P.; Stritzinger, M.;
         van der Heyden, K. J.; {\bf Vanderplas, J. T.}; Wheeler, J. C.; 
         heng, C.\\
         {\it First-year Sloan Digital Sky Survey-II supernova results:
          consistency and constraints with other intermediate-redshift
          data sets}.
          MNRAS 401:2331 (2010)
   \item LSST Science Collaboration\\
         {\it LSST Science Book}.
         arXiv:0912.0201 (2009)
   \item Kessler, Richard; Becker, Andrew C.; Cinabro, David;
         {\bf Vanderplas, Jake}; Frieman, Joshua A.; Marriner, John;
         Davis, Tamara M.; Dilday, Benjamin; Holtzman, Jon; Jha, Saurabh W.;
         Lampeitl, Hubert; Sako, Masao; Smith, Mathew; Zheng, Chen;
         Nichol, Robert C.; Bassett, Bruce; Bender, Ralf; Depoy, Darren L.;
         Doi, Mamoru; Elson, Ed; Filippenko, Alexei V.; Foley, Ryan J.;
         Garnavich, Peter M.; Hopp, Ulrich; Ihara, Yutaka; Ketzeback, William;
         Kollatschny, W.; Konishi, Kohki; Marshall, Jennifer L.;
         McMillan, Russet J.; Miknaitis, Gajus; Morokuma, Tomoki;
         Mörtsell, Edvard; Pan, Kaike; Prieto, Jose Luis; Richmond, Michael W.;
         Riess, Adam G.; Romani, Roger; Schneider, Donald P.;
         Sollerman, Jesper; Takanashi, Naohiro; Tokita, Kouichi;
         van der Heyden, Kurt; Wheeler, J. C.; Yasuda, Naoki; York, Donald.\\
         {\it First-Year Sloan Digital Sky Survey-II Supernova Results:
           Hubble Diagram and Cosmological Parameters}.
         ApJS 185:32 (2009)
   \item {\bf Vanderplas, Jake}; Connolly, Andrew.\\
         {\it Reducing the Dimensionality of Data: Locally Linear Embedding
              of Sloan Galaxy Spectra}.
         AJ 138:1365 (2009)
   \item Sollerman, J.; Mörtsell, E.; Davis, T. M.; Blomqvist, M.; Bassett, B.;
         Becker, A. C.; Cinabro, D.; Filippenko, A. V.; Foley, R. J.;
         Frieman, J.; Garnavich, P.; Lampeitl, H.; Marriner, J.; Miquel, R.;
         Nichol, R. C.; Richmond, M. W.; Sako, M.; Schneider, D. P.; Smith, M.;
         {\bf Vanderplas, J. T.}; Wheeler, J. C.\\
         {\it First-Year Sloan Digital Sky Survey-II (SDSS-II) Supernova
              Results: Constraints on Nonstandard Cosmological Models}.
         ApJ 703:1374 (2009)
   \item Kessler, Richard; Bernstein, Joseph P.; Cinabro, David;
         Dilday, Benjamin; Frieman, Joshua A.; Jha, Saurabh; Kuhlmann, Stephen;
         Miknaitis, Gajus; Sako, Masao; Taylor, Matt; {\bf Vanderplas, Jake}.
         {\it SNANA: A Public Software Package for Supernova Analysis}.\\
         PASP 121:1028 (2009)
\end{enumerate}
}